\documentclass[a4paper,11pt]{article}
\usepackage{jheppub} % for details on the use of the package, please see the JINST-author-manual
\usepackage{lineno}
\usepackage{makecell}

\usepackage{xcolor}

\usepackage{array}
\usepackage{multirow}

\usepackage[all,cmtip]{xy}

\usepackage{graphicx} % Required for inserting images
\usepackage{cancel} % For \cancel command
\usepackage{varwidth} % For cellvarwidth environment
\usepackage[normalem]{ulem}

\usepackage{tikz}
\usepackage{tikz-feynman} % Requires \usepackage[compat=1.1.0]{tikz-feynman} if using older versions
\tikzfeynmanset{compat=1.1.0}
\usepackage{amsmath} % for equation*
\usepackage{adjustbox} % for for centering tikz

\graphicspath{{figures/}}

\title{\boldmath Large $N$ Chern-Simons-matter fixed points with multiple flavors}

\author{Ofer Aharony,}
\author{Ronny Frumkin,}
\author{and Jonathan Mehl}
\affiliation{Department of Particle Physics and Astrophysics, Weizmann Institute of Science, Rehovot
7610001, Israel}

\emailAdd{ofer.aharony@weizmann.ac.il}
\emailAdd{ronny.frumkin@weizmann.ac.il}
\emailAdd{jonathan.mehl@weizmann.ac.il}

\abstract{In this paper we analyze the $2+1$d conformal fixed points arising from $SU(N_c)$ Chern-Simons-matter theories with multiple flavors $N_f > 1$ in the 't Hooft large $N_c$ limit. The multi-flavor generalization of quasi-fermionic theories (fermions or critical scalars coupled to Chern-Simons gauge fields) is straightforward, but this is not true for quasi-bosonic theories (scalars or critical fermions coupled to Chern-Simons gauge fields). The latter theories have three flavor-singlet relevant operators and also three marginal operators, that become exactly marginal for infinite $N_c$, but have a non-zero beta function at order $1/N_c$. We compute the beta functions of these couplings in various weak coupling limits, and also discuss their general structure, generalizing previous computations for $N_f=1$. We find that IR-stable fixed points of the marginal couplings exist for some values of $N_f$ and of the 't Hooft coupling $\lambda$, but not for other values, and in one case we can explicitly follow how two pairs of fixed points merge and disappear as $\lambda$ is increased. We also analyze the ``Semi-Critical'' conformal field theories that arise when fine-tuning two (rather than three) relevant operators, and compute the beta function for their (single) marginal coupling constant.}

\begin{document}
\maketitle
\flushbottom

\section{Introduction and summary of results}
$2+1$ dimensional conformal field theories arising by coupling matter to Chern-Simons gauge fields (``Chern-Simons-matter theories'') are interesting for many different reasons. Some of these theories are believed to arise in phase diagrams of condensed matter systems, as well as in $2+1$ dimensional QCD (with or without bare Chern-Simons terms). The theories of fermions coupled to Chern-Simons (CS) gauge fields are believed to be dual to those of scalars coupled to Chern-Simons gauge fields \cite{Giombi:2011kc,Aharony:2012nh,Aharony:2015mjs,Seiberg:2016gmd}, giving the simplest example of dualities between non-supersymmetric theories above $1+1$ dimensions. In the large $N_c$ limit of $SU(N_c)$ or $SO(N_c)$ theories (with fixed 't~Hooft coupling $\lambda \equiv \frac{N_c}{\kappa}$ where $\kappa$ is the CS level), many things in these theories can be computed exactly, either by resumming the planar diagrams (whose difficulty is similar to vector models rather than to matrix models), or by using their approximate high-spin symmetry \cite{Maldacena:2011jn,Maldacena:2012sf}. The large $N_c$ theories are also believed to be holographically dual to weakly coupled high-spin gravity theories on $AdS_4$, giving a relatively simple example of non-supersymmetric holography \cite{Klebanov:2002ja, Sezgin:2002rt, Giombi:2009wh,
Giombi:2011kc,  Chang:2012kt,Giombi:2016ejx}\footnote{In this paper we will discuss theories that have either only fermions or only scalars coupled to Chern-Simons gauge fields; the features and dualities mentioned above can also be generalized to theories with both scalars and fermions, some of which are supersymmetric \cite{Giveon:2008zn,Benini:2011mf,Park:2013wta,Aharony:2013dha,Jain:2013gza,Inbasekar:2015tsa,Gur-Ari:2015pca,Kapustin:2010mh,Willett:2011gp,Kapustin:2011gh,Intriligator:2013lca,Jensen:2019mga,Aharony:2019mbc}.}.

The simplest Chern-Simons-matter theories (called ``quasi-fermionic theories''  \cite{Maldacena:2011jn, Maldacena:2012sf}) arise either by coupling a fermion in the fundamental representation of $SU(N_c)$ to $SU(N_c)$ CS gauge fields (``Regular Fermion (RF) theories''), or by coupling the critical $U(N_c)$ model to such gauge fields (``Critical Boson (CB) theories'')\footnote{All the theories discussed in this paper also have $SO(N_c)$ versions, which for large $N_c$ are very similar to the $SU(N_c)$ theories in terms of our discussion here, so we will not discuss them separately.}. The two descriptions are related by strong-weak-coupling duality, so that the full family of theories is given by either the fermionic or the bosonic theory with 't~Hooft coupling $-1 < \lambda \leq 1$ (we will denote the 't~Hooft coupling in the scalar description by $\lambda_B$, and the one in the fermionic description by $\lambda_F$). These theories have a single relevant operator whose coefficient (the mass term in the fermionic theory) must be fine-tuned to flow to a fixed point (recall that Chern-Simons couplings are quantized and do not run under the renormalization group). In the large $N_c$ limit the single-trace spectrum of these theories consists of a scalar operator of dimension $\Delta = 2 + \mathcal{O}\left(\frac{1}{N_c}\right)$, and of currents of every integer spin $s$, whose anomalous dimensions (for $s>2$) go as $\frac{1}{N_c}$.

There is also a family of ``quasi-bosonic theories'' that requires additional fine-tunings. These theories arise either by coupling a scalar in the fundamental representation of $SU(N_c)$ to $SU(N_c)$ CS gauge fields (``Regular Boson (RB) theories''), or by coupling the $U(N_c)$ Gross-Neveu model to such gauge fields (``Critical Fermion (CF) theories''). These theories have two relevant operators, which in the scalar language are the $\phi^2$ and $(\phi^2)^2$ operators, and both need to be fine-tuned to flow to a fixed point. The main difference in the spectrum of these theories compared to the quasi-fermionic ones is that the single-trace scalar operator has dimension $\Delta = 1 + \mathcal{O}\left(\frac{1}{N_c}\right)$.

The ``quasi-bosonic theories'' have a $(\phi^2)^3$ ``triple-trace'' operator that is exactly marginal in the large $N_c$ limit, but has a finite beta function for finite $N_c$ (going as $\frac{1}{N_c}$ at large $N_c$). Thus, to analyze if these fixed points exist for finite large $N_c$ and how much fine-tuning is required to flow to them, one has to carefully analyze this beta function, and this was done in \cite{Aharony:2018pjn}. It was shown there that at weak coupling, either from the bosonic side ($\lambda_B \to 0$, equivalent to $|\lambda_F| \to 1$) or from the fermionic side of the duality ($\lambda_F \to 0$, equivalent to $|\lambda_B| \to 1$), the theory has one IR-stable and two UV-stable fixed points for the extra coupling, and it was conjectured that this remains true for all values of the 't~Hooft coupling $\lambda$. With this assumption, flowing to the IR-stable fixed point requires two fine-tunings of the two relevant operators.

The discussion above, and most of the analysis of these theories in the literature, was for the case of a single flavor of matter fields in the fundamental representation of $SU(N_c)$, $N_f=1$. It is straightforward to generalize these theories to $N_f>1$ flavors of fermion or scalar fields, with couplings that preserve an $SU(N_f)$ global symmetry (in addition to the $U(1)$ global symmetry of the $N_f=1$ theory).

For the ``quasi-fermionic theories'' this does not change much. These have scalar operators of dimension $\Delta \simeq 2$ in both the singlet and adjoint representations of the $SU(N_f)$ global symmetry, but they still require (for all $\lambda$) only a single fine-tuning in order to flow to the fixed point (from an $SU(N_f)$-symmetric starting point).

For the ``quasi-bosonic theories'' that will be our focus in this paper, the situation is very different. These theories have scalar operators in the singlet+adjoint representations of $SU(N_f)$ with dimensions $\Delta=1 + \mathcal{O}\left(\frac{1}{N_c}\right)$, which we can denote by ${\cal O}_S$ and ${\cal O}_A$, respectively, and this has two dramatic consequences:
\begin{itemize}
    \item These theories now have three relevant operators, schematically given by flavor-singlets coming from ${\cal O}_S$, ${\cal O}_S^2$ and ${\cal O}_A^2$. This means that we need to perform at least three fine-tunings in order to flow to these theories and not just two.
    \item These theories have (for $N_f \geq 3$) three flavor-singlet marginal operators that are all exactly marginal in the large $N_c$ limit, schematically of the form ${\cal O}_S^3$, ${\cal O}_S {\cal O}_A^2$ and ${\cal O}_A^3$. For finite large $N_c$, all of their couplings acquire beta functions, which must be computed in order to see if they have IR-stable fixed points (beyond the three relevant operators mentioned above).
\end{itemize}
Our main goal in this paper is to analyze the beta functions of the marginal couplings in these theories, generalizing the computations of \cite{Aharony:2018pjn}, and to see for which values of $\lambda$ and $N_f$ they have IR-stable fixed points. In addition to analyzing fixed values of $N_f$ as $N_c$ becomes large, it is also interesting to analyze the case of $1 \ll N_f \ll N_c$ (including the limit of large $N_c$ with fixed small $N_f/N_c$), and we will see that some computations become more tractable in this case.

As we mentioned, in order to flow to the ``quasi-bosonic theories'' for $N_f>1$ we need to tune three relevant operators, which can be thought of as a mass term and two independent four-fermi/four-scalar couplings.
We can also choose to fine-tune just one of the four-fermi/four-scalar couplings and not both of them. This leads to ``Semi-Critical theories'', in which either the dimension of the singlet scalar is $\Delta=1 + \mathcal{O}\left(\frac{1}{N_c}\right)$ and that of the adjoint scalar is $\Delta=2 + \mathcal{O}\left(\frac{1}{N_c}\right)$, or vice versa. These theories are in many ways in-between the ``quasi-fermionic'' and the ``quasi-bosonic'' theories. In particular, each class of theories has just one marginal coupling (that becomes exactly marginal for infinite $N_c$), whose beta function needs to be computed to look for IR-stable fixed points (beyond the two relevant operators of each of these theories).

We begin in section \ref{sec: theories} by defining precisely all the theories discussed above, and describing their relevant and marginal couplings, and the expected dualities and flows between them.

In section \ref{sec: Nf=1} we review the results of \cite{Aharony:2018pjn} on the $N_f=1$ theory and its fixed points.

In section \ref{sec: general case} we analyze the general form of the beta functions in the ``quasi-bosonic theories'' for all values of $\lambda$. We show that, as for $N_f=1$, the beta functions for all three marginal operators begin at order $\frac{1}{N_c}$, and at this order they are polynomials of degree three. We also analyze the large $N_f$ limit (with $N_f \ll N_c$), and show that in some scaling of the couplings (corresponding to a planar limit for $SU(N_f)$), the beta functions simplify in this limit.

In section \ref{sec: RB theory} we compute explicitly the beta functions in the theories with scalars when they are weakly coupled ($\lambda_B \to 0$). We show that, as for $N_f=1$, for $\lambda_B=0$ the free theory (with vanishing marginal deformations) is a degenerate fixed point, which splits for small finite $\lambda_B$. We find that for $N_f=2$ and for $N_f \geq 5$ there is an IR-stable fixed point for the marginal couplings at small $\lambda_B$, but that no such fixed point exists for $N_f=3,4$. At large $N_f$ we find that the IR-stable fixed point merges with another fixed point and disappears at $|\lambda_B| = 24 \pi / N_f$, such that for $24 \pi / N_f <  |\lambda_B| \ll 1$ there is no IR-stable fixed point. We classify also all the other fixed points, and discuss how they can change as we change $\lambda$.
The small $\lambda_B$ analysis is relevant also for weakly coupled fixed points at finite (not necessarily large) $N_c$, and in appendix \ref{sec: appendix finite lambda finite nc} we analyze precisely for which values of $N_c$ and $N_f$ weakly coupled IR-stable fixed points exist. 

In section \ref{sec: CF theory} we do the same in the limit where the fermionic description is weakly coupled. In this limit we find that for all $N_f > 1$ there is no IR-stable fixed point. The simplest conjecture would be that for $N_f \geq 5$ such a fixed point exists for small enough $\lambda_B$, but disappears as $|\lambda_B|$ is increased. 

In section \ref{sec: semi critical theories} we compute the beta functions for the two ``Semi-Critical theories''. These beta functions (for the single marginal coupling that exists in each of these theories) are again third order polynomials at order $\frac{1}{N_c}$. In both of these theories the $\lambda_B\to 0$ limit is not a free theory (it includes a non-trivial fixed point), so our computations in this limit depend on some unknown coefficients. In the ``Semi-Critical singlet'' theories (where the dimension of the scalar flavor-singlet $\simeq 1$ while that of the scalar flavor-adjoint $\simeq 2$) we find that for $\lambda_B \to 0$ an IR-stable fixed point (for the marginal coupling) exists at large enough $N_f$ (we cannot tell how large it needs to be), while for $\lambda_F \to 0$ there is an IR-stable fixed point for all $N_f$. The ``Semi-Critical adjoint'' theory for $N_f=2$ has no marginal coupling so there is a fixed point where we need to fine-tune just the two relevant deformations. For $N_f=3$ we find that as $\lambda_B \to 0$ the theory has one IR-stable fixed point, while for $\lambda_F \to 0$ it has two IR-stable fixed points, so presumably one of these fixed points merges with another fixed point as we increase $|\lambda_F|$. For $N_f \geq 4$ there is always an IR-stable fixed point for small $|\lambda_F|$, while for small $|\lambda_B|$ we cannot determine the existence of an IR-stable fixed point for any value of $N_f$.

We end in section \ref{sec: summary} with a summary of our results and some future directions. Several appendices contain technical results needed for computations in the main text.

\section{The theories studied in this paper \label{sec: theories}}

In this paper we study Chern-Simons-matter theories with $N_f$ massless complex scalars or Dirac fermions in the fundamental representation of an $SU(N_c)$ gauge group in $d=3$ dimensions.

We denote the scalar fields as $\phi^{c,i}$ (or $\psi^{c,i}$ for fermions), where $c=1,\cdots,N_c$ is the color index, and $i=1,\cdots,N_f$ is the flavor index. 
Here, and in the rest of this paper, we use $a,b,c,\cdots$ to denote color indices and $i,j,k,\cdots$ to denote flavor indices.

We assume that the gauge field is described by a level $\kappa$ Chern-Simons (CS) action, given in Euclidean space by 
\begin{equation}
    S_{CS}\left(A \right)=\frac{\kappa}{4\pi}\intop d^{3}x\ \left(-\frac{i}{2}\epsilon^{\mu\nu\rho}A_{\mu}^{\bar{a}}\partial_{\nu}A_{\rho}^{\bar{a}}-\frac{i}{6}\epsilon^{\mu\nu\rho}f^{\bar{a}\bar{b}\bar{c}}A_{\mu}^{\bar{a}}A_{\nu}^{\bar{b}}A_{\rho}^{\bar{c}}\right)
\end{equation}
where $\bar{a},\bar{b},\bar{c},\cdots$ are adjoint color indices.
Conventions for generators of the color group are given in Appendix \ref{sec: app finite lambda B}. We will study these theories mainly in the 't Hooft limit where $N_c,\ |\kappa|\rightarrow\infty$ with fixed $N_f$, while the 't Hooft coupling $\lambda\equiv\frac{N_c}{\kappa}$\footnote{Because $\lambda$ is the ratio of two integers 
it has no continuous renormalization group flow.} is finite and takes values\footnote{We follow the conventions of \cite{Aharony:2018pjn}, where the CS coupling is written in the dimensional reduction scheme, in which it obeys $|\kappa| \geq N_c$; it is related to the level $k$ coming from a high-energy Yang-Mills-Chern-Simons theory by $\kappa = k + N_c {\rm sign}(k)$.} between $0\le|\lambda|\le1$. When considering the possible dualities between fermion and scalar (boson) theories, we will add the super/subscript $F,B$ to indicate the level, coupling and $N_c$ of the fermion and scalar theories, respectively\footnote{The number of flavors $N_f$ will always be assumed to be the same in both theories.}. In this paper we consider only $SU(N_c)$ gauge theories, but in the 't Hooft limit that we consider they are the same as $U(N_c)$ gauge theories; dualities sometimes exchange $SU(N)$ and $U(N)$ theories \cite{Radicevic:2015yla,Aharony:2015mjs} but the distinction will not be important in this work.

The actions of the fermionic and scalar theories include kinetic terms
\begin{equation} \label{eq: bare action}
    \begin{split}
        & S_{F}(A,\psi)=\intop d^{3}x\space \bar{\psi}{}_{i,b}\cancel{D}_{ba}\psi^{i,a} \space,
        \\
        & S_{B}(A,\phi)=\intop d^{3}x\space \left(D^{ba}_{\mu}\phi{}_{i,a}\right)^{\dagger}\left(D^{\mu}_{bc}\phi^{i,c}\right) . 
    \end{split}
\end{equation}
Both theories have a $U(N_f)$ flavor symmetry, and we will assume that any additional interactions also preserve this symmetry.

In order to avoid cumbersome notation, we'll usually drop either the color or flavor index whenever the context is clear. In the following subsections we'll review the allowed interactions, and  versions of these theories that include additional auxiliary fields.

\subsection{Fermion theories}
\subsubsection{Regular and Critical Fermions} \label{RF}

The theory that includes only the fermion and the CS term is called the Regular Fermion (RF) theory
\begin{equation}\label{eq: regular fermions}
        S_{RF}(A,\psi)=S_{CS}(A)+S_{F}(A,\psi).
\end{equation}
In the 't Hooft large $N_c$ limit, the anomalous dimensions of $\psi$ and of its composites are $O(\frac{1}{N_c})$. Thus, in the large $N_c$ limit this theory has a single relevant operator -- the fermion mass term $\bar\psi \psi$, which we will fine-tune to zero -- and no marginal interactions. Upon fine-tuning the mass to zero the action \eqref{eq: regular fermions} describes a conformal theory. At least at weak coupling, one place where this theory arises is in the IR limit of the QCD-Chern-Simons theory where we add also a Yang-Mills term to \eqref{eq: regular fermions}.

To define the critical fermion theories, we introduce a new auxiliary field $\zeta^i_j$, and couple it to the fermion bilinears $M_{i}^{j}\equiv\frac{4\pi}{\kappa_F}\bar{\psi}_{c,i}\psi^{c,j}$. Note that in this normalization, at leading order in the 't Hooft large $N_c$ limit, $n$-point functions of $M$ scale as $N_c^{1-n}$. At leading order in large $N_c$, $\zeta$ has conformal dimension $\Delta=1$, so after adding this auxiliary field the large $N_c$ theory has three (for $N_f>1$) new relevant interactions, $\zeta^i_i$, $(\zeta^i_i)^2$ and $\zeta^j_i \zeta^i_j$, which need to be fine-tuned to get a conformal field theory. On the other hand, the mass term mentioned above is no longer an independent deformation since it is determined by the equation of motion of $\zeta$.
In addition, in the large $N_c$ limit there are three different types of marginal $\zeta^3$ deformations, corresponding to different ways of contracting the flavor indices:
\begin{equation} \label{eq: fermion interaction zeta}
\mathcal{L}_{int}^F=-\zeta^i_jM_{i}^{j}+
\frac{\bar{g}^F_1}{3!} \left(\zeta_i^i\right)^{3} +
\frac{\bar{g}^F_2}{2} \zeta_i^i\zeta_j^k\zeta_k^j +
\frac{\bar{g}^F_3}{3} \zeta_i^j\zeta_j^k\zeta_k^i .
\end{equation}
The resulting theory (with the three relevant operators appropriately fine-tuned) is called the Critical Fermion (CF) theory
\begin{equation}\label{eq: critical fermions}
        S_{CF}(A,\phi,\zeta)=S_{CS}(A)+S_{F}(A,\psi)+\intop d^{3}x\space\mathcal{L}_{int}^F.
\end{equation}
In the 't Hooft limit, we need to scale $\bar{g}^F_n \equiv \frac{\bar{\lambda}^F_n \cdot \left(\lambda_F\right)^3}{\left(N_c^F\right)^2}$, with $\bar{\lambda}^F_n$ finite\footnote{The factor of $\left(\lambda_F\right)^3$ in this equation is for later convenience.}. In the extreme large $N_c$ limit all three marginal deformations are exactly marginal, but for large finite $N_c$ they acquire beta functions, and our goal in this paper will be to analyze the fixed points and the flow of these beta functions (generalizing the analysis of \cite{Aharony:2018pjn} for $N_f=1$).

In some cases it's more convenient to write the interaction term in \textbf{the representation basis}, such that the $SU(N_f)$ flavor transformations are manifest. We define
\begin{equation}\label{eq: adjoint J}
M_{S}\equiv M_i^i\ ,\ \ \ \ \ \left(M_{A}\right)_{i}^{j}\equiv M_i^j-\frac{\delta_{i}^{j}}{N_{f}}M_{S},
\end{equation}
and similarly for $\zeta$. We call $M_{S}$ the singlet structure and $M_A$ the adjoint structure. To distinguish from the representation basis, we call the form \eqref{eq: fermion interaction zeta} \textbf{the index basis}. The interaction term written in the representation basis is
\begin{equation}  \label{eq: fermion interaction adjoint}
\mathcal{L}_{int}^F=
-\frac{1}{N_f}M_S\zeta_S-{\rm tr}\left(M_A\zeta_A\right)+
\frac{g^F_{SSS}}{3!}\zeta_{S}^{3}+
\frac{g^F_{SAA}}{2!}\zeta_{S}\ {\rm tr}\left(\zeta_A\zeta_A\right)+
\frac{g^F_{AAA}}{3}{\rm tr}\left(\zeta_A\zeta_A\zeta_A\right),
\end{equation}
where the traces are over the flavor indices, and the relation between the couplings in the index ($\bar{g}_n$) and in the representation ($g_n$) basis is given by
\begin{equation}\label{eq: transform to adjoint0}
    \begin{split} & g^{}_{SSS}=\bar{g}_{1}+\frac{3}{N_{f}}\bar{g}_{2}+\frac{2}{N_{f}^{2}}\bar{g}_{3}\\
 & g^{}_{SAA}=\bar{g}_{2}+\frac{2}{N_{f}}\bar{g}_{3}\\
 & g^{}_{AAA}=\bar{g}_{3}
\end{split}
\ \ \ \Leftrightarrow\ \ \ \begin{split} & \bar{g}_{1}=g^{}_{SSS}-\frac{3}{N_{f}}g^{}_{SAA}+\frac{4}{N_{f}^{2}}g^{}_{AAA}\\
 & \bar{g}_{2}=g^{}_{SAA}-\frac{2}{N_{f}}g^{}_{AAA}\\
 & \bar{g}_{3}=g^{}_{AAA}.
\end{split}
\end{equation}
Note that for the special case of $N_f=2$, the trace of three identical adjoints vanishes, so there is no $g^{}_{AAA}$ coupling in this case.

In what follows we'll use both the representation and the index basis. From now on, unless the context demands it, we'll omit the ${\rm tr}()$ for brevity, such that any multiplication of the adjoints is assumed to be traced over.

\subsubsection{Semi-Critical fermions} \label{sec: CF_SA}

The form of (\ref{eq: fermion interaction adjoint}) gives rise to a natural generalization: instead of adding auxiliary fields coupling to both $M_A$ and $M_S$, we can add a field coupling only to one of them. We call the resulting theories Semi-Critical theories (because only the singlet/adjoint degrees of freedom become critical).

The actions for these theories are given by
\begin{subequations} \label{eq: S_CF_S_and_A}
\begin{align}
    S_{CF_S}(A,\phi,\zeta_S)=&S_{CS}(A)+S_{F}(A,\psi)+
    \intop d^{3}x\left(-\frac{1}{N_f}M_S\zeta_S+
    \frac{g^F_{SSS}}{3!}\zeta_{S}^{3}\right), \label{eq: S_CF_S}\\
    S_{CF_A}(A,\phi,\zeta_A)=&S_{CS}(A)+S_{F}(A,\psi)+
    \intop d^{3}x\left(-M_A\zeta_A+
    \frac{g^F_{AAA}}{3!}\zeta_{A}^{3}\right), \label{eq: S_CF_A}
\end{align}
\end{subequations}
where we now have just a single interaction term that is marginal at large $N_c$. In the $CF_S$ theory, the singlet degree of freedom is critical and the adjoint is regular, and vice versa for $CF_A$. The $CF_S$ theory has (at large $N_c$) two relevant operators that need to be fine-tuned, $\zeta_S$ and $\zeta_S^2$. The $CF_A$ theory also has (at large $N_c$) two such operators, which are $\zeta_A^2$ and $M_S$. The existence of these fixed points at finite $N_c$ depends on having fixed points for the $\zeta^3$ couplings, and we will analyze this question below in section \ref{sec: semi critical theories}\footnote{For the special case of $N_f=2$ there is no $g_{AAA}$ coupling, so the $CF_A$ theory has no marginal operators at large $N_c$ and describes a conformal theory.}. If we ignore the marginal operators, then we can flow from the CF theory to the $CF_S$ theory by turning on the $\zeta_A^2$ deformation, or to the $CF_A$ theory by turning on the $\zeta_S^2$ deformation, while generic relevant deformations (still fine-tuning the fermion mass to zero) lead to the RF theory. Similarly, we can flow from the $CF_A$ and $CF_S$ theories to the RF theory.

\subsection{Scalar theories \label{sec: boson theories}}

\subsubsection{The Regular Boson theory\label{sec: RB}}

Naively, as in the fermionic case, one would like to define a Regular Boson theory by $S_B+S_{CS}$. This theory would have (in the large $N_c$ limit) three relevant $\phi^2$ and $\phi^4$ operators that need to be fine-tuned to reach a fixed point. But it also has three marginal deformations (at large $N_c$), which must be added to the action since for finite $N_c$ they would flow (even if we start without such interactions but only with CS interactions). As in the fermionic case, there are three different ways of contracting the $\phi$ field indices \cite{Kapoor:2021lrr}
\begin{equation} \label{eq: boson interaction}
\mathcal{L}_{int}^B=\frac{\bar{g}^B_1}{3!} \left(\bar{\phi}_{c,i}\phi^{c,i}\right)^{3} +\frac{\bar{g}^B_2}{2} \bar{\phi}_{a,i}\phi^{a,i}\bar{\phi}_{b,j}\phi^{b,k}\bar{\phi}_{c,k}\phi^{c,j} +\frac{\bar{g}^B_3}{3} \bar{\phi}_{a,i}\phi^{a,j}\bar{\phi}_{b,j}\phi^{b,k}\bar{\phi}_{c,k}\phi^{c,i} .
\end{equation}
The simplest bosonic theory that we can write,
called the Regular Boson (RB) theory, is thus
\begin{equation}\label{eq: regular bosons}
        S_{RB}(A,\phi)=S_{CS}(A)+S_{B}(A,\phi)+\intop d^{3}x\space\mathcal{L}_{int}^B.
\end{equation}

As for the CF theories, in some cases it's more convenient to write the interaction terms in the representation basis (see section \ref{RF}). We define $\tilde{M}_{j}^{i}\equiv\overline{\phi}_{c,j}\phi^{c,i}$, and using the notation of \eqref{eq: adjoint J}, we can rewrite the interaction terms in \eqref{eq: boson interaction} as
\begin{equation}
\mathcal{L}_{int}^B=\frac{g^B_{SSS}}{3!}\tilde{M}_{S}^{3}+\frac{g^B_{SAA}}{2!}\tilde{M}_{S}\ tr\left(\tilde{M}_A \tilde{M}_A\right)+\frac{g^B_{AAA}}{3}tr\left(\tilde{M}_A \tilde{M}_A \tilde{M}_A\right) .
\end{equation}
The relation between the couplings in the index basis $\left(\bar{g}_{1}^{B},\bar{g}_{2}^{B},\bar{g}_{3}^{B}\right)$ and in the representation basis $\left(g_{SSS}^{B},g_{SAA}^{B},g_{AAA}^{B}\right)$ is given by \eqref{eq: transform to adjoint0}. In the 't Hooft limit, we need to scale $\bar{g}^B_n \equiv \frac{\bar{\lambda}^B_n}{\left(N_c^B\right)^2}$, with $\bar{\lambda}^B_n$ finite\footnote{Note the difference between the definition of $\bar{\lambda}^B_n$ and the definition of $\bar{\lambda}^F_n$ below \eqref{eq: critical fermions}.}.

In order to relate this theory to the formalism used for critical theories (as in section \ref{sec: CF_SA}), we can rewrite the RB theory in a different way. We introduce two new auxiliary fields: $\sigma$, that will act as a Lagrange multiplier, and $\zeta$. We now write
\begin{equation} \label{eq: boson interaction zeta}
\mathcal{L}_{int}^B=\sigma^j_i \tilde{M}_{j}^{i}-\sigma^j_i\zeta_j^i+
\frac{\bar{g}^B_1}{3!} \left(\zeta_i^i\right)^{3} +
\frac{\bar{g}^B_2}{2} \zeta_i^i\zeta_j^k\zeta_k^j +
\frac{\bar{g}^B_3}{3} \zeta_i^j\zeta_j^k\zeta_k^i ,
\end{equation}
or equivalently in the representation basis
\begin{equation} \label{eq: RB representation}
\mathcal{L}_{int}^B=\frac{1}{N_f} \sigma_S \tilde{M}_{S}+\sigma_{A} \tilde{M}_A
-\frac{1}{N_f} \sigma_S\zeta_S
-\sigma_{A}\zeta_A
+\frac{g^B_{SSS}}{3!}\zeta_{S}^{3}
+\frac{g^B_{SAA}}{2!}\zeta_{S}\left(\zeta_A\zeta_A\right)+\frac{g^B_{AAA}}{3}\left(\zeta_A\zeta_A\zeta_A\right),
\end{equation}
where $\zeta_{A/S}$, $\sigma_{A/S}$ and $\tilde{M}_{A/S}$ were defined similarly to $M_{A/S}$ in (\ref{eq: adjoint J}). In this way, the theory with the interaction term  
includes 4 fields $S_{RB}\left(A,
\phi,\sigma,\zeta\right)$, but it is obviously equivalent to \eqref{eq: regular bosons} after integrating out $\sigma$ and $\zeta$. 

\subsubsection{Critical and Semi-Critical Boson theories} \label{sec: CB_and_semicritical}

Again, the form \eqref{eq: RB representation} gives rise to a natural generalization. Instead of adding both fields $\zeta_A$ and $\zeta_S$ to the Lagrangian, we can add either or none of them. The resulting theories are the Semi-Critical or Critical Boson (CB) theories, respectively, 
\begin{subequations} \label{eq: CB_and_semicritical}
\begin{align}
    S_{CB}(A,\phi,\sigma)=&S_{CS}(A)+S_{B}(A,\phi)+\intop d^{3}x\left(\space \frac{1}{N_f} \sigma_S \tilde{M}_{S}+\sigma_{A} \tilde{M}_A\right),\label{eq: CB}\\
    S_{CB_A}(A,\phi,\sigma,\zeta_S)=&S_{CS}(A)+S_{B}(A,\phi)+\intop d^{3}x\left(\frac{1}{N_f}\sigma_S \tilde{M}_{S}+\sigma_{A} \tilde{M}_A
    -\frac{1}{N_f}\sigma_S\zeta_S
    +\frac{g^B_{SSS}}{3!}\zeta_{S}^{3}\right),\label{eq: CBS}\\
    S_{CB_S}(A,\phi,\sigma,\zeta_A)=&S_{CS}(A)+S_{B}(A,\phi)+\intop d^{3}x\left(\frac{1}{N_f}\sigma_S \tilde{M}_{S}+\sigma_{A} \tilde{M}_A
    -\sigma_{A}\zeta_A
    +\frac{g^B_{AAA}}{3}\zeta_A^3\right). \label{eq: CBA}
\end{align}
\end{subequations}
Note that in the $CB_S$ theory the singlet degree of freedom is critical, and therefore the marginal interactions involve only the adjoint degrees of freedom\footnote{Because $\sigma_{S/A}$ acts as a Lagrange multiplier  which sets the composite operators ${\tilde M}_{S/A}$ to zero, we cannot add interactions involving these operators to the Lagrangian.} (and vice versa for $CB_A$). The CB theory has one relevant operator $\sigma_S$ (of dimension $\Delta=2$ in the large $N_c$ limit) and no marginal operators. The $CB_S$ theory has two relevant operators, $\sigma_S$ and $\zeta_A^2$, and one marginal operator in the large $N_c$ limit. Similarly, the $CB_A$ theory has two relevant operators, $\zeta_S$ and $\zeta_S^2$, and one marginal operator in the large $N_c$ limit.

One can flow from the RB theory to the $CB_S$, $CB_A$ or CB theories, by turning on $\phi^4$ couplings (or $\zeta^2$ couplings in the description \eqref{eq: boson interaction zeta}). 
We can also formally write the RB (and also $CB_{A/S}$) theory as a Legendre transform of the CB theory, as
\begin{equation} \label{eq: RB from CB}
S_{RB}\left(A,\phi,\sigma,\zeta\right)=S_{CB}\left(A,\phi,\sigma\right)+\intop d^{3}x\ \left(-\sigma^j_i\zeta_j^i+
\frac{\bar{g}^B_1}{3!} \left(\zeta_i^i\right)^{3} +
\frac{\bar{g}^B_2}{2} \zeta_i^i\zeta_j^k\zeta_k^j +
\frac{\bar{g}^B_3}{3} \zeta_i^j\zeta_j^k\zeta_k^i\right) .
\end{equation}

\subsection{Summary and expected dualities} \label{sec: dualities} 

\begin{figure}[t]
\begin{adjustbox}{center}
$
   \begin{tikzpicture}[
    >=Latex,
]
\node (RB) at (0,2.5)   {\Large $RB$};
\node (CBa) at (-1.5,0)  {\Large $CB_{A}$};
\node (CBs) at (1.5,0)   {\Large $CB_{S}$};
\node (CB) at (0,-2.5)  {\Large $CB$};

\node (CF) at ({0+8},{2.5})   {\Large $CF$};
\node (CFs) at ({-1.5+8},{0})  {\Large $CF_{S}$};
\node (CFa) at ({1.5+8},{0})   {\Large $CF_{A}$};
\node (RF) at ({0+8},{-2.5})  {\Large $RF$};

\draw[->, line width=1pt] (RB) -- (CBa)
    node[midway, xshift=-6pt, yshift=8pt] {$\zeta_A^2$};

\draw[->, line width=1pt] (RB) -- (CBs)
    node[midway, xshift=6pt, yshift=8pt] {$\zeta^2_S$};

\draw[->, line width=1pt] (CBa) -- (CB)
    node[midway, xshift=-8pt, yshift=-1pt] {$\zeta^2_S$};

\draw[->, line width=1pt] (CBs) -- (CB)
    node[midway, xshift=8pt, yshift=-0pt] {$\zeta_A^2$};

\draw[->, line width=1pt] (CF) -- (CFs)
    node[midway, xshift=-6pt, yshift=8pt] {$\zeta_A^2$};

\draw[->, line width=1pt] (CF) -- (CFa)
    node[midway, xshift=6pt, yshift=8pt] {$\zeta^2_S$};

\draw[->, line width=1pt] (CFs) -- (RF)
    node[midway, xshift=-8pt, yshift=-1pt] {$\zeta^2_S$};

\draw[->, line width=1pt] (CFa) -- (RF)
    node[midway, xshift=8pt, yshift=-0pt] {$\zeta_A^2$};

\draw[<->, double,line width=1pt, double distance=4pt] (2.3,0) -- (5.7,0);
    
\end{tikzpicture}
$
\end{adjustbox}

    \caption{The RG flows between the theories discussed in this section are depicted, with arrows labeled by the relevant operators that are turned on to generate them. The lowest relevant operator, $\zeta_S$, is fine-tuned to the fixed point throughout these flows; turning it on leads to a flow to a pure CS theory.}

    \label{fig: flow of theories diagram} 
\end{figure}
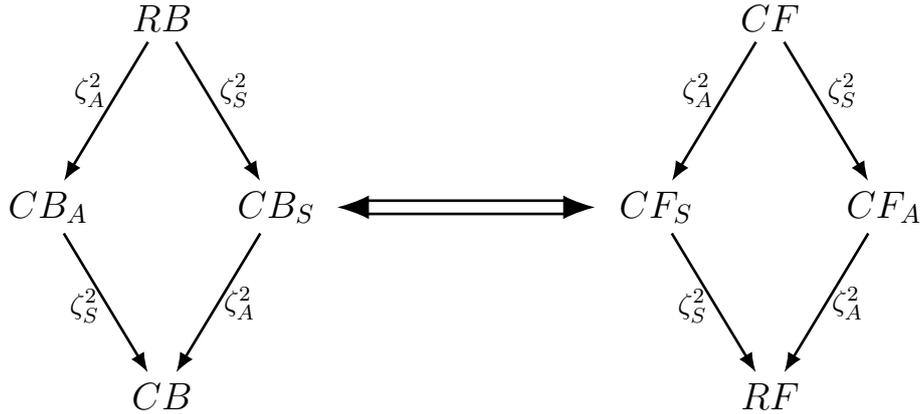

To summarize, we have presented eight theories: four scalar and four fermionic. The properties of these theories, including their interaction terms, are summarized in table \ref{tab: Theories}. All of the scalar theories and all of the fermionic theories are related, as described above, by renormalization group (RG) flows (or alternatively by Legendre transforms), which are summarized in figure \ref{fig: flow of theories diagram}\footnote{Since the flows involve multi-trace operators made from the scalar operators, the 3-point correlation functions of higher-spin operators will be the same in all of these theories at leading order in $1/N_c$; however, higher-point correlation functions, and higher-order corrections in $1/N_c$, will be different. This is similar to the relation between the RF/CB and CF/RB theories for $N_f=1$.}.

\begin{table}[t]
\centering

\begin{adjustbox}{center}

\begin{tabular}{|c|c||c|c|c|}
\hline
\multicolumn{2}{|c||}{\textbf{ \makecell{Scalars}}}
& \multicolumn{2}{c|}{\textbf{\makecell{Fermions}}} & %\begin{cellvarwidth}
\centering
\textbf{\makecell{Running \\ Marginal \\ Couplings}}
%\end{cellvarwidth}
\tabularnewline
\hline 
\hline 
$RB$ & \eqref{eq: RB representation} (see also \eqref{eq: regular bosons}) & $CF$ & \eqref{eq: fermion interaction adjoint} & \centering $g^{}_{SSS},\ g^{}_{SAA},\ g^{}_{AAA}$\tabularnewline
\hline 
$CB_{A}$ & $\frac{1}{N_f}\sigma_S \tilde{M}_{S}+\sigma_{A} \tilde{M}_A-\frac{1}{N_f}\sigma_S\zeta_S+\frac{g^B_{SSS}}{3!}\zeta_{S}^{3}$ & $CF_{S}$ & $-\frac{1}{N_f}M_{S}\zeta_{S}+\frac{g^F_{SSS}}{3!}\zeta_{S}^{3}$ & \centering $g^{}_{SSS}$\tabularnewline
\hline 
$CB_{S}$ & $\frac{1}{N_f}\sigma_S \tilde{M}_{S}+\sigma_{A} \tilde{M}_A-\sigma_{A}\zeta_A+\frac{g^B_{AAA}}{3}\zeta_A^3$ & $CF_{A}$ & $-M_{A}\zeta_{A}+\frac{g^F_{AAA}}{3}\zeta_{A}^{3}$ & \centering $g^{}_{AAA}$\tabularnewline
\hline 
$CB$ & $\frac{1}{N_f}\sigma_S \tilde{M}_{S}+\sigma_{A} \tilde{M}_A$  & $RF$ & $none$ & \centering $none$\tabularnewline
\hline 
\end{tabular}
\end{adjustbox}

\caption{The different theories considered in this paper and their interaction terms. The two theories on each row are conjectured to be dual, as explained in the text.}

\label{tab: Theories} 
\end{table}

There is strong evidence that there is a duality in the 't Hooft limit between the $CB\leftrightarrow RF$ theories \cite{Aharony:2012nh,Aharony:2015mjs,Seiberg:2016gmd,Gur-Ari:2012lgt,Bedhotiya:2015uga,Jain:2014nza,Dandekar:2014era,Yokoyama:2016sbx,Jain:2012qi,Aharony:2012ns,Jain:2013py,Takimi:2013zca,Minwalla:2015sca,Dey:2018ykx,Choudhury:2018iwf}, with $\sigma^j_i$ dual to $M^j_i$. While this has mostly been studied for $N_f=1$, the same evidence holds also for larger finite values of $N_f$ in the `t Hooft limit.
It has been conjectured \cite{Aharony:2015mjs,Seiberg:2016gmd} that this duality is true even at finite $N_c$, although this is not as well established.
Given that the different theories are related (at least at large $N_c$) by Legendre transforms, a duality between the $CB \leftrightarrow RF$  theories (the last row in table \ref{tab: Theories}) implies dualities between the $RB\leftrightarrow CF$ theories and between the $CB_{A/S}\leftrightarrow CF_{S/A}$ theories (the other rows of the table, note the exchange of subscript $A$ and $S$). Of course, the existence of the fixed points on the first three lines depends on having fixed points for their marginal couplings, which we will investigate in detail below; we conjecture that any fixed point for the scalar theories is dual to a corresponding fixed point for the fermionic theories.

The relation between the Chern-Simons couplings and level between the dual theories is\footnote{This implies that $N_c$ must also change under this duality, and in the theories above there is $N_c^B$ for scalars and $N_c^F$ for fermions.}:
\begin{equation} \label{eq: duality boson fermion lambda}
\lambda_{F}=\begin{cases}
\lambda_{B}-1 & \lambda_{B}>0\\
1+\lambda_{B} & \lambda_{B}<0
\end{cases}\ , \quad \kappa_F=-\kappa_B.
\end{equation}
This is a strong-weak duality with respect to the gauge interaction, which can help us understand the behavior of the theories at regimes where perturbation theory is no longer valid. Naively we would expect the couplings $g_n^F$ to map to $g_n^B$ under the duality, but this is subtle because the dual RF and CB theories have different contact terms, and for $N_f=1$ it was found \cite{Aharony:2018pjn} that this leads to a relative shift by a constant between $g_{SSS}^F$ and $g_{SSS}^B$ (the only coupling that exists in that case).
Similar shifts for one or more of the $g_n$ presumably are present also for $N_f>1$. Since shifting the coefficient by a constant does not change the structure of the $\beta$ function, which is our main topic in this paper, this will not affect our discussion below. In any case, we will not use the duality in most of our analysis, except when trying to interpolate the behavior of the fixed points between weak and strong coupling (in either set of variables). The theories discussed above are also conjectured to be dual to high-spin gravity theories (that are generalizations of Vasiliev's high-spin gravity, in which all the bulk fields are adjoints of $U(N_f)$), but we will not use this here.

\section{Review of the \texorpdfstring{$N_f=1$}{Nf =1} results and formalism} \label{sec: Nf=1}

The case of $N_f=1$ was extensively studied in \cite{Aharony:2018pjn}, where a formalism was developed for computing the beta function for the marginal coupling, that will also be useful for the current computations.

In this section we'll briefly review the formalism and the results for the $N_f=1$ case, and in the following sections we'll present the generalization to arbitrary $N_f$.
For the $N_f=1$ case the theories presented in section \ref{sec: theories} are significantly simplified: there is no adjoint structure (see, e.g., (\ref{eq: adjoint J})), and therefore there is only one running coupling $g^{}_{SSS}$. Moreover, the Semi-Critical theories do not exist, so we can consider only the $RB$ and $CF$ theories.

\subsection{General formalism} \label{subsec: Nf1 General}

The main idea of the formalism presented in \cite{Aharony:2018pjn} is to compute an effective action for the field $\zeta$, which can be expressed for the CF theory in terms of expectation values of $M$'s in the RF theory, and for the RB theory in terms of expectation values of $\sigma$'s in the CB theory. Then, we can use this effective action to compute the order $\frac{1}{N_c}$ correction to the 2 and 3-point correlation functions for $\zeta$, from which we derive the beta function. The effective action for $\zeta$ takes the form
\begin{equation} \label{eq: Seff n}
    S^{eff}\left(\zeta\right)=\frac{g^{}_{SSS}}{3!}\int\zeta^{3}d^{3}x-\sum_{n=2}^{\infty}\frac{1}{n!}\int d\Pi_{n}\left\langle \tilde{J}\left(-p_{1}\right)\cdots\tilde{J}\left(-p_{n}\right)\right\rangle_{
    \rm connected
    } \zeta\left(p_{1}\right)\cdots\zeta\left(p_{n}\right)
\end{equation}
with
\begin{equation} \label{eq: dPi_n definition}
d\Pi_{n}\equiv\left(2\pi\right)^{3}\delta\left(\sum_{i=1}^{n}p_{n}\right)\prod_{i=1}^{n}\frac{d^{3}p_{n}}{\left(2\pi\right)^{3}} ,
\end{equation}
and $\tilde{J}$ representing $\sigma$ (in the CB theories) or $M$ (in the RF theories).

In the CF case one can compute the expectation values for $M=\frac{4\pi}{\kappa_F}\bar{\psi}\psi$ using the Feynman rules that are derived from the RF theory \eqref{eq: regular fermions}. In the RB case the situation is more delicate, as the expectation values should be evaluated in the CB theory \eqref{eq: CB}. $\sigma$ is then also a field, so one needs first to compute the effective action for $\sigma$ and then, using this new action, to compute the effective action for $\zeta$.

It can be shown that in the 't Hooft limit the leading contribution (of order $\frac{1}{N_c}$) to the beta function come from 1-loop diagrams in $\zeta$, contributing to its 3-point function, and therefore one needs to compute the sum in (\ref{eq: Seff n}) only up to $n=5$. The result is
\begin{align} \label{eq: effectiv action Nf1}
        S^{eff}\left(\zeta\right)=&\frac{G_{2}}{2}\int\frac{d^{3}q}{\left(2\pi\right)^{3}}\left|q\right|\left|\frac{q}{\Lambda}\right|^{2\gamma}\zeta\left(q\right)\zeta\left(-q\right)\nonumber\\
        + & \frac{G_{3}}{3!}\int\frac{d^{3}q_{1}}{\left(2\pi\right)^{3}}\frac{d^{3}q_{2}}{\left(2\pi\right)^{3}}\frac{d^{3}q_{3}}{\left(2\pi\right)^{3}}\left(2\pi\right)^{3}\delta\left(q_{1}+q_{2}+q_{3}\right)\zeta\left(q_{1}\right)\zeta\left(q_{2}\right)\zeta\left(q_{3}\right)\nonumber\\
        + & \frac{\delta G_3}{3!}\int\frac{d^{3}q_{1}}{\left(2\pi\right)^{3}}\frac{d^{3}q_{2}}{\left(2\pi\right)^{3}}\frac{d^{3}q_{3}}{\left(2\pi\right)^{3}}\left(2\pi\right)^{3}\delta\left(q_{1}+q_{2}+q_{3}\right)\log\left(\frac{\Lambda}{\left|q_{1}\right|+\left|q_{2}\right|+\left|q_{3}\right|}\right)\zeta\left(q_{1}\right)\zeta\left(q_{2}\right)\zeta\left(q_{3}\right)\nonumber\\
        - & \frac{1}{4!}\int d\Pi_{4}G_{4}\left(q_{1},q_{2},q_{3},q_{4}\right) \zeta\left(q_{1}\right)\zeta\left(q_{2}\right)\zeta\left(q_{3}\right) \zeta\left(q_{4}\right)\nonumber\\
        - & \frac{1}{5!}\int d\Pi_{5}G_{5}\left(q_{1},q_{2},q_{3},q_{4},q_{5}\right)\zeta\left(q_{1}\right)\zeta\left(q_{2}\right)\zeta\left(q_{3}\right)\zeta\left(q_{4}\right)\zeta\left(q_{5}\right),
\end{align}
where $\Lambda$ is the UV cut-off used in the regularization, $\gamma$ is the anomalous dimension of ${\tilde J}$, and it was useful to write explicitly the momentum dependence of the 2 and 3 point functions. The momentum dependence of the 5 and 4-point functions is important only in a special kinematic limit, for which it is given by\footnote{More precisely we take $G_5(p,-p-p_1-p_2-p_3,p_1,p_2,p_3)$ with $|p|\gg|p_1|,|p_2|,|p_3|$.}
\begin{subequations} \label{eq: momentum regime}
\begin{equation}
  G_{4}\left(p,-p,k,-k\right)=\tilde{G}_{4}\left[-\frac{2}{\left|p\right|}+\frac{\left|k\right|}{\left|p\right|^{2}}+\frac{\left(p\cdot k\right)^{2}}{\left|p\right|^{4}\left|k\right|}+\mathcal{O}\left(\frac{k^{2}}{p^{3}}\right)\right]\ \ for\ \ p\gg k  ,
\end{equation}
\begin{equation}
G_{5}\left(p,-p,0,0,0\right)=\frac{\tilde{G}_{5}}{p^{2}}.
\end{equation}
\end{subequations}
The leading large $N_c$ scaling of the various coefficients is given by
\begin{equation} \label{gscalings}
    G_2 = O\left(\frac{1}{N_c}\right),\ \gamma = O\left(\frac{1}{N_c}\right), \ G_3 = O\left(\frac{1}{N_c^2}\right),\  \delta G_3 = O\left(\frac{1}{N_c^3}\right),\  \tilde{G}_4 = O\left(\frac{1}{N_c^3}\right),\  \tilde{G}_5 = O\left(\frac{1}{N_c^4}\right).
\end{equation}

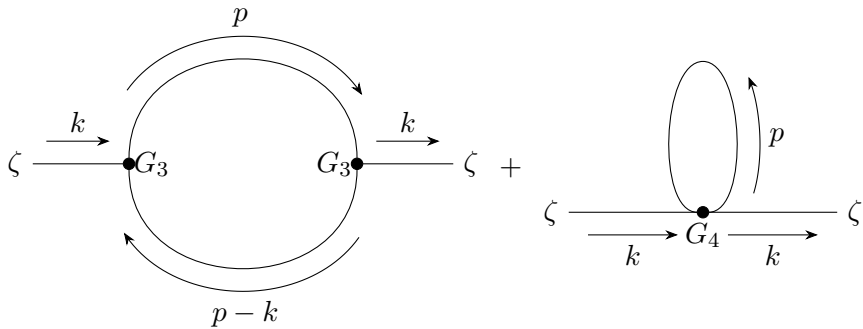
\begin{figure}[t]
    \centering
\begin{equation*}
    \vcenter{\hbox{\begin{tikzpicture}[baseline=(c.base)]
     \begin{feynman}
        \vertex (c);
        \vertex (i1) at (-3,0) {$\zeta$};
        \vertex (i2) at (3,0) {$\zeta$};
        
        \node (d1) at (-1.5,0) [dot];
        \node (d2) at (1.5,0)  [dot];

        \vertex  (d2n) at ([shift={(-0.3,0.0)}]d2) {$G_3$};
        \vertex  (d1n) at ([shift={(0.3,0.0)}]d1) {$G_3$};

        \diagram*{
            (i1) -- [momentum = {$ k $}] (d1),
            (i2) -- [reversed momentum' = {$ k $}] (d2),
            (d1) -- [half left,momentum = {$ p $}] (d2)
                 -- [half left,,momentum = {$ p - k $}] (d1),
        };
    \end{feynman}
\end{tikzpicture}}}\,+\,\vcenter{\hbox{\begin{tikzpicture}[baseline=(c.base)]
    \begin{feynman}
        \node (c) [dot];
        \vertex (h) at (0,2);
        \vertex (i1) at (-2,0) {$\zeta$};
        \vertex (i2) at (2,0) {$\zeta$};

        \vertex  (cn) at ([shift={(0.0,-0.3)}]c) {$G_4$};

        \diagram*{
            (i1) -- [momentum' = {$ k $}] (c),
            (i2) -- [reversed momentum = {$ k $}] (c),
            (c)--[ out=0,in=0,looseness=0.7,,momentum' ={[arrow shorten=0.25] $ p $ } ](h), 
            (c)--[ out=180,in=180,,looseness=0.7](h)

        };
    \end{feynman}
\end{tikzpicture}}}
\end{equation*}

    \caption{Feynman diagrams that contribute to the 2-point function of $\zeta$. The left diagram doesn't have a logarithmic divergence in the cutoff $\Lambda$ and therefore does not contribute to the anomalous dimension $\gamma^{\prime}$. In order to extract the logarithmic divergence of the right diagram, only the kinematical region of \eqref{eq: momentum regime} is needed.}
    \label{fig: gamma diagrams Nf1} 
\end{figure}

\begin{figure}[t]
    \begin{adjustbox}{center}
$
\vcenter{\hbox{\begin{tikzpicture}[baseline=(c.base)]
    \begin{feynman}
        \node (c) [dot];
        \vertex (h) at (3,0);
        \vertex (i1) at (-2,2) {$\zeta$};
        \vertex (i2) at (-2,0) {$\zeta$};
        \vertex (i3) at (-2,-2) {$\zeta$};                
        
        \diagram*{
            (i1) -- [momentum = {$ p_1 $}] (c),
            (i3) -- [ momentum' = {$ p_3 $}] (c),
            (i2) -- [ momentum' = {[arrow distance=0.1] $ p_2 $}] (c),
            (c)--[ out=90,in=90,looseness=1.5,momentum ={[arrow shorten=0.25] $ p $ } ](h), 
            (c)--[ out=270,in=270,looseness=1.5](h),
            
        };
        \vertex  (cn) at ([shift={(0.3,0.0)}]c) {$G_5$};
    \end{feynman}
\end{tikzpicture}}}
-\, 3\vcenter{\hbox{\begin{tikzpicture}[baseline=(c.base)]
    \begin{feynman}
        \vertex (c);
        \vertex (i1) at (-3,1) {$\zeta$};
        \vertex (i2) at (3,0) {$\zeta$};
        \vertex (i3) at (-3,-1) {$\zeta$};
        
        \node (d1) at (-1.5,0) [dot];
        \node (d2) at (1.5,0) [dot];

        \diagram*{
            (i1) -- [momentum = {$ p_1 $}] (d1),
            (i3) -- [ momentum' = {$ p_3 $}] (d1),
            (i2) -- [ momentum' = {$ p_2 $}] (d2),
            (d1) -- [half left,momentum ={[arrow shorten=0.25] $ p $ }] (d2)
                 -- [half left,,momentum ={[arrow shorten=0.25] $ p +p_2 $ }] (d1),
            
        };
        \vertex  (d1n) at ([shift={(0.3,0.0)}]d1) {$G_4$};
        \vertex  (d2n) at ([shift={(-0.3,0.0)}]d2) {$G_3$};

    \end{feynman}
\end{tikzpicture}}}-
    \vcenter{\hbox{\begin{tikzpicture}[baseline=(c.base)]
    \begin{feynman}
        \vertex (c);
        \vertex (i1) at (-3,0) {$\zeta$};
        \vertex (i2) at (1.5,2.6) {$\zeta$};
        \vertex (i3) at (1.5,-2.6) {$\zeta$};
        
        \node (d1) at (-1.5,0) [dot];
        \node (d2) at (0.75,1.4) [dot];
        \node (d3) at (0.75,-1.4) [dot];

        \vertex  (d1n) at ([shift={(0.0,0.3)}]d1) {$G_3$};
        \vertex  (d2n) at ([shift={(-0.3,0.1)}]d2) {$G_3$};
        \vertex  (d3n) at ([shift={(-0.3,-0.05)}]d3) {$G_3$};

        \diagram*{
            (i1) -- [momentum = {$ p_1 $}] (d1),
            (i2) -- [ momentum = {$ p_2 $}] (d2),
            (i3) -- [ momentum = {$ p_3 $}] (d3),
            (d1) -- [momentum = {$ p+p_1 $}] (d2)
                 -- [momentum = {$ p - p_3 $}] (d3)
                 -- [momentum = {$ p  $}] (d1),
            
        };
    \end{feynman}
\end{tikzpicture}
}}
$
\end{adjustbox}

    \caption{Feynman diagrams that contribute to the amputated 3-point function of $\zeta$. Together with the tree level diagram for $\delta G_3$ and the anomalous dimension, they contribute to the beta function \eqref{eq: beta Nf1}. It can be seen that for beta function computations, only the kinematic region in  \eqref{eq: momentum regime} contributes.}

    \label{fig: 3 point diagrams Nf1} 
\end{figure}
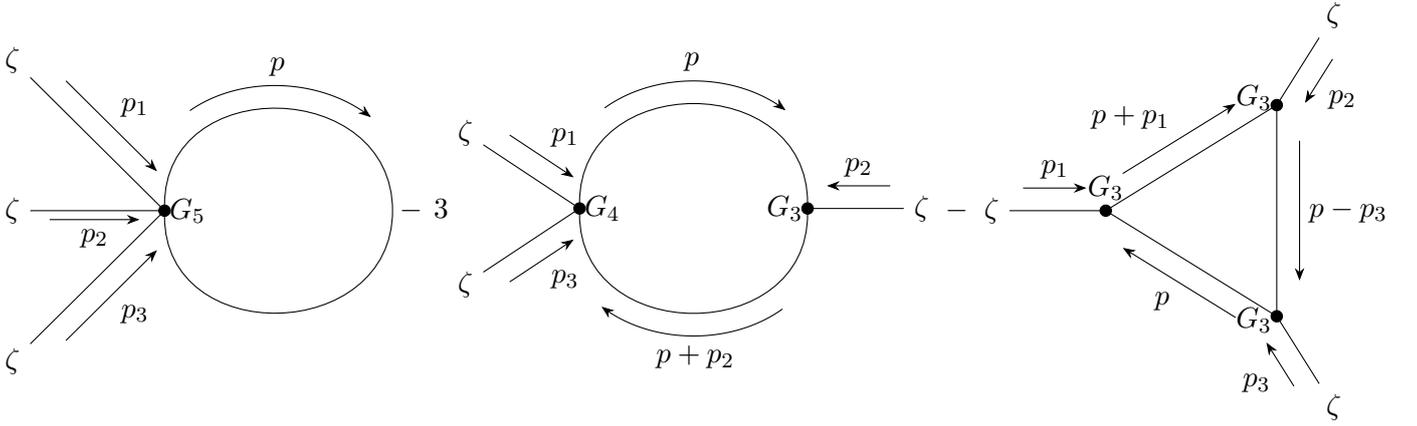

Using this action one can first compute the anomalous dimension of $\zeta$ by computing the 1-loop corrections to the 2-point correlation function, as shown in figure \ref{fig: gamma diagrams Nf1}, and find that it is given by
\begin{equation} \label{eq: anomalus dimension single flavor}
    \gamma^{\prime}=\gamma+\frac{\tilde{G}_{4}}{6\pi^{2}G_{2}^{2}}.
\end{equation}
Computing the amputated 3-point correlation function, as shown in figure \ref{fig: 3 point diagrams Nf1}, and using the Callan-Symanzik equation we can then find the beta function for $G_3$ in terms of $G_n$ and $\gamma$ as
\begin{equation}\label{eq: beta Nf1}
\beta\left(g^{}_{SSS}\right)=\left(\frac{\tilde{G}_{5}}{4\pi^{2}G_{2}}-\delta G_3\right)-G_{3}\left(-\frac{3\tilde{G}_{4}}{2\pi^{2}G_{2}^{2}}+3\gamma^{\prime}\right)-\frac{G_{3}^{3}}{2\pi^{2}G_{2}^{3}}.
\end{equation}
Note that in the large $N_c$ limit $G_3$ is of order $\frac{1}{N_c^2}$, while each term on the right-hand side (using \eqref{gscalings}) is $O\left(\frac{1}{N_c^3}\right)$, such that the beta function is suppressed by $\frac{1}{N_c}$. Equation \eqref{eq: beta Nf1} contains this leading contribution in $\frac{1}{N_c}$, exactly as a function of
the 't Hooft couplings $\lambda_{SSS}$ and $\lambda$.
$G_3$ is equal to $g^{}_{SSS}$ up to a constant (that differs for fermions and bosons), this means that the beta function for $g^{}_{SSS}$ will be a cubic polynomial, with a negative coefficient for the third power (since $G_2$ is always positive from unitarity).

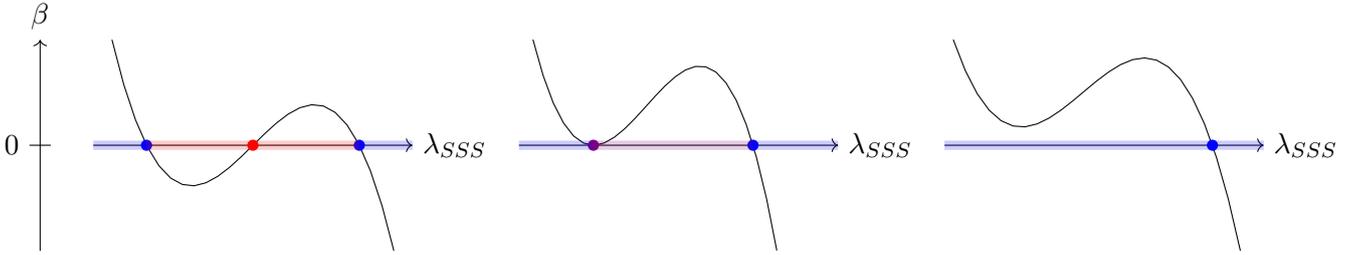
\begin{figure}[t]
\centering
\begin{adjustbox}{center}
\begin{tikzpicture}[scale=1.4]
  \draw[->] (-2,-1) -- (-2,1) node[above] {$\beta$};
  \draw (-1.9,0) -- (-2.1,0) node[left] {$0$};
  
  \draw[->] (-1.5,0) -- (1.5,0) node[right] {$\lambda_{SSS}$};
  \draw[->] (2.5,0) -- (5.5,0) node[right] {$\lambda_{SSS}$};
  \draw[->] (6.5,0) -- (9.5,0) node[right] {$\lambda_{SSS}$};

  \draw[color=black,domain=-1.3247:1.3247]  plot (\x, { \x * (1 - \x) * (1 +\x) });
  \draw[color=black,domain=2.632:4.924]  plot (\x, { -1.5*(\x-3.2)*(\x-3.2)*(\x-4.7) });
  \draw[color=black,domain=6.582:9.2805]  plot (\x, { -0.85*(\x-6.8) * (\x-7.8) * (\x-8.8)+0.5 });

  \fill[color=blue] (-1,0) circle (1.5pt);
  \fill[color=red] (0,0) circle (1.5pt);
  \fill[color=blue] (1,0) circle (1.5pt);
  \fill[color=violet] (3.2,0) circle (1.5pt);
  \fill[color=blue] (4.7,0) circle (1.5pt);
  \fill[color=blue] (9.0178,0) circle (1.5pt);

  \draw[blue, opacity=0.2, line width=3.5pt] (-1.5,0) -- (-1,0);
  \draw[blue, opacity=0.2, line width=3.5pt] (1,0) -- (1.5,0);
  \draw[red, opacity=0.2, line width=3.5pt] (-1,0) -- (1,0);

  \draw[blue, opacity=0.2, line width=3.5pt] (2.5,0) -- (3.2,0);
  \draw[blue, opacity=0.2, line width=3.5pt] (4.7,0) -- (5.5,0);
  \draw[violet, opacity=0.2, line width=3.5pt] (3.2,0) -- (4.7,0);

  \draw[blue, opacity=0.2, line width=3.5pt] (6.5,0) -- (9.5,0);

\end{tikzpicture}
\end{adjustbox}    
\caption{Illustrations of the three possibilities for the beta function in \eqref{eq: beta Nf1}. UV fixed points are denoted in blue, and IR fixed points in red. Points denoted by blue on the horizontal axis flow to (plus or minus) infinite coupling in the IR, while points denoted by red flow to the IR fixed point. The middle case corresponds to a non generic case, which under small perturbations will results in either the left or right cases. The purple point is a stable fixed point in the IR from one direction, and points denoted by purple on the horizontal axis flow in the IR to this point. }
\label{fig: beta functino nf=1 possabilities}
\end{figure}

Illustrations of the possible beta functions \eqref{eq: beta Nf1} are given in figure \ref{fig: beta functino nf=1 possabilities}. Generically, there are either one or no stable points in the IR, and either one or two fixed points in the UV.

\subsection{The beta function in the perturbative region} \label{subsec: Nf1 first order}

Some of the values of the correlation functions $G_n$ are known in the literature to all orders in the coupling $\lambda$, while others are known either only to leading order or in a specific kinematic limit (see \cite{Aharony:2018pjn} for a full summary of the known results). The values at  $\lambda_{B/F}=0$\footnote{For fermions we take the limit $\lambda_F\to 0$, using $\lambda_{SSS}^F$ as defined in section \ref{RF}.} are shown in Table \ref{tab: Gn Nf1}\footnote{The relation between the notation used here and in \cite{Aharony:2018pjn} can be read off by comparing the action \eqref{eq: effectiv action Nf1} with (3.17) there. Care must be taken in relating the signs of different factors between the fermionic and bosonic theories, e.g. in the notation of \cite{Aharony:2018pjn} $\gamma^\prime_B=\frac{\delta_B^\prime}{\kappa_B}$ but $\gamma^\prime_F=-\frac{\delta_F^\prime}{\kappa_F}$, which can be checked by requiring the duality \eqref{eq: duality boson fermion lambda} to hold.}. We can now substitute these values in \eqref{eq: beta Nf1} to find the CF and RB beta functions in these limits. 
\begin{table}[t]
\centering
\begin{tabular}{|c||c|c|}
\hline 
\textbf{Coupling} & \textbf{Bosons} & \textbf{Fermions}\tabularnewline
\hline 
\hline 
$G_{2}$ & $\frac{8}{\lambda_{B}\kappa_{B}}$ & $2\pi^{2}\frac{\lambda_{F}}{\kappa_{F}}$\tabularnewline
\hline 
$G_{3}$ & $g_{SSS}^{B}-\frac{128}{\left(N_{c}^{B}\right)^{2}}$ & $g_{SSS}^{F}$\tabularnewline
\hline 
$\delta G_3$ & see \eqref{eq: G3 G5 Nf1} & 0\tabularnewline
\hline 
$\tilde{G}_{4}$ & $\frac{2^{11}}{\lambda_{B}^{3}\kappa_{B}^{3}}$ & $\frac{\lambda_{F}}{2\kappa_{F}^{3}}\left(4\pi\right)^{4}$\tabularnewline
\hline 
$\tilde{G}_{5}$ & see \eqref{eq: G3 G5 Nf1} & 0\tabularnewline
\hline 
$\gamma$ & $-\frac{16}{3\pi^{2}\lambda_{B}\kappa_B}$ & 0\tabularnewline
\hline 
$\gamma^{\prime}$ & 0 & $\frac{16}{3\pi^{2}\lambda_{F}\kappa_F}$\tabularnewline
\hline 
\end{tabular}
    \caption{The known values of $G_n$, $\gamma$ and $\gamma^\prime$ in the limits $\lambda_{B/F}\rightarrow0$ \cite{Aharony:2018pjn,Vasiliev:1982dc,Aharony:2012nh,Yacoby:2018yvy,Turiaci:2018nua,Aharony:2018npf,Giombi:2016zwa,Jain:2019fja}. For the fermions with $\lambda_F=0$, $\tilde{G}_3$ and $\tilde{G}_5$ vanish due to parity. The vanishing of $\gamma^\prime$ for bosons and $\gamma$ for fermions in the appropriate limits can be understood because the RB and RF theories are free as $\lambda_{B/F}\rightarrow0$.
    }\label{tab: Gn Nf1}
\end{table}

The beta function for the CF theory with $\lambda_F=0$ is
\begin{equation} \label{eq: nf1 beta CF lambda 0}
    \beta\left(\lambda_{SSS}^F\right)=\frac{1}{\pi^{2}N_{c}^{F}}\left[32\lambda^F_{SSS}-\left(\lambda_{SSS}^F\right)^{3}\frac{1}{16\pi^{6}}\right].
\end{equation} 
This beta function has three roots and corresponds to the case on the left in figure \ref{fig: beta functino nf=1 possabilities}. The IR fixed point $\lambda_{SSS}^F=0$ has a finite domain of attraction $\lambda_{SSS}^F\in\left(-16\sqrt{2}\pi^3,16\sqrt{2}\pi^3\right)$, and so turning on a finite small $\lambda_F$ will not alter the number or type of fixed points. 

The situation for the bosons is somewhat more complicated. The values of $\tilde{G}_{5}$ and $\delta G_3$ are unknown in that case\footnote{It was conjectured that once $\tilde{G}_5$ is computed for $\lambda_B=0$ it will be possible to interpolate it to arbitrary values of $\lambda$ \cite{Jain:2022ajd}. \label{foot: G5 all lambda}}. However, they are related in the limit $\lambda_B=0$ by\footnote{The easiest way to see this is to note that due to the fact that the $N_f=1$ version of \eqref{eq: regular bosons} is free when $\lambda_B=0$ and $g_{SSS}^B=0$, the beta function should also vanish at this point, and this completely determines the constant part of \eqref{eq: beta Nf1}. Thus, even though $\tilde{G}_5$ is unknown, it will not play a role in this case.}
\begin{equation}\label{eq: G3 G5 Nf1}
    \frac{\tilde{G}_{5}}{4\pi^{2}G_{2}}-\delta G_3\overset{\lambda_{B}=0}{=}\frac{4096}{N_B^3\pi^2}.
\end{equation}
In this formalism, this is due to the fact that both $\tilde{G}_{5}$ and $\delta G_3$ get contributions from the same set of diagrams (see Appendix B in \cite{Aharony:2018pjn}).
Plugging this into \eqref{eq: beta Nf1} and expressing the result in term of $\lambda_{SSS}^B$ (as defined in section \ref{sec: RB}) we find\footnote{This result can also be obtained directly from \eqref{eq: regular bosons} in the usual way \cite{Pisarski:1982vz}.}
\begin{equation}  \label{eq: nf1 beta RB lambda 0}
    \beta\left(\lambda_{SSS}^B\right)=\frac{3}{8\pi^2N_{c}^{B}}\left(\lambda_{SSS}^B\right)^2\left[1-\frac{\lambda_{SSS}^B}{384}\right].
\end{equation}
This beta function has a double root at $\lambda_{SSS}^B=0$, which corresponds to the middle case in figure \ref{fig: beta functino nf=1 possabilities} (the other root is at $\lambda_{SSS}^B=384$ and is UV stable). A small perturbation making $\lambda_B$ finite can therefore either create an IR-stable point, or remove this fixed point completely. 

In order to understand which of these options is realized, one must consider finite $\lambda_B$. This was done for finite $N_c$ directly in the RB formalism \eqref{eq: regular bosons} in the region $\lambda_B^2 \sim \lambda_{SSS}^B\ll1$ \cite{Avdeev:1992jt,Aharony:2011jz,Anninos:2014hia}\footnote{This was first done in \cite{Avdeev:1992jt}. Later, \cite{Aharony:2011jz} presented a simplified version for $SO(N_c)$, which were expanded for other gauge groups (including $U(N_c)$) in \cite{Anninos:2014hia} (note a small typo in the prefactor of $\lambda^4$ in equation (9) of \cite{Aharony:2011jz}, which is fixed in equation (3.1) in \cite{Anninos:2014hia} for the $SO(N_c)$ case. The $SU(N_c)$ case is given, to leading order in $N_c$, by a factor of two compared to equation (3.1) in \cite{Anninos:2014hia}. The prefactor in (3.2) in \cite{Anninos:2014hia} for $U(N_c)$ is a mistake, as evident by explicitly summing all the diagram. See Appendix B in \cite{Anninos:2014hia}). In terms of their notation $\frac{\lambda_B}{4\pi}=\lambda$, $\lambda_{SSS}^B=\lambda_6$, and we only take leading order in $\frac{1}{N_c}$. As mentioned above, in the 't Hooft limit the beta function is the same for $SU(N_c)$ and $U(N_c)$ gauge groups, up to a factor of 2.}. The beta function for small $\lambda^{B}_{SSS}$ was found to be
\begin{equation} \label{eq: nf1 beta RB finite lambda near 0}
    \beta\left(\lambda_{SSS}^B\right)=\frac{1}{N_c^{B}\pi^2}\left[\frac{3}{2}\left(\frac{\lambda_{B}}{4\pi}\right)^{4}-\frac{5}{2}\lambda_{SSS}^B\left(\frac{\lambda_{B}}{4\pi}\right)^2+\frac{3}{8}\left(\lambda_{SSS}^B\right)^{2}\right].
\end{equation}
The $\lambda_{SSS}^B=0$ fixed point is thus split to $\lambda_{SSS}^B=\frac{2}{3}\left(\frac{\lambda_{B}}{4\pi}\right)^2$ and $\lambda_{SSS}^{B}=6\left(\frac{\lambda_{B}}{4\pi}\right)^2$, with the second one an IR-stable fixed point.

\subsection{Discussion and hypothesis in the non-perturbative regime} \label{subsec: Nf1 discussion}

Using the RB-CF duality mentioned in section \ref{sec: dualities} we can view these beta functions as the limits of the beta function of the same theory as $\lambda\rightarrow0,1$. Both beta functions have 2 UV stable fixed points and one IR-stable fixed point. This, together with other considerations, led the authors of \cite{Aharony:2018pjn} to conjecture that there is an IR-stable fixed point for \textbf{any} intermediate value of $\lambda$.

We can also see from the beta functions that the size of the region which flows to the IR-stable point stays finite even as $\lambda_B\rightarrow0$, and therefore landing on this point does not require substantial fine tuning (in figure \ref{fig: beta functino nf=1 possabilities} this corresponds to a finite length of the purple region). This region is always compact, bounded between the two UV stable points.

In the following we'll find that some of the features mentioned above carry over also to the $N_f>1$ cases. In particular, we'll find degenerate roots of the beta functions in the RB theories as $\lambda_B\rightarrow0$, so we'll need to generalize the small $\lambda_B$ computation there, while the critical (and Semi-Critical) theories do not have such a degeneracy. In contrast, the behavior of the IR-stable fixed points and the region that flows to them will be very different for $N_f>1$.

\section{Beta function structure for multiple flavors} \label{sec: general case}

Chern-Simons-matter theories with multiple flavors have several marginal interactions (see \eqref{eq: fermion interaction zeta} and \eqref{eq: boson interaction}), which leads to intricate beta functions. In this section, we analyze (to leading order in $\frac{1}{N_c}$) the generic structure of the beta functions of the marginal couplings for the CF \eqref{eq: critical fermions} and RB \eqref{eq: regular bosons} theories. 

The crucial observation is that to leading order in $\frac{1}{N_c}$, \emph{the diagrams which contribute to the beta functions in the $N_f >1$ case are the same as in the $N_f=1$ case}, the only difference is that the interaction vertices acquire additional flavor index structure\footnote{The momentum dependence of the diagrams, as well as the order of $\frac{1}{N_c}$ at which each diagram contributes, stays the same.}. We can thus use similar tools to those presented in section \ref{sec: Nf=1}, focusing on the effective action of the $\zeta^i_j$ fields.

The section is organized as follow: In section \ref{sec: multiple flavors sturctre and parametrization} we present the effective action of the $\zeta^i_j$, and parameterize its coefficients. In section \ref{sec: multiple flavors beta calculation} we present the calculation of the anomalous dimensions and of the beta functions. In section \ref{sec: multiple falvor beta function comments} we analyze the structure of the beta functions, and general features which can be extracted from their form. Finally, in section \ref{sec: general case large nf limit}, we discuss the limit of a large
number of flavors $N_f \rightarrow \infty$ (but still with $N_f \ll N_c$).

\subsection{Effective action for multiple flavors} \label{sec: multiple flavors sturctre and parametrization}
\subsubsection{General structure}

Our first step, in both the CF and RB theories, is to write the path integral in terms of the $\zeta^i_j$ fields only, integrating out all other fields. Generalizing the $N_f=1$ case above, the effective action of $\zeta^i_j$ takes the form
\begin{equation} \label{eq: S zeta multiple flavor step 1}
    \begin{split}
        S^{eff}\left(\zeta_{i}^{j}\right)=&\int d^{3}x\left[\frac{\bar{g}_{1}}{3!}\left(\zeta_{i}^{i}\right)^{3}+\frac{\bar{g}_{2}}{2}\zeta_{i}^{i}\zeta_{j}^{k}\zeta_{k}^{j}+\frac{\bar{g}_{3}}{3}\zeta_{i}^{j}\zeta_{j}^{k}\zeta_{k}^{i}\right] \\
        &-\sum_{n=2}^{\infty}\frac{1}{n!}\int d\Pi_{n}\left\langle \tilde{J}_{i_{1}}^{j_{1}}\left(-p_{1}\right)\cdots\tilde{J}_{i_{n}}^{j_{n}}\left(-p_{n}\right)\right\rangle_{\rm connected} \zeta_{j_{1}}^{i_{1}}\left(p_{1}\right)\cdots\zeta_{j_{n}}^{i_{n}}\left(p_{n}\right),
    \end{split}
\end{equation}
where $\tilde{J}_{i}^{j}$ represents $\sigma_{i}^{j}$ (in the scalar case) or $M_{i}^{j}$ (in the fermionic case).

In the 't Hooft limit, the leading contribution to the beta functions of the marginal couplings is of order $\frac{1}{N_c}$. As in the $N_f=1$ case, we then need to expand the sum in \eqref{eq: S zeta multiple flavor step 1} only up to $n=5$, so that it can be rewritten as
\begin{equation} \label{eq: S zeta multiple flavor step 2}
    \begin{split}
        S^{eff}\left(\zeta\right)	&=\frac{1}{2}\int\frac{d^{3}q}{\left(2\pi\right)^{3}}\left|q\right|\left(\left(G_{2}\right)_{i_{1},i_{2}}^{j_{1},j_{2}}+\left(\delta G_{2}\right)_{i_{1},i_{2}}^{j_{1},j_{2}}\log\left(\frac{q}{\Lambda}\right)\right)\zeta_{j_{1}}^{i_{1}}\left(q\right)\zeta_{j_{2}}^{i_{2}}\left(-q\right) \\
	&+\frac{1}{3!}\int d\Pi_{3}\left(G_{3}\right)_{i_{1},i_{2},i_{3}}^{j_{1},j_{2},j_{3}}\left(p_{1},p_{2},p_{3}\right)\zeta_{j_{1}}^{i_{1}}\left(p_{1}\right)\zeta_{j_{2}}^{i_{2}}\left(p_{2}\right)\zeta_{j_{3}}^{i_{3}}\left(p_{3}\right)\\
	&+\frac{1}{3!}\int d\Pi_{3}\left(\delta G_{3}\right)_{i_{1},i_{2},i_{3}}^{j_{1},j_{2},j_{3}}\left(p_{1},p_{2},p_{3}\right)\log\left(\frac{\Lambda}{\left|p_{1}\right|+\left|p_{2}\right|+\left|p_{3}\right|}\right)\zeta_{j_{1}}^{i_{1}}\left(p_{1}\right)\zeta_{j_{2}}^{i_{2}}\left(p_{2}\right)\zeta_{j_{3}}^{i_{3}}\left(p_{3}\right)\\
	&-\frac{1}{4!}\int d\Pi_{4}\left(G_{4}\right)_{i_{1},i_{2},i_{3},i_{4}}^{j_{1},j_{2},j_{3},j_{4}}(p_i) \zeta_{j_{1}}^{i_{1}}\left(p_{1}\right)\zeta_{j_{2}}^{i_{2}}\left(p_{2}\right)\zeta_{j_{3}}^{i_{3}}\left(p_{3}\right)\zeta_{j_{4}}^{i_{4}}\left(p_{4}\right) \\
    &-\frac{1}{5!}\int d\Pi_{5}\left(G_{5}\right)_{i_{1},i_{2},i_{3},i_{4},i_{5}}^{j_{1},j_{2},j_{3},j_{4},j_{5}}(p_i) \zeta_{j_{1}}^{i_{1}}\left(p_{1}\right)\zeta_{j_{2}}^{i_{2}}\left(p_{2}\right)\zeta_{j_{3}}^{i_{3}}\left(p_{3}\right)\zeta_{j_{4}}^{i_{4}}\left(p_{4}\right)\zeta_{j_{5}}^{i_{5}}\left(p_{5}\right),
    \end{split}
\end{equation}
where the $\bar{g}_n$ couplings in \eqref{eq: S zeta multiple flavor step 1} are inserted into $\left(G_{3}\right)_{i_{1},i_{2},i_{3}}^{j_{1},j_{2},j_{3}}$ in the second line of \eqref{eq: S zeta multiple flavor step 2}. This is the generalization of \eqref{eq: effectiv action Nf1}\footnote{The anomalous dimension $\gamma$ in \eqref{eq: effectiv action Nf1} is expressed here as $\delta G_2= 2 G_2 \gamma$ for $N_f=1$.}.

The coefficients $\left(G_{n}\right)_{i_{1},\cdots,i_{n}}^{j_{1},\cdots,j_{n}}\left(p_{1},\cdots,p_{n}\right)$ in \eqref{eq: S zeta multiple flavor step 2} can be greatly simplified in our limits, since for each coefficient we consider only its leading order in  $\frac{1}{N_c}$. In the following discussion, we concentrate on their parametrization in the CF theory, with the conclusions being the same for the RB theory.

\subsubsection{Parameterizing \texorpdfstring{$G_2$ and $\delta G_2$}{G2 and dG2}}

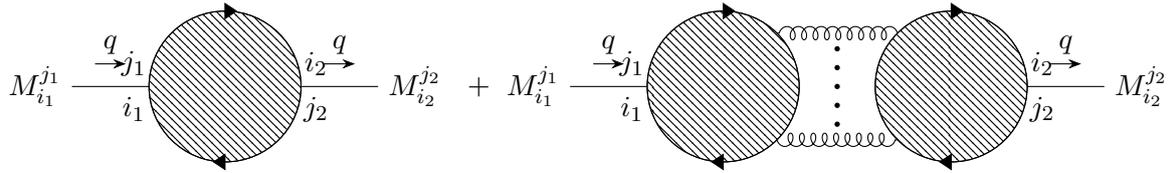
\begin{figure}[t]
    \begin{adjustbox}{center}
\begin{adjustbox}{center}
$
    \vcenter{\hbox{\begin{tikzpicture}[baseline=(c.base)]
  \draw (0,0) circle (1);
  \filldraw[black] (0,1.1) -- (0,0.9) -- (0.15,1) -- cycle;
  \filldraw[black] (0,-1.1) -- (0,-0.9) -- (-0.15,-1) -- cycle;
  \begin{feynman}
    \vertex [blob,minimum size=2cm] (c) at (0,0) {};
    \vertex (i1) at (-2.5,0) {$M^{j_1}_{i_1}$};
    \vertex (i2) at (2.5,0) {$M^{j_2}_{i_2}$};
        
    \vertex (d1) at (-1,0);
    \vertex (d2) at (1,0);
    \diagram*{
            (i1) -- [reversed momentum' ={[arrow shorten=0.7] $ q $ }] (d1),
            (i2) -- [momentum ={[arrow shorten=0.7] $ q $ }] (d2),
        };
  \end{feynman}
  \node at ([shift={(-0.2,-0.3)}]d1) {\(i_1\)};
  \node at ([shift={(-0.2,0.3)}]d1) {\(j_1\)};
  \node at ([shift={(0.2,0.3)}]d2) {\(i_2\)};
  \node at ([shift={(0.2,-0.3)}]d2) {\(j_2\)};
\end{tikzpicture}
}}+\vcenter{\hbox{\begin{tikzpicture}[baseline=(c.base)]
  \draw (0,0) circle (1);
  \filldraw[black] (0,1.1) -- (0,0.9) -- (0.15,1) -- cycle;
  \filldraw[black] (0,-1.1) -- (0,-0.9) -- (-0.15,-1) -- cycle;

  \draw (3,0) circle (1);
  \filldraw[black] (3,1.1) -- (3,0.9) -- (3.15,1) -- cycle;
  \filldraw[black] (3,-1.1) -- (3,-0.9) -- (2.85,-1) -- cycle;
  
  \begin{feynman}
    \vertex [blob,minimum size=2cm] (c) at (0,0) {};
    \vertex [blob,minimum size=2cm] (c2) at (3,0) {};

    \vertex (i1) at (-2.5,0) {$M^{j_1}_{i_1}$};
    \vertex (i2) at (5.5,0) {$M^{j_2}_{i_2}$};
        
    \vertex (d1) at (-1,0);
    \vertex (d2) at (4,0);
    
    \vertex (g11) at ({1*cos(45)},{1*sin(45)});
    \vertex (g21) at ({3+cos(135)},{sin(135)});

    \vertex (g12) at ({1*cos(-45)},{1*sin(-45)});
    \vertex (g22) at ({3+cos(225)},{sin(225)});
    \diagram*{
            (i1) -- [reversed momentum' ={[arrow shorten=0.7] $ q $ }] (d1),
            (i2) -- [momentum ={[arrow shorten=0.7] $ q $ }] (d2),
            (g21) -- [gluon] (g11),
            (g12) -- [gluon] (g22),
        };
  \end{feynman}
  \node at ([shift={(-0.2,-0.3)}]d1) {\(i_1\)};
  \node at ([shift={(-0.2,0.3)}]d1) {\(j_1\)};
  \node at ([shift={(0.2,0.3)}]d2) {\(i_2\)};
  \node at ([shift={(0.2,-0.3)}]d2) {\(j_2\)};

    \fill (1.5,0.5) circle (1pt);
    \fill (1.5,0.25) circle (1pt);
    \fill (1.5,0) circle (1pt);
    \fill (1.5,-0.25) circle (1pt);
    \fill (1.5,-0.5) circle (1pt);
    
\end{tikzpicture}
}}
$
\end{adjustbox}
\end{adjustbox}

    \caption{Two `topologies' contributing to the 2-point function \eqref{eq: G2 general}. The shaded area includes possible gluon lines and interactions (the diagram on the right is schematic and actually there can be many different types of gluon structures). The diagrams on the right start at one higher order in $\frac{1}{N_c}$ in the 't Hooft large $N_c$ limit.} 

    \label{fig: g2 for general case different topologies} 
\end{figure}

The two-point function of two mesons in the CF theory is given by a sum of diagrams with two distinct `topologies', shown in figure \ref{fig: g2 for general case different topologies}. In one type of diagrams (on the left-hand side), the fermion line which starts with flavor index $j_1$ ($j_2$), connects to the flavor index $i_2$ ($i_1$); this results in a contribution to the two-point function with the flavor structure $\delta_{i_2}^{j_1}\delta_{i_1}^{j_2}$. Note that the only fermion interactions in the RF theory are with gluons, and those preserve the flavor index, and so it is valid to assign a flavor to a fermion line. Diagrams like this start at order $\frac{1}{N_c}$, when the gluon diagram filling the disk in figure \ref{fig: g2 for general case different topologies} is planar and there are no extra fermion loops \cite{coleman1988aspects}.
For the second type of diagrams (on the right-hand side in the figure), the fermion line connects each meson to itself,
resulting in the structure $\delta_{i_1}^{j_1}\delta_{i_2}^{j_2}$. These diagrams start with ``annulus diagrams'' of order $\frac{1}{N_c^2}$.

Since we are interested only in the leading order result of $\frac{1}{N_c}$, we can take the structure of the two-point function to be of the form\footnote{Note that here, and in the following, we leave the dependence on $\lambda_{F/B}$ implicit in the parametrization.}
\begin{equation} \label{eq: G2 general}
\left(G_{2}\right)_{i_{1},i_{2}}^{j_{1},j_{2}}=G_{2}\delta_{i_{2}}^{j_{1}}\delta_{i_{1}}^{j_{2}} ,
\end{equation}
where by dimensional analysis, $G_{2}$ is constant\footnote{Note that $G_2$ is actually known for any value of $\lambda_{B/F}$. This follows form the fact that only the planar diagrams (left diagram in figure \ref{fig: g2 for general case different topologies}) contribute to $G_2$ in leading order, and this diagram is the same as for the single flavor case, which is known in the literature \cite{Aharony:2012nh,Aharony:2018pjn}.\label{foot: G2}}.

However, we are interested also in $\delta G_2$, which is suppressed compared to $G_2$ by an extra factor of $\frac{1}{N_c}$. Thus, both types of diagrams can contribute to $\delta G_2$, and
we parameterize it by
\begin{equation} \label{eq: deltaG2 general}
    \left(\delta G_{2}\right)_{i_{1},i_{2}}^{j_{1},j_{2}}=2G_{2}\gamma_{A}\delta_{i_{1}}^{j_{2}}\delta_{i_{2}}^{j_{1}}+2\frac{G_{2}}{N_{f}}\left(\gamma_{S}-\gamma_{A}\right)\delta_{i_{1}}^{j_{1}}\delta_{i_{2}}^{j_{2}} ,
\end{equation}
where we explain the subscripts in $\gamma_{A/S}$ below. Once again, $\gamma_{A/S}$ are constants and do not depend on the momentum, by dimensional analysis.

The general classification by `topologies' for the diagrams can be extended to $n>2$. More concretely, for any $n$, the leading contribution at large $N_c$ comes only from diagrams where all the fermion lines exiting the mesons are connected (in some cyclic order) along a single fermion loop, as in figure \ref{fig: g4 for general case different topologies}.
A diagram in which there are several connected pieces gives a subleading contribution, with each added `connected' piece reducing the contribution of this diagram by a factor of $\frac{1}{N_c}$.

Before continuing to $n>2$, we have two comments. First, since \eqref{eq: G2 general} is diagonal at leading order in $\frac{1}{N_c}$, it is easy to find the propagator of the $\zeta$ fields to leading order in $\frac{1}{N_c}$,
\begin{equation} \label{eq: general G2 -1}
\left(G_{2}^{-1}\right)_{j_{1},j_{2}}^{i_{1},i_{2}}\left(q\right)\equiv\left\langle \zeta_{j_{1}}^{i_{1}}\left(q\right)\zeta_{j_{2}}^{i_{2}}\left(-q\right)\right\rangle=\frac{1}{\left|q\right|G_{2}}\delta_{j_{2}}^{i_{1}}\delta_{j_{1}}^{i_{2}} .
\end{equation}
We will use \eqref{eq: general G2 -1} in evaluating the Feynman diagrams below.

Second, the theory preserves the $SU\left( N_f \right)$ flavor symmetry, and quantities multiplying $\zeta_{j_{1}}^{i_{1}}$ can be written in terms of the representation basis. For two $\zeta$ fields
\begin{equation} \label{eq: projectio to SA to fields}
\left(a\delta_{i_{1}}^{j_{2}}\delta_{i_{2}}^{j_{1}}+b\delta_{i_{1}}^{j_{1}}\delta_{i_{2}}^{j_{2}}\right)\zeta_{j_{1}}^{i_{1}}\zeta_{j_{2}}^{i_{2}} = a\left(\zeta_{A}\right)_{j}^{i}\left(\zeta_{A}\right)_{i}^{j}+\left(\frac{a}{N_{f}}+b\right)\zeta_{S}\zeta_{S} ,
\end{equation}
and so using the parametrization in \eqref{eq: G2 general} and \eqref{eq: deltaG2 general}, the first line  of the effective action \eqref{eq: S zeta multiple flavor step 2} can be written as
\begin{equation}
\left(\left(G_{2}\right)_{i_{1},i_{2}}^{j_{1},j_{2}}+\left(\delta G_{2}\right)_{i_{1},i_{2}}^{j_{1},j_{2}}\log\left(\frac{q}{\Lambda}\right)\right)\zeta_{j_{1}}^{i_{1}}\zeta_{j_{2}}^{i_{2}} = G_{2}\left(\frac{q}{\Lambda}\right)^{2\gamma_{A}}\left(\zeta_{A}\right)_{j}^{i}\left(\zeta_{A}\right)_{i}^{j}+\frac{G_{2}}{N_{f}}\left(\frac{q}{\Lambda}\right)^{2\gamma_{S}}\zeta_{S}\zeta_{S} .
\end{equation}
We see that $\gamma_{A}$ and $\gamma_{S}$ can be identified with the `classical' anomalous dimension of the $\zeta_{A}$ and $\zeta_{S}$ fields (classical in the sense that $\zeta$ loops have not been taken into account).

\subsubsection{Parameterizing \texorpdfstring{$G_3$ and $\delta G_3$}{G3 and dG3}}

For $n=3$, we have three different possible structures of flavor indices. The general $G_3$ can therefore be written as
\begin{equation} \label{eq: G3 general}
\left(G_{3}\right)_{i_{1},i_{2},i_{3}}^{j_{1},j_{2},j_{3}}=\left[\begin{array}{c}
\bar{G}_{3,1}\left(\delta_{i_{1}}^{j_{1}}\delta_{i_{2}}^{j_{2}}\delta_{i_3}^{j_{3}}\right)+\\
\bar{G}_{3,2}\left(\delta_{i_{1}}^{j_{1}}\delta_{i_{2}}^{j_{3}}\delta_{i_{3}}^{j_{2}}+\delta_{i_{2}}^{j_{2}}\delta_{i_{1}}^{j_{3}}\delta_{i_{3}}^{j_{1}}+\delta_{i_{3}}^{j_{3}}\delta_{i_{1}}^{j_{2}}\delta_{i_{2}}^{j_{1}}\right)+\\
\bar{G}_{3,3}\left(\delta_{i_{1}}^{j_{2}}\delta_{i_{2}}^{j_{3}}\delta_{i_{3}}^{j_{1}}+\delta_{i_{1}}^{j_{3}}\delta_{i_{3}}^{j_{2}}\delta_{i_{2}}^{j_{1}}\right)
\end{array}\right].
\end{equation}
Following the previous discussion, the leading order large $N_c$ contributions from the correlation functions of mesons appear only in $\bar{G}_{3,3}$. We thus find
\begin{equation} \label{eq: general G3 coeff}
\bar{G}_{3,1}=\bar{g}_{1},\quad\bar{G}_{3,2}=\bar{g}_{2},\quad\bar{G}_{3,3}=\bar{g}_{3}-\left\langle \tilde{J}_{1}^{2}\left(-p_{1}\right)\tilde{J}_{2}^{3}\left(-p_{2}\right)\tilde{J}_{3}^{1}\left(-p_{3}\right)\right\rangle _{{\rm leading}}.
\end{equation}
For $\delta G_3$, we are interested in coefficients with a factor of $\frac{1}{N_c}$ compared to $G_3$, and so we can write
\begin{equation}  \label{eq: general delta G3}
    \left(\delta G_{3}\right)_{i_{1},i_{2},i_{3}}^{j_{1},j_{2},j_{3}}=\delta\bar{G}_{3,2}\left(\delta_{i_{1}}^{j_{1}}\delta_{i_{2}}^{j_{3}}\delta_{i_{3}}^{j_{2}}+\delta_{i_{2}}^{j_{2}}\delta_{i_{1}}^{j_{3}}\delta_{i_{3}}^{j_{1}}+\delta_{i_{3}}^{j_{3}}\delta_{i_{1}}^{j_{2}}\delta_{i_{2}}^{j_{1}}\right)+\delta\bar{G}_{3,3}\left(\delta_{i_{1}}^{j_{2}}\delta_{i_{2}}^{j_{3}}\delta_{i_{3}}^{j_{1}}+\delta_{i_{1}}^{j_{3}}\delta_{i_{3}}^{j_{2}}\delta_{i_{2}}^{j_{1}}\right),
\end{equation}
where we implicitly used $\delta \bar{G}_{3,1}=0$ since its calculation contains three disconnected pieces, so it not at the order in $\frac{1}{N_c}$ that we are interested in.

The parametrization adopted in \eqref{eq: general G3 coeff} and \eqref{eq: general delta G3} is natural within the index basis, which is more convenient for explicit calculations. However, for the purposes of presentation and gaining conceptual insight, it is more advantageous to write the beta functions in the representation basis, with the transformation given in \eqref{eq: transform to adjoint0}. To facilitate the understanding of the results, in the following we use $G_{SSS},G_{SAA},G_{AAA}$ instead of $\bar{G}_{3,1},\bar{G}_{3,2},\bar{G}_{3,3}$  and $\delta G_{SSS},\delta G_{SAA},\delta G_{AAA}$ instead of $\delta \bar{G}_{3,1}, \delta \bar{G}_{3,2}, \delta \bar{G}_{3,3}$.

\subsubsection{Parameterizing \texorpdfstring{$G_4$ and $G_5$}{G4 and G5}}

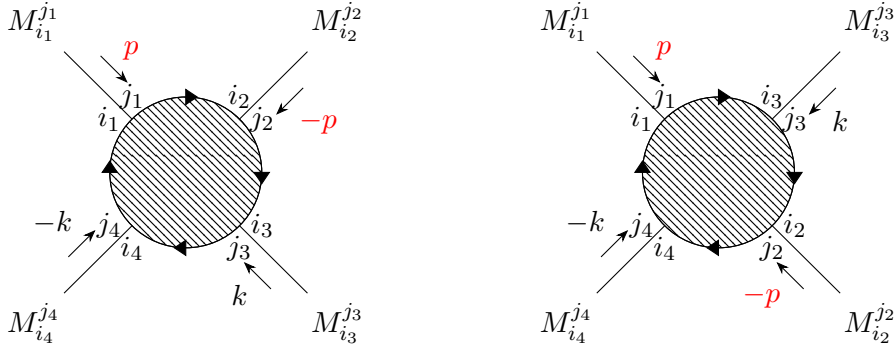
\begin{figure}[t]
\begin{adjustbox}{center}
$
    \vcenter{\hbox{\begin{tikzpicture}[baseline=(c.base)]
  \draw (0,0) circle (1);
  \filldraw[black] (0,1.1) -- (0,0.9) -- (0.15,1) -- cycle;
  \filldraw[black] (0,-1.1) -- (0,-0.9) -- (-0.15,-1) -- cycle;
  \filldraw[black] (1.1,0) -- (0.9,0) -- (1,-0.15) -- cycle;
  \filldraw[black] (-1.1,0) -- (-0.9,0) -- (-1,0.15) -- cycle;

  \begin{feynman}
    \vertex [blob,minimum size=2cm] (c) at (0,0) {};
    \vertex (i1) at ({-2},{2}) {$M^{j_1}_{i_1}$};
    \vertex (i2) at ({2},{2}) {$M^{j_2}_{i_2}$};
    \vertex (i3) at ({2},{-2}) {$M^{j_3}_{i_3}$};
    \vertex (i4) at ({-2},{-2}) {$M^{j_4}_{i_4}$};
        
    \vertex (d1) at ({-sqrt(1/2)},{sqrt(1/2)}) ;
    \vertex (d2) at ({sqrt(1/2)},{sqrt(1/2)}) ;
    \vertex (d3) at ({sqrt(1/2)},{-sqrt(1/2)}) ;
    \vertex (d4) at ({-sqrt(1/2)},{-sqrt(1/2)}) ;
    \diagram*{
            (i1) -- [reversed momentum' ={[arrow shorten=0.7] $ \color{red}{p} $ }] (d1),
            (i2) -- [reversed momentum' ={[arrow shorten=0.7] $ \color{red}{-p} $  }] (d2),
            (i3) -- [reversed momentum' ={[arrow shorten=0.7] $ k $ }] (d3),
            (i4) -- [reversed momentum' ={[arrow shorten=0.7] $ -k $ }] (d4),
        };
  \end{feynman}
  \node at ([shift={(-0.3,0)}]d1) {\(i_1\)};
  \node at ([shift={(0,0.3)}]d1) {\(j_1\)};
  \node at ([shift={(0,0.3)}]d2) {\(i_2\)};
  \node at ([shift={(0.3,0)}]d2) {\(j_2\)};
  \node at ([shift={(0.3,0)}]d3) {\(i_3\)};
  \node at ([shift={(0,-0.3)}]d3) {\(j_3\)};
  \node at ([shift={(0,-0.3)}]d4) {\(i_4\)};
  \node at ([shift={(-0.3,0)}]d4) {\(j_4\)};
\end{tikzpicture}
}}\quad \quad \quad \quad \quad \vcenter{\hbox{\begin{tikzpicture}[baseline=(c.base)]
  \draw (0,0) circle (1);
  \filldraw[black] (0,1.1) -- (0,0.9) -- (0.15,1) -- cycle;
  \filldraw[black] (0,-1.1) -- (0,-0.9) -- (-0.15,-1) -- cycle;
  \filldraw[black] (1.1,0) -- (0.9,0) -- (1,-0.15) -- cycle;
  \filldraw[black] (-1.1,0) -- (-0.9,0) -- (-1,0.15) -- cycle;

  \begin{feynman}
    \vertex [blob,minimum size=2cm] (c) at (0,0) {};
    \vertex (i1) at ({-2},{2}) {$M^{j_1}_{i_1}$};
    \vertex (i2) at ({2},{2}) {$M^{j_3}_{i_3}$};
    \vertex (i3) at ({2},{-2}) {$M^{j_2}_{i_2}$};
    \vertex (i4) at ({-2},{-2}) {$M^{j_4}_{i_4}$};
        
    \vertex (d1) at ({-sqrt(1/2)},{sqrt(1/2)}) ;
    \vertex (d2) at ({sqrt(1/2)},{sqrt(1/2)}) ;
    \vertex (d3) at ({sqrt(1/2)},{-sqrt(1/2)}) ;
    \vertex (d4) at ({-sqrt(1/2)},{-sqrt(1/2)}) ;
    \diagram*{
            (i1) -- [reversed momentum' ={[arrow shorten=0.7] $ \color{red}{p} $ }] (d1),
            (i2) -- [reversed momentum' ={[arrow shorten=0.7] $ k $ }] (d2),
            (i3) -- [reversed momentum' ={[arrow shorten=0.7] $ \color{red}{-p} $ }] (d3),
            (i4) -- [reversed momentum' ={[arrow shorten=0.7] $ -k $ }] (d4),
        };
  \end{feynman}
  \node at ([shift={(-0.3,0)}]d1) {\(i_1\)};
  \node at ([shift={(0,0.3)}]d1) {\(j_1\)};
  \node at ([shift={(0,0.3)}]d2) {\(i_3\)};
  \node at ([shift={(0.3,0)}]d2) {\(j_3\)};
  \node at ([shift={(0.3,0)}]d3) {\(i_2\)};
  \node at ([shift={(0,-0.3)}]d3) {\(j_2\)};
  \node at ([shift={(0,-0.3)}]d4) {\(i_4\)};
  \node at ([shift={(-0.3,0)}]d4) {\(j_4\)};
\end{tikzpicture}
}}
$
\end{adjustbox}

    \caption{Two momentum structures that contribute to the 4-point function in the kinematic limit discussed in the text. Either the two external lines of momentum $\pm p$ are near each other in the cyclic order (left), or they are separated by another external line (right).}

    \label{fig: g4 for general case different topologies} 
\end{figure}

For $n=4$ we need only the leading order results, which means only diagrams with one `connected' piece, and only in a specific momentum limit (we want $\left(G_{4}\right)_{i_{1},i_{2},i_{3},i_{4}}^{j_{1},j_{2},j_{3},j_{4}}\left(p,-p,k,-k\right)$ in the limit $|p|\gg|k|$). By dimensional analysis, at the leading and first subleading orders in $\frac{1}{|p|}$, there can be only three momentum terms which contribute: $\frac{1}{\left|p\right|},\frac{\left|k\right|}{\left|p\right|^{2}},\frac{\left(p\cdot k\right)^{2}}{\left|p\right|^{4}\left|k\right|}$, as in \eqref{eq: momentum regime}\footnote{There can be in general also contributions of the type $\frac{k\cdot p}{|p|^3}$ or $\frac{k\cdot p}{|p|^2|k|}$, however, they will cancel when integrating over $p$.}. However, here there is a slight complication: once we fix the momenta of the different operators, different flavor structures correspond to different diagrams. This is evident in the examples in figure \ref{fig: g4 for general case different topologies}. The diagram on the left is proportional to $\delta_{i_{2}}^{j_{1}}\delta_{i_{3}}^{j_{2}}\delta_{i_{4}}^{j_{3}}\delta_{i_{1}}^{j_{4}}$, and the diagram on the right to  $\delta_{i_{3}}^{j_{1}}\delta_{i_{2}}^{j_{3}}\delta_{i_{4}}^{j_{2}}\delta_{i_{1}}^{j_{4}}$. Each structure will have a different momentum dependence, and so we assign different coefficients to them. There is a symmetry of taking $p\rightarrow-p$ and $k\rightarrow-k$, and so in total there are only 2 options -- either $p$ is `near' $-p$ (left diagram) or $p$ is `far' from $-p$ (right diagram). We conclude that the general structure of $G_4$ is\footnote{\label{footnote: on g4 and symetrized}Note that in the calculation below, when we use the 4-point function we integrate over all values of $p$, and odd functions of $p$ in $G_4$ will not contribute to the beta function. Thus, we do not use $\left(G_{4}\right)_{i_{1},i_{2},i_{3},i_{4}}^{j_{1},j_{2},j_{3},j_{4}}\left(p,-p,k,-k\right)$ per se, but the symmetrized combination
\begin{equation*}
    \frac{1}{2}\left[\left(G_{4}\right)_{i_{1},i_{2},i_{3},i_{4}}^{j_{1},j_{2},j_{3},j_{4}}\left(-p,p,k,-k\right)+\left(G_{4}\right)_{i_{1},i_{2},i_{3},i_{4}}^{j_{1},j_{2},j_{3},j_{4}}\left(p,-p,k,-k\right)\right] .
\end{equation*}
This, instead of $\left(G_{4}\right)_{i_{1},i_{2},i_{3},i_{4}}^{j_{1},j_{2},j_{3},j_{4}}\left(p,-p,k,-k\right)$ alone, is what the parametrization in \eqref{eq: g4 general case} stands for. Indeed, we calculate in Appendix \ref{sec: appendix CB correlations} the symmetrized quantity for the CB theory with $\lambda_B=0$, and show that this quantity is well defined and free from IR divergences, rather than the non-symmetrized $G_4$.
}
\begin{equation} \label{eq: g4 general case}
\begin{adjustbox}{center}
$
\left(G_{4}\right)_{i_{1},i_{2},i_{3},i_{4}}^{j_{1},j_{2},j_{3},j_{4}}\left(p,-p,k,-k\right)=\frac{1}{\left|p\right|}\left(\begin{array}{c}
\left(\begin{array}{c}
G_{4,1,N}\left(\delta_{i_{2}}^{j_{1}}\delta_{i_{3}}^{j_{2}}\delta_{i_{4}}^{j_{3}}\delta_{i_{1}}^{j_{4}}+\delta_{i_{2}}^{j_{1}}\delta_{i_{4}}^{j_{2}}\delta_{i_{3}}^{j_{4}}\delta_{i_{1}}^{j_{3}}+\delta_{i_{1}}^{j_{2}}\delta_{i_{3}}^{j_{1}}\delta_{i_{4}}^{j_{3}}\delta_{i_{2}}^{j_{4}}+\delta_{i_{1}}^{j_{2}}\delta_{i_{4}}^{j_{1}}\delta_{i_{3}}^{j_{4}}\delta_{i_{2}}^{j_{3}}\right)\\
+G_{4,1,F}\left(\delta_{i_{3}}^{j_{1}}\delta_{i_{2}}^{j_{3}}\delta_{i_{4}}^{j_{2}}\delta_{i_{1}}^{j_{4}}+\delta_{i_{4}}^{j_{1}}\delta_{i_{2}}^{j_{4}}\delta_{i_{3}}^{j_{2}}\delta_{i_{1}}^{j_{3}}\right)
\end{array}\right) +\\
\frac{\left|k\right|}{\left|p\right|}\left(\begin{array}{c}
G_{4,2,N}\left(\delta_{i_{2}}^{j_{1}}\delta_{i_{3}}^{j_{2}}\delta_{i_{4}}^{j_{3}}\delta_{i_{1}}^{j_{4}}+\delta_{i_{2}}^{j_{1}}\delta_{i_{4}}^{j_{2}}\delta_{i_{3}}^{j_{4}}\delta_{i_{1}}^{j_{3}}+\delta_{i_{1}}^{j_{2}}\delta_{i_{3}}^{j_{1}}\delta_{i_{4}}^{j_{3}}\delta_{i_{2}}^{j_{4}}+\delta_{i_{1}}^{j_{2}}\delta_{i_{4}}^{j_{1}}\delta_{i_{3}}^{j_{4}}\delta_{i_{2}}^{j_{3}}\right)\\
+G_{4,2,F}\left(\delta_{i_{3}}^{j_{1}}\delta_{i_{2}}^{j_{3}}\delta_{i_{4}}^{j_{2}}\delta_{i_{1}}^{j_{4}}+\delta_{i_{4}}^{j_{1}}\delta_{i_{2}}^{j_{4}}\delta_{i_{3}}^{j_{2}}\delta_{i_{1}}^{j_{3}}\right)
\end{array}\right) +\\
\frac{\left(p\cdot k\right)^{2}}{\left|p\right|^{3}\left|k\right|}\left(\begin{array}{c}
G_{4,3,N}\left(\delta_{i_{2}}^{j_{1}}\delta_{i_{3}}^{j_{2}}\delta_{i_{4}}^{j_{3}}\delta_{i_{1}}^{j_{4}}+\delta_{i_{2}}^{j_{1}}\delta_{i_{4}}^{j_{2}}\delta_{i_{3}}^{j_{4}}\delta_{i_{1}}^{j_{3}}+\delta_{i_{1}}^{j_{2}}\delta_{i_{3}}^{j_{1}}\delta_{i_{4}}^{j_{3}}\delta_{i_{2}}^{j_{4}}+\delta_{i_{1}}^{j_{2}}\delta_{i_{4}}^{j_{1}}\delta_{i_{3}}^{j_{4}}\delta_{i_{2}}^{j_{3}}\right)\\
+G_{4,3,F}\left(\delta_{i_{3}}^{j_{1}}\delta_{i_{2}}^{j_{3}}\delta_{i_{4}}^{j_{2}}\delta_{i_{1}}^{j_{4}}+\delta_{i_{4}}^{j_{1}}\delta_{i_{2}}^{j_{4}}\delta_{i_{3}}^{j_{2}}\delta_{i_{1}}^{j_{3}}\right)
\end{array}\right)
\end{array}\right)   ,
$
\end{adjustbox}
\end{equation}
where the subscript $N$ stands for `Near' and $F$ for `Far'\footnote{For the $N_f=1$ case the 4-point function is known exactly for any value of $\lambda$ \cite{Kalloor:2019xjb,Turiaci:2018nua}, and this fixes some specific combinations of the $N_f>1$ coefficients as described in \eqref{eq: g4g5 parametrization multiple to single} below. It is possible that using the same techniques one can  compute each of these terms separately, for $G_{4,1,N/F}$ this was done in \cite{Turiaci:2018nua}.}.

The $n=5$ case is similar to the $n=4$ case. We need only the leading order in $\frac{1}{N_c}$, and only for $\left(G_{5}\right)_{i_{1},i_{2},i_{3},i_{4},i_{5}}^{j_{1},j_{2},j_{3},j_{4},j_{5}}\left(p,-p,0,0,0 \right)$ (namely, the other momenta are small compared to $p$). As in the $n=4$ case, there are two structures, where the $p$ meson is `near' or `far' from the $-p$ meson. The only momentum contribution which we care about has the form  $\frac{1}{\left|p\right|^2}$, and so we can write\footnote{Note that as described in footnote \ref{footnote: on g4 and symetrized} for $G_4$, we use only the symmetrized version of $G_5$.}
\begin{equation} \label{eq: genral g5}
    \left(G_{5}\right)_{i_{1},i_{2},i_{3},i_{4},i_{5}}^{j_{1},j_{2},j_{3},j_{4},j_{5}}\left(p,-p,0,0,0\right)=\frac{1}{\left|p\right|^{2}}\left(\begin{array}{c}
G_{5,N}\left(\delta_{i_{2}}^{j_{1}}\delta_{i_{3}}^{j_{2}}\delta_{i_{4}}^{j_{3}}\delta_{i_{5}}^{j_{4}}\delta_{i_{1}}^{j_{5}}+\left(3,4,5\right)+\left(1\leftrightarrow2\right)\right)+\\
G_{5,F}\left(\delta_{i_{3}}^{j_{1}}\delta_{i_{2}}^{j_{3}}\delta_{i_{4}}^{j_{2}}\delta_{i_{5}}^{j_{4}}\delta_{i_{1}}^{j_{5}}+\left(3,4,5\right)+\left(1\leftrightarrow2\right)\right)
\end{array}\right),
\end{equation}
where all the non-written permutations are given by permutations of the indices $\{3,4,5\}$ and exchanging $1\leftrightarrow2$. All in all, we have 24 terms, corresponding to the number of permutations of the 5 mesons up to cyclic permutations.

\subsection{Calculating the beta function} \label{sec: multiple flavors beta calculation}

The calculation of the beta function proceeds in a similar fashion to the calculation reviewed in section \ref{sec: Nf=1}; one has to compute the diagrams in figures \ref{fig: gamma diagrams Nf1} and \ref{fig: 3 point diagrams Nf1}, adding the appropriate flavor structures.

For the 2-point function, as in the $N_f=1$ case, the logarithmic UV divergence comes only from the right diagram of figure \ref{fig: gamma diagrams Nf1}
\begin{equation} \label{eq: general 2point 1 loop}
\begin{split}
\left(\delta\Gamma_{2}\right)_{i_{1},i_{2}}^{j_{1},j_{2}}\left(q\right)= & -\frac{1}{2}\intop\frac{d^{3}p}{\left(2\pi\right)^{3}}\left(G_{2}^{-1}\right)_{j_{3},j_{4}}^{i_{3},i_{4}}\left(p\right)\left(G_{4}\right)_{i_{3},i_{4},i_{1},i_{2}}^{j_{3},j_{4},j_{1},j_{2}}\left(p,-p,q,-q\right)\\
= & \left(\frac{N_{f}\left(3G_{4,2,N}+G_{4,3,N}\right)}{3\pi^{2}G_{2}}\delta_{i_{2}}^{j_{1}}\delta_{i_{1}}^{j_{2}}+\frac{3G_{4,2,F}+G_{4,3,F}}{6\pi^{2}G_{2}}\delta_{i_{1}}^{j_{1}}\delta_{i_{2}}^{j_{2}}\right)\left|q\right|\log\left(\frac{q}{\Lambda}\right),
\end{split}
\end{equation}
where we used $\int\frac{d^{3}p}{\left(2\pi\right)^{3}}\frac{\left|q\right|}{\left|p\right|^{3}} = \frac{\left|q\right|}{2\pi^{2}}\log\left(\Lambda\right) $ and $ \int\frac{d^{3}p}{\left(2\pi\right)^{3}}\frac{\left(p\cdot q\right)^{2}}{\left|p\right|^{5}\left|q\right|} = \frac{\left|q\right|}{6\pi^{2}}\log\left(\Lambda\right)$.

For the 3-point function, we need to take into account all the diagrams in figure \ref{fig: 3 point diagrams Nf1}. The computation is more lengthy but straightforward and we find (the dots are to abbreviate the  full structures presented in \eqref{eq: G3 general})
\begin{equation}
\left(\delta\Gamma_{3}\right)_{i_{1},i_{2},i_{3}}^{j_{1},j_{2},j_{3}}=\left[\delta\bar{\Gamma}_{3,1}\left(\delta_{i_{1}}^{j_{1}}\delta_{i_{2}}^{j_{2}}\delta_{i_3}^{j_{3}}\right)+\delta\bar{\Gamma}_{3,2}\left(\delta_{i_{1}}^{j_{1}}\delta_{i_{2}}^{j_{3}}\delta_{i_{3}}^{j_{2}}+\cdots\right)+\delta\bar{\Gamma}_{3,3}\left(\delta_{i_{1}}^{j_{2}}\delta_{i_{2}}^{j_{3}}\delta_{i_{3}}^{j_{1}}+\cdots\right)\right]\log(\Lambda),
\end{equation}
where
\begin{multline}
    \delta\bar{\Gamma}_{3,1}=\Bigg( \frac{3\bar{G}_{3,2}G_{4,1,F}}{2\pi^{2}G_{2}^{2}}+\frac{\bar{G}_{3,1}^{3}N_{f}^{3}+\bar{G}_{3,2}^{3}\left(N_{f}^{2}+14\right)+9\bar{G}_{3,1}^{2}\bar{G}_{3,2}N_{f}^{2}}{2\pi^{2}G_{2}^{3}}\\
    +\frac{24\bar{G}_{3,1}\bar{G}_{3,2}^{2}N_{f}+6\bar{G}_{3,3}\left(\bar{G}_{3,2}^{2}N_{f}+\bar{G}_{3,1}^{2}N_{f}+4\bar{G}_{3,1}\bar{G}_{3,2}\right)+6\bar{G}_{3,3}^{2}\bar{G}_{3,2}}{2\pi^{2}G_{2}^{3}} \Bigg) \log \left( \Lambda \right) ,
\end{multline}
\begin{multline}
    \delta\bar{\Gamma}_{3,2}=\Bigg(-\frac{G_{5,F}}{\pi^{2}G_{2}}+\frac{2\bar{G}_{3,3}\left(G_{4,1,F}+G_{4,1,N}\right)+\bar{G}_{3,1}\left(G_{4,1,F}+2G_{4,1,N}\right)+2\bar{G}_{3,2}N_{f}G_{4,1,N}}{2\pi^{2}G_{2}^{2}}\\
    +\frac{2\bar{G}_{3,3}^{2}\left(\bar{G}_{3,2}N_{f}+2\bar{G}_{3,1}\right)+2\bar{G}_{3,2}\bar{G}_{3,3}\left(2\bar{G}_{3,1}N_{f}+7\bar{G}_{3,2}\right)+\bar{G}_{3,2}^{2}N_{f}\left(\bar{G}_{3,1}N_{f}+4\bar{G}_{3,2}\right)+2\bar{G}_{3,3}^{3}}{2\pi^{2}G_{2}^{3}} \Bigg) \log \left( \Lambda \right),
\end{multline}
and
\begin{multline}
    \delta\bar{\Gamma}_{3,3}=\Bigg(-\frac{3N_{f}G_{5,N}}{2\pi^{2}G_{2}}+\frac{\bar{G}_{3,3}\left(\bar{G}_{3,3}^{2}N_{f}+3\bar{G}_{3,2}^{2}N_{f}+12\bar{G}_{3,2}\bar{G}_{3,3}\right)}{2\pi^{2}G_{2}^{3}}\\
    +\frac{3\left(\bar{G}_{3,2}\left(G_{4,1,F}+2G_{4,1,N}\right)+\bar{G}_{3,3}N_{f}G_{4,1,N}\right)}{2\pi^{2}G_{2}^{2}} \Bigg) \log \left( \Lambda \right) .
\end{multline}

We can now use the above results to write the regulated 1-loop correction to the inverse of the $\zeta_i^j$ propagator and to the amputated 3-point function:
\begin{equation}
   \left(\Gamma_{2}\right)_{i_{1},i_{2}}^{j_{1},j_{2}}\left(q\right) \equiv \left|q\right|\left(G_{2}\right)_{i_{1},i_{2}}^{j_{1},j_{2}}+\left|q\right|\left(\delta G_{2}\right)_{i_{1},i_{2}}^{j_{1},j_{2}}\log\left(\frac{q}{\Lambda}\right)+\left(\delta\Gamma_{2}\right)_{i_{1},i_{2}}^{j_{1},j_{2}}\left(q\right),
\end{equation}
\begin{equation}
    	\left\langle \zeta_{j_{1}}^{i_{1}}\left(q_1\right)\zeta_{j_{2}}^{i_{2}}\left(q_2\right)\zeta_{j_{3}}^{i_{3}}\left(q_3\right)\right\rangle_{amputated}=\left(G_{3}\right)_{i_{1},i_{2},i_{3}}^{j_{1},j_{2},j_{3}}+\left(\delta G_{3}\right)_{i_{1},i_{2},i_{3}}^{j_{1},j_{2},j_{3}}\log\left(\Lambda\right)+\left(\delta\Gamma_{3}\right)_{i_{1},i_{2},i_{3}}^{j_{1},j_{2},j_{3}},
\end{equation}
where we emphasize that $\delta \Gamma_2$ and $\delta \Gamma_3$ have a factor of $\log (\Lambda)$ inside them.

To proceed with the beta functions calculation, one has to renormalize the $\zeta$ fields. To preserve the $SU\left( N_f \right)$ flavor symmetry of the theory, the renormalization of all the fields in the adjoint $\zeta_A$ structure must be the same, and thus there are two renormalization conditions -- one for $\zeta_S$ and one for all the fields in $\zeta_A$. 

Thus, we need to project the equations above into the representation basis. For the two-point function this is done using \eqref{eq: projectio to SA to fields}, so together with \eqref{eq: G2 general}, \eqref{eq: deltaG2 general} and \eqref{eq: general 2point 1 loop},  we get that
\begin{equation}
    \left(\Gamma_{2}\right)_{i_{1},i_{2}}^{j_{1},j_{2}}\left(q\right)\zeta_{j_{1}}^{i_{1}}\left(q\right)\zeta_{j_{2}}^{i_{2}}\left(-q\right)
    =G_2 \left(\frac{q}{\Lambda}\right)^{2\gamma_{A}'}\left|q\right|\left(\zeta_{A}\right)_{j}^{i}\left(q\right)\left(\zeta_{A}\right)_{i}^{j}\left(-q\right)+\frac{G_{2}}{N_{f}}\left|q\right|\left(\frac{q}{\Lambda}\right)^{2\gamma_{S}'}\zeta_{S}\left(q\right)\zeta_{S}\left(-q\right)
\end{equation}
with the one loop corrected anomalous dimensions
\begin{equation}    \label{eq: general case anomalus A and S}
\gamma_{A}'=\gamma_{A}+\frac{N_{f}\left(3G_{4,2,N}+G_{4,3,N}\right)}{6\pi^{2}G_{2}^{2}}, \quad \gamma_{S}'=\gamma_{S}+\frac{N_{f}\left(3G_{4,2,F}+G_{4,3,F}+6G_{4,2,N}+2G_{4,3,N}\right)}{12\pi^{2}G_{2}^{2}} .
\end{equation}

Proceeding in the same manner as in section \ref{sec: Nf=1} and \cite{Aharony:2018pjn}, and using \eqref{eq: transform to adjoint0} to relate the 3-point structures in the index basis to the representation basis, we find that the beta functions of the $G_{SSS},\ G_{SAA},\ G_{AAA}$ couplings are

\begin{equation} \label{eq: beta multyflavour}
    \begin{split}
        \beta_{G_{SSS}}&=\frac{1}{\pi^{2}G_{2}}\frac{3\left(G_{5,F}+G_{5,N}\right)}{N_{f}}-\delta G_{SSS}
        %\\&
        -G_{SSS}\left(3\gamma_{S}^{\prime}+\frac{3\left(G_{4,1,F}+2G_{4,1,N}\right)}{2\pi^{2}G_{2}^{2}N_{f}}\right)\\&-G_{SAA}\left(\frac{3\left(N_{f}^{2}-1\right)\left(G_{4,1,F}+2G_{4,1,N}\right)}{2\pi^{2}G_{2}^{2}N_{f}^{2}}\right)
        %\\&
        -\frac{1}{2\pi^{2}G_{2}^{3}}\left(N_{f}^{3}G_{SSS}^{3}+\left(N_{f}^{2}-1\right)G_{SAA}^{3}\right) ,\\
        \beta_{G_{SAA}}&=\frac{1}{\pi^{2}G_{2}}\left(G_{5,F}+3G_{5,N}\right)-\delta G_{SAA}
        %\\&
        -G_{SSS}\left(\frac{G_{4,1,F}+2G_{4,1,N}}{2\pi^{2}G_{2}^{2}}\right)\\&-G_{SAA}\left(\gamma_{S}^{\prime}+2\gamma_{A}^{\prime}+\frac{3G_{4,1,F}+2\left(N_{f}^{2}+3\right)G_{4,1,N}}{2\pi^{2}G_{2}^{2}N_{f}}\right)\\&-G_{AAA}\left(\frac{\left(N_{f}^{2}-4\right)\left(G_{4,1,F}+2G_{4,1,N}\right)}{\pi^{2}G_{2}^{2}N_{f}^{2}}\right)\\&-\frac{1}{2\pi^{2}G_{2}^{3}}\left(N_{f}^{2}G_{SSS}G_{SAA}^{2}+N_{f}G_{SAA}^{3}+2\frac{\left(N_{f}^{2}-4\right)}{N_{f}}G_{SAA}G_{AAA}^{2}\right) ,\\
        \beta_{G_{AAA}}&=\frac{3}{2\pi^{2}G_{2}}N_{f}G_{5,N}-\delta G_{AAA}
        %\\&
        -G_{SAA}\left(\frac{3\left(G_{4,1,F}+2G_{4,1,N}\right)}{2\pi^{2}G_{2}^{2}}\right)\\&-G_{AAA}\left(3\gamma_{A}^{\prime}+\frac{3\left(N_{f}^{2}-4\right)G_{4,1,N}-6G_{4,1,F}}{2\pi^{2}G_{2}^{2}N_{f}}\right)\\&-\frac{1}{2\pi^{2}G_{2}^{3}}\left(3N_{f}G_{SAA}^{2}G_{AAA}+\frac{\left(N_{f}^{2}-12\right)}{N_{f}}G_{AAA}^{3}\right) .
    \end{split}
\end{equation}
Those serve as the analog of \eqref{eq: beta Nf1} in the single flavor case. 

As a check, we note that this computation reproduces the single flavor result presented in section \ref{sec: Nf=1}. For $N_f=1$, there is only the $SSS$ structure, and indeed $\beta_{G_{SSS}}$  depends on $G_{SSS}$ only. Comparing the parametrization of $\tilde{G}_4$ and $\tilde{G}_5$ \eqref{eq: momentum regime} in the single flavor case and the parametrization of \eqref{eq: g4 general case} and \eqref{eq: genral g5} we find
\begin{equation} \label{eq: g4g5 parametrization multiple to single}
    \text{for }N_{f}=1:\begin{array}{c}
\tilde{G}_{5}=12G_{5,N}+12G_{5,F},\\
-2\tilde{G}_{4}=4G_{4,1,N}+2G_{4,1,F},\\
\tilde{G}_{4}=4G_{4,2,N}+2G_{4,2,F},\\
\tilde{G}_{4}=4G_{4,3,N}+2G_{4,3,F} .
\end{array}
\end{equation}
Substituting \eqref{eq: g4g5 parametrization multiple to single} into equations \eqref{eq: general case anomalus A and S} and \eqref{eq: beta multyflavour} reproduces the $N_f=1$ results in \eqref{eq: anomalus dimension single flavor} and \eqref{eq: beta Nf1}.

We can find the beta functions  \eqref{eq: beta multyflavour} exactly (at leading order in $\frac{1}{N_c}$) for the RB theory with $\lambda_B=0$ and the CF theory with $\lambda_F=0$. The correlation functions in those cases are calculated analytically in Appendices \ref{sec: appendix CB correlations} and \ref{sec: CF appendix}. The results are summarized in table \ref{tab: beta coefficients}. We study those cases in detail in sections \ref{sec: RB theory} and \ref{sec: CF theory}.

\begin{table}[t]
    \centering
\begin{tabular}{|c|c||c|c|}
\hline 
\textbf{Element} & \textbf{coupling} & \textbf{Bosons} & \textbf{Fermions}\tabularnewline
\hline 
\hline 
\multicolumn{2}{|c||}{$G_{2}$} & $\frac{8}{\lambda_{B}\kappa_{B}}$ & $2\pi^{2}\frac{\lambda_{F}}{\kappa_{F}}$\tabularnewline
\hline 
 & $G_{SSS}$ & $g_{1}^{B}-\frac{128}{N_{f}^{2}\left(N_{c}^{B}\right)^{2}}$ & $g_{1}^{F}$\tabularnewline
\cline{2-4}
$G_{3}$ & $G_{SAA}$ & $g_{2}^{B}-\frac{128}{N_{f}\left(N_{c}^{B}\right)^{2}}$ & $g_{2}^{F}$\tabularnewline
\cline{2-4}
 & $G_{AAA}$ & $g_{3}^{B}-\frac{64}{\left(N_{c}^{B}\right)^{2}}$ & $g_{3}^{F}$\tabularnewline
\hline 
 & $\delta G_{SSS}$ & $\frac{4096\left(3s_{5,F}+3s_{5,N}-1\right)}{\left(N_{c}^{B}\right)^{3}N_{f}\pi^{2}}$ & $0$\tabularnewline
\cline{2-4}
$\delta G_{3}$ & $\delta G_{SAA}$ & $\frac{2048\left(2s_{5,F}+6s_{5,N}-1\right)}{\left(N_{c}^{B}\right)^{3}\pi^{2}}$ & $0$\tabularnewline
\cline{2-4}
 & $\delta G_{AAA}$ & $\frac{512N_f\left(12s_{5,N}-1\right)}{\left(N_{c}^{B}\right)^{3}\pi^{2}}$ & $0$\tabularnewline
\hline 
 & $G_{4,1,N}$ & $-\frac{1}{8}\frac{8^{4}}{\lambda_{B}^{3}\kappa_{B}^{3}}$ & $-\frac{1}{8}\left(4\pi\right)^{4}\frac{\lambda_{F}}{\kappa_{F}^{3}}$\tabularnewline
\cline{2-4}
 & $G_{4,2,N}$ & 0 & 0\tabularnewline
\cline{2-4}
$G_{4}$ & $G_{4,3,N}$ & $\frac{1}{8}\frac{8^{4}}{\lambda_{B}^{3}\kappa_{B}^{3}}$ & $\frac{1}{8}\left(4\pi\right)^{4}\frac{\lambda_{F}}{\kappa_{F}^{3}}$\tabularnewline
\cline{2-4}
 & $G_{4,1,F}$ & $-\frac{1}{4}\frac{8^{4}}{\lambda_{B}^{3}\kappa_{B}^{3}}$ & $-\frac{1}{4}\left(4\pi\right)^{4}\frac{\lambda_{F}}{\kappa_{F}^{3}}$\tabularnewline
\cline{2-4}
 & $G_{4,2,F}$ & $\frac{1}{4}\frac{8^{4}}{\lambda_{B}^{3}\kappa_{B}^{3}}$ & $\frac{1}{4}\left(4\pi\right)^{4}\frac{\lambda_{F}}{\kappa_{F}^{3}}$\tabularnewline
\cline{2-4}
 & $G_{4,3,F}$ & 0 & 0\tabularnewline
\hline 
$G_{5}$ & $G_{5,N/F}$ & $\frac{8^{5}}{\left(N_{c}^{B}\right)^{4}}s_{5,N/F}$ & $0$\tabularnewline
\hline 
\multicolumn{2}{|c||}{$\gamma_{S}$} & $-\frac{1}{N_{c}^{B}}\frac{16N_{f}}{3\pi^{2}}$ & 0\tabularnewline
\hline 
\multicolumn{2}{|c||}{$\gamma_{S}^{\prime}$} & $0$ & $\frac{1}{N_{c}^{F}}\frac{16N_{f}}{3\pi^{2}}$\tabularnewline
\hline 
\multicolumn{2}{|c||}{$\gamma_{A}$} & $-\frac{1}{N_{c}^{B}}\frac{4N_{f}}{3\pi^{2}}$ & 0\tabularnewline
\hline 
\multicolumn{2}{|c||}{$\gamma_{A}^{\prime}$} & $0$ & $\frac{1}{N_{c}^{F}}\frac{4N_{f}}{3\pi^{2}}$\tabularnewline
\hline 
\end{tabular}
    \caption{Summary of the results of Appendices \ref{sec: appendix CB correlations} (bosons) and \ref{sec: CF appendix} (fermions). These are the coefficients that appear in \eqref{eq: beta multyflavour} when $\lambda_{B/F}\rightarrow0$ and they are used to derive \eqref{eq: beta function fermions}, and as an alternative method to derive \eqref{eq: beta function free boson}. \eqref{eq: transform to adjoint0} is used to transform from the index basis in the text to the representation basis in this table. $s_{5,N/F}$ are defined in \eqref{eq: s5 effective RB lambda=0} (see also \eqref{eq: s5Nf}) and their values are unknown. 
    }
    \label{tab: beta coefficients}
\end{table}

\subsection{Analyzing the beta function structure} \label{sec: multiple falvor beta function comments}
\begin{figure}[t]
\begin{adjustbox}{center}
$
    \vcenter{\hbox{\begin{tikzpicture}[baseline=(c.base)]
  \draw (0,0) circle (1.5);
  
  \begin{feynman}
    
    \vertex (i1) at ({-2.5},{0}) {$S$};
    \vertex (i2) at ({2.5*cos(60)},{2.5*sin(60)}) {$S$};
    \vertex (i3) at ({2.5*cos(-60)},{2.5*sin(-60)}) {$S$};
        
    \node (d1) at ({-1.5},{0})  [dot];
    \node (d2) at ({1.5*cos(60)},{1.5*sin(60)})  [dot];
    \node (d3) at ({1.5*cos(-60)},{1.5*sin(-60)})  [dot];

    \vertex  (dn1) at ([shift={(0.5,0.0)}]d1) {$G_{SSS}$};
    \vertex  (dn2) at ([shift={(-0.2,-0.3)}]d2) {$G_{SSS}$};
    \vertex  (dn3) at ([shift={(-0.2,0.3)}]d3) {$G_{SSS}$};

    \vertex  (d1r) at ({1.7},{0}) {$S$};
    \vertex  (d2r) at ({-1.7*cos(60)},{-1.7*sin(60)}) {$S$};
    \vertex  (d3r) at ({-1.7*cos(-60)},{-1.7*sin(-60)}) {$S$};
    \diagram*{
            (i1) --  (d1),
            (i2) --  (d2),
            (i3) --  (d3),
        };
  \end{feynman}
\end{tikzpicture}
}}\quad \quad \quad \quad \quad \vcenter{\hbox{\begin{tikzpicture}[baseline=(c.base)]
  \draw (0,0) circle (1.5);
  \draw[blue] (0,0) circle (1.5);
  \begin{feynman}
    
    \vertex (i1) at ({-2.5},{0}) {$S$};
    \vertex (i2) at ({2.5*cos(60)},{2.5*sin(60)}) {$S$};
    \vertex (i3) at ({2.5*cos(-60)},{2.5*sin(-60)}) {$S$};
        
    \node (d1) at ({-1.5},{0})  [dot];
    \node (d2) at ({1.5*cos(60)},{1.5*sin(60)})  [dot];
    \node (d3) at ({1.5*cos(-60)},{1.5*sin(-60)})  [dot];

    \vertex  (dn1) at ([shift={(0.5,0.0)}]d1) {$G_{S\color{blue}{AA}}$};
    \vertex  (dn2) at ([shift={(-0.2,-0.3)}]d2) {$G_{S\color{blue}{AA}}$};
    \vertex  (dn3) at ([shift={(-0.2,0.3)}]d3) {$G_{S\color{blue}{AA}}$};

    \vertex  (d1r) at ({1.7},{0}) {$\color{blue}{A}$};
    \vertex  (d2r) at ({-1.7*cos(60)},{-1.7*sin(60)}) {$\color{blue}{A}$};
    \vertex  (d3r) at ({-1.7*cos(-60)},{-1.7*sin(-60)}) {$\color{blue}{A}$};
    \diagram*{
            (i1) --  (d1),
            (i2) --  (d2),
            (i3) --  (d3),
        };
  \end{feynman}

\end{tikzpicture}
}}
$
\end{adjustbox}

    \caption{The structures contributing to the cubic terms in $\beta_{G_{SSS}}$: either all the interactions are $SSS$ vertices and the internal lines are $\zeta_S$, or they are all $SAA$ vertices and the internal lines are $\zeta_A$.}

    \label{fig: beta1 cubic term posabilities} 
\end{figure}
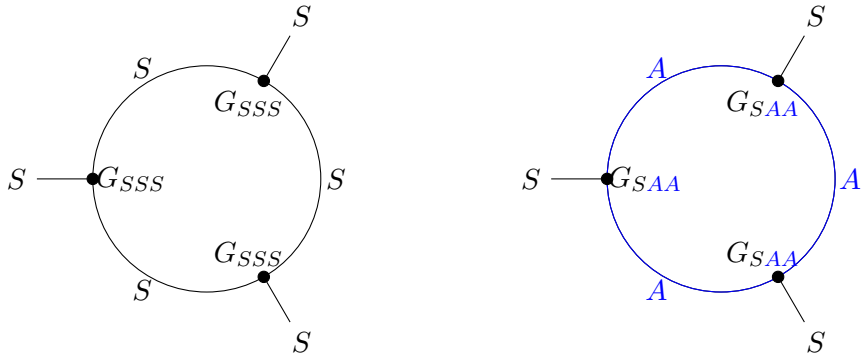

In the following sections \ref{sec: RB theory} and \ref{sec: CF theory} we compute the beta functions in two specific limits. 
Here, we wish to emphasize several general aspects of the beta functions presented in (\ref{eq: beta multyflavour}):
\begin{enumerate}
    \item We are interested in the beta functions in terms of the finite 't Hooft couplings $\lambda_{SSS},\lambda_{SAA},\lambda_{AAA}$ (see section \ref{sec: dualities}). The connection between them and $G_{SSS},G_{SAA},G_{AAA}$ is given by shifting by some constants (provided in the index basis in \eqref{eq: general G3 coeff}) and rescaling. Substituting the explicit dependence on $N_c$, one gets that the beta functions for the finite 't Hooft couplings scale as $\sim\frac{1}{N_c}$. Furthermore, as emphasized in the paragraph below equation \eqref{eq: beta Nf1}, at leading order in $\frac{1}{N_c}$, the beta functions found are \emph{exact} to all orders in $\lambda_{SSS},\lambda_{SAA},\lambda_{AAA}$ \cite{Aharony:2018pjn}.
    \item The beta functions for $\lambda_{SSS},\lambda_{SAA},\lambda_{AAA}$ are 3 polynomials of degree 3. To find the fixed points of this system, we set all the beta functions to zero. Such a system of real equations can have in general between 1 and 27 real solutions (see section \ref{sec: RB complex analysis} for more details), and so these are the minimal and maximal possible numbers of fixed points in the system (at large $N_c$).
    \item Note that the beta functions in \eqref{eq: beta multyflavour} have a special form: no $G_{AAA}$ dependence in $\beta_{G_{SSS}}$ and no $G_{SSS}$ dependence in $\beta_{G_{AAA}}$.
    \item One way to find the fixed points is to relate them to a single complicated equation in one variable. 
    The property mentioned in the previous item implies that one can solve third-order equations for $G_{SSS}$ and for $G_{AAA}$ in terms of $G_{SAA}$. One can then substitute each of the 9 possible options into $\beta_{G_{SAA}}$, and get an equation just for $G_{SAA}$. 
    \item The cubic terms in each of the beta functions (the last term in each equation) are the same for all our theories, up to an overall factor of $G_2^3$. These terms are not affected by shifting the $\lambda_n$ by constants, and $G_2$ is always positive by unitarity.
    Thus, the analysis of the flow for very large values of $\lambda_{SSS},\lambda_{SAA},\lambda_{AAA}$ (when the cubic part dominates the beta function) will be the same in all our theories and for any values of $\lambda_F$ or $\lambda_B$. We will analyze this flow in sections \ref{sec: RB nf2 large lambda i} and \ref{sec: RB nf general large lambda i}. \label{item4.3: cubic are universal}
    \item     We mentioned above the simplification of the beta functions for $N_f=1$, where only the $G_{SSS}$ coupling exists, and $G_{SAA}$ does not appear in its beta function because of factors of $(N_f^2-1)$. There is a similar simplification for $N_f=2$, where the structure associated with $G_{AAA}$ does not exist, and indeed it does not appear in the other two beta functions; in  $\beta_{G_{SAA}}$ this is because its coefficients are proportional to  $\left(N_f^2-4\right)$.
    
    \label{item4.3: nf-1 factor near lambda1}
    \item The general structure of the beta functions is similar to the $N_f=1$ case in \eqref{eq: beta Nf1}, with constant terms, linear terms, and cubic terms, but no quadratic term, because they arise from the same diagrams just with different flavor indices. 
    The $\delta G_3$ terms, as well as the $G_5$ Feynman diagram (the left diagram in figure \ref{fig: 3 point diagrams Nf1}), lead to the constant parts. The renormalization of the two-point functions (manifested by the anomalous dimensions) and the $G_4-G_3$ Feynman diagram (the middle diagram in figure \ref{fig: 3 point diagrams Nf1}) lead to the linear parts, and the $G_3^3$ Feynman diagram (the right diagram in figure \ref{fig: 3 point diagrams Nf1}) corresponds to the cubic part. It is easy to see from the diagrams why the specific linear and cubic terms that we have in \eqref{eq: beta multyflavour} arise, see for instance figure \ref{fig: beta1 cubic term posabilities}. The $\lambda_n$ couplings are shifted from $G_n$ by constants, so their beta functions will contain some quadratic terms, but otherwise they will have the same properties.
    \label{item4.3: structre of monimials for g3}
    \item For each $N_f\ge3$, there are 10 parameters that determine the beta functions and thus the fixed points and the flows between them\footnote{The prefactors of the cubic terms are equal in all of the beta functions, and so it can be set to 1 by rescaling.}: the 3 constant terms in each beta function and the 7 prefactors of the linear terms. 
    
\end{enumerate}

\subsection{The large \texorpdfstring{$N_f$}{Nf} limit} \label{sec: general case large nf limit}

In our analysis, we consider the limit $ N_c \to \infty $, retaining only the leading non-trivial terms of order $ \frac{1}{N_c} $ in the beta functions. While the previous analysis was conducted for a fixed number of flavors $ N_f $, we now extend it to the large $ N_f $ limit, assuming $ N_f \ll N_c $.

In the $N_f\rightarrow \infty$ limit we have (on top of the $N_c$-scalings discussed above)
\begin{equation} \label{eq: scaling of Non G parameters}
\begin{split}
    \gamma_{A}^{\prime},\gamma_{S}^{\prime},\delta G_{AAA}&\sim N_{f},\\
    \delta G_{SAA}&\sim1,\\\delta G_{SSS}&\sim\frac{1}{N_{f}} ,
\end{split}    
\end{equation}
with the rest of the parameters having no scaling under $N_f$. (See table \ref{tab: beta coefficients} for examples of these parameter scalings in free theories.)

It is interesting to ask what are the possible $N_f$ scalings of the fixed points associated with \eqref{eq: beta multyflavour}, in the limit $N_f\rightarrow\infty$ and for constant $\lambda$. In other words, denoting the couplings as
\begin{equation} \label{eq: large nf general scaling}
     G_{SSS}=\frac{Y_{SSS}}{N_{f}^{a}}+\mathcal{O}\left(\frac{1}{N_{f}^{a+\epsilon}}\right),\quad G_{SAA}=\frac{Y_{SAA}}{N_{f}^{b}}+\mathcal{O}\left(\frac{1}{N_{f}^{b+\epsilon}}\right),\quad G_{AAA}=\frac{Y_{AAA}}{N_{f}^{c}}+\mathcal{O}\left(\frac{1}{N_{f}^{c+\epsilon}}\right),
\end{equation}
what are the possible values of $\left(a,b,c\right)$ with $\epsilon>0$.

Plugging \eqref{eq: large nf general scaling} into \eqref{eq: beta multyflavour} and using the scalings \eqref{eq: scaling of Non G parameters}, we can write the beta functions as sums of terms with different powers of $N_f$ that depend on $a,\ b$ and $c$. Generically (i.e, if no special relation between the different components of \eqref{eq: beta multyflavour} holds) we expect to get a root of the beta functions (i.e, a fixed point) only when at least two terms in the sum that contribute to the leading $N_f$ power\footnote{Which terms are leading and which are subleading, as well as what is the leading power, depends on $a,\ b$ and $c$.} are the same. We find that this condition holds only for
\begin{equation} \label{eq: allowd scalings}
    \left(a,b,c\right):\ \ \left(\frac{2}{3},\frac{1}{3},0\right),\ \left(1,1,0\right),\ \left(2,1,0\right) ,
\end{equation}
and any other leading scaling is forbidden.

In sections \ref{sec: RB large nf} and \ref{sec: CF large nf}, we will see explicit examples of solutions of the types  $\left(\frac{2}{3},\frac{1}{3},0\right)$ and $\left(2,1,0\right)$\footnote{The type $\left(1,1,0\right)$ requires $\gamma_S^{\prime}<0$, which does not happen for the values of $\lambda_{B/F}$ considered in this paper.}.
The $\left(2,1,0\right)$ solution is particularly interesting, as we discuss in the following.

\subsubsection{The (2,1,0) scaling} \label{sec: general case large nf limit (2,1,0)}

Since our gauge-invariant operators are $N_f\times N_f$ matrices, this scaling is the natural one associated with
the standard large $N_f$ limit of matrix models, with single-$SU(N_f)$-trace terms dominating, and other terms suppressed by powers of $N_f$. The large $N_f$ limit is most natural to write in the index basis, in which it is natural to scale
\begin{equation} \label{eq: large nf coupling definition specific scaling}
    {\bar g_1}= \frac{\bar{Y}_1}{N_f^2}, \quad {\bar g}_2= \frac{\bar{Y}_2}{N_f}, \quad {\bar g}_3= \bar{Y}_3,
\end{equation}
with the $\bar{Y}_i$'s fixed in the large $N_f$ limit.
Using the relation \eqref{eq: transform to adjoint0} between the index and representation basis, this is equivalent to taking
\begin{equation} \label{eq: large nf coupling definition spesific scaling}
    G_{SSS}= \frac{Y_{SSS}}{N_f^2}, \quad G_{SAA}= \frac{Y_{SAA}}{N_f}, \quad G_{AAA}= Y_{AAA} ,
\end{equation}
which is seen to correspond to the third solution in \eqref{eq: allowd scalings}. In this limit the beta functions simplify, and we find as expected from general large-$N_f$ arguments that the flow of the single-trace coupling (which is the dominant contribution to $Y_{AAA}$) decouples from the flow of the other couplings, and the flow of the double-trace (the dominant contribution to $Y_{SAA}$) decouples from the triple-trace.

One can show that this is the case explicitly in the parametrization chosen using \eqref{eq: scaling of Non G parameters}. The resulting beta functions, at leading order in $1/N_f$, are
\begin{equation} \label{eq: general case beta nf to infinity}
\begin{split}\frac{\beta_{Y_{SSS}}}{N_{f}}= & \left(\frac{3\left(G_{5,F}+G_{5,N}\right)}{\pi^{2}G_{2}}-N_{f}\delta G_{SSS}\right)-Y_{SSS}\left(3\frac{\gamma_{S}^{\prime}}{N_{f}}\right)\\
 & -Y_{SAA}\left(\frac{3\left(G_{4,1,F}+2G_{4,1,N}\right)}{2\pi^{2}G_{2}^{2}}\right)-\frac{1}{2\pi^{2}G_{2}^{3}}Y_{SAA}^{3}+\mathcal{O}\left(\frac{1}{N_{f}}\right),\\
\frac{\beta_{Y_{SAA}}}{N_{f}}= & \left(\frac{G_{5,F}+3G_{5,N}}{\pi^{2}G_{2}}-\delta G_{SAA}\right)-Y_{SAA}\left(\frac{\gamma_{S}^{\prime}}{N_{f}}+2\frac{\gamma_{A}^{\prime}}{N_{f}}+\frac{G_{4,1,N}}{\pi^{2}G_{2}^{2}}\right)\\
 & -Y_{AAA}\left(\frac{G_{4,1,F}+2G_{4,1,N}}{\pi^{2}G_{2}^{2}}\right)-\frac{1}{2\pi^{2}G_{2}^{3}}2Y_{SAA}Y_{AAA}^{2}+\mathcal{O}\left(\frac{1}{N_{f}}\right),\\
\frac{\beta_{Y_{AAA}}}{N_{f}}= & \left(\frac{3G_{5,N}}{2\pi^{2}G_{2}}-\frac{\delta G_{AAA}}{N_{f}}\right)-Y_{AAA}\left(3\frac{\gamma_{A}^{\prime}}{N_{f}}+\frac{3 G_{4,1,N}}{2\pi^{2}G_{2}^{2}}\right)-\frac{1}{2\pi^{2}G_{2}^{3}}Y_{AAA}^{3}+\mathcal{O}\left(\frac{1}{N_{f}}\right).
\end{split}
\end{equation}
Note that all the terms on the right-hand side are $O(1)$ in the large $N_f$ limit. Upon rewriting \eqref{eq: general case beta nf to infinity} in terms of the finite 't Hooft couplings $ y_n = N_c^2 Y_n $, the beta functions for $ y_n $ become proportional to $ \frac{N_f}{N_c} \ll 1 $. We see that
indeed, the flow of $Y_{AAA}$ does not depend on the other couplings, and the flow of $Y_{SAA}$ does not depend on $Y_{SSS}$. 

In searching for fixed points, it is possible in this limit to solve the equations iteratively: we first solve $ \beta_{Y_{AAA}}=0$, which is a cubic polynomial in $Y_{AAA}$, that generically has $1$ or $3$ zeros. Next, we can substitute this solution to  $\beta_{Y_{SAA}}=0$ to get $Y_{SAA}$, and repeat the process to find $Y_{SSS}$. Since $\beta_{Y_{SAA}}$ and $\beta_{Y_{SSS}}$ are linear in $Y_{SAA}$ and $Y_{SSS}$ respectively, we find that there are always between 1 and 3 solutions with the scaling \eqref{eq: large nf coupling definition spesific scaling}.

The argument above holds unless the linear terms in the leading order equation \eqref{eq: general case beta nf to infinity} happen to vanish. If for one of the solutions of $Y_{AAA}$, the prefactor of $Y_{SAA}$ in its beta function vanishes, which happens when
\begin{equation} \label{eq: condition for degeneracy for large nf}
    \left(\frac{3G_{5,N}}{2\pi^{2}G_{2}}-\frac{\delta G_{AAA}}{N_{f}}\right)^{2}=-\pi^{2}G_{2}^{3}\left(2\frac{\gamma_{A}^{\prime}}{N_{f}}+\frac{G_{4,1,N}}{\pi^{2}G_{2}^{2}}+\frac{\gamma_{S}^{\prime}}{N_{f}}\right)\left(2\frac{\gamma_{A}^{\prime}}{N_{f}}+\frac{G_{4,1,N}}{\pi^{2}G_{2}^{2}}-\frac{1}{2}\frac{\gamma_{S}^{\prime}}{N_{f}}\right)^{2} ,
\end{equation}
or the prefactor of $Y_{SSS}$ in its beta function vanishes, which happens when $\gamma_{S}^{\prime}=0$, then it is not enough to look at the leading order terms \eqref{eq: general case beta nf to infinity} to find all possible solutions with the leading scaling of  \eqref{eq: large nf coupling definition spesific scaling}. We will see examples of these degeneracies in sections \ref{sec: RB large nf} and \ref{sec: CF large nf}. However, for generic values of the 't Hooft coupling $\lambda$ we do not expect such degeneracies to hold.

\section{Regular boson theories with vanishing and small \texorpdfstring{$\lambda_B$}{lambda}} \label{sec: RB theory}

In this section, we analyze in detail the number of fixed points and the flows coming from the beta functions of the marginal couplings $\lambda_{SSS}^{B},\lambda_{SAA}^{B},\lambda_{AAA}^{B}$, in the weakly coupled limit of the RB theories defined in section \ref{sec: RB}. We will ignore the subscript and superscript $B$, as we concentrate only on scalar theories in this section.

We first calculate the exact beta functions at leading order in $\frac{1}{N_c}$ for the `gluon free' theory with $\lambda=0$. As for the single flavor case (see \eqref{eq: nf1 beta RB lambda 0} and figure \ref{fig: beta functino nf=1 possabilities}), the free point $\lambda_{SSS}=\lambda_{SAA}=\lambda_{AAA}=0$ is a degenerate fixed point in this case. To observe its splitting to several fixed points at finite $\lambda$, we calculate the beta function for $\lambda\ll1$\footnote{A subtlety in this definition arises for large $N_f$, which we discuss in section \ref{sec: small lambda calculation}.} in the vicinity of this point, and show that an IR-stable fixed point emerges.

The structure of this section is as follow: in section \ref{sec: RB clacualte beta function summary} we compute the beta function in two regimes of interest, when $\lambda=0$ and when $\lambda^2\sim\lambda_{n}\ll1$. We then analyze the fixed points structure and the flows for $N_f=2$ and for $N_f\ge3$, in sections \ref{sec: RB nf=2 analysis} and \ref{sec: RB nf>=3 analysis}, respectively. In section \ref{sec: RB large nf}, we comment on the solutions for fixed points at large values of $N_f$. We conclude in section  \ref{sec: RB general analysis complex topology} with a general discussion of the possible behavior of the fixed points as we change $\lambda$.

\subsection{Calculating the beta functions} \label{sec: RB clacualte beta function summary}

In this section we outline the calculations of the beta functions for the $\lambda=0$ case and for the perturbative $\lambda\ll1$ case. The full details of the calculations, including conventions for the Lie algebra generators, renormalization conditions, and the contribution of each Feynman diagram, are presented in Appendix \ref{sec: app finite lambda B}.  
\subsubsection{The \texorpdfstring{$\lambda=0$}{lambda=0} case}
\begin{figure}[t]
\begin{adjustbox}{center}
$
    \vcenter{\hbox{
    \begin{tikzpicture}[baseline=(c.base)]

  \begin{feynman}
    
    \node (i1) at ({0},{1.5}) [dot];
    \node (i2) at ({1.5*cos(-30)},{1.5*sin(-30)}) [dot];
    \node (i3) at ({1.5*cos(210)},{1.5*sin(210)}) [dot];

    \vertex (c) at (0,0);

    \vertex (e1) at ({2.7*cos(90+15)},{2.7*sin(90+15)});
    \vertex (e2) at ({2.7*cos(90-15)},{2.7*sin(90-15)});

    \vertex (e3) at ({2.7*cos(-30+15)},{2.7*sin(-30+15)});
    \vertex (e4) at ({2.7*cos(-30-15)},{2.7*sin(-30-15)});

    \vertex (e5) at ({2.7*cos(210+15)},{2.7*sin(210+15)});
    \vertex (e6) at ({2.7*cos(210-15)},{2.7*sin(210-15)});

    \diagram*{
            (e1) -- [fermion] (i1)  -- [fermion] (e2),
            (e3) -- [fermion] (i2)  -- [fermion] (e4),
            (e5) -- [fermion] (i3)  -- [fermion] (e6),

            (i1)-- [fermion, quarter left,reversed momentum = $k-q_3$] (i2),
            (i1)-- [anti fermion,reversed momentum'={[arrow shorten=0.25,arrow distance=0.15,label distance=-2] $ q_3 $ }] (i2),

            (i2)-- [fermion, quarter left,reversed momentum = $k-q_2$] (i3),
            (i2)-- [anti fermion,reversed momentum'={[arrow shorten=0.25,arrow distance=0.15,label distance=-2] $ q_2 $ }] (i3),

            (i3)-- [fermion, quarter left,reversed momentum = $k-q_1$] (i1),
            (i3)-- [anti fermion,reversed momentum'={[arrow shorten=0.25,arrow distance=0.15,label distance=-2] $ q_1 $ } ] (i1),

        };

    \vertex (n) at (0,-2.4) {\large $\left( A0 \right)$};
  \end{feynman}
  
\end{tikzpicture}
}}\quad \vcenter{\hbox{
\begin{tikzpicture}[baseline=(c.base)]
    \begin{feynman}
        \vertex (c);

        \vertex (e1) at (-2.5,2) ;
        \vertex (e2) at (-2.5,0) ;
        \vertex (e3) at (-2.5,-2) ;
        \vertex (e4) at (2.5,2) ;
        \vertex (e5) at (2.5,0) ;
        \vertex (e6) at (2.5,-2) ;

        \node (il) at (-1.5,0) [dot];
        \node (ir) at (1.5,0) [dot];

        \diagram*{
            (e1) -- [fermion] (il),
            (e2) -- [fermion] (il),
            (e3) -- [anti fermion] (il),

            (e4) -- [anti fermion] (ir),
            (e5) -- [anti fermion] (ir),
            (e6) -- [fermion] (ir),

            (il) -- [fermion, half left,momentum = {$ q $}] (ir),
            (il) -- [fermion,,momentum = {$ k $}] (ir),
            (il) -- [anti fermion, half right,reversed momentum = {$ q+k $}] (ir),

        };

        \vertex (n) at (0,-2.4) {\large $\left( A8a \right)$};
    \end{feynman}
\end{tikzpicture}
}}
$
\end{adjustbox}

\caption{The only two diagrams contributing to the beta function of RB theories when $\lambda=0$, at leading order in the 't Hooft $\frac{1}{N_c}$ expansion, when the external legs have zero momentum.}
\label{fig: two feyman diagrams for nf=1 lambda=0}
\end{figure}
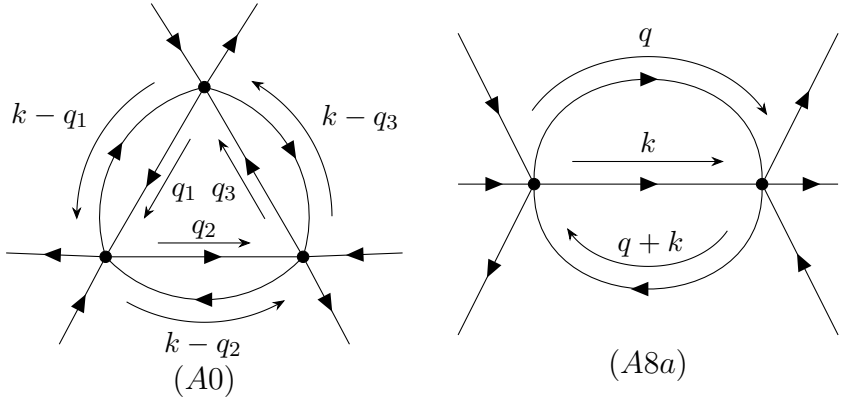

The RB theory \eqref{eq: regular bosons} with $\lambda=0$ has the simple Lagrangian
\begin{equation}
    \mathcal{L}=\partial_{\mu}\bar{\phi}{}_{i,a}\partial_{\mu}\phi^{i,a}+\frac{\bar{g}_{1}^{B}}{3!}\left(\bar{\phi}_{c,i}\phi^{c,i}\right)^{3}+\frac{\bar{g}_{2}^{B}}{2}\bar{\phi}_{a,i}\phi^{a,i}\bar{\phi}_{b,j}\phi^{b,k}\bar{\phi}_{c,k}\phi^{c,j}+\frac{\bar{g}_{3}^{B}}{3}\bar{\phi}_{a,i}\phi^{a,j}\bar{\phi}_{b,j}\phi^{b,k}\bar{\phi}_{c,k}\phi^{c,i} ,
\end{equation}
which contains only six-scalar vertices. In the 't Hooft limit we are working in, there are only two diagrams which contribute to the 6-point correlation function, corresponding to two and three interaction vertices. The diagrams are presented in figure \ref{fig: two feyman diagrams for nf=1 lambda=0}. We leave the proof that in this formalism, those are the only two diagrams that contribute in the leading 't Hooft limit, to Appendix \ref{sec: app lambdab=0}.

Evaluating the diagrams, the beta functions are found to be\footnote{The result for the $SO(N_c)$ case, which includes an additional factor of two compared to \eqref{eq: beta function free boson}, was also computed in \cite{Kapoor:2021lrr}, though with a different normalization of the couplings.} 
\begin{equation} \label{eq: beta function free boson}
    \begin{split}    2^{10}\pi^{2}N_{c}\beta_{SSS}=&2^{6}\left(6N_{f}\lambda_{SSS}^{2}+6\frac{\left(N_{f}^{2}-1\right)}{N_{f}}\lambda_{SAA}^{2}\right)-\left(N_{f}^{3}\lambda_{SSS}^{3}+\left(N_{f}^{2}-1\right)\lambda_{SAA}^{3}\right), \\
    2^{10}\pi^{2}N_{c}\beta_{SAA}=&2^{6}\left(4N_{f}\lambda_{SSS}\lambda_{SAA}+8\lambda_{SAA}^{2}+4\frac{\left(N_{f}^{2}-4\right)}{N_{f}}\lambda_{SAA}\lambda_{AAA}+4\frac{\left(N_{f}^{2}-4\right)}{N_{f}^{2}}\lambda_{AAA}^{2}\right)
    \\-&\left(N_{f}^{2}\lambda_{SSS}\lambda_{SAA}^{2}+N_{f}^{1}\lambda_{SAA}^{3}+2\frac{\left(N_{f}^{2}-4\right)}{N_{f}}\lambda_{SAA}\lambda_{AAA}^{2}\right), \\
    2^{10}\pi^{2}N_{c}\beta_{AAA}=&2^{6}\left(3N_{f}\lambda_{SAA}^{2}+\frac{3\left(N_{f}^{2}-12\right)}{N_{f}}\lambda_{AAA}^{2}+12\lambda_{SAA}\lambda_{AAA}\right)\\-&\left(3N_{f}\lambda_{SAA}^{2}\lambda_{AAA}+\frac{\left(N_{f}^{2}-12\right)}{N_{f}}\lambda_{AAA}^{3}\right) \, .
    \end{split}
\end{equation}

It is also possible to evaluate the beta function in this case directly using the formalism presented in section \ref{sec: general case}. In Appendix \ref{sec: appendix CB correlations} we calculate the correlation functions for RB theories with $\lambda=0$. Substituting the results summarized in table \ref{tab: beta coefficients} in \eqref{eq: beta multyflavour} reproduces \eqref{eq: beta function free boson}.

The beta functions \eqref{eq: beta function free boson} are a simple special case of the general possible form of the beta functions for RB theories, which is presented in \eqref{eq: beta multyflavour}. The reduction to the $N_f=1$ and $N_f=2$ cases is straightforward by ignoring $\beta_{SAA}$ and/or $\beta_{AAA}$ (as manifested by the $N_f^2-1$ and $N_f^2-4$ factors, see item \ref{item4.3: nf-1 factor near lambda1} in section \ref{sec: multiple falvor beta function comments}). Also, one can explicitly check that redefining the couplings by $\lambda_{SSS}\rightarrow\lambda_{SSS}+\frac{128}{N_{f}^{2}},\lambda_{SAA}\rightarrow\lambda_{SAA}+\frac{128}{N_{f}},\lambda_{AAA}\rightarrow\lambda_{AAA}+64$ removes the quadratic terms in all of the equations and brings them to the form of section \ref{sec: general case}\footnote{Note that without the general formalism presented in section \ref{sec: general case}, this property is rather surprising, since unlike the one coupling case in which it is always possible to redefine the variable in a cubic polynomial to remove the quadratic part, this is not true for a generic system of three cubic polynomials with three variables.}.

In sections \ref{sec: RB nf=2 analysis} and \ref{sec: RB nf>=3 analysis} we present a full analysis of the fixed point structure and the flow which results from the beta functions \eqref{eq: beta function free boson}. Note that in the $\lambda=0$ theory 
the beta functions do not contain any constants or linear terms. An immediate consequence of this is that there is a degenerate fixed point at $\left(\lambda_{SSS},\lambda_{SAA},\lambda_{AAA} \right) = \left(0,0,0 \right)$, which is simply the free massless scalar theory, and that will either split or vanish when turning on a finite $\lambda$ (the three dimensional analog of figure \ref{fig: beta functino nf=1 possabilities}). To determine what happens to this point, we'll turn on a small but finite $\lambda$, and use perturbation theory to compute the beta functions near this free point $\lambda_n=0$.

\subsubsection{Finite \texorpdfstring{$\lambda 
\ll1$}{lambda<<1} in the vicinity of the free theory \label{sec: small lambda calculation}}

In order to calculate the beta functions at finite $\lambda \ll 1$ near $\lambda_n=0$, one has to take into account the full action of the RB theory \eqref{eq: regular bosons}.

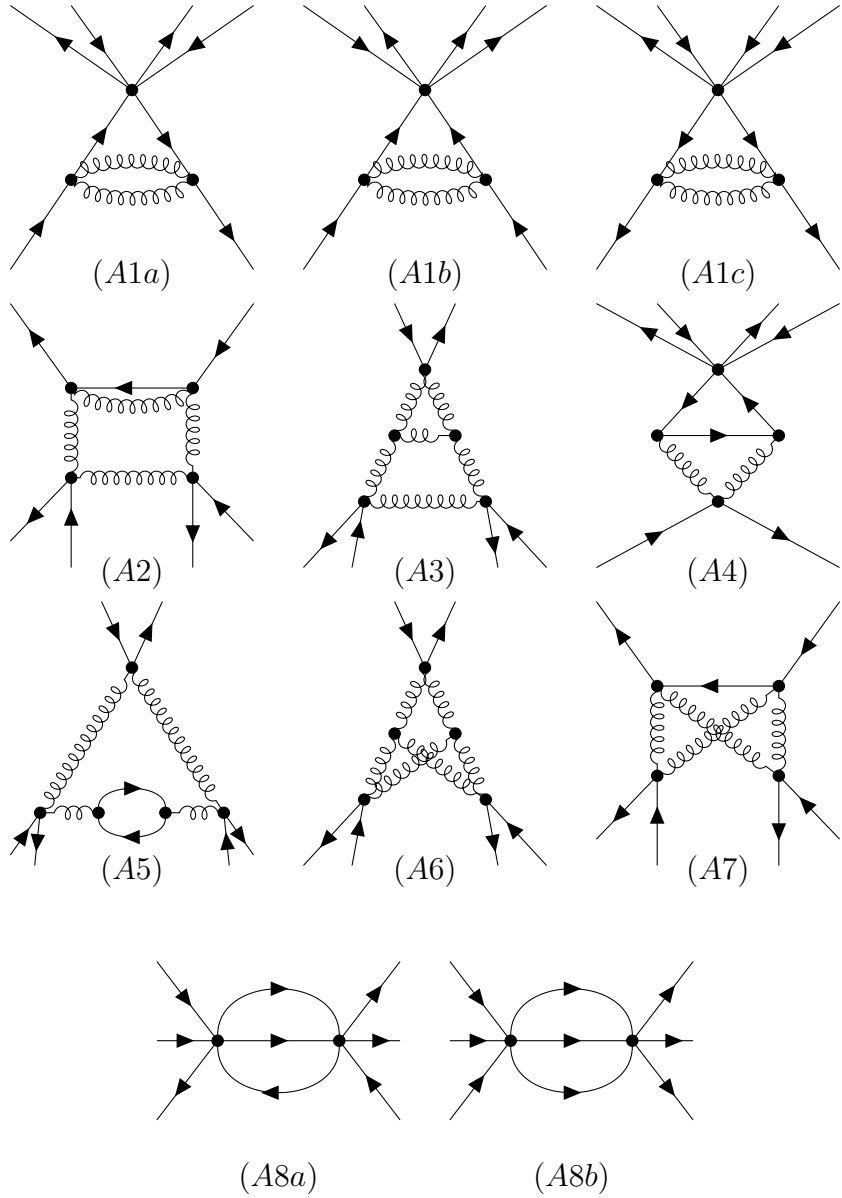
\begin{figure}[t]
\begin{adjustbox}{center}
$
    \vcenter{\hbox{
\begin{tikzpicture}[baseline=(c.base),xscale=0.8, yscale=0.7]
    \begin{feynman}
        \vertex (c);
        \vertex (e1) at (-2,0) ;
        \vertex (e2) at (2,0) ;

        \vertex (e3) at (-2,5) ;
        \vertex (e4) at (-1,5) ;
        \vertex (e5) at (1,5) ;
        \vertex (e6) at (2,5) ;
        
        \node (iu) at (0,3.4) [dot];
        \node (idl) at (-1,1.7) [dot];
        \node (idr) at (1,1.7) [dot];

        \diagram*{
            (e3) -- [anti fermion] (iu),
            (e4) -- [fermion] (iu),
            (e5) -- [anti fermion] (iu),
            (e6) -- [fermion] (iu),
            (e1) -- [fermion] (idl) -- [fermion] (iu),
            (e2) -- [anti fermion] (idr) -- [anti fermion] (iu),
            (idl) -- [gluon, out=45,in=135,looseness=0.7] (idr),
            (idr) -- [gluon, out=225,in=315,looseness=0.7] (idl), 
        };

        \vertex (n) at (0,-0.1) {\large $\left( A1a \right)$};
    \end{feynman}
\end{tikzpicture}    
}}\quad \vcenter{\hbox{
\begin{tikzpicture}[baseline=(c.base),xscale=0.8, yscale=0.7]
    \begin{feynman}
        \vertex (c);
        \vertex (e1) at (-2,0) ;
        \vertex (e2) at (2,0) ;

        \vertex (e3) at (-2,5) ;
        \vertex (e4) at (-1,5) ;
        \vertex (e5) at (1,5) ;
        \vertex (e6) at (2,5) ;
        
        \node (iu) at (0,3.4) [dot];
        \node (idl) at (-1,1.7) [dot];
        \node (idr) at (1,1.7) [dot];

        \diagram*{
            (e3) -- [fermion] (iu),
            (e4) -- [anti fermion] (iu),
            (e5) -- [anti fermion] (iu),
            (e6) -- [anti fermion] (iu),
            (e1) -- [fermion] (idl) -- [fermion] (iu),
            (e2) -- [fermion] (idr) -- [fermion] (iu),
            (idl) -- [gluon, out=45,in=135,looseness=0.7] (idr),
            (idr) -- [gluon, out=225,in=315,looseness=0.7] (idl), 
        };

        \vertex (n) at (0,-0.1) {\large $\left( A1b \right)$};
    \end{feynman}
\end{tikzpicture}
}}\quad \vcenter{\hbox{
\begin{tikzpicture}[baseline=(c.base),xscale=0.8, yscale=0.7]
    \begin{feynman}
        \vertex (c);
        \vertex (e1) at (-2,0) ;
        \vertex (e2) at (2,0) ;

        \vertex (e3) at (-2,5) ;
        \vertex (e4) at (-1,5) ;
        \vertex (e5) at (1,5) ;
        \vertex (e6) at (2,5) ;
        
        \node (iu) at (0,3.4) [dot];
        \node (idl) at (-1,1.7) [dot];
        \node (idr) at (1,1.7) [dot];

        \diagram*{
            (e3) -- [anti fermion] (iu),
            (e4) -- [fermion] (iu),
            (e5) -- [fermion] (iu),
            (e6) -- [fermion] (iu),
            (e1) -- [anti fermion] (idl) -- [anti fermion] (iu),
            (e2) -- [anti fermion] (idr) -- [anti fermion] (iu),
            (idl) -- [gluon, out=45,in=135,looseness=0.7] (idr),
            (idr) -- [gluon, out=225,in=315,looseness=0.7] (idl), 
        };

        \vertex (n) at (0,-0.1) {\large $\left( A1c \right)$};
    \end{feynman}
\end{tikzpicture}
}}
$
\end{adjustbox}
%\vspace{0.5cm} % spacing between figures

\begin{adjustbox}{center}
$\vcenter{\hbox{
\begin{tikzpicture}[baseline=(c.base),xscale=0.8, yscale=0.7]
    \begin{feynman}
        \vertex (c);
        \vertex (e1) at (-2,5) ;
        \vertex (e2) at (2,5) ;

        \vertex (e3) at (-2,0.5) ;
        \vertex (e4) at (-1,0) ;
        \vertex (e5) at (1,0) ;
        \vertex (e6) at (2,0.5) ;
        
        \node (iul) at (-1,3.4) [dot];
        \node (iur) at (1,3.4) [dot];
        \node (idl) at (-1,1.7) [dot];
        \node (idr) at (1,1.7) [dot];

        \diagram*{
            (e1) -- [anti fermion] (iul) -- [anti fermion] (iur) -- [anti fermion] (e2),

            (e3) -- [anti fermion] (idl) -- [anti fermion] (e4),
            (e5) -- [anti fermion] (idr) -- [anti fermion] (e6),

            (iur)-- [gluon] (idr) -- [gluon] (idl) -- [gluon] (iul),
            (iur) -- [gluon, out=225,in=315,looseness=0.7] (iul), 
        };

        \vertex (n) at (0,-0.1) {\large $\left( A2 \right)$};
    \end{feynman}
\end{tikzpicture}    
}}\quad \vcenter{\hbox{
\begin{tikzpicture}[baseline=(c.base),xscale=0.8, yscale=0.7]
    \begin{feynman}
        \vertex (c);
        \vertex (e1) at (-0.5,5) ;
        \vertex (e2) at (0.5,5) ;

        \vertex (e3) at (-2,0) ;
        \vertex (e4) at (-1.2,0) ;
        \vertex (e5) at (1.2,0) ;
        \vertex (e6) at (2,0) ;
        
        \node (iu) at (0,3.75) [dot];
        \node (iml) at (-0.5,2.5) [dot];
        \node (imr) at (0.5,2.5) [dot];
        \node (idl) at (-1,1.25) [dot];
        \node (idr) at (1,1.25) [dot];

        \diagram*{
            (e1) -- [fermion] (iu) -- [fermion] (e2),

            (e3) -- [anti fermion] (idl) -- [anti fermion] (e4),
            (e5) -- [anti fermion] (idr) -- [anti fermion] (e6),

            (idl)-- [gluon] (iml) -- [gluon] (iu) -- [gluon] (imr) -- [gluon] (idr),
            (idl)-- [gluon] (idr),
            (iml)-- [gluon] (imr),
            
        };

        \vertex (n) at (0,-0.1) {\large $\left( A3 \right)$};
    \end{feynman}
\end{tikzpicture}
}}\quad \vcenter{\hbox{
\begin{tikzpicture}[baseline=(c.base),xscale=0.8, yscale=0.7]
    \begin{feynman}
        \vertex (c);
        \vertex (e1) at (-2,0) ;
        \vertex (e2) at (2,0) ;

        \vertex (e3) at (-2,5) ;
        \vertex (e4) at (-1,5) ;
        \vertex (e5) at (1,5) ;
        \vertex (e6) at (2,5) ;
        
        \node (iu) at (0,3.75) [dot];
        
        \node (iml) at (-1,2.5) [dot];
        \node (imr) at (1,2.5) [dot];

        \node (id) at (0,1.25) [dot];

        \diagram*{
            (e1) -- [fermion] (id) -- [fermion] (e2),
        
            (e3) -- [anti fermion] (iu),
            (e4) -- [fermion] (iu),
            (e5) -- [anti fermion] (iu),
            (e6) -- [fermion] (iu),

            (iu) -- [fermion] (iml) -- [fermion] (imr) -- [fermion] (iu),

            (iml) -- [gluon] (id) -- [gluon] (imr),

        };

        \vertex (n) at (0,-0.1) {\large $\left( A4 \right)$};
    \end{feynman}
\end{tikzpicture}
}}
$
\end{adjustbox}
%\vspace{0.5cm} % spacing between figures

\begin{adjustbox}{center}
$\vcenter{\hbox{
\begin{tikzpicture}[baseline=(c.base),xscale=0.8, yscale=0.7]
    \begin{feynman}
        \vertex (c);
        \vertex (e1) at (-0.5,5) ;
        \vertex (e2) at (0.5,5) ;

        \vertex (e3) at (-2,0.2) ;
        \vertex (e4) at (-1.6,0) ;
        \vertex (e5) at (1.6,0) ;
        \vertex (e6) at (2,0.2) ;
        
        \node (iu) at (0,3.75) [dot];
        \node (il) at (-1.51,1) [dot];
        \node (ir) at (1.51,1) [dot];
        \node (icl) at (-0.55,1) [dot];
        \node (icr) at (0.55,1) [dot];

        \diagram*{
            (e1) -- [fermion] (iu) -- [fermion] (e2),

            (e3) -- [fermion] (il) -- [fermion] (e4),
            (e5) -- [fermion] (ir) -- [fermion] (e6),

            (icl)-- [gluon] (il) -- [gluon] (iu) -- [gluon] (ir) -- [gluon] (icr),
            (icl) -- [fermion, half left,looseness=1.1] (icr) -- [fermion, half left,looseness=1.1] (icl), 

        };

        \vertex (n) at (0,-0.1) {\large $\left( A5 \right)$};
    \end{feynman}
\end{tikzpicture}
}}\quad \vcenter{\hbox{
\begin{tikzpicture}[baseline=(c.base),xscale=0.8, yscale=0.7]
    \begin{feynman}
        \vertex (c);
        \vertex (e1) at (-0.5,5) ;
        \vertex (e2) at (0.5,5) ;

        \vertex (e3) at (-2,0) ;
        \vertex (e4) at (-1.2,0) ;
        \vertex (e5) at (1.2,0) ;
        \vertex (e6) at (2,0) ;
        
        \node (iu) at (0,3.75) [dot];
        \node (iml) at (-0.5,2.5) [dot];
        \node (imr) at (0.5,2.5) [dot];
        \node (idl) at (-1,1.25) [dot];
        \node (idr) at (1,1.25) [dot];

        \diagram*{
            (e1) -- [fermion] (iu) -- [fermion] (e2),

            (e3) -- [anti fermion] (idl) -- [anti fermion] (e4),
            (e5) -- [anti fermion] (idr) -- [anti fermion] (e6),

            (idl)-- [gluon] (iml) -- [gluon] (iu) -- [gluon] (imr) -- [gluon] (idr),
            (iml)-- [gluon] (idr),
            (idl)-- [gluon] (imr),
            
        };

        \vertex (n) at (0,-0.1) {\large $\left( A6 \right)$};
    \end{feynman}
\end{tikzpicture}
}}\quad \vcenter{\hbox{
\begin{tikzpicture}[baseline=(c.base),xscale=0.8, yscale=0.7]
    \begin{feynman}
        \vertex (c);
        \vertex (e1) at (-2,5) ;
        \vertex (e2) at (2,5) ;

        \vertex (e3) at (-2,0.5) ;
        \vertex (e4) at (-1,0) ;
        \vertex (e5) at (1,0) ;
        \vertex (e6) at (2,0.5) ;
        
        \node (iul) at (-1,3.4) [dot];
        \node (iur) at (1,3.4) [dot];
        \node (idl) at (-1,1.7) [dot];
        \node (idr) at (1,1.7) [dot];

        \diagram*{
            (e1) -- [anti fermion] (iul) -- [anti fermion] (iur) -- [anti fermion] (e2),

            (e3) -- [anti fermion] (idl) -- [anti fermion] (e4),
            (e5) -- [anti fermion] (idr) -- [anti fermion] (e6),

            (iul)-- [gluon] (idl) -- [gluon] (iur),
            (iul)-- [gluon] (idr) -- [gluon] (iur), 
        };

        \vertex (n) at (0,-0.1) {\large $\left( A7 \right)$};
    \end{feynman}
\end{tikzpicture}
}}
$
\end{adjustbox}
%\vspace{0.5cm} % spacing between figures

\begin{adjustbox}{center}
$\vcenter{\hbox{
\begin{tikzpicture}[baseline=(c.base),xscale=0.8, yscale=0.7]
    \begin{feynman}
        \vertex (c);
        \vertex (tempver) at (0,5) {};
        \vertex (e1) at (-2,4) ;
        \vertex (e2) at (-2,2.5) ;
        \vertex (e3) at (-2,1) ;
        \vertex (e4) at (2,4) ;
        \vertex (e5) at (2,2.5) ;
        \vertex (e6) at (2,1) ;

        \node (il) at (-1,2.5) [dot];
        \node (ir) at (1,2.5) [dot];

        \diagram*{
            (e1) -- [fermion] (il),
            (e2) -- [fermion] (il),
            (e3) -- [anti fermion] (il),

            (e4) -- [anti fermion] (ir),
            (e5) -- [anti fermion] (ir),
            (e6) -- [fermion] (ir),

            (il) -- [fermion, half left] (ir),
            (il) -- [fermion] (ir),
            (il) -- [anti fermion, half right] (ir),
            
        };

        \vertex (n) at (0,-0.1) {\large $\left( A8a \right)$};
    \end{feynman}
\end{tikzpicture}   
}}\quad \vcenter{\hbox{
\begin{tikzpicture}[baseline=(c.base),xscale=0.8, yscale=0.7]
    \begin{feynman}
        \vertex (c);
        \vertex (tempver) at (0,5) {};
        \vertex (e1) at (-2,4) ;
        \vertex (e2) at (-2,2.5) ;
        \vertex (e3) at (-2,1) ;
        \vertex (e4) at (2,4) ;
        \vertex (e5) at (2,2.5) ;
        \vertex (e6) at (2,1) ;

        \node (il) at (-1,2.5) [dot];
        \node (ir) at (1,2.5) [dot];

        \diagram*{
            (e1) -- [fermion] (il),
            (e2) -- [fermion] (il),
            (e3) -- [fermion] (il),

            (e4) -- [anti fermion] (ir),
            (e5) -- [anti fermion] (ir),
            (e6) -- [anti fermion] (ir),

            (il) -- [fermion, half left] (ir),
            (il) -- [fermion] (ir),
            (il) -- [fermion, half right] (ir),

        };

        \vertex (n) at (0,-0.1) {\large $\left( A8b \right)$};
    \end{feynman}
\end{tikzpicture}
}}
$

\end{adjustbox}

    \caption{Diagrams that contribute to the 2-loop correction to the 6-point vertex. Each external line carries a color and a flavor index.}

    \label{fig: diagrmas A} 
\end{figure}

\begin{figure}[t]
\begin{adjustbox}{center}
$
    \vcenter{\hbox{
    \begin{tikzpicture}[baseline=(c.base)]
    \begin{feynman}
        \vertex (c);
        \vertex (e1) at (-2,0) ;
        \vertex (e2) at (2,0) ;
        \vertex (tempveru) at (0,1.7) {};
        \vertex (tempverd) at (0,-1.5) {};

        \node (il) at (-1,0) [dot];
        \node (im) at (0,0) [dot];
        \node (ir) at (1,0) [dot];
        \node (iu) at (0,1) [dot];

        \diagram*{
            (e1) -- [fermion] (il) -- [fermion] (im)-- [fermion] (ir)  -- [fermion] (e2),

            (il)-- [gluon, quarter left] (iu), 
            (im)-- [gluon] (iu),
            (ir)-- [gluon,quarter right] (iu),

        };
        \vertex (n) at (0,-1.3) {\large $\left( B1 \right)$};
    \end{feynman}
\end{tikzpicture}
}}\quad \vcenter{\hbox{
\begin{tikzpicture}[baseline=(c.base)]
    \begin{feynman}
        \vertex (c);
        \vertex (e1) at (-2,0) ;
        \vertex (e2) at (2,0) ;
        \vertex (tempveru) at (0,1.7) {};
        \vertex (tempverd) at (0,-1.5) {};

        \node (il) at (-0.8,0) [dot];
        \node (ir) at (0.8,0) [dot];
        
        \diagram*{
            (e1) -- [fermion] (il) -- [fermion] (ir)  -- [fermion] (e2),

            (il)-- [gluon, half left,looseness=2] (ir), 
            (il)-- [gluon, half left,looseness=0.7] (ir), 

        };
        \vertex (n) at (0,-1.3) {\large $\left( B2 \right)$};
    \end{feynman}
\end{tikzpicture}
}}\quad \vcenter{\hbox{
\begin{tikzpicture}[baseline=(c.base)]
    \begin{feynman}
        \vertex (c);
        \vertex (e1) at (-2,0) ;
        \vertex (e2) at (2,0) ;
        \vertex (tempveru) at (0,1.7) {};
        \vertex (tempverd) at (0,-1.5) {};

        \node (ill) at (-1.2,0) [dot];
        \node (il) at (-0.4,0) [dot];
        \node (ir) at (0.4,0) [dot];
        \node (irr) at (1.2,0) [dot];
        
        \diagram*{
            (e1) -- [fermion] (ill) -- [fermion] (il)  -- [fermion] (ir)  -- [fermion] (irr)  -- [fermion] (e2),

            (ill)-- [gluon, half left,looseness=1.5] (ir), 
            (il)-- [gluon, half right,looseness=1.5] (irr), 

        };
        \vertex (n) at (0,-1.3) {\large $\left( B3 \right)$};
    \end{feynman}
\end{tikzpicture}
}}\quad \vcenter{\hbox{
\begin{tikzpicture}[baseline=(c.base)]
    \begin{feynman}
        \vertex (c);
        \vertex (e1) at (-2,0) ;
        \vertex (e2) at (2,0) ;
        \vertex (tempveru) at (0,1.7) {};
        \vertex (tempverd) at (0,-1.5) {};

        \node (il) at (-1.5,0) [dot];
        \node (iul) at (-0.7,1) [dot];
        \node (iur) at (0.7,1) [dot];
        \node (ir) at (1.5,0) [dot];
        
        \diagram*{
            (e1) -- [fermion] (il)  -- [fermion] (ir)  -- [fermion] (e2),

            (il)  -- [gluon, quarter left] (iul),
            (ir)  -- [gluon, quarter right] (iur),

            (iul) -- [fermion, half left,looseness=1.1] (iur) -- [fermion, half left,looseness=1.1] (iul),

        };
        \vertex (n) at (0,-1.3) {\large $\left( B4 \right)$};
    \end{feynman}
\end{tikzpicture}
}}
$
\end{adjustbox}

    \caption{Diagrams that contribute to the 2-loop correction to the scalar propagator.}

    \label{fig: B diagrams} 
\end{figure}
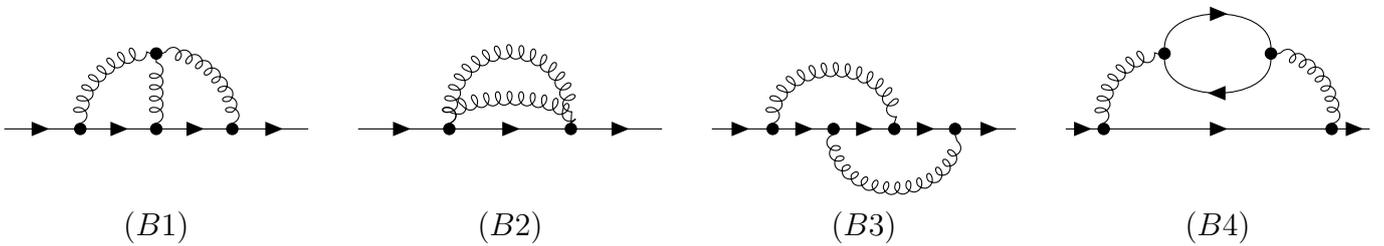

This calculation, for a Lagrangian involving both relevant terms and fermions (see equation 23 in~\cite{Avdeev:1992jt}), was first performed in~\cite{Avdeev:1992jt} (see equation 34), for a $SU(N_c)$ gauge group with finite $N_c$ and one flavor. It involved a very large number of Feynman diagrams. In~\cite{Aharony:2011jz}\footnote{The calculation in~\cite{Aharony:2011jz} was done for the $SO(N_c)$ gauge group with $N_f = 1$.} (and later in~\cite{Banerjee:2013nca,Anninos:2014hia}), the authors showed that in the region $\lambda_{n} \sim \lambda^2 \ll 1$, there are 11\footnote{In~\cite{Aharony:2011jz}, there are 8 diagrams contributing to the 6-point vertex for real scalars. For complex scalars, the orientation of the scalar lines splits diagrams A1 and A8.} diagrams that contribute to the 6-point vertex (figure~\ref{fig: diagrmas A}), and 4 that contribute to the propagator (figure~\ref{fig: B diagrams}).
Adding flavors does not alter the number of diagrams, but only adds flavor indices to each scalar line.

The full calculation of the 2-loop beta function is presented in Appendix \ref{sec: appndiex RB lambda finite}, with the result in the representation basis\footnote{Note that the quadratic term in \eqref{eq: beta small lambdab} is the same as in \eqref{eq: beta function free boson} since in both, the only diagram that contributes to the quadratic term in the 't Hooft limit is $A8a$.}

\begin{equation} \label{eq: beta small lambdab}
    \begin{split}
        16\pi^{2}&N_{c}\beta_{SSS}=\left(\frac{\lambda}{4\pi}\right)^4\frac{24 }{N_f}-\left(\frac{\lambda}{4\pi}\right)^2\frac{8  \left(\lambda_{SSS} N_f \left(2 N_f^2+3\right)+3 \lambda_{SAA} \left(N_f^2-1\right)\right)}{N_f^2}\\+&6 \lambda_{SSS}^2 N_f+\frac{6 \lambda_{SAA}^2 \left(N_f^2-1\right)}{N_f}, \\
        16\pi^{2}&N_{c}\beta_{SAA}=12 \left(\frac{\lambda}{4\pi}\right)^4
        -\left(\frac{\lambda}{4\pi}\right)^2\frac{4  \left(N_f \left(2 \lambda_{SSS} N_f+3 \lambda_{SAA} \left(N_f^2+2\right)\right)+4 \lambda_{AAA} \left(N_f^2-4\right)\right)}{N_f^2}\\
        +&\frac{4 \left(\lambda_{SAA} N_f^2 \left(\lambda_{SSS} N_f+2 \lambda_{SAA}\right)+\lambda_{SAA} \lambda_{AAA} \left(N_f^2-4\right) N_f+\lambda_{AAA}^2 \left(N_f^2-4\right)\right)}{N_f^2}, \\
        16\pi^{2}&N_{c}\beta_{AAA}=3 N_f \left(\frac{\lambda}{4\pi}\right)^4+\left(\frac{\lambda}{4\pi}\right)^2 \left(-10 \lambda_{AAA} N_f+\frac{48 \lambda_{AAA}}{N_f}-24 \lambda_{SAA}\right)\\+&3 \left(\lambda_{SAA}^2 N_f+\frac{\lambda_{AAA}^2 \left(N_f^2-12\right)}{N_f}+4 \lambda_{AAA} \lambda_{SAA}\right) .
    \end{split}
\end{equation}
As a consistency check, we see that for $N_f=1$, $\beta_{SSS}$ reduces to \eqref{eq: nf1 beta RB finite lambda near 0},
and there is a $N_f^2-4$ factor in front of $\lambda_{AAA}$ in $\beta_{SAA}$ (so that for $N_f=2$, $\beta_{AAA}$ decouples from $\beta_{SSS}$ and $\beta_{SAA}$).

As in the $N_f=1$ case, we analyze the splitting of the degenerate free fixed point at $\lambda=0$ using \eqref{eq: beta small lambdab}. The resulting fixed points are located at couplings of order $\lambda^2$, i.e.  $\lambda_{SSS},\lambda_{SAA},\lambda_{AAA}\sim\lambda^2 \ll 1$. This hierarchy will justify the use of perturbation theory to the order considered in the derivation of \eqref{eq: beta small lambdab}, if all next order terms will be smaller. In \eqref{eq: beta function free boson} the cubic terms come with factors proportional to powers of $N_f$ (up to $N_f^3$), therefore this perturbative approach is fully justified only for $\lambda\ll\frac{1}{N_f^2}$. For finite $N_f$ this subtlety is of little importance, and in section \ref{sec: Trucking the IR fixed point at large nf} we'll discuss this regime in the case of large $N_f$.

We note that the computations presented in Appendix \ref{sec: appndiex RB lambda finite} were carried out without invoking any specific assumptions regarding $ N_c $ or $ N_f $. The 't Hooft limit  
was imposed only in the final step. This allows for the perturbative analysis to be extended to finite values of $ N_c $ as well. In Appendix \ref{sec: appendix finite lambda finite nc}, we investigate whether an infrared (IR)-stable fixed point persists for finite values of $ N_c $.

\subsection{The  \texorpdfstring{$N_f=2$}{Nf=2} case} \label{sec: RB nf=2 analysis}

We can now analyze the fixed point structure and the flow of the beta functions in \eqref{eq: beta function free boson} and \eqref{eq: beta small lambdab}. We start with the $N_f=2$ case, in which we have only $\beta_{SSS}$ and $\beta_{SAA}$. 

\subsubsection{Analysis for \texorpdfstring{$\lambda=0$}{lambda=0}}

\begin{figure}[t]
\centering
\includegraphics[width=0.49\linewidth]{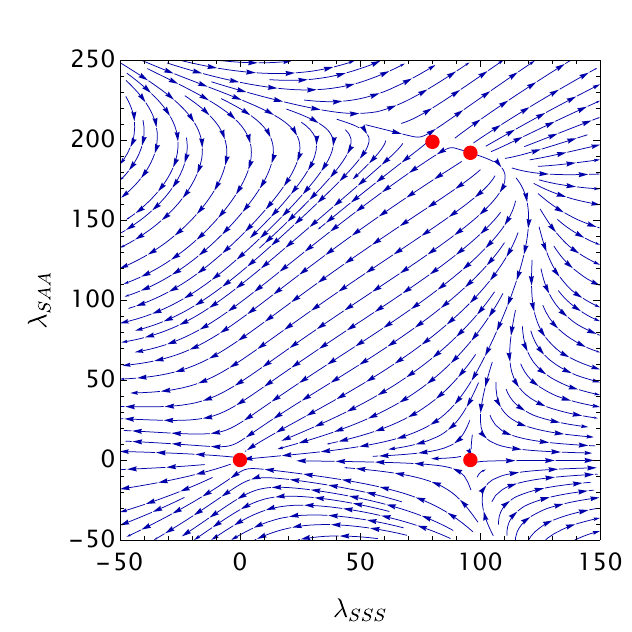}
\includegraphics[width=0.49\linewidth]{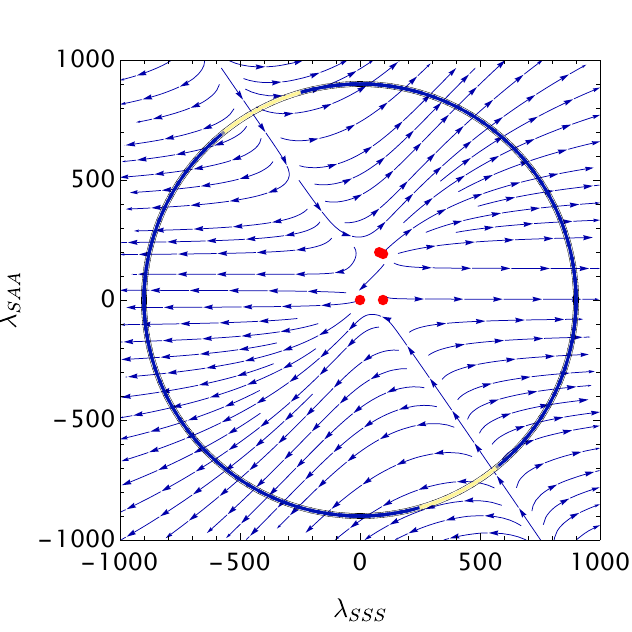}

\caption{ The flow to the IR of (minus) \eqref{eq: beta function free boson} for $N_f=2$ and $\lambda=0$. The fixed points \eqref{eq: RB nf=2 lmabda=0 fixed points} are shown in red.
{\bf Left:} The flow near the fixed points. {\bf Right:} The flows ``at infinity", 
the color of the circle encodes whether the flow goes outward (blue), or inward (yellow). Note that the angular size of the outward and inward regions stays finite as $\lambda_n\rightarrow\infty$.
}

\label{fig: large L 2d}
\end{figure}

For $\lambda=0$, we find from \eqref{eq: beta function free boson} that there are 4 fixed points
\begin{equation} \label{eq: RB nf=2 lmabda=0 fixed points}
\left(\lambda_{SSS},\lambda_{SAA}\right):\left(0,0\right),\left(\sim80,\sim199\right),\left(96,0\right),\left(96,192\right).
\end{equation}
Diagonalization of the derivative matrix of the beta function at these points $\partial_i\beta_j$ shows that (see section \ref{sec: RB morse theory} for more details)
\begin{itemize}
    \item The free point $\left(0,0\right)$ has two zero eigenvalues and is thus degenerate. 
    \item The $\left(\sim80,\sim199\right)$ and $\left(96,0\right)$ points have both positive and negative eigenvalues, which means that they are mixed points and not UV nor IR-stable.
    \item The $\left(96,192\right)$ point has only positive eigenvalues, so it is UV-stable 
    \cite{Kapoor:2021lrr}.
\end{itemize}
The RG flow of the couplings is shown in the streamplot in the left panel of figure \ref{fig: large L 2d}, which shows the behaviors of the fixed points mentioned above  
(note that arrows represent flows to the IR).

\subsubsection{Analysis of large \texorpdfstring{$\lambda_n$}{lambdai}} \label{sec: RB nf2 large lambda i}
An interesting aspect of the RG flow is its behavior for large values of $\lambda_n$. In this region, we consider only the cubic terms in \eqref{eq: beta function free boson}, and, as we discuss in section \ref{sec: general case} (see item \ref{item4.3: cubic are universal} in section \ref{sec: multiple falvor beta function comments}), this behavior is universal for any value of $\lambda$.

We compute the direction of the flow (inward or outward) predicted from the cubic terms in the right panel of figure \ref{fig: large L 2d}, where we plot a circle centered at the origin with large radius, on top of a stream plot like that in the left panel. 

For finite $\lambda$, angles for which the flow points inward can either land at an IR-stable point, or turn around and flow to infinity. Angles for which the flow points outward will only flow to infinity. This simple test restricts the possible domains of attraction of a possible IR fixed point.

\subsubsection{Analysis for \texorpdfstring{$\lambda \ll 1$}{lambda l1}}
We can now use the perturbative results obtained in \eqref{eq: beta small lambdab} to see what happens to the degenerate free fixed point when turning on a finite $\lambda$. Solving the equations, we find four fixed points at
\begin{equation} \label{eq: nf=2 lmabda finite 4 fixed point near 0}
    \begin{split}
        \left(\lambda_{SSS},\lambda_{SAA}\right)	: \ & \left(\frac{1}{6}\left(\frac{\lambda}{4\pi}\right)^{2},\frac{1}{3}\left(\frac{\lambda}{4\pi}\right)^{2}\right),\left(\frac{3}{2}\left(\frac{\lambda}{4\pi}\right)^{2},3\left(\frac{\lambda}{4\pi}\right)^{2}\right), \\ & \left(\frac{1}{42}\left(131-2\sqrt{22}\right)\left(\frac{\lambda}{4\pi}\right)^{2},\frac{1}{7}\left(5-2\sqrt{22}\right)\left(\frac{\lambda}{4\pi}\right)^{2}\right), \\
	&\left(\frac{1}{42}\left(2\sqrt{22}+131\right)\left(\frac{\lambda}{4\pi}\right)^{2},\frac{1}{7}\left(2\sqrt{22}+5\right)\left(\frac{\lambda}{4\pi}\right)^{2}\right).
    \end{split}
\end{equation}
The first point $\left(\frac{1}{6}\left(\frac{\lambda}{4\pi}\right)^{2},\frac{1}{3}\left(\frac{\lambda}{4\pi}\right)^{2}\right)$  is UV-stable, the last point is IR-stable, and the middle two points are mixed. We summarize the number of fixed points of any type for $\lambda \ll1$ in table \ref{tab: nf2 fixed points RB}.

\begin{figure}[t]
\centering
\includegraphics[width=0.45\linewidth]{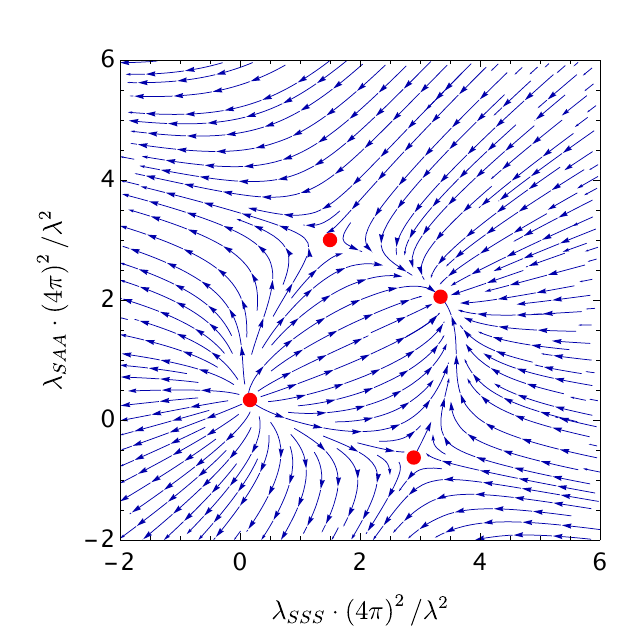}
\includegraphics[width=0.45\linewidth]{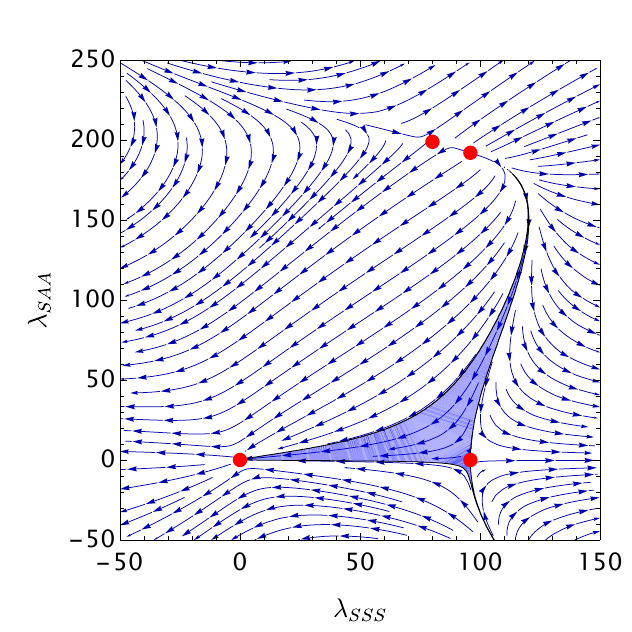}

\caption{\textbf{Left:} Stream plot of flow to the IR of (minus) \eqref{eq: beta small lambdab} for $N_f=2$ and $\lambda=0.2$. The fixed points \eqref{eq: nf=2 lmabda finite 4 fixed point near 0} are shown in red.
{\bf Right:} 
The basin of attraction of the IR fixed point, overlaid on the beta function flows shown in figure \ref{fig: large L 2d}.
}

\label{fig: nf2 spliting of fixed point} 
\end{figure}

The stream plot zoomed on the points in \eqref{eq: nf=2 lmabda finite 4 fixed point near 0} is given in the left panel of figure \ref{fig: nf2 spliting of fixed point}. As can be seen from the figure, couplings inside the `polygon' created by the perturbative fixed points flow to the IR fixed point. This shows that the area in the coupling space which flows to the IR-stable point is bounded from below by $\sim\lambda^4$. 

To find the global structure of the flows to the IR fixed point, we solved equations \eqref{eq: beta function free boson} and \eqref{eq: beta small lambdab} together, by adding the linear and constant terms in \eqref{eq: beta small lambdab} to \eqref{eq: beta function free boson}. Since $\lambda \ll 1$, this has a minor effect on the flow far from $\lambda_n=0$, 
but it splits the degenerate fixed point into the four points in \eqref{eq: nf=2 lmabda finite 4 fixed point near 0}.

In the right panel of figure \ref{fig: nf2 spliting of fixed point} we show the basin of attraction to the IR fixed point for $\lambda = 0.2$. We see from this figure that there is deformation of the UV-stable fixed point at $\left(96,192\right)$ which flows to the IR fixed point, and passes near the mixed fixed point $\left(96,0\right)$. We also see that unlike the single flavor case in section \ref{sec: Nf=1}, here the domain of attraction is non-compact, namely there is a flow to the IR-stable fixed point from infinity. We have checked numerically that the size of the basin of attraction of the IR-stable point decreases continuously to 0 as $\lambda$ decreases. This stands in contrast to the single flavor case, in which the IR-stable fixed point's basin of attraction stays finite even as $\lambda\rightarrow0$, and is compact.

\begin{table}[t]
\centering
\begin{tabular}{|c|c|c|c|c|}
\hline 
 &  $\gamma$ & $\text{ind}_{p}=\left(-1\right)^{\gamma}$ &  Num. of points  & $\sum\text{ind}_{p}$\tabularnewline
\hline 
\hline 
IR fixed points & $0$ & $+1$ & $1$ & $1$\tabularnewline
\hline 
Mixed fixed points & $1$ & $-1$ & $4$ & $-4$\tabularnewline
\hline 
UV fixed points & $2$ & $+1$ & $2$ & $2$\tabularnewline
\hline 
\hline 
Total &  &  & $7$ & $-1$\tabularnewline
\hline 
\end{tabular}
    \caption{Summary of the $N_f=2$ fixed points with $\lambda \ll1$. We denote by $\gamma$ the number of relevant directions at each point. The importance of $\text{ind}_{p}$ is discussed in section \ref{sec: RB morse theory}.
    }
    \label{tab: nf2 fixed points RB}
\end{table}

\subsection{The  \texorpdfstring{$N_f\ge3$}{Nf g= 3} case} \label{sec: RB nf>=3 analysis}

In this section we analyze the fixed points structure and the flows for the $N_f \ge 3 $ case. Unlike $N_f=2$, now the flow is three dimensional.

\subsubsection{Analysis for \texorpdfstring{$\lambda=0$}{lambda=0}} \label{subsec: Nf>3 lambda=0}
The number of fixed points predicted by the beta functions \eqref{eq: beta function free boson} depends on the number of flavors $N_f$, and can be found directly by solving the equations for each $N_f$. We summarize the number of fixed points, and the number of relevant directions of each such fixed point (obtained by diagonalizing the derivative matrix of the beta function $\partial_i\beta_j$ at each point, and counting the number of negative eigenvalues), in table \ref{tab: free_bosonic_fixed_points}.

Out of all the fixed points, there are three fixed points with simple expressions \cite{Kapoor:2021lrr}
\begin{equation} \label{eq: fixed points RB nf=3 always}
\left(\lambda_{SSS},\lambda_{SAA},\lambda_{AAA}\right):\left(0,0,0\right),\left(\frac{384}{N_{f}^{2}},0,0\right),\left(\frac{384}{N_{f}^{2}},\frac{384}{N_{f}},192\right) .
\end{equation}
The additional 4 fixed points that exist for $N_f\ge 10$ converge to the point $\left(0,0,192\right)$ as we take $N_f\rightarrow \infty$.  We elaborate on this limit in section \ref{sec: RB large nf}.

Unlike the cases $N_f=1,2$, for $N_f\ge 3$ there are two fixed points in which the derivative matrix of the beta function $\partial_i\beta_j$ is a degenerate matrix. For the point $\left(0,0,0\right)$ we have $\partial_i\beta_j=0$, and so it is degenerate in all of the directions. For the point $\left(\frac{384}{N_{f}^{2}},0,0\right)$ there is one zero, one positive, and one negative eigenvalue.  The point $\left(\frac{384}{N_{f}^{2}},\frac{384}{N_{f}},192\right)$ is a mixed fixed point.

\begin{table}[t]
\centering
\begin{tabular}{|c||c|c|c|c|c||c|c|}
\hline 
$N_{f}$ & 
\centering
\makecell{$\gamma=0$ \\ IR-stable}
 & $\gamma=1$ & $\gamma=2$ & 
\centering
\makecell{$\gamma=3$ \\ UV stable}
 & 
\centering
\makecell{Degenerate \\ points}
&
\makecell{Total num. \\ of fixed points}
 & 
\centering
$\sum\text{ind}_{p}$
\tabularnewline
\hline 
\hline 
\textbf{3} & 0 & 1 & 4 & 0 & 2 & 7 & 3\tabularnewline
\hline 
\textbf{4-9} & 0 & 0 & 2 & 1 & 2 & 5 & 1\tabularnewline
\hline 
\textbf{10$\leq$} & 0 & 0 & 3 & 2 & 2 & 7  & 1\tabularnewline
\hline 
\end{tabular}
\caption{Summary of the $N_f>2$ fixed points with $\lambda=0$. We denote by $\gamma$ the number of relevant directions at each point. In our case, the degenerate points do not contribute to the sum in the last column, whose importance is explained in section \ref{sec: RB morse theory}. The fact that this sum differs between $N_f=3$ and $N_f>3$ is also discussed in section \ref{sec: RB morse theory}.}
\label{tab: free_bosonic_fixed_points}
\end{table}

\subsubsection{Analysis of large \texorpdfstring{$\lambda_n$}{lambdai}\label{sec: RB nf general large lambda i}}

One can perform a similar analysis to the one done in section \ref{sec: RB nf2 large lambda i} and analyze the RG flow for large couplings, considering only the (universal) cubic terms.
Since the RG flow in this case is three dimensional, the origin is circled by a sphere $S^2$ instead of $S^1$ as in the $N_f =2$ case.

\begin{figure}[t]
\centering
\includegraphics[width=0.45\linewidth]{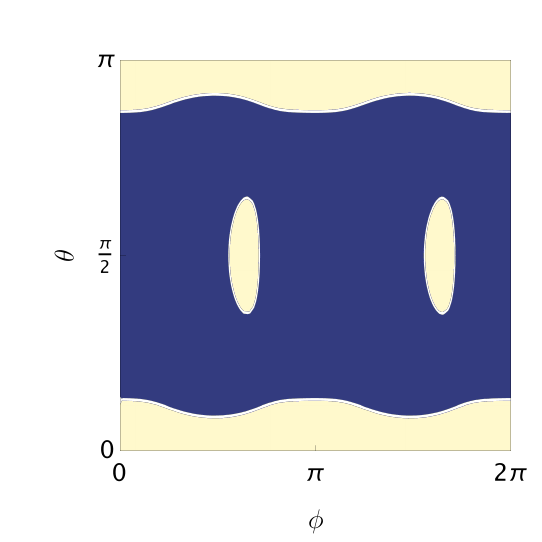}
\includegraphics[width=0.45\linewidth]{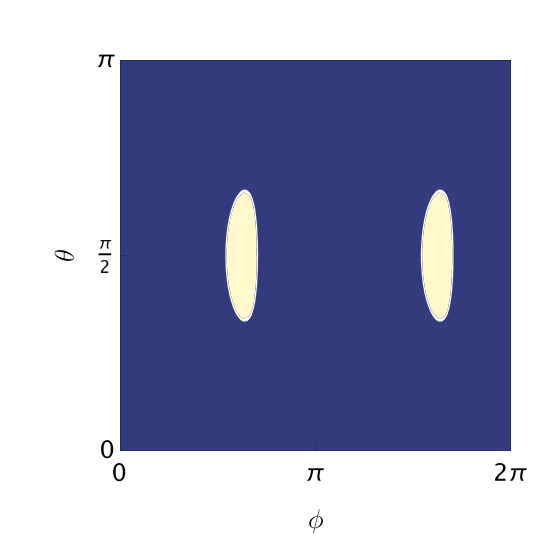}
\includegraphics[width=0.45\linewidth]{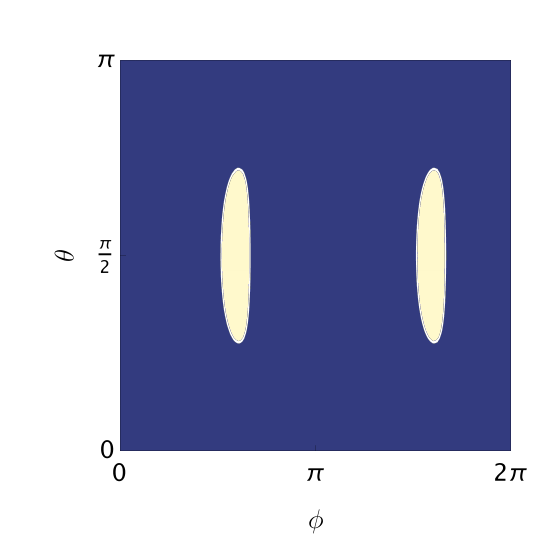}
\includegraphics[width=0.45\linewidth]{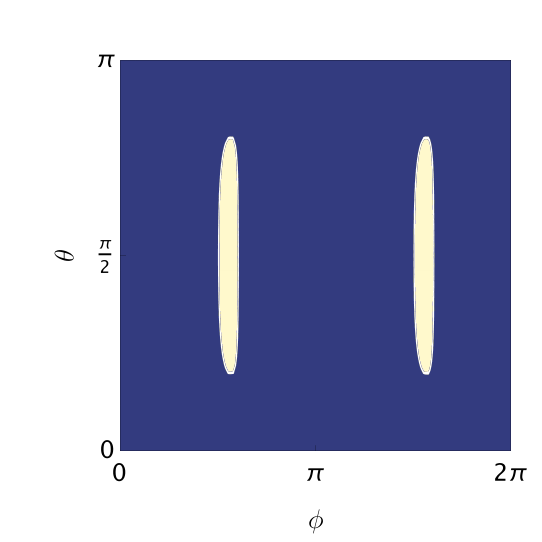}

\caption{
Direction of the flow to the IR of (minus) the cubic term in \eqref{eq: beta function free boson} for $N_f=3$ (\textbf{upper left}), for $N_f=4$ (\textbf{upper right}), for $N_f=10$ (\textbf{bottom left}), and for $N_f=50$ (\textbf{bottom right}). \textcolor{blue}{Blue-} The direction is outward (to infinity), \textcolor{brown}{Yellow-} The direction is inward.
}
\label{fig: large L 3d} 
\end{figure}

We parametrize the different couplings on the sphere as
\begin{equation}
    \lambda_{AAA}=R\cos\left(\theta\right), \lambda_{SAA}=R\sin\left(\theta\right)\sin\left(\phi\right), \lambda_{SSS}=R\sin\left(\theta\right)\cos\left(\phi\right) ,
\end{equation}
with $R\rightarrow\infty$. The direction of the flow, as a function of $\theta \in \left[ 0, \pi  \right]$ and $\phi\in \left[ 0, 2\pi  \right]$, for different values of $N_f$, is shown in figure \ref{fig: large L 3d}.  We see that there is a qualitative difference between the behavior for $N_f=3$ and for $N_f>3$ (which are all similar to each other), which is that for $\theta\sim0,\pi$ the flow for $N_f=3$ is inward, unlike the $N_f>3$ case. This stems from the factor of $\left(N_f^2-12\right)$ in front of $\lambda_{AAA}^3$ in $\beta_{AAA}$ in \eqref{eq: beta function free boson}, which flips sign between $N_f=3$ and $N_f=4$.

\subsubsection{Analysis for \texorpdfstring{$\lambda \ll 1$}{lambda l1}} \label{sec: RB lambda finite nf ge 3}

We now turn to study what happens to the degenerate \textit{free} fixed point when we turn on a small coupling $\lambda$.

Solving the beta functions \eqref{eq: beta small lambdab} for various values of $N_f$, we find that the free point splits into 6 points (with the exception of $N_f=4$), where there is always a UV-stable fixed point, and for $N_f\ge 5$ there is also an IR-stable fixed point. Two of the points have simple solutions (the second point is for any $N_f \ge 4$):
\begin{equation} \label{eq: RB finite lambda analytic}
    \begin{split}
        \left(\lambda_{SSS},\lambda_{SAA},\lambda_{AAA}\right):&\left(\frac{6}{N_{f}^{2}}\left(\frac{\lambda}{4\pi}\right)^{2},\frac{6}{N_{f}}\left(\frac{\lambda}{4\pi}\right)^{2},3\left(\frac{\lambda}{4\pi}\right)^{2}\right),\\&\left(\frac{2}{3N_{f}^{2}}\left(\frac{\lambda}{4\pi}\right)^{2},\frac{2}{3N_{f}}\left(\frac{\lambda}{4\pi}\right)^{2},\frac{1}{3}\left(\frac{\lambda}{4\pi}\right)^{2}\right) ,
    \end{split}
\end{equation}
where the first point in \eqref{eq: RB finite lambda analytic} is mixed, while the second point is UV-stable. We summarize the number of fixed points and their types in table 
\ref{tab: pertubed_bosonic_fixed_points}. As apparent from \eqref{eq: beta function free boson}, the coupling values scale as $\lambda^2$.

For $N_f \ge 5$, We find that the 3D polygon generated by the points is in the basin of attraction of the IR-stable fixed point, so the area of the basin of attraction grows at least as $\lambda^{6}$. Furthermore, we also find that there is a direction from infinity that flows to the IR-stable fixed point. Like in the $N_f=2$ case, it seems that the area of the basin of attraction goes to zero smoothly as $ \lambda \rightarrow 0 $.

The methods presented above are insufficient to determine what happens to the degenerate fixed point $\left(\frac{384}{N_{f}^{2}},0,0\right)$, which may be eliminated by the small-$\lambda$ corrections or split into several non-degenerate fixed points. The beta functions \eqref{eq: beta small lambdab} were calculated in perturbation theory around the free fixed point, and not around a point for which one of the couplings ($\lambda_{SSS}$) is finite. However, since already for $\lambda=0$ this fixed point had an IR-unstable direction, it is clear that it cannot lead to an IR-stable fixed point at small $\lambda$. In section \ref{sec: Trucking the IR fixed point at large nf}, we show that in the large $N_f$ limit in which the CS coupling scales as $\lambda \sim \frac{1}{N_f}$, this point can be treated perturbatively, and that it splits for small $\lambda$. 

\begin{table}[t]
\centering
\begin{tabular}{|c||c|c|c|c||c|}
\hline 
$N_{f}$ & \makecell{$\gamma=0$ \\ IR-stable} & $\gamma=1$ & $\gamma=2$ & \makecell{$\gamma=3$ \\ UV stable} &
\makecell{Total num. \\ of fixed Points} \tabularnewline
\hline 
\hline 
\textbf{3} & 0 & 2 & 3 & 1 & 6\tabularnewline
\hline 
\textbf{4} & 0 & 1 & 2 & 1 & 4\tabularnewline
\hline 
\textbf{5$\leq$} & 1 & 2 & 2 & 1 & 6\tabularnewline
\hline 
\end{tabular}
\caption{Summary of  $N_f>2$ and $\lambda\ll1$ fixed points near $\left(0,0,0\right)$. We denote by $\gamma$ the number relevant directions at each point. Note that there are no longer any degenerate fixed points in that area. Also note that the sum $\sum\text{ind}_{p}$ for these points is $0$ for any $N_f$, this is because all these points came from a splitting of a degenerate fixed point, as 
explained in section \ref{sec: RB morse theory}.}
\label{tab: pertubed_bosonic_fixed_points}
\end{table}

\subsection{The limit  \texorpdfstring{$N_f\rightarrow\infty$}{Nf}} \label{sec: RB large nf}

In this section we focus on the limit $N_f \rightarrow \infty$.  We start by taking the limit $N_f \rightarrow\infty$ of the beta functions \eqref{eq: beta function free boson} and \eqref{eq: beta small lambdab}, and show that the derivation of \eqref{eq: beta small lambdab} is inconsistent with this limit. Then, we take a double scaling limit, in which we can use the calculation above, and we show that the IR-stable fixed point merges with another fixed point and disappears as $\lambda$ is increased. Finally, we look concretely at the $(2,1,0)$ scaling.

All the solutions found in this section are presented to leading relevant order in $\frac{1}{N_f}$, we drop the $\mathcal{O}\left(N_f^{-\alpha}\right)$ notation for convenience whenever explicit solutions are discussed.

\subsubsection{The \texorpdfstring{$\lambda=0$}{lambda=0}  case} \label{sec: RB large nf lambda =0}

It is interesting to check how the fixed point solutions found above behave in the large $N_f\rightarrow \infty$. A naive approach would be to scale the parameters according to the scalings of \eqref{eq: allowd scalings}, and solve at leading order in the large $N_f$ expansion (e.g. substituting into \eqref{eq: general case beta nf to infinity} the values in table \ref{tab: beta coefficients}).

However, this does not give us all the possible solutions with that leading order scaling of $N_f$. The reason is that the leading order of the $N_f$ expansion might vanish, or give us degenerate answers (see section \ref{sec: 210} for a concrete example of degeneracy that forces us to use the next orders). Thus, we need to use the subleading orders of $\frac{1}{N_f}$ to determine the couplings, and those depend on the subleading scalings of the running couplings $\lambda_n$.

Therfore, in order to apply perturbation theory correctly, one has to write the couplings as a series
\begin{equation} \label{eq: Nf perturbation definition}
    \begin{split}
    \lambda_{SSS}&=\frac{1}{N_{f}^{a}}\left(y^{}_{SSS}+\frac{z^{}_{SSS}}{N_{f}}+\frac{w^{}_{SSS}}{N_{f}^{2}}+\frac{x^{}_{SSS}}{N_{f}^{3}}\cdots\right),\\
    \lambda_{SAA}&=\frac{1}{N_{f}^{b}}\left(y^{}_{SAA}+\frac{z^{}_{SAA}}{N_{f}}+\frac{w^{}_{SAA}}{N_{f}^{2}}+\frac{x^{}_{SAA}}{N_{f}^{3}}\cdots\right),\\
    \lambda_{AAA}&=\frac{1}{N_{f}^{c}}\left(y^{}_{AAA}+\frac{z^{}_{AAA}}{N_{f}}+\frac{w^{}_{AAA}}{N_{f}^{2}}+\frac{x^{}_{AAA}}{N_{f}^{3}}\cdots\right),
    \end{split}
\end{equation}
with $a,\ b,\ c$ as in \eqref{eq: allowd scalings}, and to solve the vanishing of the beta functions \eqref{eq: beta function free boson} order by order in $\frac{1}{N_f}$ for $y_n,z_n,w_n,x_n,\cdots $\footnote{Note that higher-order corrections of $ N_f $ do not necessarily appear with powers of $ \frac{1}{N_f} $. In the $\left(\frac{2}{3},\frac{1}{3},0\right)$ scaling, they may also involve fractional powers, such as $ \frac{1}{N_f^{1/3}} $. This is seen explicitly in \eqref{eq: RB nf to infinity lambda 0 sol second scaling}.
}. Doing so, we find that at leading order in $\frac{1}{N_f}$, the three analytic solutions presented in \eqref{eq: fixed points RB nf=3 always}, two additional UV-stable fixed points
\begin{equation} \label{eq: RB nf to infinity lambda 0 sol}
\begin{split}
 \left(\lambda_{SSS},\lambda_{SAA},\lambda_{AAA}\right): &
  \left(\frac{384\sqrt{2}}{N_{f}^{2}},\frac{384}{N_{f}}-\frac{768\left(-1+\sqrt{2}\right)}{N_{f}^{3}},192+\frac{3072\left(-1+\sqrt{2}\right)}{N_{f}^{4}}\right), \\
  &\left(\frac{-384\sqrt{2}}{N_{f}^{2}},\frac{384}{N_{f}}+\frac{768\left(1+\sqrt{2}\right)}{N_{f}^{3}},192-\frac{3072\left(1+\sqrt{2}\right)}{N_{f}^{4}}\right),
\end{split}
\end{equation}
have the $\left(2,1,0\right)$ scaling,
and two additional mixed fixed points have the $\left(\frac{2}{3},\frac{1}{3},0\right)$ scaling
\begin{equation} \label{eq: RB nf to infinity lambda 0 sol second scaling}
    \left(\lambda_{SSS},\lambda_{SAA},\lambda_{AAA}\right):    \left(\frac{-64\sqrt{6}}{N_{f}^{2/3}},\frac{64\sqrt{6}}{N_{f}^{1/3}},192-\frac{256}{N_{f}^{2/3}}\right),\left(\frac{64\sqrt{6}}{N_{f}^{2/3}},\frac{-64\sqrt{6}}{N_{f}^{1/3}},192-\frac{256}{N_{f}^{2/3}}\right) .
\end{equation}
Note that the points in \eqref{eq: RB nf to infinity lambda 0 sol} and \eqref{eq: RB nf to infinity lambda 0 sol second scaling} have additional higher order corrections in all of their components. The way they are written here is to emphasize that different subleading orders in several couplings  can affect the leading order of others. For example, in the first point in \eqref{eq: RB nf to infinity lambda 0 sol}, we have $\lambda_{AAA}=192+\frac{3072\left(-1+\sqrt{2}\right)}{N_{f}^{4}}+\mathcal{O}\left( \frac{1}{N_f^5} \right)$, and  the correction to the leading order of  $\lambda_{AAA}$ (i.e. $\frac{3072\left(-1+\sqrt{2}\right)}{N_{f}^{4}}$) affects the leading values of $y^{}_{SSS}$. 

In total we find 7 solutions, as indeed expected from table \ref{tab: free_bosonic_fixed_points}.

\subsubsection{Very small \texorpdfstring{$\lambda$}{lambda}}

As in the $\lambda=0$ case, we can apply the $N_f \rightarrow \infty$ limit to \eqref{eq: beta small lambdab} and find analytically the leading order (in $\frac{1}{N_f}$) dependence of the fixed points arising from the splitting of the $\lambda_{SSS}=\lambda_{SAA}=\lambda_{AAA}=0$ degenerate point. 

In section \ref{sec: RB lambda finite nf ge 3}, we saw that \eqref{eq: beta small lambdab} predicted that there are six such fixed points for $N_f \ge5$ (see table \ref{tab: pertubed_bosonic_fixed_points}), with two of them given exactly by \eqref{eq: RB finite lambda analytic}, both of which have the scaling $\left(2,1,0\right)$.

Using the same methods as in section \ref{sec: RB large nf lambda =0}, we can find the leading scalings of the additional fixed points. There is one additional mixed ($\gamma=1$) fixed point with leading $\left(2,1,0\right)$ scaling
\begin{equation} \label{eq: RB finite lambda largenf 2,1,0}
        \left(\lambda_{SSS},\lambda_{SAA},\lambda_{AAA}\right):\left(\frac{294}{25}\frac{1}{N_{f}^{2}}\left(\frac{\lambda}{4\pi}\right)^{2},-\frac{18}{5}\frac{1}{N_{f}}\left(\frac{\lambda}{4\pi}\right)^{2},\left(3+\frac{1}{N_{f}^{2}}\frac{576}{25}\right)\left(\frac{\lambda}{4\pi}\right)^{2}\right)  . 
\end{equation}
These are only 3 points. In order to find the other 3 we must relax the assumption that 
the leading powers in \eqref{eq: Nf perturbation definition} are constrained by \eqref{eq: allowd scalings} (we'll explain what this relaxation means momentarily). Doing this, we find
two points with leading $\left(0,0,0\right)$ scaling
\begin{equation} 
\label{eq: RB finite lambda largenf 0,0,0}
\begin{split} 
 \left(\lambda_{SSS},\lambda_{SAA},\lambda_{AAA}\right): & \left(\frac{2}{3}\left(5-\sqrt{5}\right)\left(\frac{\lambda}{4\pi}\right)^{2},-\frac{2}{3}\sqrt{6\sqrt{5}-10}\left(\frac{\lambda}{4\pi}\right)^{2},\frac{1}{3}\left(2\sqrt{5}-1\right)\left(\frac{\lambda}{4\pi}\right)^{2}\right)
  , \\
  & \left(\frac{2}{3}\left(5-\sqrt{5}\right)\left(\frac{\lambda}{4\pi}\right)^{2},\frac{2}{3}\sqrt{6\sqrt{5}-10}\left(\frac{\lambda}{4\pi}\right)^{2},\frac{1}{3}\left(2\sqrt{5}-1\right)\left(\frac{\lambda}{4\pi}\right)^{2}\right),
\end{split}
\end{equation}
(both mixed with $\gamma=2,1$, respectively), and an IR-stable fixed point with the leading $\left(0,1,0\right)$ scaling
\begin{equation} \label{eq: IR fixed point large nf wrong scaling}
        \left(\lambda_{SSS},\lambda_{SAA},\lambda_{AAA}\right):\ \left(\frac{8}{3}\left(\frac{\lambda}{4\pi}\right)^{2},\frac{2}{N_{f}}\left(\frac{\lambda}{4\pi}\right)^{2},3\left(\frac{\lambda}{4\pi}\right)^{2}\right) . 
\end{equation}
Overall we found 6 points as indeed was expected.

While the points in  \eqref{eq: RB finite lambda analytic} and \eqref{eq: RB finite lambda largenf 2,1,0} have one of the three allowed scalings presented in \eqref{eq: allowd scalings}, the points in \eqref{eq: RB finite lambda largenf 0,0,0} and \eqref{eq: IR fixed point large nf wrong scaling} have different scalings. 
This ``contradiction" is resolved by a careful analysis of the different limits; \eqref{eq: allowd scalings} was derived under the assumption of taking $N_f\rightarrow\infty$ at a constant $\lambda$, while \eqref{eq: beta small lambdab} assumes that $\lambda$ is the smallest parameter (and, in particular, $\lambda \ll \frac{1}{N_f^2}$).

Indeed, one can check explicitly for the IR fixed point in \eqref{eq: IR fixed point large nf wrong scaling} that the approximation which leads to \eqref{eq: beta small lambdab} breaks. Since $\lambda_{SSS}\sim \lambda^2$, the cubic term in $\beta_{SSS}$ (see \eqref{eq: beta multyflavour} and \eqref{eq: beta function free boson}), which always exists, has the scaling of $N_f^3 \lambda_{SSS}^3 \sim N_f^3 \lambda^6$, while the linear term in \eqref{eq: beta small lambdab} scales as $ \lambda^2 N_f \lambda_{SSS} \sim N_f \lambda^4$. Thus, for $\lambda \sim \frac{1}{N_f}$, the cubic term is of the same order of magnitude as some of the terms in  \eqref{eq: beta small lambdab}, and it is unjustified to ignore it. One can do the same scaling analysis for different terms in \eqref{eq: beta small lambdab}, and show that for the cubic terms to be smaller than all of them, we must demand  $\lambda \ll \frac{1}{N_f^2}$.

We therefore expect that as we increase $\lambda$ for constant, but large, values of $N_f$\footnote{Or equivalently, increase $N_f$ for constant small $\lambda$.} the 3 fixed points in \eqref{eq: RB finite lambda largenf 0,0,0} and \eqref{eq: IR fixed point large nf wrong scaling} that have ``wrong scalings'' will merge in pairs with other fixed points (or with themselves) and disappear. More concretely, for large enough $N_f$, we expect that (for each of these fixed points) there exist $l_0,\alpha>0$ with $\lambda_{crit}\simeq \frac{4\pi l_0}{N_f^\alpha}$ such that these fixed points merge with others at $\lambda=\lambda_{crit}$.

\subsubsection{Tracking the IR-stable fixed point} \label{sec: Trucking the IR fixed point at large nf}

In this section, we show explicitly how the IR-stable fixed point \eqref{eq: IR fixed point large nf wrong scaling} merges with another fixed point, as we increase $\lambda$ beyond the perturbative regime in which \eqref{eq: beta small lambdab} holds. In order to see this we take a double scaling limit in which $\lambda$ scales as $\lambda = \frac{4\pi l_0}{N_f}$, with $l_0$ a constant which does not scale with $N_f$.

The appropriate beta functions that describe this phenomenon are given by adding the cubic term in \eqref{eq: beta function free boson} to \eqref{eq: beta small lambdab}. By itself this will not yield  a well defined expression in perturbation theory.
It is not the beta function at leading order in $\lambda$, as there are also 
$\lambda^2$ corrections to the quadratic and cubic terms. Nor is it the leading order beta function in $\frac{1}{N_f}$, as it mixes different orders of $\frac{1}{N_f}$.

However, it is still useful in the double scaling limit, as it captures the leading order contribution in $\frac{1}{N_f}$ of each term separately\footnote{Consider for example, the constant term in $\beta_{SSS}$. It scales as $\frac{\lambda^4}{N_f} \sim\frac{1}{N_f^5}$. All higher corrections in $\frac{1}{N_f}$ of this term will scale with at least $\mathcal{O} \left( \frac{1}{N_f^6} \right) $. 

It is possible to prove this statement by looking at the general case directly. We can compute this term by taking \eqref{eq: beta multyflavour}, and shifting $G_{SAA}$ and $G_{SSS}$ according to \eqref{eq: general G3 coeff}. We find
\begin{equation*}
    \begin{split}\text{Constant term in }\beta_{SSS} & =6\frac{\gamma_{S}^{\prime}}{N_{f}^{2}}\left(\bar{G}_{3,3,\bar{g}_{3}=0}\right)-\delta G_{SSS}+\frac{1}{N_{f}}\frac{3\left(G_{5F}+G_{5N}\right)}{\pi^{2}G_{2}}\\
 & +\frac{1}{N_{f}}\frac{3\left(G_{4,1,F}+2G_{4,1,N}\right)\left(\bar{G}_{3,3,\bar{g}_{3}=0}\right)}{\pi^{2}G_{2}^{2}}+\frac{1}{N_{f}}\frac{4\left(\bar{G}_{3,3,\bar{g}_{3}=0}\right)^{3}}{\pi^{2}G_{2}^{3}},
\end{split}
\end{equation*}
where we denoted $\bar{G}_{3,3,\bar{g}_{3}=0}\equiv\left\langle \tilde{J}_{1}^{2}\left(-p_{1}\right)\tilde{J}_{2}^{3}\left(-p_{2}\right)\tilde{J}_{3}^{1}\left(-p_{3}\right)\right\rangle _{{\rm leading}}$.
By going over the sum, and using the $N_f \rightarrow \infty$ scaling laws presented in section \ref{sec: general case large nf limit} (and the fact that $\bar{G}_{3,3,\bar{g}_{3}=0}$ has no scaling with $N_f$) we see that for constant $\lambda$, this term must be proportional to $\frac{1}{N_f}$ for large $N_f$. Thus, if we expand the constant term in $\beta_{SSS}$ as a series in $\lambda$ and $\frac{1}{N_f}$
\begin{equation*}
    \text{Constant term in }\beta_{SSS}=\sum_{i\in2\mathbb{N}}f_{i}\left(N_{f}\right)\lambda^{i} ,
\end{equation*}
we must have for all $i$ that $f_{i}\left(N_{f}\right)$ scales as at most as $\frac{1}{N_f}$. Since we take a double scaling limit in which $\lambda \sim \frac{1}{N_f}$, the term $\frac{\lambda^4}{N_f}$ has the highest order in $\frac{1}{N_f}$ (as there are no terms with $\lambda^0$ or $\lambda^2$; otherwise, they would have appeared in \eqref{eq: beta small lambdab}).

We can now go term by term in \eqref{eq: beta multyflavour} and show that it gives the leading order in $\frac{1}{N_f}$, in a similar fashion.
}.
Now (in accordance with \eqref{eq: IR fixed point large nf wrong scaling}) we scale
\begin{equation} \label{eq: scaling limits for couplings in the double scaling limit for the IR}
    \lambda_{SSS}=\frac{y^{}_{SSS}}{N_{f}^{2}},\quad\lambda_{SAA}=\frac{y^{}_{SAA}}{N_{f}^{3}},\quad\lambda_{AAA}=\frac{y^{}_{AAA}}{N_{f}^{2}} .
\end{equation}
The beta functions (that contain the terms from both \eqref{eq: beta function free boson} and \eqref{eq: beta small lambdab}) in this scaling become
\begin{equation} \label{eq: RB beta function for the IR fixed point}
    \begin{split}
    \pi^{2}N_{c}N_{f}^{3}\beta_{SSS}= & -\frac{y^{}_{SSS}\left(y_{SSS}^{2}-384y^{}_{SSS}+1024l_0^{2}\right)}{1024}+\mathcal{O}\left(\frac{1}{N_{f}}\right),\\
\pi^{2}N_{c}N_{f}^{4}\beta_{SAA}= & \frac{y^{}_{SSS}+y^{}_{AAA}-3l_0^{2}}{4}y^{}_{SAA}+\frac{3l_0^{4}+y_{AAA}^{2}-2l_0^{2}y^{}_{SSS}-4l_0^{2}y^{}_{AAA}}{4}+\mathcal{O}\left(\frac{1}{N_{f}}\right),\\
\pi^{2}N_{c}N_{f}^{3}\beta_{AAA}= & \frac{3\left(y^{}_{AAA}-\frac{l_0^{2}}{3}\right)\left(y^{}_{AAA}-3l_0^{2}\right)}{16}+\mathcal{O}\left(\frac{1}{N_{f}}\right) ,
\end{split}
\end{equation}
which now include all the leading order contributions.

From the first of these equations we see that the possible values of $y^{}_{SSS}$ at a fixed point are 
\begin{equation}
    y^{}_{SSS}=0,\ 32\left(6-\sqrt{36-l_0^{2}}\right),\ 32\left(6+\sqrt{36-l_0^{2}}\right),
\end{equation}
and those of $y^{}_{AAA}$ are
\begin{equation}
    y^{}_{AAA}=\frac{1}{3}l_0^{2},\ 3l_0^{2} ,
\end{equation}
and for each choice $y^{}_{SAA}$ is determined (except for the choice $y_{SSS}=0, y_{AAA}=3l_0^2$, for which its beta function vanishes at leading order).

The solutions with $y^{}_{SSS}=32\left(6+\sqrt{36-l_0^{2}}\right)$, which are
\begin{equation}
    \left(\frac{32\left(6+\sqrt{36-l_{0}^{2}}\right)}{N_{f}^{2}},\frac{2\left(l_{0}^{2}-24\sqrt{36-l_{0}^{2}}+144\right)}{3N_{f}^{3}},\frac{l_{0}^{2}}{3N_{f}^{2}}\right)\ ,\ \ \left(\frac{32\left(6+\sqrt{36-l_{0}^{2}}\right)}{N_{f}^{2}},\frac{2l_{0}^{2}}{N_{f}^{3}},\frac{3l_{0}^{2}}{N_{f}^{2}}\right),
\end{equation}
correspond to two fixed points which originate from the degenerate point at $\left(\frac{384}{N_{f}^{2}},0,0\right)$\footnote{Indeed $32\left(6+\sqrt{36-l_0^{2}}\right) \rightarrow 384$ in the limit $l_0\rightarrow0$. This also justifies our claim in the end of section \ref{sec: RB lambda finite nf ge 3}.}. The other solutions describe the splitting of the free degenerate fixed point $\left(0,0,0\right)$, one of them is the IR fixed point, which corresponds to $\left(\frac{32\left(6-\sqrt{36-l_{0}^{2}}\right)}{N_{f}^{2}},\frac{2l_{0}^{2}}{N_{f}^{3}},\frac{3l_{0}^{2}}{N_{f}^{2}}\right)$\footnote{For very small $l_0$, this indeed reproduces \eqref{eq: IR fixed point large nf wrong scaling}, and one can check explicitly the derivative matrix of \eqref{eq: RB beta function for the IR fixed point} and see that this is a fixed point.}. Note that there is a degeneracy for $\beta_{SAA}$ when $y^{}_{SSS}=0,\ y^{}_{AAA}=3l_0^2$. This means that in order to determine the fixed points for this case we must go to subleading corrections in $N_f$, which will result in three different points (this is inferred by counting, as we showed in section \ref{sec: RB lambda finite nf ge 3} that there are in total six fixed points)\footnote{It is instructive to revisit the fixed points found in the regime $\lambda \ll 1$ and identify their corresponding solutions in the language used here. The second point in \eqref{eq: RB finite lambda analytic}, which is a UV fixed point, corresponds to $y_{SSS} = 0$ and $y_{AAA} = \frac{1}{3} l_0^2$. This is confirmed by examining the beta functions in \eqref{eq: RB beta function for the IR fixed point}, where the structure of the derivative matrix indicates a UV fixed point.

The first point in \eqref{eq: RB finite lambda analytic}, as well as the point in \eqref{eq: RB finite lambda largenf 2,1,0}, both correspond to the degenerate solution $y_{SSS} = 0$, $y_{AAA} = 3l_0^2$. There is another point corresponds to the degenerate solution, as well as to the case of $y^{}_{SSS}=32\left(6-\sqrt{36-l_0^{2}}\right),y^{}_{AAA}=\frac{1}{3}$. Those two correspond (by elimination of other possibilities) to the two fixed points in \eqref{eq: RB finite lambda largenf 0,0,0}. However, this correspondence is not immediately evident from the expression in \eqref{eq: RB finite lambda largenf 0,0,0}, as that result is valid only in the regime $\lambda \ll \frac{1}{N_f^2}$, while in this section we consider a different scaling.}.

As we increase $l_0$ from $0$ to $6$, the two points with $y^{}_{SSS}=32\left(6-\sqrt{36-l_0^{2}}\right)$ and the two points with $y^{}_{SSS}=\ 32\left(6+\sqrt{36-l_0^{2}}\right)$ get closer to each other. At $l_0=6$ those points merge to a degenerate point, and for $l_0>6$ they no longer exist. 

\begin{figure}[t]
    \centering
    \includegraphics[width=0.6\linewidth]{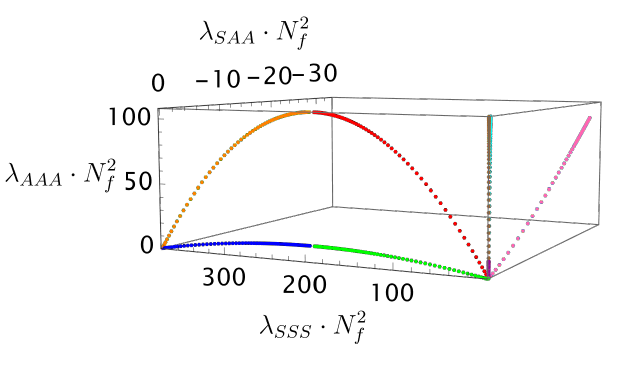}

    \caption{ The fixed points of \eqref{eq: RB beta function for the IR fixed point} (including next order corrections that break the degeneracy) at different values of $0<l_0<6$. The degenerate fixed points at $\left(0,0,0\right)$ and $\left(\frac{384}{N_f^2},0,0\right)$  (for $\lambda=0$) split and some of the resulting fixed points merge and disappear at $l_0=6$. The IR-stable fixed point is denoted in red, the other fixed points discussed in the text appear as well (note that the purple, brown and cyan fixed points appear very close to one another).
    }
    \label{fig: RB large nf merging of fixed points} 
\end{figure}

Thus, we managed to show analytically that the IR-stable fixed point disappears, at least at large values of $N_f$, at $\lambda_{crit}\simeq \frac{24 \pi}{N_f}$. Since the perturbative parameter in the equations is $\left( \frac{\lambda}{4 \pi} \right)^2 =\frac{l_0^2}{N_f^2}$, we can trust this analysis as long as $\frac{l_0^2}{N_f^2} \ll 1$. A visualization of the merging of points is presented in figure \ref{fig: RB large nf merging of fixed points} for $N_f=1000$ and different values of $\lambda$. We see that the two degenerate points for $\lambda=0$ split, and 2 pairs of points merge at the critical value of $\lambda$.

\subsubsection{The \texorpdfstring{$(2,1,0)$}{(2,1,0)} case\label{sec: 210}}

The $(2,1,0)$ scaling possesses a specific physical significance, and so here we examine this case in greater detail. For this scaling, the beta functions at leading order in $1/N_f$ and at $\lambda=0$ are given by (substituting into \eqref{eq: general case beta nf to infinity} the values in table \ref{tab: beta coefficients})
\begin{equation} \label{eq: beta functinos of y}
\begin{split}
    \pi^{2}N_{c}\frac{\beta_{y^{}_{SSS}}}{N_{f}}&=-\frac{\left(y^{}_{SAA}-384\right)y_{SAA}^{2}}{1024}+\mathcal{O}\left(\frac{1}{N_{f}}\right),\\\pi^{2}N_{c}\frac{\beta_{y^{}_{SAA}}}{N_{f}}&=\frac{y^{}_{AAA}\left(128y^{}_{AAA}-y^{}_{SAA}\left(y^{}_{AAA}-128\right)\right)}{512}+\mathcal{O}\left(\frac{1}{N_{f}}\right),\\\pi^{2}N_{c}\frac{\beta_{y^{}_{AAA}}}{N_{f}}&=-\frac{\left(y^{}_{AAA}-192\right)y_{AAA}^{2}}{1024}+\mathcal{O}\left(\frac{1}{N_{f}}\right) ,
\end{split}
\end{equation}
where the $y_n$ are given by (see table \ref{tab: beta coefficients} and the definition of $Y_n$ in \eqref{eq: large nf coupling definition spesific scaling})
\begin{equation}
    y^{}_{SSS}=N_{c}^{2}Y_{SSS}+128,\quad y^{}_{SAA}=N_{c}^{2}Y_{SAA}+128,\quad y^{}_{AAA}=N_{c}^{2}Y_{AAA}+64 .
\end{equation}

We see that the degeneracies discussed in section \ref{sec: general case large nf limit (2,1,0)} apply in this case for \eqref{eq: beta functinos of y}, since both $\gamma_S^{\prime}=0$, and the condition \eqref{eq: condition for degeneracy for large nf} is satisfied. First, solving $\beta_{y^{}_{AAA}}=0$, we find $y^{}_{AAA}=0$ or $y^{}_{AAA}=192$. In the $y^{}_{AAA}=0$ case, $\beta_{y^{}_{SAA}}$ automatically vanishes, and we cannot determine $y^{}_{SAA}$ from the equation $\beta_{y^{}_{SAA}}=0$.
Second, $y^{}_{SSS}$ does not appear in the leading order beta functions, but instead
the equation $\beta_{y^{}_{SSS}}=0$ becomes an additional condition for $y^{}_{SAA}$. Solving for $y^{}_{SAA}$, we find the following possibilities\footnote{Note that the solution $ \left(384,0\right)$ in \eqref{eq: Infinite solutions} disappears at the next order in $\frac{1}{N_f}$ (i.e, there are no perturbed solutions of the form \eqref{eq: Nf perturbation definition} around this solution), and so there are no solutions with these values.}
\begin{equation} \label{eq: Infinite solutions}
    \left(y^{}_{SAA},y^{}_{AAA}\right):\ \left(0,0\right),\ \left(384,192\right),\ \left(384,0\right) .
\end{equation}

To remove the degeneracies mentioned above, we must go to the next order of $\frac{1}{N_f}$, and apply the approach presented by \eqref{eq: Nf perturbation definition}. Doing so, we find in total 5 points - the 3 points in \eqref{eq: fixed points RB nf=3 always} and the 2 points in \eqref{eq: RB nf to infinity lambda 0 sol}. 

We see explicitly how a degeneracy adds possible solutions. In the generic case described in \eqref{sec: general case large nf limit (2,1,0)}, there should be only one fixed point with $y^{}_{AAA}=192$, corresponding to the single root in $\beta_{AAA}$ in \eqref{eq: beta functinos of y}. However, we find explicitly three such solutions - the last point in \eqref{eq: fixed points RB nf=3 always}, and the two points \eqref{eq: RB nf to infinity lambda 0 sol}.

As we turn on finite $\lambda$, our generic expectation is that the degeneracies of \eqref{eq: general case beta nf to infinity}, which appear for the $\lambda=0$ case in \eqref{eq: beta functinos of y}, will disappear.  It is easy to see that this is the case for at least one of the degeneracies: in the region $\lambda \ll \frac{1}{N_f^2}$ we have $\gamma_{S}^{\prime}/N_f \sim \lambda^2$, and this adds a linear term in $\lambda_{SSS}$ in $\beta_{SSS}$. Indeed, taking the  $N_f \rightarrow\infty$ limit of \eqref{eq: large nf coupling definition spesific scaling} in equation \eqref{eq: beta small lambdab} we find
\begin{equation} \label{eq: beta small lambdab large nf}
    \begin{split}
        \pi^{2}N_{c}\frac{\beta_{y^{}_{SSS}}}{N_{f}}&=\frac{1}{8}\left(-8y^{}_{SSS}+3\left(y^{}_{SAA}-2\left(\frac{\lambda}{4\pi}\right)^{2}\right)^{2}\right)+\mathcal{O}\left(\frac{1}{N_{f}},\lambda^{2}\right),\\
        \pi^{2}N_{c}\frac{\beta_{y^{}_{SAA}}}{N_{f}}&=\frac{1}{4}\left(y^{}_{AAA}-3\left(\frac{\lambda}{4\pi}\right)^{2}\right)\left(y^{}_{SAA}+y^{}_{AAA}-\left(\frac{\lambda}{4\pi}\right)^{2}\right)+\mathcal{O}\left(\frac{1}{N_{f}},\lambda^{2}\right),\\
        \pi^{2}N_{c}\frac{\beta_{y^{}_{AAA}}}{N_{f}}&=\frac{3}{16}\left(y^{}_{AAA}-3\left(\frac{\lambda}{4\pi}\right)^{2}\right)\left(y^{}_{AAA}-\frac{1}{3}\left(\frac{\lambda}{4\pi}\right)^{2}\right)+\mathcal{O}\left(\frac{1}{N_{f}},\lambda^{2}\right) ,
    \end{split}
\end{equation}
and there is a linear term of $y^{}_{SSS}$ in $\beta_{SSS}$.

The second degeneracy is not removed in \eqref{eq: beta small lambdab large nf}. We see that for $y^{}_{AAA}= 3\left(\frac{\lambda}{4\pi}\right)^{2}$, we have $\beta_{y^{}_{AAA}}=\beta_{y^{}_{SAA}}=0$ at leading order of $\frac{1}{N_f}$. There are thus two possibilities: the first is that this degeneracy is removed when going to the next order of perturbation theory in $\lambda$. The second is that the condition \eqref{eq: condition for degeneracy for large nf} persists to all values of $\lambda$ (in the large $N_f$ limit).

Whether the degeneracy is removed or not, we can use \eqref{eq: beta small lambdab}, \eqref{eq: beta functinos of y}, and \eqref{eq: beta small lambdab large nf} to search for the possible points with the scaling \eqref{eq: Nf perturbation definition} for $(a,b,c)=(2,1,0)$ and $\lambda \ll \frac{1}{N_f^2}$. Solving for $y^{}_{AAA}$, the three possibilities are:
\begin{enumerate}
    \item $y^{}_{AAA} = 192 +\mathcal{O} \left( \lambda^2 \right)$: For this solution, there is no degeneracy of $\beta_{SAA}$, and so by \eqref{eq: beta functinos of y} we get $y^{}_{SAA}=384+\mathcal{O} \left( \lambda^2 \right)$. Since the degeneracy in $y^{}_{SSS}$ is removed, there will be for finite $\lambda\ll1$ only one point with the leading scaling of $\left(2,1,0\right)$. One might think that since $\gamma_S^{\prime}\sim \mathcal{O}\left(\lambda^2 \right)$, the solution of $y^{}_{SSS}$ will be $y^{}_{SSS}\sim \mathcal{O}\left(\lambda^{-2} \right)$, but this is not true because the other terms in the beta function also vanish as $\lambda \to 0$ for this fixed point and depend on even powers of $\lambda$. 
    In any case, one of the 3 points with $y^{}_{AAA}=192$ and $y^{}_{SAA}=384$ (that is, the two points in \eqref{eq: RB nf to infinity lambda 0 sol} and the last point in \eqref{eq: fixed points RB nf=3 always}) will preserve the scaling of $\left(2,1,0\right)$ as we turn on $\lambda$, even beyond perturbation theory, while the other two points will presumably get a different scaling.
    \item $y^{}_{AAA}=\frac{1}{3}\left(\frac{\lambda}{4\pi}\right)^{2}$: For this solution, there is no degeneracy in \eqref{eq: beta small lambdab large nf} and so there will be only one solution with the scaling of $\left(2,1,0\right)$, and we already found this point at small $\lambda$: the second point in \eqref{eq: RB finite lambda analytic}.
    \item $y^{}_{AAA}=3\left(\frac{\lambda}{4\pi}\right)^{2}$: For this solution, there is a degeneracy in \eqref{eq: beta small lambdab large nf}, and so, at least at leading order in $\lambda^2$, there might be several points with this scaling. The possible points are the first point in \eqref{eq: RB finite lambda analytic} and the point in \eqref{eq: RB finite lambda largenf 2,1,0}. From this analysis only, it is also possible that the fixed point $\left(\frac{384}{N_{f}^{2}},0,0\right)$, splits under perturbation theory into two points with $y^{}_{AAA}=3\left(\frac{\lambda} {4\pi}\right)^{2}$, and gives two more candidates for fixed points with the scaling of $(2,1,0)$. However, from the analysis in section \ref{sec: Trucking the IR fixed point at large nf} we see that those two points have different scalings. 
    If the degeneracy is removed at higher orders in $\lambda$, then one of the points necessarily loses the scaling of $\left(2,1,0\right)$ as we go to higher orders of perturbation theory in $\lambda$. If the degeneracy remains then possibly both can have the scaling.
\end{enumerate}

\subsection{General discussion -- elimination, splitting, and merging of fixed points} \label{sec: RB general analysis complex topology}

Our objective in this section is twofold - first, we wish to gain additional understanding of the structure of solutions to the vanishing of the beta functions \eqref{eq: beta function free boson} and \eqref{eq: beta small lambdab}. Second, we develop tools to analyze what can happen to the fixed points beyond the perturbative region of small $\lambda$. 

In section \ref{sec: RB complex analysis}, we take a `static' approach by focusing on the fixed points as roots of a system of polynomials. This will involve mostly elementary complex analysis. In section \ref{sec: RB morse theory}, we look at the flow induced by the beta functions, and classify the points according to their Poincaré–Hopf index. This gives topological constraints for the possible splittings and mergings of fixed points.

\subsubsection{Complex analysis perspective} \label{sec: RB complex analysis}

We showed in section \ref{sec: general case} that the general form of the beta functions of the marginal couplings $\lambda_n$ is exactly (at order $\frac{1}{N_c}$) a system of cubic polynomials in the couplings. As such, searching for fixed points of the flow is equivalent to finding the roots of a system of three polynomials of degree three in three variables (with coefficients that depend on $\lambda$).

By Bézout's Theorem, the total number of zeros (including degeneracies) over the complex plane of three polynomials of degree three in three variables is equal to or lower than $27$\footnote{This is sometimes known as the Bézout bound. The exact statement of Bézout's Theorem for $n$ homogeneous polynomials in $n+1$ variables is that the number of solutions on the projective plane is the multiplication of their degrees. Our polynomials are not homogeneous, but using homogenization (i.e, defining $P^h\left(x_0,x_1,\cdots\right)=x_0^dP\left(\frac{x_1}{x_0},\cdots\right)$ with $d$ the degree of $p$) we can apply the theorem, any solution with $x_0\ne0$ corresponds to a complex solution in our case, while solutions with $x_0=0$ are solutions ``at infinity" and therefore the statement in the text is only an upper bound (that is generically saturated). If the polynomials have a common component, the number of solutions might be uncountable.}.
Since all the coefficients of the polynomials are real, solutions with an imaginary part must come in pairs (i.e. if $\left(\lambda_{SSS},\lambda_{SAA},\lambda_{AAA}\right)$ is a complex root of the beta functions, so is $\left(\lambda_{SSS}^{*},\lambda_{SAA}^{*},\lambda_{AAA}^{*}\right)$). Thus, the number of real solutions (including degeneracies) must be odd.

To emphasize this point, we work out explicitly the example of \eqref{eq: beta function free boson} with $N_f=3$. The algebraic degeneracy of the point $\left(0,0,0\right)$ is $2^3=8$, while for the point $\left(\frac{384}{N_{f}^{2}},0,0\right)$ the degeneracy is $2$\footnote{One can see that by Taylor expanding around these points. Around the free fixed point all the beta functions are polynomials of degree 2, while around $\left(\frac{384}{N_{f}^{2}},0,0\right)$, $\beta_{SSS/SAA}$ are linear and $\beta_{AAA}$ is of degree 2.}.
Adding the other $5$ real non degenerate fixed points, one finds in total $2+8+5=15$ real solutions. Indeed, solving \eqref{eq: beta function free boson} with $N_f=3$ over the complex plane, one finds 12 solutions with non-zero imaginary part. This analysis extends to all $N_f\ge 3$ (with different numbers of complex solutions).

We now turn to see how the solutions change under continuous changes in the coefficients of the polynomials, which in our case corresponds to turning on a finite $\lambda$. As a consequence of the pairing of points with imaginary part, real roots can't be eliminated on their own (that is, lifted from the real line to the complex plane), but must do so in pairs. That is to say, for a point to gain an imaginary part and be lifted from the real line, there must be another point with the same real part, which gains the opposite imaginary part. For this process to occur, the two points must have the same real part, before the small perturbation lifts them. This means that the only way a point can be eliminated in the process of continuously changing the polynomial coefficients, is by first merging with another point, and only after this merger the resulting degenerate point can be eliminated (see figure \ref{fig: beta functino nf=1 possabilities} for a pictorial example in one variable).

The conclusion of the analysis above, is that from the perspective of complex analysis, fixed points cannot be added or removed arbitrarily, but only in pairs, and then there must be a critical value of $\lambda$ for which there is a degenerate point.
At such a degenerate point the CFT has a marginal deformation.

A similar analysis can be done to investigate the $N_f\rightarrow\infty$ limit, and in particular the $\left(2,1,0\right)$ scaling. As discussed in section \ref{sec: 210}, generically (i.e, assuming that \eqref{eq: condition for degeneracy for large nf} does not hold, and $\gamma_S^{\prime}\neq 0 $) \eqref{eq: general case beta nf to infinity} has exactly 3 solutions (including multiplicity), but \eqref{eq: beta multyflavour} can, and indeed does, have many more solutions (see tables \ref{tab: free_bosonic_fixed_points} and \ref{tab: pertubed_bosonic_fixed_points}). It is therefore interesting to ask what happens to these solutions as $N_f\rightarrow\infty$?

To answer this question, we define $Y_n$ as in \eqref{eq: large nf coupling definition spesific scaling} and look at $\delta\equiv\frac{1}{N_f}$ as a continuous variable. For $\delta=0$, and $\lambda$ now held constant at some non-zero value, we get equation \eqref{eq: general case beta nf to infinity} exactly (without higher order corrections). It is therefore evident from the discussion above that (assuming generic behavior) only 3 solutions can have the scaling \eqref{eq: large nf coupling definition specific scaling}: if two solutions merge into one in that limit, it is counted as a multiplicity and the resulting point will be degenerate. Every other solution must either merge and be eliminated at finite $N_f$, or go to infinity in the $Y_n$ coordinates.

\subsubsection{Topological perspective} \label{sec: RB morse theory}

There is a deep connection between the local properties of the flow around the fixed points, and the overall topology of the flow. We can use such connections to restrain if and how fixed points can split, merge, or be eliminated. We start by looking at the process of degenerate point splitting, and then use the Poincaré–Hopf theorem to gain insight into global restrictions on the fixed point structure.

For each fixed point we evaluate the derivative matrix of the beta functions $\partial_i\beta_j$ for this point. The eigenvectors and eigenvalues of the $\partial_i\beta_j$ matrix characterize the structure of the flow around this fixed point. Each positive eigenvalue corresponds to an irrelevant deformation (when decreasing the energy scale) in the direction of its corresponding eigenvector, and each negative eigenvalue corresponds to a relevant deformation. If one or more of the eigenvalues is zero, we call this point a \emph{degenerate} point (the point is degenerate in that sense, if and only if it is algebraically degenerate).

Perturbations of a given flow can either split or eliminate degenerate fixed points (as discussed above), while non-degenerate fixed points are protected under infinitesimal changes of the flow\footnote{To see how perturbations can eliminate a fixed point with only one degenerate direction, it is instructive to look at the example $\beta_x=x,\beta_y=y^2$, with a degenerate fixed point at $\left(0,0 \right)$ and a perturbation that gives $\beta_x=x,\beta_y=y^2+\epsilon$, and eliminates the fixed point}. Note that due to continuity, if the original degenerate point had non-degenerate directions (i.e, non-zero eigenvalues), the signs of the eigenvalues in these directions stays the same for the resulting fixed points after the splitting.

There are also global constraints on the types and numbers of fixed points. As discussed in point \ref{item4.3: cubic are universal} of section \ref{sec: multiple falvor beta function comments}, the flow at large values of $\lambda_{n}$ is the same for every value of $\lambda$. 
One can then assign a ``Poincaré-Hopf index" to the flow at infinity, that is constant for any value of $\lambda$. The index of a point ${\rm ind}_p$ is defined as the degree of the map $u: \partial D \rightarrow S^{n-1}$, where $D$ is a closed ball around the fixed point, and $u$ is the normalized flow\footnote{Intuitively, it counts how many times the function $u$ wraps the $S^{n-1}$ manifold. To define this quantity at infinity one must compactify the plane.}.

The Poincaré–Hopf theorem asserts that the sum of the Poincaré–Hopf indices of all the fixed points + infinity is constant. Since in our case the index at infinity is constant for every $\lambda$, so is the sum of all the fixed points (not including infinity). We summarize this as
\begin{equation} \label{eq: Poincare–Hopf theory}
    \sum_{p\ fixed \ point}\text{ind}_{p}=\text{Constant in $\lambda$} .
\end{equation}
The Poincaré–Hopf index of a nondegenerate fixed point is the sign of the determinant of the $\partial_i\beta_j$ matrix, i.e $\left(-1\right)^\gamma$ where $\gamma$ is the number of negative eigenvalues \cite{Milnor1965,guillemin2010differential}. Finding the Poincaré–Hopf index of a degenerate fixed point is in general more complicated, however, in our case the index of the degenerate points is zero\footnote{We can see from \eqref{eq: beta function free boson} that in the vicinity of the free fixed point $\beta_{SSS} \ge0$ so the map $u: \partial D \rightarrow S^{2}$, of the normalized flow is not surjective, and so the degree of the map is zero (i.e. it does not wrap $S^2$). Similar arguments also hold for the other degenerate fixed point for $N_f>4$. For $N_f=3$ one can apply the Eisenbud–Levine–Khimshiashvili formula.}.

The fact that this sum is related to the flow at infinity shows why it is different for the cases $N_f=3$ and $N_f>3$ (see table \ref{tab: free_bosonic_fixed_points}): this is because the flow at infinity of these 2 cases is topologically very different, as was shown in figure \ref{fig: large L 3d}.

Another implication of \eqref{eq: Poincare–Hopf theory} is that if several points merge to a single degenerate point, then the sum of their indices is the index of the resulting point. In particular, if they annihilate, the sum of their indices must be 0. Indeed, we can check explicitly using tables \ref{tab: nf2 fixed points RB} and \ref{tab: pertubed_bosonic_fixed_points} that in our case the sum of the indices of the non-degenerate points that split from the degenerate one is 0.

The implications extracted from \eqref{eq: Poincare–Hopf theory} persist as we go beyond the perturbative regime in $\lambda$. Note that the opposite of merging can also occur -- it may be that for some finite value of $\lambda$, a new degenerate fixed point will appear, and it will then split into additional non-degenerate fixed points.

To give a concrete example of the consequences of the type of analysis above, we can compute the sum of indices for $N_f=2$ as shown in table \ref{tab: nf2 fixed points RB}, and find it is $-1$. Since for $N_f=2$, UV and IR-stable fixed points both contribute positively to this sum, this means that \emph{for any value of $\lambda$} there must be at least one mixed point.

\section{The Critical Fermion theory with \texorpdfstring{$\lambda_F=0$}{lambdaf=0}} \label{sec: CF theory}

In this section, we'll analyze the CF theory described in section \ref{RF} in the particular limit where $\lambda_F=0$ (this is the same as the $SU(N_c)$-singlet sector of the Gross-Neveu model). First, in section \ref{sec: CF calculating beta function}, we'll outline how to compute the beta function for this theory. Then, in section \ref{sec: CF fixed points and flow}, we'll analyze the flows due to the beta functions at large $N_c$, and in particular the absence of IR-stable fixed points. Finally, in section \ref{sec: CF large nf}, we will consider the large $N_f$ limit. Since this section only deals with fermion theories we'll drop the sub/superscript $F$ in all the equations that follow. 

Unlike the RB theory, we'll find that there are no degenerate fixed points for $\lambda=0$ that can split into IR-stable points, and therefore we will not need to carry out any computations for small $\lambda$.

\subsection{Computation of the \texorpdfstring{$\beta$}{beta} functions} \label{sec: CF calculating beta function}

As in section \ref{sec: general case}, we compute the effective action for $\zeta^i_j$ in terms of the correlation functions of the meson operators $M_{i}^{j}$. These correlation functions are computed in the RF theory with $\lambda=0$ (a free fermion theory) and they are therefore relatively simple. 
In fact, in this limit the 2, 3 and 5-point correlation functions are trivial to compute:
\begin{itemize}
    \item The leading term of the \textbf{two-point} function, $G_2$, is equal to its value for the $N_f=1$ case in section \ref{sec: Nf=1}, and given by $G_2=2\pi^2\frac{N_c}{\kappa_F^2}$ (see footnote \ref{foot: G2}).

    \item  Since the RF theory is free when $\lambda\rightarrow0$, we find that the anomalous dimensions are $\gamma_S=0$ and $\gamma_A=0$.

    \item The 3 and 5-point correlation functions must vanish. This follows from parity invariance, as the RF theory \eqref{eq: regular fermions} for $\lambda=0$ contains only parity conserving terms. Under parity, the mesons transform to themselves with a minus sign, and so all correlation functions with an odd number of mesons are equal to minus themselves, and so they need to be zero. In particular, this means that $\bar{G}_{3,n}=\bar{g}_{n}$ (see \eqref{eq: general G3 coeff}), and therefore $G_{SSS}=g^{}_{SSS}$, $G_{SAA}=g^{}_{SAA}$ and $G_{AAA}=g^{}_{AAA}$.

\end{itemize}

The only non-trivial correlation function is the 4-point correlation function, which (as explained in section \ref{sec: general case}) we only need to know in a specific kinematical limit. The computation is a straightforward, yet cumbersome, loop integral. We leave the full derivation to Appendix \ref{sec: CF appendix}. The result for the leading orders in $k/p$ is
\begin{equation} \label{eq: fermions CF lambda=0 4 point}
    \left(G_{4}\right)_{i_{1},i_{2},i_{3},i_{4}}^{j_{1},j_{2},j_{3},j_{4}}\left(p,-p,k,-k\right)=-\left(\frac{4\pi}{\kappa}\right)^{4}N_{c}\left(\begin{array}{c}
\left(\delta_{i_{2}}^{j_{1}}\delta_{i_{3}}^{j_{2}}\delta_{i_{4}}^{j_{3}}\delta_{i_{1}}^{j_{4}}+\delta_{i_{4}}^{j_{1}}\delta_{i_{3}}^{j_{4}}\delta_{i_{2}}^{j_{3}}\delta_{i_{1}}^{j_{2}}\right)\left(\frac{1}{8\left|p\right|}+\frac{k\cdot p}{8p^{3}}-\frac{k\cdot p}{8\left|k\right|p^{2}}-\frac{\left(k\cdot p\right)^{2}}{8\left|k\right|p^{4}}\right)+\\
\left(\delta_{i_{2}}^{j_{1}}\delta_{i_{4}}^{j_{2}}\delta_{i_{3}}^{j_{4}}\delta_{i_{1}}^{j_{3}}+\delta_{i_{3}}^{j_{1}}\delta_{i_{4}}^{j_{3}}\delta_{i_{2}}^{j_{4}}\delta_{i_{1}}^{j_{2}}\right)\left(\frac{1}{8\left|p\right|}-\frac{k\cdot p}{8p^{3}}+\frac{k\cdot p}{8\left|k\right|p^{2}}-\frac{\left(k\cdot p\right)^{2}}{8\left|k\right|p^{4}}\right)+\\
\left(\delta_{i_{3}}^{j_{1}}\delta_{i_{2}}^{j_{3}}\delta_{i_{4}}^{j_{2}}\delta_{i_{1}}^{j_{4}}+\delta_{i_{4}}^{j_{1}}\delta_{i_{2}}^{j_{4}}\delta_{i_{3}}^{j_{2}}\delta_{i_{1}}^{j_{3}}\right)\left(\frac{1}{4\left|p\right|}-\frac{\left|k\right|}{4p^{2}}\right)
\end{array}\right),
\end{equation}
where the third line is different because the large-momentum meson operators are not adjacent to each other in the cyclic ordering of the mesons around the fermion loop\footnote{As mentioned in footnote \ref{footnote: on g4 and symetrized}, we use only the symmetrized version of the 4-point function \eqref{eq: fermions CF lambda=0 4 point}, Under symmetrization, \eqref{eq: fermions CF lambda=0 4 point} has the form of the general parametrization in \eqref{eq: g4 general case}.}.

The results are summarized in table \ref{tab: beta coefficients}.  Inserting these values into \eqref{eq: beta multyflavour}, we find the beta functions
\begin{equation}\label{eq: beta function fermions}
    \begin{split}
        16\pi^{8}N_{c}\cdot\beta_{SSS}=&256\pi^{6}\left(3\lambda_{SAA}\frac{N_{f}^{2}-1}{N_{f}^2}-\lambda_{SSS}\frac{N_{f}^{2}-3}{N_f}\right)-\left(N_{f}^{3}\lambda_{SSS}^{3}+\left(N_{f}^{2}-1\right)\lambda_{SAA}^{3}\right), \\
16\pi^{8}N_{c}\cdot\beta_{SAA}=&256\pi^{6}\left(\lambda_{SSS}+2\lambda_{AAA}\frac{N_{f}^{2}-4}{N_{f}^2}+\lambda_{SAA}\frac{3}{N_f}\right) \\
&-\left(N_{f}^{2}\lambda_{SSS}\lambda_{SAA}^{2}+N_{f}\lambda_{SAA}^{3}+2\frac{N_{f}^{2}-4}{N_{f}}\lambda_{SAA}\lambda_{AAA}^{2}\right),  \\
16\pi^{8}N_{c}\cdot\beta_{AAA}=&128\pi^{6}\left(\lambda_{AAA}\frac{N_{f}^{2}-12}{N_f}+6\lambda_{SAA}\right)-\left(3N_{f}\lambda_{SAA}^{2}\lambda_{AAA}+\frac{N_{f}^{2}-12}{N_{f}}\lambda_{AAA}^{3}\right).
    \end{split}
\end{equation}
Note that the couplings $\lambda_n$ are parity-odd in this case, so parity implies that only odd powers of the couplings can appear in their beta functions. This implies that
\begin{itemize}
\item
$\left( 0,0,0 \right)$ is always a fixed point, and unlike the RB case with $\lambda_{B}=0$ in \eqref{eq: beta function free boson}, there are linear terms in $\lambda_n$ which prevent this point from being degenerate. Thus, turning on a very small but finite $\lambda$ (which would break the parity symmetry) does not change the behavior of this fixed point.

\item 
Except for the fixed point at $\left(0,0,0\right)$, all other fixed points come in pairs, and the two points in the pair are of the same type (they have the same number of relevant/irrelevant directions). In particular, this means that if the point at $\left(0,0,0\right)$ is non-degenerate as we argued above, then the sum of the Poincaré-Hopf indices will be odd.
\end{itemize}

\subsection{Analysis of the flow} \label{sec: CF fixed points and flow}

The beta functions found above can be compared to the $N_f=1$ case. In that case, $\beta_{SSS}$ decouples from the other equations, it is only dependent on $\lambda_{SSS}$ and it is equal to the result given in \eqref{eq: nf1 beta CF lambda 0}, which has an IR-stable fixed point. 

\begin{figure}[t]
    \centering
    \includegraphics[width=0.5\linewidth]{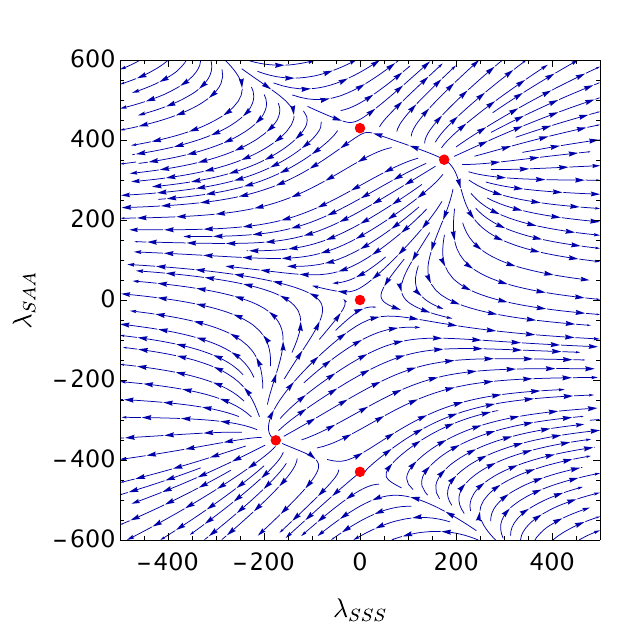}

    \caption{ Stream plot of the flow to the IR of (minus) the fermion beta functions \eqref{eq: beta function fermions} for $N_f=2$. The fixed points are shown in Red.
    }
    \label{fig: stream plot fermion nf=2} 
\end{figure}

For the case $N_f=2$ we get that $\beta_{SSS}$ and $\beta_{SAA}$ are independent of $\lambda_{AAA}$, as is expected in this case and as was also observed for the scalar theories and for the general case. The flow to the IR for this case is seen in figure \ref{fig: stream plot fermion nf=2}. The fixed points can all be found analytically as
\begin{equation}
\left(\lambda_{SSS},\lambda_{SAA}\right):\ \pm \left(4\sqrt{2}\pi^{3},8\sqrt{2}\pi^{3}\right),\ \pm \left(0,8\sqrt{3}\pi^{3}\right)
\end{equation}
(and the extra point $\left( 0,0\right)$). The first point is UV-stable, and both the second and the one at $\left( 0,0\right)$ are mixed fixed points. There is no IR-stable fixed point for this number of flavors.

The absence of an IR-stable fixed point continues as we take $N_f\ge 3$. For $N_f \ge 3$ there are $5$ or $9$ fixed points. Two of the points are given analytically by
\begin{equation} \label{eq: Fermion Nf3 analytic fixed points}
\left(\lambda_{SSS},\lambda_{SAA},\lambda_{AAA}\right):\pm \left(\frac{16\sqrt{2}\pi^{3}}{N_{f}^{2}},\frac{16\sqrt{2}\pi^{3}}{N_{f}},8\sqrt{2}\pi^{3}\right).
\end{equation}
For $N_f > 3$ the points \eqref{eq: Fermion Nf3 analytic fixed points} are UV-stable points, while for $N_f=3$ they have one degenerate direction and two relevant directions\footnote{It is possible to show analytically that the Poincaré-Hopf index of these points at $N_f=3$ is zero, as expected for a simple degeneration (where higher order derivatives do not vanish). To do so, we shift the couplings to be zero at this point, and change variables to $v_i$ such that
\begin{equation*}
    \left(\begin{array}{c}
v_{1}\\
v_{2}\\
v_{3}
\end{array}\right)=\left(\begin{array}{ccc}
\frac{9}{\sqrt{2977}} & -\frac{36}{\sqrt{2977}} & \frac{40}{\sqrt{2977}}\\
\frac{4}{\sqrt{17}} & \frac{1}{\sqrt{17}} & 0\\
-\frac{40}{\sqrt{50609}} & \frac{160}{\sqrt{50609}} & 9\sqrt{\frac{17}{2977}}
\end{array}\right)\left(\begin{array}{c}
\lambda_{SSS}\\
\lambda_{SAA}\\
\lambda_{AAA}
\end{array}\right) .
\end{equation*}
In these variables, the beta function of $v_1$ at quadratic order is $\beta_{v_{1}}=-144\sqrt{\frac{2}{2977}}\pi^{3}\left(9\lambda_{SSS}^{2}+8\lambda_{SAA}^{2}\right)$ where $\lambda_{SSS}$ and $\lambda_{SAA}$ are linear combinations of the $v_i$'s. Thus, $\beta_{v_{1}}\le0$ in the vicinity of this point, and the map $u$ is not surjective, and so the index is zero.
\label{footnote: CF degenerate point nf=3}}. Therefore, for $N_f=3$, under a small perturbation to finite $\lambda$, these points can either disappear, or split to one mixed and one UV-stable point, but not to an IR-stable fixed point. The summary of the fixed points and their type for $N_f \ge 3$ is found in table \ref{tab: fermion_fixed_points}.

Finally, we note that as discussed in point \ref{item4.3: cubic are universal} of section \ref{sec: multiple falvor beta function comments}, the cubic terms are the same as in the RB theory (up to a common pre-factor). Therefore the large $\lambda_n$ analysis done in sections \ref{sec: RB nf2 large lambda i} and \ref{sec: RB nf general large lambda i} holds in this case as well. This also implies that the sum of the Poincaré-Hopf indices, defined in section \ref{sec: RB morse theory}, should be the same as in the RB theories, as can be verified by comparing table \ref{tab: fermion_fixed_points} with table \ref{tab: free_bosonic_fixed_points}.

\begin{table}[t]
\centering
\begin{tabular}{|c||c|c|c|c|c||c|c|}
\hline 
$N_{f}$ & 
\centering
\makecell{$\gamma=0$ \\ IR-stable}
 & $\gamma=1$ & $\gamma=2$ & 
\centering
\makecell{$\gamma=3$ \\ UV stable}
 & 
\centering
\makecell{Degenerate \\ points}
&
\makecell{Total num. \\ of fixed points}
 & 
\centering
$\sum\text{ind}_{p}$
\tabularnewline
\hline 
\hline 
\textbf{3} & 0 & 0 & 3 & 0 & 2 & 5 & 3\tabularnewline
\hline 
\textbf{4-13} & 0 & 0 & 3 & 2 & 0 & 5 & 1 \tabularnewline
\hline 
\textbf{14$\leq$} & 0 & 2 & 5 & 2 & 0 & 9 & 1\tabularnewline
\hline 
\end{tabular}
\caption{Summary of the $N_f\geq 3$ fixed points in the fermionic theory with $\lambda=0$. We denote by $\gamma$ the number of relevant directions at each point. The degenerate points do not contribute to the sum in the last column (see footnote \ref{footnote: CF degenerate point nf=3}). The fact that this sum differs between $N_f=3$ and $N_f>3$, and is equal to the same sum in the RB theory (see table \ref{tab: free_bosonic_fixed_points}), is discussed in section \ref{sec: RB morse theory}.}
\label{tab: fermion_fixed_points}
\end{table}

\subsection{Solutions in the limit  \texorpdfstring{$N_f\rightarrow\infty$}{Nf}} \label{sec: CF large nf}

We can now take the $N_f \rightarrow \infty$ limit, in which it is possible to find analytically (to leading order) all nine fixed points which appear for $N_f \ge 14$. Three of the points were already found: the $\left(0,0,0 \right)$ point, and the two points given by \eqref{eq: Fermion Nf3 analytic fixed points} with the $\left(2,1,0 \right)$ scaling law.

Using the methods presented in section \ref{sec: RB large nf lambda =0}, we find six additional points. Two mixed fixed points with the leading scaling of $\left(2,1,0 \right)$ are given by
\begin{equation} \label{eq: Fermion large Nf analytic fixed points scaling 2,1,0}
    \left(\lambda_{SSS},\lambda_{SAA},\lambda_{AAA}\right):\ \pm \left(\frac{48\sqrt{6}\pi^{3}}{N_{f}^{2}},-\frac{16\sqrt{6}\pi^{3}}{N_{f}},0+\frac{96\sqrt{6}\pi^{3}}{N_{f}^{2}}\right) ,
\end{equation}
and four mixed fixed points with the leading scaling of $\left(\frac{2}{3},\frac{1}{3},0\right)$ are given by
\begin{equation}
    \begin{split}
        \left(\lambda_{SSS},\lambda_{SAA},\lambda_{AAA}\right):  \ \pm \left(\frac{-16\pi^{3}}{N_{f}^{2/3}},\frac{16\pi^{3}}{N_{f}^{1/3}},-8\sqrt{2}\pi^{3}\right),
        % \\ &
        \ 
          \pm \left(\frac{-16\pi^{3}}{N_{f}^{2/3}},\frac{16\pi^{3}}{N_{f}^{1/3}},8\sqrt{2}\pi^{3}\right).
    \end{split}
\end{equation}

Note that, although the formal scaling of the points in \eqref{eq: Fermion large Nf analytic fixed points scaling 2,1,0} is $\left(2,1,2 \right)$, this is not a contradiction to the allowed scaling laws presented in \eqref{eq: allowd scalings}, since we can classify it under the $\left(2,1,0 \right)$ scaling class. This is just a special case in which the leading order result for $\lambda_{AAA}$ happens to be zero.

In a similar fashion, it is not clear from the $\lambda=0$ analysis what is the large $N_f$ scaling of the free fixed point at the origin, and one has to check how it behaves for non-zero $\lambda$ in order to determine this.

\subsubsection{The \texorpdfstring{$\left(2,1,0\right)$}{(2,1,0)} case}

We see that there are 5 fixed points which we can classify under the scalings laws of $\left(2,1,0\right)$: The free point $(0,0,0)$, the two points in \eqref{eq: Fermion Nf3 analytic fixed points} and the two points in \eqref{eq: Fermion large Nf analytic fixed points scaling 2,1,0}. As explained in section \ref{sec: general case large nf limit (2,1,0)}, this is possible only if there is a degeneracy in the leading order (in $\frac{1}{N_f}$) terms in the beta functions.

Indeed, taking the limit $N_f\rightarrow \infty$ with the scaling of \eqref{eq: large nf coupling definition spesific scaling}, the beta functions are\footnote{In \eqref{eq: beta large nf 2,1,0} we use $y_n=N_{c}^2 Y_n$, as in \eqref{eq: Nf perturbation definition}.}
\begin{equation} \label{eq: beta large nf 2,1,0}
    \begin{split}
        16\pi^{8}N_{c}\frac{\beta_{y^{}_{SSS}}}{N_{f}}& =\left(256\pi^{6}\left(y^{}_{SSS}-3y^{}_{SAA}\right)+y^{3}_{SAA}\right)+\mathcal{O}\left(\frac{1}{N_{f}}\right),\\
        16\pi^{8}N_{c}\frac{\beta_{y^{}_{SAA}}}{N_{f}}& =2y^{}_{AAA}\left(y^{}_{SAA}y^{}_{AAA}-256\pi^{6}\right)+\mathcal{O}\left(\frac{1}{N_{f}}\right),\\
        16\pi^{8}N_{c}\frac{\beta_{y^{}_{AAA}}}{N_{f}}& =y^{}_{AAA}\left(y^2_{AAA}-128\pi^{6}\right)+\mathcal{O}\left(\frac{1}{N_{f}}\right).
    \end{split}
\end{equation}

We see from \eqref{eq: beta large nf 2,1,0} that there are 3 possible solutions for $y^{}_{AAA}$, two of them are described by \eqref{eq: Fermion Nf3 analytic fixed points} and are non-degenerate. The third solution $y^{}_{AAA}=0$ is degenerate (the condition \eqref{eq: condition for degeneracy for large nf} is satisfied for the $CF$ theory with $\lambda=0$), and one has to expand to next order in $\frac{1}{N_f}$ to find which solutions for $y^{}_{SSS}$ and $y^{}_{SAA}$ are consistent with this value of $y^{}_{AAA}$. This process results in three points (the free point, and the two points given by \eqref{eq: Fermion large Nf analytic fixed points scaling 2,1,0}).

As discussed in section \ref{sec: 210}, once we turn on a finite $\lambda$ there are two options: either the degeneracy is lifted and then only one\footnote{Note that non-zero $\lambda$ breaks the $\lambda_n\rightarrow-\lambda_n$ symmetry, and therefore it is possible that only one of the non-trivial solutions loses its scaling.} of these solutions keeps its $N_f$ scaling to all orders in $\lambda$, or it is not lifted and then it is also possible that two or all of them keep the scaling.

\section{Semi-Critical theories} \label{sec: semi critical theories}

In this section we analyze the Semi-Critical theories, presented in sections \ref{sec: CF_SA} and \ref{sec: CB_and_semicritical}, in the limits $\lambda_{F}=0$ and $\lambda_{B}=0$.
We start by presenting the general formalism of how to calculate the beta functions for them, and then analyze the results in these limits.

\subsection{Calculation of the beta functions}

The beta functions of the Semi-Critical theories defined in \eqref{eq: S_CF_S_and_A} and \eqref{eq: CB_and_semicritical} are much simpler than those of the RB and CF theories, since there is only one marginal coupling constant in each of them. 

However, one has to be wary, as the calculation of the beta functions does not directly follow from \eqref{eq: beta multyflavour}. This is due to the fact that the Lagrange multiplier which sets the field combinations $\zeta_A$ or $\zeta_S$ to zero, should be applied \emph{before} calculating the 1-loop diagrams in section \ref{sec: multiple flavors beta calculation}, and not after.

Consider, for example, the simple $G_5$ diagram on the left of figure \ref{fig: 3 point diagrams Nf1}, which contributes to the constant terms of the beta functions. The contribution of this diagram to $\beta_{G_{SSS}}$ in  \eqref{eq: beta multyflavour} comes from setting the three external legs to $\zeta_S$, and integrating over all the possible $\zeta^i_j$ fields propagating in the loop diagram. This is \emph{not} the same calculation that needs to be done in the $CF_S$ and $CB_A$ theories, since in these theories the combinations of fields in $\zeta_A$ do not propagate at all, and so the internal loop contains only the $\zeta_S$ structure (see also figure \ref{fig: beta1 cubic term posabilities}).

This has important consequences for the semiclassical boson theories with $\lambda_B=0$. In \eqref{eq: beta function free boson}, there is no constant term in the beta function, as the contributions of all the terms exactly match to cancel each other\footnote{This is of course no accident, but results from the fact that the RB theory is free when $\lambda_B=\lambda_{n}^B=0$. However, in the Semi-Critical theories at $\lambda_B=0$ one sector is a critical bosonic theory, so they are not free.}. In particular, the contribution of the $G_5$ diagram in figure \ref{fig: 3 point diagrams Nf1} cancels the term in $\delta G_3$ which originated from calculating a diagram with the same topology (see Appendix B.2.3 in \cite{Aharony:2018pjn} for the full details, Appendix \ref{subsec: App subleading G3} here, or compare the $s_{5,N/F}$ terms in $\delta G_3$ in table \ref{tab: beta coefficients} to the constants in \eqref{eq: beta multyflavour}).

To apply the Lagrange multiplier properly and project out the vanishing modes in the effective action \eqref{eq: S zeta multiple flavor step 2} before calculating the 1-loop diagrams, we define the projectors to the singlet and adjoint representation
\begin{equation}
P_{j^\prime i}^{i^\prime j}=\begin{cases}
\frac{1}{N_{f}}\delta_{j^\prime}^{i^\prime}\delta_{i}^{j} & S\\
\delta_{i}^{i^\prime}\delta_{j^\prime}^{j}-\frac{1}{N_{f}}\delta_{j^\prime}^{i^\prime}\delta_{i}^{j} & A
\end{cases}
\end{equation}
and use them on the $\zeta$ fields. This is equivalent to applying the projections on the $G_n$'s in \eqref{eq: S zeta multiple flavor step 2}. The end result is that the action of the Semi-Critical theories is similar in form to \eqref{eq: S zeta multiple flavor step 2}, with either only $\zeta_{S}$ fields or only $\zeta_{A}$ fields, and with the prefactors\footnote{Equivalently one can project only the propagator.}
\begin{equation}
    \left( G_{n}^{SC} \right)_{i_{1},i_{2},\cdots}^{j_{1},j_{2},\cdots}=\left(G_{n}\right){}_{i_{1}^{\prime},i_{2}^{\prime},\cdots}^{j_{1}^{\prime},j_{2}^{\prime},\cdots}P_{j_{1}^{\prime}i_{1}}^{i_{1}^{\prime}j_{1}}P_{j_{2}^{\prime}i_{2}}^{i_{2}^{\prime}j_{2}}\cdots .
\end{equation}
The superscript $SC$ stands for Semi-Critical effective action.

The rest of the computations go on exactly as described in section \ref{sec: multiple flavors beta calculation} but now with the $G_n$ replaced by $G_{n}^{SC}$.

For the $CB_{A}$ and $CF_{S}$ Semi-Critical theories, in which the adjoint field combinations vanish $\zeta_A=0$ (see table \ref{tab: Theories}), we find that the one loop corrected anomalous dimension of the $\zeta_S$ field is 
\begin{equation} \label{eq: anomalus S for Semi-Critical}
\gamma_{S}^{\prime}=\gamma_{S}+\frac{3G_{4,2,F}+G_{4,3,F}+6G_{4,2,N}+2G_{4,3,N}}{12\pi^{2}G_{2}^{2}N_{f}} ,
\end{equation}
and the beta function is
\begin{equation} \label{eq: beta for Semi-Critical S}
    \begin{split} 
        \beta_{G_{SSS}}&=\frac{1}{\pi^{2}G_{2}}\frac{3\left(G_{5,F}+G_{5,N}\right)}{N_{f}^{3}}-\delta G_{SSS}
        \\&-G_{SSS}\left(3\gamma_{S}^{\prime}+\frac{3\left(G_{4,1,F}+2G_{4,1,N}\right)}{2\pi^{2}G_{2}^{2}N_{f}}\right)
        %\\&
        -\frac{1}{2\pi^{2}G_{2}^{3}}N_{f}^{3}G_{SSS}^{3} .
    \end{split}
\end{equation}

Note that \eqref{eq: anomalus S for Semi-Critical} and \eqref{eq: beta for Semi-Critical S} reproduce the $N_f=1$ results in section \ref{sec: Nf=1}, as one would expect, since for $N_f=1$, the Semi-Critical singlet theories are the same as the one flavor theory (there is no adjoint structure to project out).

Also, the prefactor of the cubic and linear terms in \eqref{eq: beta for Semi-Critical S} are the same as in \eqref{eq: beta multyflavour} (ignoring the change in $\gamma_{S}^{\prime}$), as the only diagrams that contribute to $\beta_{G_{SSS}}$ and that are proportional only to the $G_{SSS}$ coupling in the non projected theories, contain only $\zeta_S$ fields as internal propagators. The anomalous dimension and the constant term in \eqref{eq: beta for Semi-Critical S} change however, as the closed loops in the right diagram in figure \ref{fig: gamma diagrams Nf1} and the left diagrams in figure \ref{fig: 3 point diagrams Nf1} contain internal propagation of $\zeta_A$ in the non projected theories.

For the $CB_{S}$ and $CF_{A}$ Semi-Critical theories, in which the singlet field combination vanishes $\zeta_S=0$, we find that the one-loop corrected anomalous dimension of the $\zeta_A$ field is
\begin{equation} \label{eq: anomalus A for Semi-Critical}
\gamma_{A}^{\prime}=\gamma_{A}+\frac{3G_{4,2,F}+G_{4,3,F}-2\left(N_{f}^{2}-1\right)\left(3G_{4,2,N}+G_{4,3,N}\right)}{12\pi^{2}G_{2}^{2}N_{f}} ,
\end{equation}
and the beta function is
\begin{equation} \label{eq: beta for Semi-Critical A}
    \begin{split} 
        \beta_{G_{AAA}}&=\frac{3\left(N_{f}^{2}-1\right)G_{5,N}-3G_{5,F}}{2\pi^{2}G_{2}N_{f}}-\delta G_{AAA}\\&-G_{AAA}\left(3\gamma_{A}^{\prime}+\frac{3\left(N_{f}^{2}-4\right)G_{4,1,N}-6G_{4,1,F}}{2\pi^{2}G_{2}^{2}N_{f}}\right)
        %\\&
        -\frac{1}{2\pi^{2}G_{2}^{3}}\frac{N_{f}^{2}-12}{N_{f}}G_{AAA}^{3} .
    \end{split}
\end{equation}

In addition to the similar comments made above regarding the cubic, linear and constant coefficients, note that for $N_f\rightarrow \infty$, the leading order results of \eqref{eq: anomalus A for Semi-Critical}, \eqref{eq: beta for Semi-Critical A} and of \eqref{eq: general case anomalus A and S},  \eqref{eq: beta multyflavour}  (setting $G_{SAA}=0$ in $\beta_{G_{AAA}}$) are the same, respectively. This stems from the fact that in $\zeta_A$ there are $N_f^2-1$ degrees of freedom and in $\zeta_S$ only 1, so projecting it out doesn't change the leading large $N_f$ behavior.

More remarkably, for $N_f=3$ the prefactor of the cubic term is positive for every $\lambda$ (see a similar discussion in item \ref{item4.3: cubic are universal} of section \ref{sec: general case}). This means that there is at least 1 (and up to 2) IR-stable fixed points \emph{for every value of $\lambda$}, and the basin of attraction of these points is all of $\lambda_{AAA}$. 

The conclusion of this section is that one can use the results obtained for the correlation functions of the mesons $G_n$ in the $CB$ and $RF$ theories to compute also the beta function in the Semi-Critical theories. In the following, we will use the results summarized in table \ref{tab: beta coefficients} (which were obtained in Appendices \ref{sec: appendix CB correlations} and \ref{sec: CF appendix}) to explicitly calculate the beta functions in the scalar and fermionic theories with $\lambda_{B/F}=0$.

\subsection{Semi-Critical Singlet theories}

In this section, we apply the results obtained for the singlet theories in \eqref{eq: anomalus S for Semi-Critical} and \eqref{eq: beta for Semi-Critical S} to the scalar and fermionic theories with $\lambda_{B}=0$ and $\lambda_{F}=0$, respectively.

\subsubsection{Scalar theories \texorpdfstring{$CB_{A}$}{CBA}}
For $CB_{A}$ theories with $\lambda_{B}=0$, the anomalous dimension is (the super-script $B$ stands for Boson)
\begin{equation} \label{eq: anomalus diemsnion Semi-Critical RB}
    \gamma_{S}^{B\prime}=-\frac{16\left(N_{f}^{2}-1\right)}{3\pi^{2}N_{c}^{B}N_{f}} ,
\end{equation}
and the beta function is
\begin{equation} \label{eq: beta funtion Semi-Critical singler RB}
2^{10}\pi^{2}N_{c}^{B}\beta_{S}^{B}=2^{21}\frac{\left(N_{f}^{2}-1\right)\left(1-6\left(s_{5,F}+s_{5,N}\right)\right)}{N_{f}^{3}}+2^{14}\frac{N_{f}^{2}-1}{N_{f}}\lambda_{SSS}+384N_{f}\lambda_{SSS}^{2}-N_{f}^{3}\lambda_{SSS}^{3} .
\end{equation}
We see explicitly that for the case $N_f=1$, $\gamma_{S}^{B\prime}$ and $\beta_{S}^{B}$ reproduce the result of \eqref{eq: nf1 beta RB lambda 0}. 
Although the beta function depends on $s_{5,N/F}$, which are unknown in the literature, the dependence is only through the sum $s_{5,N}+s_{5,F}$, which may be easier to compute than each one separately, as the sum appears also in the $N_f=1$ theory.

We already observe an interesting behavior from the anomalous dimension \eqref{eq: anomalus diemsnion Semi-Critical RB}. Specifically, since $\gamma_{S}^{B\prime} \neq 0$ for $N_f > 1$, the two-point correlation function of the mesons $\tilde{M}_S$ no longer resembles that of a free theory. As a result, even if $\lambda_{SSS} = 0$ corresponds to a fixed point, the 
resulting theory
is not equivalent to a free theory.

The beta function \eqref{eq: beta funtion Semi-Critical singler RB} is a cubic polynomial, and thus has generically either 1 or 3 zeros (see figure \ref{fig: beta functino nf=1 possabilities} and the discussion in section \ref{sec: Nf=1}). Because of the unknown quantity, we cannot find whether in general these theories have an IR-stable fixed point, however we can find the range of its values for which such an IR-stable fixed point appears. By evaluating the discriminant of the polynomials we find an IR-stable fixed point if
\begin{equation} \label{eq: Semi-Critical rf singlet condition IR}
    \frac{9N_{f}^{2}-\sqrt{3}\left(N_{f}^{2}+2\right)^{\frac{3}{2}}}{27\left(N_{f}^{2}-1\right)}< s_{5,F}+s_{5,N}<\frac{9N_{f}^{2}+\sqrt{3}\left(N_{f}^{2}+2\right)^{\frac{3}{2}}}{27\left(N_{f}^{2}-1\right)} .
\end{equation}
Since $s_{5,N}+s_{5,F}$ is just a number which is currently unknown, and the right and left sides in  condition \eqref{eq: Semi-Critical rf singlet condition IR} go to positive/negative infinity as $N_f\rightarrow\infty$, there exists a value of $N_f^{*}$ from which there will be an IR-stable fixed point for any $N_f> N_f^{*} $.

The conclusion of this section is that for large enough $N_f$, the $CB_A$ theory has a non-trivial IR-stable fixed point for $\lambda_B=0$.

\subsubsection{Fermionic theories  \texorpdfstring{$CF_S$}{CFS}}

For the $CF_S$ theories with $\lambda_F=0$, the anomalous dimension is (the super-script $F$ stands for Fermions)
\begin{equation}
    \gamma_{S}^{F\prime}=\frac{16}{3\pi^{2}N_{c}N_{f}^{F}},
\end{equation}
and the beta function is
\begin{equation} \label{eq: beta funtion Semi-Critical singlet CF}
    16\pi^{8}N_{c}^{F}\cdot\beta_{S}^{F}=\frac{512\pi^{6}}{N_{f}}\lambda_{SSS}-N_{f}^{3}\lambda_{SSS}^{3}.
\end{equation}
Once again, we see that for $N_f=1$, the beta function reproduces the beta function \eqref{eq: nf1 beta CF lambda 0} of the CF theory.

The beta function \eqref{eq: beta funtion Semi-Critical singlet CF} always exhibits an IR-stable fixed point at $\lambda_{SSS}=0$, for any $N_f$.

\subsubsection{Conclusions and duality}

We will now analyze these results in light of the dualities presented in section \ref{sec: dualities} that relate the fermion theory to the scalar theory. We found that for large enough values of $N_f$ there is a single IR-stable fixed point for both the perturbative regions of the scalar and fermion theories (i.e, in small neighborhoods of $\lambda_B=0,1$), as is the case for $N_f=1$. It is then natural to conjecture that such an IR-stable fixed point also exists for every intermediate value of $\lambda$.

For small $N_f$ the situation is more complicated because of the unknown sum $s_{5,F}+s_{5,N}$. We know that in this case there is a fixed point for $|\lambda_B|\simeq 1$ (i.e. $\lambda_F\simeq 0$) but it is yet unknown whether there is one for $\lambda_B\simeq 0$.

\subsection{Semi-Critical Adjoint theories}

In this section, we repeat the analysis in the previous section for the adjoint theories, by using \eqref{eq: anomalus A for Semi-Critical} and \eqref{eq: beta for Semi-Critical A}. We emphasize that those theories are defined only for $N_f\ge3$ (more precisely, the theory exists also for $N_f=2$, but it does not have a marginal coupling in this case).

\subsubsection{Scalar theories \texorpdfstring{$CB_{S}$}{CBS}}

For the $CB_{S}$ theories with $\lambda_{B}=0$, the anomalous dimension is
\begin{equation}
    \gamma_{A}^{B\prime}=-\frac{16}{3\pi^{2}N_{c}^{B}N_{f}},
\end{equation}
and the beta function is
\begin{equation} \label{eq: beta funtion Semi-Critical adjoint RB}
    2^{10}\pi^{2}N_{c}^{B}\beta_{A}^{B}=2^{21}\frac{1-3\left(s_{5,F}+s_{5,N}\right)}{N_{f}}+\frac{2^{16}}{N_{f}}\lambda_{AAA}+192\frac{N_{f}^{2}-12}{N_{f}}\lambda_{AAA}^{2}-\frac{N_{f}^{2}-12}{N_{f}}\lambda_{AAA}^{3} .
\end{equation}
We explicitly see that, as expected, for $N_f\rightarrow\infty$ the beta function \eqref{eq: beta funtion Semi-Critical adjoint RB} reproduces the coefficients of the beta function $\beta_{AAA}$ in \eqref{eq: beta function free boson}, and, as in the singlet theory, also here the theory is not free even for $\lambda_{AAA}=0$.

The beta function \eqref{eq: beta funtion Semi-Critical adjoint RB} results in one IR-stable fixed point if either $N_f=3$ (since the beta function in that case is monotonically increasing), or if
\begin{equation} \label{eq: Semi-Critical cba condition IR}
\frac{9N_{f}^{2}-\sqrt{3\frac{\left(3N_{f}^{2}-20\right)^{3}}{N_{f}^{2}-12}}}{108}< s_{5,F}+s_{5,N}<\frac{9N_{f}^{2}+\sqrt{3\frac{\left(3N_{f}^{2}-20\right)^{3}}{N_{f}^{2}-12}}}{108} \quad \ and\ \  N_f\ge4.
\end{equation}
Unlike the singlet case, the left hand side of \eqref{eq: Semi-Critical cba condition IR} is bounded from below as $N_f\rightarrow\infty$ by $\frac{1}{3}$, and so if $s_{5,F}+s_{5,N}>\frac{1}{3}$, there always exists a value of $N_f^{*}$ such that there will be an IR-stable fixed point for any $N_f> N_f^{*} $, but if $s_{5,F}+s_{5,N}<\frac{1}{3}$ there is no IR-stable fixed point for large enough $N_f$.

\subsubsection{Fermionic theories  \texorpdfstring{$CF_A$}{CFA}}

For the $CF_A$ theories with $\lambda_F=0$, the anomalous dimension is
\begin{equation}
\gamma_{A}^{F\prime}=\frac{4\left(N_{f}^{2}-4\right)}{3\pi^{2}N_{c}^{F}N_{f}} ,
\end{equation}
and the beta function is
\begin{equation} \label{eq: beta funtion Semi-Critical adjoint CF}
    16\pi^{8}N_{c}^{F}\cdot\beta_{A}^{F} = 128\pi^{6}\frac{N_{f}^{2}-10}{N_{f}}\lambda_{AAA}-\frac{N_{f}^{2}-12}{N_{f}}\lambda_{AAA}^{3} .
\end{equation}
Once again, we see that in the $N_f\rightarrow\infty$ limit the beta function \eqref{eq: beta funtion Semi-Critical adjoint CF} reproduces the coefficients of  $\beta_{AAA}$ in \eqref{eq: beta function fermions} for that limit. This beta function has 2 IR-stable fixed points (and one UV-stable) for $N_f=3$, and a single IR-stable point at $\lambda_{AAA}=0$ for any $N_f>3$.

\subsubsection{Conclusions and duality}

Again we analyze these results in light of the dualities of section \ref{sec: dualities}. As was noted below \eqref{eq: beta for Semi-Critical A}, for $N_f=3$ there is an IR-stable point for every value of $\lambda_B$. However, we see that for $\lambda_B=0$ there is only one such point, while for $|\lambda_B|=1$ there are two. Thus, in this case an IR-stable point \emph{emerges} as one increases $\lambda_B$. This theory is also the only system we find with two IR-stable fixed points.

For $N_f>3$, at $|\lambda_B|=1$ there is an IR-stable fixed point. Whether this point emerges for some critical value of $\lambda_B$ or exists for any value of $\lambda_B$ is unknown, due to the unknown factor $s_{5,F}+s_{5,N}$. In particular, if $\frac{1}{3}<s_{5,F}+s_{5,N}<\frac{2}{3}$ then there would be an IR-stable fixed point for all the (perturbative regions of the) Semi-Critical theories (singlet and adjoint) and for every value of $N_f$.

\section{Conclusions and future directions} \label{sec: summary}

The main result of this paper is the computation of the beta functions for the marginal couplings in the RB, CF, and Semi-Critical theories, for a general flavor number $N_f$, at first non-trivial order in $\frac{1}{N_c}$. For small $|\lambda|$ (in one of the descriptions of these theories) we compute the beta functions explicitly, and we also analyze their general structure for all values of $\lambda$. Using the dualities described in section \ref{sec: dualities}, our explicit computations describe the beta functions of the RB/CF theory in the limits $|\lambda_B| \to 0,1$. Adopting this viewpoint, we describe the beta functions in the rest of this section in terms of $\lambda_B$.

For the RB/CF theory, we find that the beta function behavior for general values of $N_f$ is qualitatively different from the $N_f=1$ case described in \cite{Aharony:2018pjn}. For $|\lambda_B|=1$ there are no IR-stable fixed points for any $N_f>1$, in contrast to the $N_f=1$ case. Moreover, we find that for $\lambda_B\rightarrow0$ there is an IR-stable fixed point for any $N_f\ne3,4$, which means that the structure of the fixed points changes as we go between $0<\lambda_B<1$. In particular, this means that there is a critical value $\lambda_{B,crit}$ for which the IR-stable fixed point merges with another fixed point and then disappears. 

We were able to find this value, and to show the merging explicitly, for large (but finite) values of $N_f$. For small $N_f$ this merger happens outside the perturbative regime, so its analysis requires an expression for the beta functions beyond the perturbative regime. 

As in the $N_f=1$ case, we were able to construct a general expression for the beta functions, as a third degree polynomial in the marginal couplings, which depends on some unknown $n$-point correlation functions of meson operators. 
Nevertheless this formal expression gives meaningful results that are true for any $\lambda_B$, such as the
large $\lambda_n$ behavior, and restrictions on types and numbers of possible fixed points.

We also managed to find general restrictions in the large $N_f$ limit, in which we showed that the fixed points must scale in some way with $N_f$, and we showed that for the natural scaling of $\left(2,1,0 \right)$ there are generically three fixed points. However, we showed that this is not the case for  $|\lambda_B|=0,1$, and we conjectured that this is resolved as we go to higher orders of perturbation theory in $\lambda_{B,F}$\footnote{Alternatively, if one finds that the beta functions \emph{do} degenerate, then this would imply a non-trivial relation between the 5 and 3-point functions and the anomalous dimension, see \eqref{eq: condition for degeneracy for large nf}.}.

For both Semi-Critical theories, we found there is always an IR-stable fixed point for $|\lambda_B|=1$. For $\lambda_B=0$ the situation is more complex, as the beta function depends on the unknown values of the 5-point function. An exception is the $N_f=3$ Adjoint ($CF_A/CB_S$) theories, for which we were able to prove that there exists an IR-stable fixed point \emph{for every value of $\lambda_B$}. It is therefore evident that finding the 5-point function would greatly contribute to the understanding of the Semi-Critical theories as well as the RB/CF theories. 

One obvious direction for future work is the computation of the $n$-point correlation functions of the meson operators at large $N_c$, which enter the beta functions of the theory. A complete understanding of these correlation functions for arbitrary values of $\lambda$ would enable the determination of the beta functions even in the strongly coupled regime. Some conjectures about the structures of these correlation functions for arbitrary values of $\lambda$ appeared in \cite{Jain:2022ajd}.

A particular concrete computation that can be pursued involves the 5-point function of mesons. 
It would be interesting to compute it even just at $\lambda_B=0$, since this would tell us if the Semi-Critical theories possess IR fixed points in that case. If they do, then one could conjecture that the fixed points survive for all $\lambda$; while if they don't, then it would mean that new fixed points must appear as $\lambda_B$ is increased, and those theories could serve as simple candidates for scenarios in which fixed-point merging occurs.

In addition, computing the correlation functions of the CF theory at finite but small $\lambda_F \ll 1$, or evaluating the beta function of the RB theory at small $\lambda_B \ll 1$ to the next order in perturbation theory, would allow us to test whether the degeneracy we identified in the $N_f \rightarrow \infty$ limit for the $\left(2,1,0\right)$ scaling at weak coupling stops holding (i.e. the condition \eqref{eq: condition for degeneracy for large nf} is not satisfied) at higher orders in perturbation theory, as we propose, or whether this limit introduces a new type of symmetry in the theory, that implies \eqref{eq: condition for degeneracy for large nf}.

In this paper we have not used the approximate higher-spin symmetry of the CS-matter theories, and it would be interesting to generalize the analysis of \cite{Maldacena:2012sf} to the case of higher $N_f$, where in addition to the energy-momentum tensor there is another spin-2 $SU(N_f)$-adjoint operator that becomes conserved in the large-$N_c$ limit (and similarly for all other spins). Proving that the large $N_c$ 3-point functions of the ``quasi-fermionic'' theories in this case are still uniquely determined by one parameter (corresponding to $\lambda$), and that the ``quasi-bosonic'' ones are determined up to four parameters (corresponding to $\lambda$ and the three marginal deformations), would provide more evidence for the $N_f>1$ dualities between the scalar and fermion theories. It would be particularly interesting to compute the anomalous dimension of the extra spin-2 flavor-adjoint operator in all of these theories at leading order in $1/N_c$.

Additional computations that would be interesting to do for general values of $N_f$ involve the phase diagram of the theory as a function of its three relevant operators, that can be accessed by computing the large-$N_c$ effective potential (as done for $N_f=1$ in \cite{Aharony:2018pjn,Dey:2018ykx}). The naive expectation is that by tuning two of the three relevant operators one should be able to flow from the RB/CF theories to the Semi-Critical Adjoint or Semi-Critical Scalar theories, while by tuning one operator one should be able to flow to the CB/RF theory. However, this may not be true for all values of the marginal couplings, and in addition some values of these couplings may correspond to unstable theories (as found for $N_f=1$ in \cite{Aharony:2018pjn}). 

It would be interesting to understand what happens to all these fixed points when $N_f$ increases to become of order $N_c$ or larger, though it is no longer known how to sum all the large-$N_c$ diagrams in this case. In addition, it would be interesting to understand the flows from QCD-Chern-Simons theories to all these theories; currently we do not have the tools to sum all the planar diagrams involved, even in the large $N_c$ limit for small $N_f$, but it may still be possible to understand these flows, at least at small $\lambda$.

Our analysis in this paper can be generalized to many other theories, and in particular to theories with both scalars and fermions. For some values of the couplings one has supersymmetric theories with various amounts of supersymmetry, and the beta functions for this case were computed for all values of $N_f$ in \cite{Aharony:2019mbc}, but there should also be many non-supersymmetric fixed points, and it would be interesting to know if some of them are IR-stable.

\acknowledgments
 The authors would like to thank Netanel Barel, Sachin Jain, Trivko Kukolj, Jiangyuan Qian and Shimon Yankielowicz for useful
discussions. This work was supported in part by ISF grant no. 2159/22, by Simons Foundation grant 994296 (Simons Collaboration on Confinement and QCD Strings), by the Minerva foundation with funding from the Federal German Ministry for Education and Research, and by the German Research Foundation through a German-Israeli Project Cooperation (DIP) grant ``Holography and the Swampland''. OA is the Samuel Sebba Professorial Chair of Pure and Applied Physics.

\appendix

\section{Calculations for finite \texorpdfstring{$\lambda_B$}{lambda }} \label{sec: app finite lambda B}

In this appendix we'll compute the beta function of the RB theory in the weak 't Hooft coupling limit. First, we'll lay out our conventions for the CS action and the $SU(N_c)$ generators, from which we'll derive the Feynman rules of the theory. Then, we'll compute the beta function to first non-vanishing order (in the 't Hooft large $N_c$ expansion) when $\lambda_B=0$. The result is valid to all orders in the $\phi^6$ couplings $\lambda_n^B$. 

We then move on to perturbative computations when both $\lambda_B$ and $\lambda^B_n$ are small but non-zero. We compute the beta function to first non-vanishing order in both of these couplings, for any value of $N_c$ and $N_f$.
Since this appendix deals only with the RB theory, we drop the sub/superscript $B$ in all of the following equations (e.g., write $\lambda$ instead of $\lambda_B$).

\subsection{The action and Feynman rules}
\subsubsection{Conventions for generators}

In the calculation below, we use a convention in which the generators of the $SU(N_c)$ algebra are anti-hermitian, which is $\left(T_{ab}^{\bar{a}}\right)^{\dagger}=-T_{ab}^{\bar{a}}$, orthogonal with respect to the Killing form $\text{tr}\left(T^{\bar{a}}T^{\bar{b}}\right)\propto\delta^{\bar{a}\bar{b}}$, and with the structure constants defined by $\left[T^{\bar{b}},T^{\bar{c}}\right]=-f^{\bar{a}\bar{b}\bar{c}}T^{\bar{a}}$.

The generators of $SU(N_c)$ obey \cite{Aharony:2011jz,Anninos:2014hia}
\begin{equation}
    T_{ij}^{\bar{a}}T_{kl}^{\bar{a}}=C_{3}I_{ij;kl};\quad I_{ij;kl}\equiv \delta_{il}\delta_{kj}-\frac{1}{N_{c}}\delta_{ij}\delta_{kl} ,
\end{equation}
where $C_3<0$, and so
\begin{equation}
\text{tr}\left(T^{\bar{a}}T^{\bar{b}}\right)=C_{3}\delta^{\bar{a}\bar{b}}; \quad f^{\bar{a}\bar{b}\bar{c}}=-\frac{1}{C_{3}}\text{tr}\left(T^{\bar{a}}\left[T^{\bar{b}},T^{\bar{c}}\right]\right).
\end{equation}
In the following, we use the normalization $C_3=-1$.

\subsubsection{The Chern-Simons action}

The normalization for the Chern-Simons action which we use is
\begin{equation}
    S_{CS}=\frac{\kappa}{4\pi}\intop d^{3}x\left(-\frac{i}{2}\epsilon^{\mu\nu\rho}A_{\mu}^{\bar{a}}\partial_{\nu}A_{\rho}^{\bar{a}}-\frac{i}{6}\epsilon^{\mu\nu\rho}f^{\bar{a}\bar{b}\bar{c}}A_{\mu}^{\bar{a}}A_{\nu}^{\bar{b}}A_{\rho}^{\bar{c}}\right).
\end{equation}
In this normalization, the level $\kappa$ for the $SU(N_c)$ gauge group is an integer. We now rescale the gauge field $A\rightarrow \sqrt{\frac{4\pi}{\left|\kappa\right|}}A$ such that
\begin{equation}
    S=\text{sign}\left(\kappa\right)\intop d^{3}x\left(\frac{-i}{2}\epsilon^{\mu\nu\rho}A_{\mu}^{\bar{a}}\partial_{\nu}A_{\rho}^{\bar{a}}-\frac{i}{6}\sqrt{\frac{4\pi}{\left|\kappa\right|}}f^{\bar{a}\bar{b}\bar{c}}A_{\mu}^{\bar{a}}A_{\nu}^{\bar{b}}A_{\rho}^{\bar{c}}\right) . 
\end{equation}

The full action also contains gauge-fixing and ghost terms. The latter will not play a role in our calculations (they only appear at higher orders, see \cite{Aharony:2011jz}). We work in the Landau gauge, in which the gluon propagator is given by \cite{Aharony:2011jz,Anninos:2014hia}
\begin{equation}
    \left\langle A_{\mu}^{\bar{a}}\left(p\right)A_{\nu}^{\bar{b}}\left(-p\right)\right\rangle =-\delta^{\bar{a}\bar{b}}\epsilon_{\mu\nu\lambda}\frac{p^{\lambda}}{p^{2}} .
\end{equation}
The non-linear term in the Chern-Simons action generates an interaction vertex with no momentum dependence
\begin{equation}
    \left\langle A_{\mu}^{\bar{a}}A_{\nu}^{\bar{b}}A_{\rho}^{\bar{c}}\right\rangle_{vertex} =i\sqrt{\frac{4\pi}{\left|\kappa\right|}}\epsilon_{\mu\nu\rho}f^{\bar{a}\bar{b}\bar{c}}.
\end{equation}

\subsubsection{The scalar kinetic term}

The kinetic term of the scalars in the bosonic theory \eqref{eq: bare action} can be written as
\begin{equation} \label{eq: kinetic bosonic theory}
\begin{split}
    S_{B}(A,\phi)=&\intop d^{3}x\space \left(D^{ba}_{\mu}\phi{}_{i,a}\right)^{\dagger}\left(D^{\mu}_{bc}\phi^{i,c}\right) \\
    =& \intop d^{3}x\ \left( \partial_{\mu}\bar{\phi}_{a,i}\partial_{\mu}\phi_{a,i}+T_{ba}^{\bar{a}}A_{\mu}^{\bar{a}}\left(\partial_{\mu}\bar{\phi}_{b,i}\phi_{a,i}-\bar{\phi}_{b,i}\partial_{\mu}\phi_{a,i}\right)-\left\{ T^{\bar{b}},T^{\bar{a}}\right\} _{cb}A_{\mu}^{\bar{b}}A_{\mu}^{\bar{a}}\bar{\phi}_{c,i}\phi_{b,i} \right) ,
\end{split}
\end{equation}
where the covariant derivative is given by $D_{\mu}=\partial_{\mu}+T^{\bar{a}}A_{\mu}^{\bar{a}}$.
The propagator of the scalar is given by
\begin{equation}
    \left\langle \phi_{a,i}\left(p\right)\bar{\phi}_{b,j}\left(-p\right)\right\rangle =\frac{\delta_{a,b}\delta_{i,j}}{p^{2}} ,
\end{equation}
and the interaction terms which follows from \eqref{eq: kinetic bosonic theory}, after rescaling $A\rightarrow \sqrt{\frac{4\pi}{\left|\kappa\right|}} A$, are
\begin{equation}
    \begin{split}
        \left\langle A_{\mu}^{\bar{a}}\bar{\phi}_{a,i}\left(p_{1}\right)\phi_{b,j}\left(p_{2}\right)\right\rangle_{vertex} &=i\sqrt{\frac{4\pi}{\left|\kappa\right|}}\left(p_{1}-p_{2}\right)_{\mu}T_{ab}^{\bar{a}}\delta_{i,j}, \\
        \left\langle A_{\mu}^{\bar{a}}A_{\nu}^{\bar{b}}\bar{\phi}_{a,i}\phi_{b,j}\right\rangle_{vertex} &=\frac{4\pi}{\left|\kappa\right|}\left\{ T^{\bar{a}},T^{\bar{b}}\right\} _{ab}\delta_{\mu\nu}\delta_{i,j}.
    \end{split}
\end{equation}

\subsubsection{The scalar self interaction term}

The last term in the RB action is the scalar self-interactions, given in \eqref{eq: boson interaction}, which we repeat here for clarity
\begin{equation} \label{eq: copy of boson self interaction}
    \mathcal{L}_{int}=\frac{\bar{g}_1}{3!} \left(\bar{\phi}_{c,i}\phi^{c,i}\right)^{3} +\frac{\bar{g}_2}{2} \bar{\phi}_{a,i}\phi^{a,i}\bar{\phi}_{b,j}\phi^{b,k}\bar{\phi}_{c,k}\phi^{c,j} +\frac{\bar{g}_3}{3} \bar{\phi}_{a,i}\phi^{a,j}\bar{\phi}_{b,j}\phi^{b,k}\bar{\phi}_{c,k}\phi^{c,i} .
\end{equation}
This leads to the following vertex:
\begin{equation} \label{eq: feynman diagram vertex 6 point function}
    \begin{split}
        \left\langle \bar{\phi}_{a_{1},i_{1}}\phi^{a_{2},i_{2}}\bar{\phi}_{a_{3},i_{3}}\phi^{a_{4},i_{4}}\bar{\phi}_{a_{5},i_{5}}\phi^{a_{6},i_{6}}\right\rangle_{vertex} =&-\bar{g}_1\left(\delta_{a_{1}a_{2}}\delta_{a_{3}a_{4}}\delta_{a_{5}a_{6}}\delta_{i_{1}i_{2}}\delta_{i_{3}i_{4}}\delta_{i_{5}i_{6}}+\left(\text{5 terms}\right)\right) \\
        &-\bar{g}_2\left(\delta_{a_{1}a_{2}}\delta_{a_{3}a_{4}}\delta_{a_{5}a_{6}}\delta_{i_{1}i_{2}}\delta_{i_{3}i_{6}}\delta_{i_{5}i_{4}}+\left(\text{17 terms}\right)\right) \\
        &-\bar{g}_3\left(\delta_{a_{1}a_{2}}\delta_{a_{3}a_{4}}\delta_{a_{5}a_{6}}\delta_{i_{1}i_{4}}\delta_{i_{3}i_{6}}\delta_{i_{5}i_{2}}+\left(\text{11 terms}\right)\right).
    \end{split}
\end{equation}

\subsubsection{Renormalization Conditions}

The renormalization conditions chosen in this paper are such that, at the energy scale we are working with, the connected six-point correlation functions of $\phi$ are given by
\begin{equation} \label{eq: tree level correlation function indices}
\begin{array}{c}
\bar{\Gamma}_{1}^{\left(6\right)}\equiv\left\langle \bar{\phi}_{a,i}\phi^{a,i}\bar{\phi}_{b,j}\phi^{b,j}\bar{\phi}_{c,k}\phi^{c,k}\right\rangle _{\text{amp}}=-\bar{g}_{1}\\
\bar{\Gamma}_{2}^{\left(6\right)}\equiv\left\langle \bar{\phi}_{a,i}\phi^{a,i}\bar{\phi}_{b,j}\phi^{b,k}\bar{\phi}_{c,k}\phi^{c,j}\right\rangle _{\text{amp}}=-\bar{g}_{2}\\
\bar{\Gamma}_{3}^{\left(6\right)}\equiv\left\langle \bar{\phi}_{a,i}\phi^{a,j}\bar{\phi}_{b,j}\phi^{b,k}\bar{\phi}_{c,k}\phi^{c,i}\right\rangle _{\text{amp}}=-\bar{g}_{3}
\end{array},\quad\text{no sum over all indices,\ \  }\begin{array}{c}
i\neq j\neq k\\
a\neq b\neq c
\end{array},
\end{equation}
where the subscript $\text{amp}$ indicates a connected amputated diagram, and all momenta are taken to lie at the same scale.

In the 't Hooft limit one can contract the color indices to get the \emph{leading} $N_c$ expressions
\begin{equation} \label{eq: tree level corelation function indices leading order}
\begin{array}{c}
\left\langle \bar{\phi}_{a,i}\phi^{a,i}\bar{\phi}_{b,j}\phi^{b,j}\bar{\phi}_{c,k}\phi^{c,k}\right\rangle _{\text{amp}}=-N_{c}^{3}\bar{g}_{1}\\
\left\langle \bar{\phi}_{a,i}\phi^{a,i}\bar{\phi}_{b,j}\phi^{b,k}\bar{\phi}_{c,k}\phi^{c,j}\right\rangle _{\text{amp}}=-N_{c}^{3}\bar{g}_{2}\\
\left\langle \bar{\phi}_{a,i}\phi^{a,j}\bar{\phi}_{b,j}\phi^{b,k}\bar{\phi}_{c,k}\phi^{c,i}\right\rangle _{\text{amp}}=-N_{c}^{3}\bar{g}_{3}
\end{array},\quad\text{no sum over flavor indices, }i\neq j\neq k.
\end{equation}
Alternatively, one can consider the six-point correlation function in terms of the representation basis. The renormalization conditions for the three mesons (at leading order in $\frac{1}{N_c}$) are
\begin{align} \label{eq: tree level corelation function representation}
\Gamma_{SSS}^{\left(6\right)}&\equiv\left\langle \tilde{M}_{S}\tilde{M}_{S}\tilde{M}_{S}\right\rangle _{\text{amp}}=-N_{c}^{3}N_{f}^{3}g^{}_{SSS} , \nonumber \\
\Gamma_{SAA}^{\left(6\right)}&\equiv\left\langle \tilde{M}_{S}\left(\tilde{M}_{A}\right)_{j}^{i}\left(\tilde{M}_{A}\right)_{i}^{j}\right\rangle _{\text{amp}}=-N_{c}^{3}N_{f}\left(N_{f}^{2}-1\right)g^{}_{SAA},\\
\nonumber\Gamma_{AAA}^{\left(6\right)}&\equiv\left\langle \left(\tilde{M}_{A}\right)_{j}^{i}\left(\tilde{M}_{A}\right)_{k}^{j}\left(\tilde{M}_{A}\right)_{i}^{k}\right\rangle _{\text{amp}}=-N_{c}^{3}\frac{\left(N_{f}^{2}-1\right)\left(N_{f}^{2}-4\right)}{N_{f}}g^{}_{AAA}.
\end{align}
At leading order, choosing \eqref{eq: tree level corelation function representation} is equivalent to choosing \eqref{eq: tree level corelation function indices leading order}, as can be shown by projecting \eqref{eq: tree level corelation function indices leading order} to the representation basis, and using \eqref{eq: transform to adjoint0}. 

\subsection{The \texorpdfstring{$\lambda = 0$}{lambda = 0} case} \label{sec: app lambdab=0}

In this section we calculate the $\beta$ function of the scalar self-interaction terms at leading order in $\frac{1}{N_c}$, in the 't Hooft limit where the couplings scale as $g_n\sim\frac{1}{N_c^2}$. 
The computation is a simple generalization of the single flavor $N_f=1$ case studied long ago in \cite{Pisarski:1982vz} (for real scalars). First, we present the answer for the $N_f=1$ case, and then we generalize this to multiple flavors.

Note that in this section we use a momentum cutoff regularization, unlike in the next section, in which dimensional regularization is used.

\subsubsection{Calculation for $N_f=1$}

For the $\lambda=0$, $N_f=1$ case, the Euclidean Lagrangian with UV momentum cutoff $\Lambda$ is
\begin{equation}
\mathcal{L}=\partial_{\mu}\bar{\phi}\partial_{\mu}\phi+\frac{g^{}_{SSS}}{3!}\left(\bar{\phi}\phi\right)^{3}|_{\Lambda},
\end{equation}
where the color summations are implicit (as was noted before, for $N_f=1$ only the $g^{}_{SSS}$ structure exists). The connection between the regularized and bare scalar field is given by $\phi_{B}=\sqrt{Z}\phi$, and the Lagrangian is
\begin{equation}
\mathcal{L}=\partial_{\mu}\bar{\phi}\partial_{\mu}\phi+\frac{g^{}_{SSS}}{3!}\left(\bar{\phi}\phi\right)^{3}+\left(Z-1\right)\partial_{\mu}\bar{\phi}\partial_{\mu}\phi+\frac{1}{3!}\left(Z^{3}g^{}_{SSS,B}-g^{}_{SSS}\right)\left(\bar{\phi}_{B}\phi_{B}\right)^{3} ,
\end{equation}
where the second term provides the tree-level six point correlation function \eqref{eq: tree level correlation function indices}, and the last two terms serve as counter-terms. 

There are only two diagrams that contribute to the beta function $\beta_{SSS}$ at leading order, shown in figure \ref{fig: two feyman diagrams for nf=1 lambda=0}. The first, which is a 4-loop diagram, contributes to the zero-momentum amputated correlator $\bar\Gamma^{(6)}_1$ a UV divergence
\begin{equation}
\begin{split}
    A0 =& \left(-g^{}_{SSS}\right)^{3}N_{c}^{3}\intop\frac{d^{3}q_{1}}{\left(2\pi\right)^{3}}\frac{d^{3}q_{2}}{\left(2\pi\right)^{3}}\frac{d^{3}q_{3}}{\left(2\pi\right)^{3}}\frac{d^{3}k}{\left(2\pi\right)^{3}}\frac{1}{q_{2}^{2}\left(k-q_{2}\right)^{2}q_{3}^{2}\left(k-q_{3}\right)^{2}q_{1}^{2}\left(k-q_{1}\right)^{2}} \\
    =&-g_{SSS}^{3}N_{c}^{3}\frac{1}{2^{10}\pi^{2}}\ln(\Lambda) .
\end{split}
\end{equation}
The second, which is a two-loop diagram, contributes
\begin{equation}
    A8a=6g_{SSS}^{2}N_{c}\intop\frac{d^{3}q}{\left(2\pi\right)^{3}}\frac{d^{3}k}{\left(2\pi\right)^{3}}\frac{1}{q^{2}k^{2}\left(q+k\right)^{2}}=6\cdot g_{SSS}^{2}N_{c}\frac{1}{16\pi^{2}}\ln(\Lambda) .
\end{equation}
At leading order, there are no diagrams which contribute to $Z-1$, and so we can set $Z=1$. We find from the renormalization condition that:
\begin{equation} \label{eq: counterterm and loop for nf=1 lam=0 case}
    -\left(Z^{3}g^{}_{SSS,B}-g^{}_{SSS}\right)+A0+A8a=0
\end{equation}
and so, by taking a derivative with respect to $\Lambda$,
\begin{equation}
2^{10}\pi^{2}\beta_{g^{}_{SSS}}=2^{6}\cdot6N_{c}g_{SSS}^{2}-g_{SSS}^{3}N_{c}^{3} .
\end{equation}
Finally, in the 't Hooft limit $\lambda_{SSS}\equiv g^{}_{SSS} N_c^2$ is a finite quantity, and by writing the $\beta$ function for $\lambda_{SSS}$ we get
\begin{equation} \label{eq: beta function  regular boson  lammbdab=0 nf=1}
    2^{10}\pi^{2}\beta_{SSS}=\frac{2^{6}\cdot6\lambda_{SSS}^{2}-\lambda_{SSS}^{3}}{N_{c}}.
\end{equation}
The form of the $\frac{1}{N_c}$ leading term of the $\beta$ function is cubic in the coupling, which must be the case, as proved in \cite{Aharony:2018pjn} and reviewed in section \ref{sec: Nf=1} of this paper. 
The proof that only the two diagrams contribute to the $\beta$-function proceeds as follows. To calculate the $\beta$ function, we consider amputated diagrams with $E = 6$ external legs, as well as possible contributions from the field strength renormalization $Z$. We are interested in contributions to the 6-point correlation functions (see \eqref{eq: tree level correlation function indices}) that scale like $\log (\Lambda)$, without any additional powers of the cutoff $\Lambda$, that is, diagrams that do not contain factors of $\Lambda$. Thus, we ignore diagrams with self-connected vertices, and so for a diagram with $E$ external legs, $I$ internal legs, and $V$ 6-point interaction vertices, we have the relation
\begin{equation}
    E+2I=6V .
\end{equation}

For a diagram with $E=6$ external legs, we have the relation $I=3V-3$. A factor of $N_c$ comes when summing over the color index of two internal lines, so given $I$ internal lines, the maximum power of $N_c$ a diagram can produce is $\left\lfloor \frac{I}{2}\right\rfloor $. The vertices contribute $N_c^{-2V}$ (since $g=\frac{\lambda}{N_c^2}$), so the total power of $N_c$ of a Feynman diagram is (the plus two is since $\beta_{SSS} = N_c^2 \beta_{g_{SSS}}$)
\begin{equation}
    \text{\# maximal power of $N_c$} = \left\lfloor \frac{I}{2}\right\rfloor -2V+2=\left\lfloor \frac{-V+1}{2}\right\rfloor ,
\end{equation}
and so there will be contributions to subleading order only for $V=2$ or $V=3$. Furthermore, in order to obtain the maximal power of $N_c$, all the available propagators of $\phi$ must be paired to give factors of $N_c$. This leaves only the two diagrams drawn in figure \ref{fig: two feyman diagrams for nf=1 lambda=0}\footnote{An example with ``unpaired'' propagators is diagram $A8b$, shown in figure \ref{fig: diagrmas A}. The contribution of this diagram is subleading to the $A8a$ diagram.}.

Next, we claim that there is no contribution of the field strength renormalization $Z$ at the leading order in $\frac{1}{N_c}$. For such diagrams, the number of external lines is $E=2$ and so the number of internal lines is given by the equation $I=3V-1$. By the same power counting, the maximal power of $N_c$ is $\left\lfloor \frac{I}{2}\right\rfloor -2V=\left\lfloor \frac{-V-1}{2}\right\rfloor$. However, the only possibility to get a contribution at order $\frac{1}{N_c}$ is by taking $V=1$, and this cannot be the case, since we do not allow interaction vertices connected to themselves. This completes the proof.

\subsubsection{Calculation for $N_f>1$}

The calculation for the $N_f>1$ case follows directly from the $N_f=1$ case. The only difference is that the vertex \eqref{eq: feynman diagram vertex 6 point function} also contains a flavor structure. The integrals over the internal loop momenta are the same, and so one has only to calculate the prefactor which comes from the symmetry factor and the flavor structure. 

We find it illuminating to work with the renormalization condition in the representation basis \eqref{eq: tree level corelation function representation}. The analog of \eqref{eq: counterterm and loop for nf=1 lam=0 case} is (for $n=SSS,SAA,AAA$)
\begin{equation} \label{eq: counterterm and loop for nf>1 lam=0 case}
    \#_{tree,n}\left(g_{n,B}-g_{n}\right)=\#_{2loop,n}\frac{N_{c}}{16\pi^{2}}\ln(\Lambda)-\#_{4loop,n}\frac{N_{c}^{3}}{2^{10}\pi^{2}}\ln(\Lambda),
\end{equation}
where the coefficients which correspond to the different flavor structures are given in table \ref{tab: RB lmabda=0 factors for beta}.

\begin{table}[t]
\begin{tabular}{|c||c|c|c|}
\hline 
$i$ &  $\#_{tree,n}$ & $\#_{2loop,n}$ & $\#_{4loop,n}$\tabularnewline
\hline 
\hline 
$SSS$ &  $N_{f}^{3}$ & $\begin{array}{c}6N_{f}^{4}g_{SSS}^{2}+\\6N_{f}^{2}\left(N_{f}^{2}-1\right)g_{SAA}^{2}\end{array}$ & $N_{f}^{3}\left(g_{SSS}^{3}N_{f}^{3}+g_{SAA}^{3}\left(N_{f}^{2}-1\right)\right)$\tabularnewline
\hline 
$\frac{SAA}{N_{f}^{2}-1}$ & $N_{f}$ & $\begin{array}{c}
4N_{f}^{2}g^{}_{SSS}g^{}_{SAA}+\\
8N_{f}^{}g_{SAA}^{2}+\\
4\left(N_{f}^{2}-4\right)g^{}_{SAA}g^{}_{AAA}+\\
4\frac{N_{f}^{2}-4}{N_f}g_{AAA}^{2}
\end{array}$ & $\begin{array}{c}
N_{f}^{3}g^{}_{SSS}g_{SAA}^{2}+\\N_{f}^{2}g_{SAA}^{3}+\\
2\left(N_{f}^{2}-4\right)g^{}_{SAA}g_{AAA}^{2}
\end{array}$\tabularnewline
\hline 
$\frac{AAA}{\left(N_{f}^{2}-1\right)\left(N_{f}^{2}-4\right)}$ & $\frac{1}{N_{f}}$ & $\begin{array}{c}
3g_{SAA}^{2}+\\3\frac{N_{f}^{2}-12}{N_f^2}g_{AAA}^{2}+\\
\frac{12}{N_f}g^{}_{SAA}g^{}_{AAA}
\end{array}$ & $\begin{array}{c}
3g_{SAA}^{2}g^{}_{AAA}+\\
\frac{N_{f}^{2}-12}{N_f^2}g_{AAA}^{3}
\end{array}$\tabularnewline
\hline 
\end{tabular}

\caption{The prefactors which correspond to the flavor structure part in the calculation of the $\beta$ functions in \eqref{eq: counterterm and loop for nf>1 lam=0 case}. Note the factor of $N_f^2-1$ and $\left(N_{f}^{2}-1\right)\left(N_{f}^{2}-4\right)$ common to all the $SAA$ and $AAA$ contributions respectively. This is to be expected because for $N_f=1$ (respectively $2$) both (respectively the $AAA$) structures do not exist.}
\label{tab: RB lmabda=0 factors for beta}
\end{table}

Taking the derivative with respect to the UV cutoff $\Lambda$, we find that the $\beta$ functions are
\begin{equation} \label{eq: beta function free boson copy}
    \begin{split}    2^{10}\pi^{2}N_{c}\beta_{SSS}=&2^{6}\left(6N_{f}\lambda_{SSS}^{2}+6\frac{\left(N_{f}^{2}-1\right)}{N_{f}}\lambda_{SAA}^{2}\right)-\left(N_{f}^{3}\lambda_{SSS}^{3}+\left(N_{f}^{2}-1\right)\lambda_{SAA}^{3}\right), \\
    2^{10}\pi^{2}N_{c}\beta_{SAA}=&2^{6}\left(4N_{f}\lambda_{SSS}\lambda_{SAA}+8\lambda_{SAA}^{2}+4\frac{\left(N_{f}^{2}-4\right)}{N_{f}}\lambda_{SAA}\lambda_{AAA}+4\frac{\left(N_{f}^{2}-4\right)}{N_{f}^{2}}\lambda_{AAA}^{2}\right)
    \\-&\left(N_{f}^{2}\lambda_{SSS}\lambda_{SAA}^{2}+N_{f}^{1}\lambda_{SAA}^{3}+2\frac{\left(N_{f}^{2}-4\right)}{N_{f}}\lambda_{SAA}\lambda_{AAA}^{2}\right), \\
    2^{10}\pi^{2}N_{c}\beta_{AAA}=&2^{6}\left(3N_{f}\lambda_{SAA}^{2}+\frac{3\left(N_{f}^{2}-12\right)}{N_{f}}\lambda_{AAA}^{2}+12\lambda_{SAA}\lambda_{AAA}\right)\\-&\left(3N_{f}\lambda_{SAA}^{2}\lambda_{AAA}+\frac{\left(N_{f}^{2}-12\right)}{N_{f}}\lambda_{AAA}^{3}\right) \, ,
    \end{split}
\end{equation}
which is equation \eqref{eq: beta function free boson} in the main text.

The advantage of working with the couplings in the representation basis, compared to the index basis, is that the contributions of the Feynman diagrams to each of the beta functions is more transparent. 
In complete analogy to the beta functions \eqref{eq: beta multyflavour} in the main text (explained in section \ref{sec: multiple falvor beta function comments} in item \ref{item4.3: structre of monimials for g3}), one can extract the possible monomials which can appear in the beta function by treating each pair of particle $\phi$ and anti particle $\bar{\phi}$ entering a vertex of type \eqref{eq: feynman diagram vertex 6 point function} as a composite meson $\tilde{M}$. For example, the contribution of the four-loop diagram to $\Gamma^{6}_{SSS}$ involves three scalar mesons, each entering the diagram at a different vertex. The internal loops correspond to exchanging mesons between the vertices, which in our example, can be either all scalar mesons, or all adjoint mesons, but not both. As a result, the only two cubic structures which contribute to $\beta_{SSS}$ are $\lambda_{SSS}^3$ and $\lambda_{SAA}^3$ (as one cannot construct a scenario in which there is $\lambda_{SSS}^2\lambda_{SAA}$ or $\lambda_{SSS}\lambda_{SAA}^2$). This way, each two lines in diagram $A0$ in figure \ref{fig: beta1 cubic term posabilities} are viewed in the same way as a $\zeta$ line in the $G_3^3$ diagrams in figure \ref{fig: 3 point diagrams Nf1}. The same reasoning can be applied for the other two couplings $\beta_{SAA}$ and $\beta_{AAA}$ and for the two-loop diagram. This shows which monomials of the couplings can appear in the $\beta$ function.
Also, each $\lambda_{SAA}$ in $\beta_{SSS}$ is accompanied by a factor of $\left(N_f^2-1\right)$, and each $\lambda_{AAA}$ in $\beta_{SAA}$ is accompanied by a factor of $\left(N_f^2-4\right)$, which makes the reduction to the $N_f=1$ and $N_f=2$ cases natural (as discussed in item \ref{item4.3: nf-1 factor near lambda1} in section \ref{sec: multiple falvor beta function comments}).

\newpage

\subsection{The $\lambda\ll1$ case} \label{sec: appndiex RB lambda finite}

Turning on a small $\lambda$ we can also perturbatively compute the beta function, in the regime where also $g_{n}$ are small and of order $\frac{\lambda^2}{N_c^2}$ (without necessarily taking the 't Hooft large $N_c$ limit).

To second order in both $g_{n}$ and $\lambda^2$ there are 11 diagrams that contribute to the 6-point vertex (figure \ref{fig: diagrmas A}) and 4 that contribute to the propagator (figure \ref{fig: B diagrams}) \cite{Aharony:2011jz}. The momentum structure of these diagrams is the same as in the $N_f=1$ case, so we can use the known results \cite{Aharony:2011jz,Anninos:2014hia} to extract the divergent part in \textbf{dimensional regularization} with $d=3-\epsilon$, as is shown in the first column of table \ref{tab:properties of A diagrams}. The symmetry factor for these diagrams is also shown in that table.
\begin{table}[t]
    \centering
\begin{tabular}{|c|c|c|c|c|}
\hline 
Diagram 
& \makecell{Momentum \\ Integral} 
& \makecell{Symmetry \\ Factor}  
& \makecell{Overcounting \\ Factor} 
& \makecell{Factors \\ of $4\pi/\left|\kappa\right|$} 
\tabularnewline
\hline 
\hline 
A1a & $-\frac{1}{32\pi^{2}}$ & $\frac{1}{2}$ & 4 & $\left(4\pi/\left|\kappa\right|\right)^2$\tabularnewline
\hline 
A1b/c & $-\frac{1}{32\pi^{2}}$ & $\frac{1}{2}$ & 12 & $\left(4\pi/\left|\kappa\right|\right)^2$\tabularnewline
\hline 
A2 & $\frac{1}{32\pi^{2}}$ & 1 & 1 & $\left(4\pi/\left|\kappa\right|\right)^4$\tabularnewline
\hline 
A3 & $\frac{1}{32\pi^{2}}$ & 1 & 2 & $\left(4\pi/\left|\kappa\right|\right)^4$\tabularnewline
\hline 
A4 & $-\frac{1}{16\pi^{2}}$ & 1 & 4 & $\left(4\pi/\left|\kappa\right|\right)^2$\tabularnewline
\hline 
A5 & $\frac{1}{32\pi^{2}}$ & 1 & 2 & $\left(4\pi/\left|\kappa\right|\right)^4$\tabularnewline
\hline 
A6 & $\frac{3}{64\pi^{2}}$ & $\frac{1}{2}$ & 6 & $\left(4\pi/\left|\kappa\right|\right)^4$\tabularnewline
\hline 
A7 & $\frac{3}{64\pi^{2}}$ & 1 & 2 & $\left(4\pi/\left|\kappa\right|\right)^4$\tabularnewline
\hline 
A8a & $\frac{1}{32\pi^{2}}$ & $\frac{1}{2}$ & 4 & $1$\tabularnewline
\hline 
A8b & $\frac{1}{32\pi^{2}}$ & $\frac{1}{2}$ & 36 & $1$ \tabularnewline
\hline 
\end{tabular}
    \caption{The factors contributing to the diagrams of figure \ref{fig: diagrmas A}, as explained in the text. The momentum integral column is the coefficient of $\frac{1}{\epsilon}$. The last column indicates the factors of $4\pi/\left|\kappa\right|$ which come from the gauge field interactions. For color and flavor structures see table \ref{tab:very big table}.}
    \label{tab:properties of A diagrams}
\end{table}

Each of the diagrams of figure \ref{fig: diagrmas A} also has a color and flavor structure (some sum of delta functions in the external indices). In order to extract the correct contribution to the different delta function structures of \eqref{eq: feynman diagram vertex 6 point function} one must sum over every permutation of the external color-flavor indices \textbf{that gives a different diagram}. This is not the full $3!\cdot3!$ possible permutations (e.g. permuting two of the upper scalars in diagram A1b will not result in a different diagram). We call the number of permutations of external indices that leaves the diagram the same an ``Overcounting Factor'' and it is given in the second to right column of table \ref{tab:properties of A diagrams}.
\begin{table}[t]
    \centering
\begin{tabular}{|c|c|c|}
\hline 
\makecell{Contribution \\ to:} & Diagram & Contribution\tabularnewline
\hline 
\hline 
\multirow{11}{*}{$\delta\bar{\Gamma}_{1}^{\left(6\right)}$} & A1a & $3\bar{g}_{1}N_{c}^{2}+6\bar{g}_{2}N_{c}+\frac{18\bar{g}_{1}}{N_{c}^{2}}-\frac{24\bar{g}_{2}}{N_{c}}-3\bar{g}_{1}$\tabularnewline
\cline{2-3}
 & A1b/c & $\frac{6\bar{g}_{1}}{N_{c}^{2}}+3\bar{g}_{2}N_{c}-\frac{12\bar{g}_{2}}{N_{c}}+3\bar{g}_{1}$\tabularnewline
\cline{2-3}
 & A2 & $\frac{96}{N_{c}^{4}}+12$\tabularnewline
\cline{2-3}
 & A3 & $-12$\tabularnewline
\cline{2-3}
 & A4 & see \eqref{eq: A4 A8}\tabularnewline
\cline{2-3}
 & A5 & $-\frac{24N_{f}}{N_{c}^{3}}$\tabularnewline
\cline{2-3}
 & A6 & $-2$\tabularnewline
\cline{2-3}
 & A7 & $\frac{48}{N_{c}^{4}}+\frac{48}{N_{c}^{2}}+6$\tabularnewline
\cline{2-3}
 & A8a & see \eqref{eq: A4 A8}\tabularnewline
\cline{2-3}
 & A8b & $\bar{g}_{1}^{2}+3\bar{g}_{2}^{2}+2\bar{g}_{3}^{2}$\tabularnewline
\hline 
\hline 
\multirow{11}{*}{$\delta\bar{\Gamma}_{2}^{\left(6\right)}$} & A1a & $3\bar{g}_{2}N_{c}^{2}+2\left(\bar{g}_{1}+2\bar{g}_{3}\right)N_{c}+\frac{18\bar{g}_{2}}{N_{c}^{2}}-\frac{8\left(\bar{g}_{1}+2\bar{g}_{3}\right)}{N_{c}}-3\bar{g}_{2}$\tabularnewline
\cline{2-3}
 & A1b/c & $\frac{6\bar{g}_{2}}{N_{c}^{2}}+\left(\bar{g}_{1}+2\bar{g}_{3}\right)N_{c}-\frac{4\left(\bar{g}_{1}+2\bar{g}_{3}\right)}{N_{c}}+3\bar{g}_{2}$\tabularnewline
\cline{2-3}
 & A2 & $10N_{c}-\frac{96}{N_{c}^{3}}-\frac{16}{N_{c}}$\tabularnewline
\cline{2-3}
 & A3 & $-10N_{c}-\frac{8}{N_{c}}$\tabularnewline
\cline{2-3}
 & A4 & see \eqref{eq: A4 A8}\tabularnewline
\cline{2-3}
 & A5 & $\frac{24N_{f}}{N_{c}^{2}}+6N_{f}$\tabularnewline
\cline{2-3}
 & A6 & $-2N_{c}$\tabularnewline
\cline{2-3}
 & A7 & $6N_{c}-\frac{64}{N_{c}^{3}}-\frac{32}{N_{c}}$\tabularnewline
\cline{2-3}
 & A8a & see \eqref{eq: A4 A8}\tabularnewline
\cline{2-3}
 & A8b & $2\bar{g}_{2}\left(\bar{g}_{1}+2\bar{g}_{3}\right)$\tabularnewline
\hline 
\hline 
\multirow{11}{*}{$\delta\bar{\Gamma}_{3}^{\left(6\right)}$} & A1a & $3\bar{g}_{3}N_{c}^{2}+6\bar{g}_{2}N_{c}+\frac{18\bar{g}_{3}}{N_{c}^{2}}-\frac{24\bar{g}_{2}}{N_{c}}-3\bar{g}_{3}$\tabularnewline
\cline{2-3}
 & A1b/c & $3\bar{g}_{2}N_{c}-\frac{12\bar{g}_{2}}{N_{c}}+\frac{6\bar{g}_{3}}{N_{c}^{2}}+3\bar{g}_{3}$\tabularnewline
\cline{2-3}
 & A2 & $3N_{c}^{2}+\frac{144}{N_{c}^{2}}-48$\tabularnewline
\cline{2-3}
 & A3 & $36-3N_{c}^{2}$\tabularnewline
\cline{2-3}
 & A4 & $\frac{3\left(N_{c}^{2}-4\right)\left(\bar{g}_{3}N_{f}+2\bar{g}_{2}\right)}{N_{c}}$\tabularnewline
\cline{2-3}
 & A5 & $\frac{-36N_{f}}{N_{c}}+3N_{c}N_{f}$\tabularnewline
\cline{2-3}
 & A6 & 4\tabularnewline
\cline{2-3}
 & A7 & $\frac{96}{N_{c}^{2}}-12$\tabularnewline
\cline{2-3}
 & A8a & see \eqref{eq: A4 A8}\tabularnewline
\cline{2-3}
 & A8b & $3\bar{g}_{2}^{2}+\bar{g}_{3}\left(2\bar{g}_{1}+\bar{g}_{3}\right)$\tabularnewline
\hline 
\end{tabular}
    \caption{Color and flavor structures that contribute to the 2-loop corrections to \eqref{eq: tree level correlation function indices}, from the diagrams in figure \ref{fig: diagrmas A}, after accounting for symmetry and overcounting factors (but not the momentum integrals).}
    \label{tab:very big table}
\end{table}

The computations of the color and flavor structures of these diagrams are straightforward given the Feynman rules above. We can then extract the corrections to the beta functions using the renormalization conditions in \eqref{eq: tree level correlation function indices}. The results are shown in table \ref{tab:very big table}, except for the contributions to the $A4$ and $A8a$ diagrams that are too long to fit in the table, and \newpage 

\noindent are given by \thispagestyle{empty}
\begin{align} \label{eq: A4 A8} 
A4_{1}	&=\frac{6\left(\bar{g}_{1}\left(N_{c}^{3}N_{f}-N_{c}N_{f}+N_{c}^{2}+2\right)+\bar{g}_{2}\left(N_{c}^{2}N_{f}+2N_{c}^{3}-2N_{c}+2N_{f}\right)+\bar{g}_{3}\left(N_{c}^{2}+2\right)\right)}{N_{c}^{2}}, \nonumber\\ 
A4_{2}	&=2\left(N_{c}\left(2\bar{g}_{2}N_{f}+\bar{g}_{1}+3\bar{g}_{3}\right)+\frac{2\bar{g}_{3}N_{f}+4\bar{g}_{2}}{N_{c}^{2}}-\frac{5\bar{g}_{2}N_{f}+4\bar{g}_{1}+6\bar{g}_{3}}{N_{c}}+\bar{g}_{3}N_{f}+2\bar{g}_{2}\right), \nonumber \\ 
A8a_{1}	&=3\left(\bar{g}_{1}^{2}\left(2N_{c}N_{f}+7\right)+8\bar{g}_{1}\left(\bar{g}_{2}\left(N_{c}+N_{f}\right)+\bar{g}_{3}\right)+\bar{g}_{2}^{2}\left(2N_{c}N_{f}+19\right)+4\bar{g}_{2}\bar{g}_{3}\left(N_{c}+N_{f}\right)+4\bar{g}_{3}^{2}\right),\nonumber \\ 
A8a_{2}	&=2\left(7\bar{g}_{2}^{2}\left(N_{c}+N_{f}\right)+\bar{g}_{1}\bar{g}_{2}\left(2N_{c}N_{f}+17\right)+2\bar{g}_{3}\bar{g}_{2}\left(N_{c}N_{f}+16\right)+3\bar{g}_{3}^{2}\left(N_{c}+N_{f}\right)+4\bar{g}_{1}\bar{g}_{3}\left(N_{c}+N_{f}\right)\right),\nonumber \\ 
A8a_{3}	&=3\left(\bar{g}_{2}^{2}\left(N_{c}N_{f}+15\right)+8\bar{g}_{3}\bar{g}_{2}\left(N_{c}+N_{f}\right)+\bar{g}_{3}\left(\bar{g}_{3}\left(N_{c}N_{f}+9\right)+6\bar{g}_{1}\right)\right).
\end{align}

\begin{table}[t]
    \centering
\begin{tabular}{|c|c|c|c|}
\hline 
Diagram & \makecell{Momentum \\ Integral} & \makecell{Symmetry \\ Factor}  & \makecell{Total Color- \\ Flavor Factor} \tabularnewline
\hline 
\hline 
B1 & $\frac{1}{24\pi^{2}}$ & 1 & $N_{c}^{2}-1$\tabularnewline
\hline 
B2 & $\frac{1}{96\pi^{2}}$ & $\frac{1}{2}$ & $\frac{N_{c}^{4}-3N_{c}^{2}+2}{N_{c}^{2}}$\tabularnewline
\hline 
B3 & $\frac{1}{12\pi^{2}}$ & 1 & $\frac{1}{N_{c}^{2}}-1$\tabularnewline
\hline 
B4 & $\frac{1}{24\pi^{2}}$ & 1 & $\frac{N_{c}^{2}-1}{N_{c}}N_{f}$\tabularnewline
\hline 
\end{tabular}
    \caption{The factors contributing to the diagrams of figure \ref{fig: B diagrams}, as explained in the text. The momentum integral column is the coefficient of $\frac{1}{\epsilon}$, the color structure is after multiplication by the symmetry factor. All the diagrams are of order $\left(\frac{4\pi}{\left|\kappa\right|}\right)^2$.}
    \label{tab: B table}
\end{table}

We also need the contribution that comes from the corrections to the propagators. This is simpler, as the overall structure is just $\delta_{ij}\delta_{ab}$ and there are no overcounting factors. The results are shown in table \ref{tab: B table}.

Summing all the contributions we can get the beta functions for $\bar{g}_{1/2/3}$. By the Callan-Symanzik equation we have
\begin{multline}
\beta_{\bar{g}_i}=2\sum_{A\ diagrams}
\left(\begin{array}{c}
Factors\\
of\ 4\pi/\left|\kappa\right|
\end{array}\right)\times
\left(\begin{array}{c}
Momentum\\
Integration
\end{array}\right)\times\left(\begin{array}{c}
Symmetry\\
Factor
\end{array}\right)\times\frac{3!\cdot3!}{Overcounting\ Factor}\times\delta\bar{\Gamma}_{i}^{\left(6\right)} \\
-6\times \left(\frac{4\pi}{\left|\kappa\right|}\right)^2 \times \bar{g}_i\times\sum_{B\ diagrams}\left(\begin{array}{c}
Momentum\\
Integration
\end{array}\right)\times\left(\begin{array}{c}
Symmetry\\
Factor
\end{array}\right)\times\left(\text{factors in table}\ \ref{tab: B table}\right).
\end{multline}

In the paper, we are interested in the 't Hooft limit $N_c \rightarrow \infty$, with the finite 't Hooft parameters $\bar{g}_n = \frac{\bar{\lambda}_n}{\left(N_c\right)^2}$ and $\lambda=\frac{N_c}{\kappa}$.  In this limit, the beta functions are (at leading order in $\frac{1}{N_c}$)
\begin{align} \label{eq: beta small lambda indices}
16\pi^{2}N_{c}\beta_{\overline{1}}&=	6 \left(\bar{\lambda }_1^2 N_f+\bar{\lambda }_2 \left(\bar{\lambda }_2 N_f+2 \bar{\lambda }_3\right)+4 \bar{\lambda }_2 \bar{\lambda }_1\right)-4 \left(\frac{\lambda}{4\pi}\right)^2 \left(4 \bar{\lambda }_1 N_f+9 \bar{\lambda }_2\right), \\
16\pi^{2}N_{c}\beta_{\overline{2}}&=	4 \bar{\lambda }_2 \left(\bar{\lambda }_1+\bar{\lambda }_3\right) N_f+14 \bar{\lambda }_2^2+2 \bar{\lambda }_3 \left(4 \bar{\lambda }_1+3 \bar{\lambda }_3\right)-4 \left(\frac{\lambda}{4\pi}\right)^2 \left(3 \bar{\lambda }_2 N_f+2 \bar{\lambda }_1+5 \bar{\lambda }_3\right)+6 \left(\frac{\lambda}{4\pi}\right)^4, \\
16\pi^{2}N_{c}\beta_{\overline{3}}&=	3 \left(\bar{\lambda }_2^2 N_f+\bar{\lambda }_3^2 N_f+8 \bar{\lambda }_3 \bar{\lambda }_2\right)-2 \left(\frac{\lambda}{4\pi}\right)^2 \left(5 \bar{\lambda }_3 N_f+12 \bar{\lambda }_2\right)+3 N_f \left(\frac{\lambda}{4\pi}\right)^4.
\end{align}
As in the case $\lambda=0$, it is more convenient to write this in the representation basis. Using the conversion \eqref{eq: transform to adjoint0} we find
\begin{equation} \label{eq: beta small lambdab copy}
    \begin{split}
        16\pi^{2} N_{c}\beta_{SSS}=&\left(\frac{\lambda}{4\pi}\right)^4\frac{24 }{N_f}-\left(\frac{\lambda}{4\pi}\right)^2\frac{8  \left(\lambda_{SSS} N_f \left(2 N_f^2+3\right)+3 \lambda_{SAA} \left(N_f^2-1\right)\right)}{N_f^2}\\+&6 \lambda_{SSS}^2 N_f+\frac{6 \lambda_{SAA}^2 \left(N_f^2-1\right)}{N_f}, \\
        16\pi^{2}N_{c}\beta_{SAA}=&12 \left(\frac{\lambda}{4\pi}\right)^4
        -\left(\frac{\lambda}{4\pi}\right)^2\frac{4  \left(N_f \left(2 \lambda_{SSS} N_f+3 \lambda_{SAA} \left(N_f^2+2\right)\right)+4 \lambda_{AAA} \left(N_f^2-4\right)\right)}{N_f^2}\\
        +&\frac{4 \left(\lambda_{SAA} N_f^2 \left(\lambda_{SSS} N_f+2 \lambda_{SAA}\right)+\lambda_{SAA} \lambda_{AAA} \left(N_f^2-4\right) N_f+\lambda_{AAA}^2 \left(N_f^2-4\right)\right)}{N_f^2}, \\
        16\pi^{2}N_{c}\beta_{AAA}=&3 N_f \left(\frac{\lambda}{4\pi}\right)^4+\left(\frac{\lambda}{4\pi}\right)^2 \left(-10 \lambda_{AAA} N_f+\frac{48 \lambda_{AAA}}{N_f}-24 \lambda_{SAA}\right)\\+&3 \left(\lambda_{SAA}^2 N_f+\frac{\lambda_{AAA}^2 \left(N_f^2-12\right)}{N_f}+4 \lambda_{AAA} \lambda_{SAA}\right) ,
    \end{split}
\end{equation}
which is equation \eqref{eq: beta small lambdab} we use in the main text. We note that as expected from the general proof in the paper, and as shown in the  $N_f=1$ case \cite{Aharony:2011jz}, the highest power of $N_c^2$ cancels between diagrams $A1a$, $A2$, $B1$ and $B2$, and we are left with a leading contribution to the beta function of order $\frac{1}{N_c}$.

\subsubsection{IR-stable points at finite $N_c$} \label{sec: appendix finite lambda finite nc}

The results presented in table~\ref{tab:very big table} were obtained using perturbation theory in $\sqrt{\frac{4\pi}{\left|\kappa\right|}}$, with $ g_{n} \sim \frac{\lambda^2}{N_c^2} $, and are valid for arbitrary $ N_c $ and $ N_f $. Although this is not the main focus of the paper, the calculations performed can also be applied to analyze systems with finite $ N_c $.

\begin{figure}[t]
\centering
\includegraphics[width=0.45\linewidth]{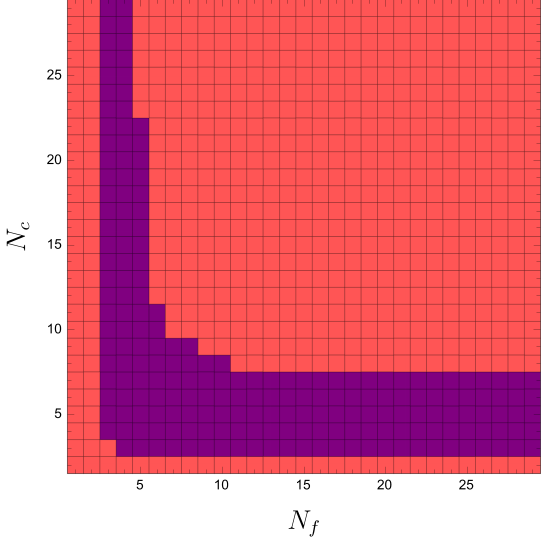}

\caption{
Existence of a weakly coupled IR-stable point for different values of $N_c\ge2$ and $N_f\ge1$. \textcolor{violet}{\textbf{Violet-}} no IR-stable fixed point \textcolor{red}{\textbf{Red-}} one IR-stable fixed point. The large $N_c$ behavior that was derived in the main text appears here for $N_c>22$.
}
\label{fig: finite Nc} 
\end{figure}

One particular question of interest is whether there exists an IR-stable fixed point for the marginal running couplings $ g_{n}$. The results, for different values of $N_c$ and $N_f$, are shown in figure \ref{fig: finite Nc}.

As we see from the figure, the existence of of an IR fixed point depends both on the number of colors and on the number of flavors. For the $N_f=1,2$ (or $N_c=2$) cases, there exist only 1 or 2 marginal couplings, we see that an IR-stable fixed point exists for any $N_c$ ($N_f$).

For the cases with all three marginal couplings, the existence of the IR-stable fixed point depends on the specific values chosen. An interesting result is that for $N_f=3,4$ (with the exception of $N_c=N_f=3$) there is no IR fixed point for any $N_c$ (we see this explicitly for $N_c\le30$ and is in accordance with the result for $N_c \rightarrow\infty$ we saw in the main text, see table \ref{tab: pertubed_bosonic_fixed_points}).

Another interesting observation from the figure is that for $3\le N_c\le6$, there is no IR-stable fixed point for any value of $N_f\ge3$ (with the exception of $N_c=N_f=3$).

\section{Correlation functions for the critical boson 
theory at \texorpdfstring{$\lambda = 0$}{lambda = 0}} \label{sec: appendix CB correlations}
In this appendix, we calculate correlation functions of the $\sigma^j_i$ operators, for the critical boson theory \eqref{eq: CB} with $\lambda=0$. As described in \cite{Aharony:2018pjn} and in the main text, in order to calculate the $\beta_{\lambda_n}$ functions at leading order in $\frac{1}{N_c}$, we need the leading expansion in terms of $\frac{1}{N_c}$ for the four-point and five-point correlation functions, as well as the leading and first subleading terms of the two-point and three-point functions. 

To calculate the desired correlation functions, we do the following\footnote{We closely follow Appendix B of \cite{Aharony:2018pjn}, extending their results to multiple flavors.}:
\begin{itemize}
    \item In section \ref{subsec: App efective sigma action}, we integrate out the scalar fields $\phi, \bar{\phi}$ and obtain an effective action for the $\sigma$ field. While the formal expression for the effective action is an infinite series, we truncate it to obtain a polynomial in $\sigma$. The coefficients of this expansion, denoted by $S_n$, correspond to the amputated correlation functions.
    \item In section \ref{subsec: App CB s4 s5}, we present the amputated tree level $n$-point correlation functions of $\sigma$, $G_n$, and derive an expression for the amputated correlation functions of four $\sigma$ fields, in a particular limit of external momenta.
    \item In section \ref{subsec: App gn}, we summarize the tree-level correlation functions of $\sigma$ in the limit that we use in the paper, and in this appendix afterwards.
    \item In section \ref{subsec: App subleading G2} we calculate the first sub-leading correction to the two-point correlation function of $\sigma$, which depends on $\ln(\Lambda)$. From this, we extract the anomalous dimensions $\gamma_{S/A}$ of the singlet $\sigma_S$ and adjoint $\sigma_A$ operators.
    \item In section \ref{subsec: App subleading G3}, we present the first sub-leading correction to the three-point correlation function of $\sigma$, which in the main text we denote as $\delta G_3$.
\end{itemize}

\subsection{Effective action for \texorpdfstring{$\sigma$}{sigma}} \label{subsec: App efective sigma action}

In the $\lambda=0$ limit, the gauge fields do not interact with the $\phi$ and $\sigma$ fields in the theory, and can be trivially integrated out. The scalar fields $\phi$ can also be integrated out, providing the effective action for $\sigma$
\begin{equation} \label{eq S eff J}
\begin{split}
    & \intop D\phi D\bar\phi e^{-\intop\left(\sum_{i}\partial_{\mu}\bar{\phi}\partial_{\mu}\phi+\sigma_{i}^{\ j}\bar{\phi}_{j}\phi^{i}\right)}=e^{-S_{eff}\left(\sigma_{i}^{\ j}\right)}, \\
    & S_{eff}\left(\sigma_{i}^{\ j}\right)=\sum_{n=2}^{\infty}\frac{1}{n!}\intop d\Pi_{n}\left(S_{n}\right)_{j_{1}\cdots j_{n}}^{i_{1}\cdots i_{n}}\left(p_{1},\cdots,p_{n}\right)\sigma_{i_{1}}^{\ j_{1}}\left(p_{1}\right)\cdots \sigma_{i_{n}}^{\ j_{n}}\left(p_{n}\right),
\end{split}    
\end{equation}
where $d\Pi_n$ is given by \eqref{eq: dPi_n definition}.

The $\left(S_{n}\right)_{j_{1}\cdots j_{n}}^{i_{1}\cdots i_{n}}\left(p_{1},\cdots,p_{n}\right)$ coefficients of the effective action can be found (at least formally) by taking the derivatives with respect to $\sigma^j_i$ on both sides of \eqref{eq S eff J}. For each $n$, $\left(S_{n}\right)_{j_{1}\cdots j_{n}}^{i_{1}\cdots i_{n}}\left(p_{1},\cdots,p_{n}\right)$ can be calculated by summing over all the connected diagrams of the $\phi$'s. For the purpose of our calculations, we need only the terms with $n\le5$. 

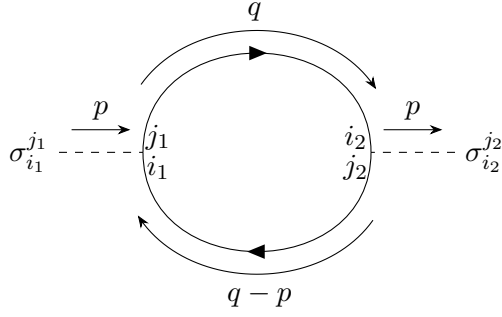
\begin{figure}[t]
    \centering
    \begin{tikzpicture}[baseline=(c.base)]
    \begin{feynman}
        \vertex (c);
        \vertex (i1) at (-3,0) {$\sigma^{j_1}_{i_1}$};
        \vertex (i2) at (3,0) {$\sigma^{j_2}_{i_2}$};
        
        \vertex (d1) at (-1.5,0);
        \vertex (d2) at (1.5,0);

        \diagram*{
            (i1) -- [scalar,momentum = {$ p $}] (d1),
            (i2) -- [scalar,reversed momentum' = {$ p $}] (d2),
            (d1) -- [fermion,half left,momentum = {$ q $}] (d2)
                 -- [fermion,half left,,momentum = {$ q - p $}] (d1),
            
        };

        \node at ([shift={(0.2,-0.2)}]d1) {\(i_1\)};
        \node at ([shift={(0.2,0.2)}]d1) {\(j_1\)};
        \node at ([shift={(-0.2,0.2)}]d2) {\(i_2\)};
        \node at ([shift={(-0.2,-0.2)}]d2) {\(j_2\)};
    \end{feynman}
\end{tikzpicture}

    \caption{Feynman diagram for calculating $S_2$. Broken lines denote $\sigma$ fields and solid lines the scalars $\phi$.}
    \label{fig: appendix diagram s2} 
\end{figure}

For $n=2$, there is only one connected diagram, and one finds (see figure \ref{fig: appendix diagram s2})
\begin{equation}
    \left(S_{2}\right)_{j_{1}j_{2}}^{i_{1}i_{2}} \left(p\right)=\delta_{j_{2}}^{i_{1}}\delta_{j_{1}}^{i_{2}}\tilde{S}_{2}(p); \quad \tilde{S}_{2}(p)= -N_c\intop\frac{d^{3}q}{\left(2\pi\right)^{3}}\frac{1}{q^{2}\left(q-p\right)^{2}}= -N_c\frac{1}{8\left|p\right|}.
\end{equation}

\begin{figure}[t]
    \centering
    \begin{equation*}
    \vcenter{\hbox{\begin{tikzpicture}[baseline=(c.base)]
    \begin{feynman}
        \vertex (c);
        \vertex (i1) at (-3,0) {$\sigma^{j_1}_{i_1}$};
        \vertex (i2) at (1.5,2.6) {$\sigma^{j_2}_{i_2}$};
        \vertex (i3) at (1.5,-2.6) {$\sigma^{j_3}_{i_3}$};
        
        \vertex (d1) at (-1.5,0);
        \vertex (d2) at (0.75,1.4);
        \vertex (d3) at (0.75,-1.4);

        \diagram*{
            (i1) -- [scalar,momentum = {$ p_1 $}] (d1),
            (i2) -- [scalar, momentum = {$ p_2 $}] (d2),
            (i3) -- [scalar, momentum = {$ p_3 $}] (d3),
            (d1) -- [fermion,momentum = {$ q+p_1 $}] (d2)
                 -- [fermion,momentum = {$ q - p_3 $}] (d3)
                 -- [fermion,momentum = {$ q  $}] (d1),
            
        };

        \node at ([shift={(0.0,-0.2)}]d1) {\(i_1\)};
        \node at ([shift={(0.0,0.2)}]d1) {\(j_1\)};
        \node at ([shift={(-0.2,0.2)}]d2) {\(i_2\)};
        \node at ([shift={(0.3,-0.1)}]d2) {\(j_2\)};
        \node at ([shift={(0.2,0.1)}]d3) {\(i_3\)};
        \node at ([shift={(-0.2,-0.1)}]d3) {\(j_3\)};
    \end{feynman}
\end{tikzpicture}}}\,+\,\vcenter{\hbox{\begin{tikzpicture}[baseline=(c.base)]
    \begin{feynman}
        \vertex (c);
        \vertex (i1) at (-3,0) {$\sigma^{j_1}_{i_1}$};
        \vertex (i2) at (1.5,2.6) {$\sigma^{j_3}_{i_3}$};
        \vertex (i3) at (1.5,-2.6) {$\sigma^{j_2}_{i_2}$};
        
        \vertex (d1) at (-1.5,0);
        \vertex (d2) at (0.75,1.4);
        \vertex (d3) at (0.75,-1.4);

        \diagram*{
            (i1) -- [scalar,momentum = {$ p_1 $}] (d1),
            (i2) -- [scalar, momentum = {$ p_3 $}] (d2),
            (i3) -- [scalar, momentum = {$ p_2 $}] (d3),
            (d1) -- [fermion,momentum = {$ q+p_1 $}] (d2)
                 -- [fermion,momentum = {$ q - p_2 $}] (d3)
                 -- [fermion,momentum = {$ q  $}] (d1),
            
        };

        \node at ([shift={(0.0,-0.2)}]d1) {\(i_1\)};
        \node at ([shift={(0.0,0.2)}]d1) {\(j_1\)};
        \node at ([shift={(-0.2,0.2)}]d2) {\(i_3\)};
        \node at ([shift={(0.3,-0.1)}]d2) {\(j_3\)};
        \node at ([shift={(0.2,0.1)}]d3) {\(i_2\)};
        \node at ([shift={(-0.2,-0.1)}]d3) {\(j_2\)};
    \end{feynman}
\end{tikzpicture}}}
\end{equation*}

    \caption{Feynman diagrams for calculating $S_3$, with the same notations as in figure \ref{fig: appendix diagram s2}.}
    \label{fig: appendix diagram s3} 
\end{figure}
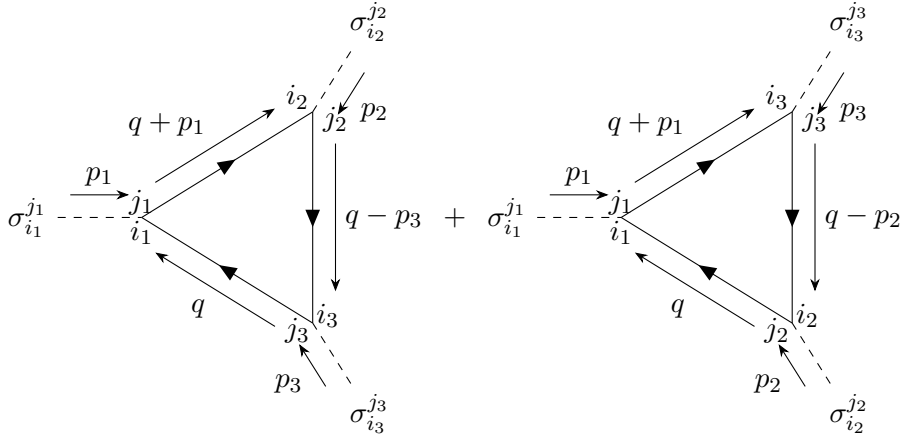

For $n=3$, there are two connected diagrams, as shown in figure \ref{fig: appendix diagram s3}: one in which the $i_1$ index is contracted with $j_2$ to provide $\delta_{j_2}^{i_1}$ and one in which we have $\delta_{j_3}^{i_1}$. It is evident (as can be shown by explicit calculation) that the momentum dependence in both cases is the same, and we find
\begin{equation}
    \begin{split}
    & \left(S_{3}\right)_{j_{1},j_{2},j_{3}}^{i_{1},i_{2},i_{3}}\left(p_{1},p_{2},p_{3}\right)=\tilde{S}_{3}\left(\delta_{j_{2}}^{i_{1}}\delta_{j_{3}}^{i_{2}}\delta_{j_{1}}^{i_{3}}+\delta_{j_{3}}^{i_{1}}\delta_{j_{2}}^{i_{3}}\delta_{j_{1}}^{i_{2}}\right), \\
    & \tilde{S}_{3}\left(p_{1},p_{2},p_{3}\right)=N_c\intop\frac{d^{3}q}{\left(2\pi\right)^{3}}\frac{1}{q^{2}\left(q+p_{1}\right)^{2}\left(q-p_{3}\right)^{2}}=\frac{N_c}{8}\frac{1}{\left|p_{1}\right|\left|p_{2}\right|\left|p_{3}\right|} .
    \end{split}
\end{equation}

\begin{figure}[t]
    \centering
    \begin{tikzpicture}[baseline=(c.base)]
    \begin{feynman}
        \vertex (c);
        \vertex (i1) at (-2,2) {$\sigma^{j_1}_{i_1}$};
        \vertex (i2) at (2,2) {$\sigma^{j_2}_{i_2}$};
        \vertex (i3) at (2,-2) {$\sigma^{j_3}_{i_3}$};
        \vertex (i4) at (-2,-2) {$\sigma^{j_4}_{i_4}$};
        
        \vertex (d1) at (-1,1);
        \vertex (d2) at (1,1);
        \vertex (d3) at (1,-1);
        \vertex (d4) at (-1,-1);

        \diagram*{
            (i1) -- [scalar,momentum = {$ p_1 $}] (d1),
            (i2) -- [scalar,momentum = {$ p_2 $}] (d2),
            (i3) -- [scalar, momentum = {$ p_3 $}] (d3),
            (i4) -- [scalar, momentum = {$ p_4 $}] (d4),
            (d1) -- [fermion,momentum = {$ q+p_1 $}] (d2)
                 -- [fermion,momentum = {$ q + p_1+p_2 $}] (d3)
                 -- [fermion,momentum = {$ q - p_4 $}] (d4)
                 -- [fermion,momentum = {$ q  $}] (d1),
            
        };

        \node at ([shift={(0.2,-0.5)}]d1) {\(i_1\)};
        \node at ([shift={(0.5,-0.2)}]d1) {\(j_1\)};
        \node at ([shift={(-0.5,-0.2)}]d2) {\(i_2\)};
        \node at ([shift={(-0.2,-0.5)}]d2) {\(j_2\)};
        \node at ([shift={(-0.2,0.5)}]d3) {\(i_3\)};
        \node at ([shift={(-0.5,0.2)}]d3) {\(j_3\)};
        \node at ([shift={(0.5,0.2)}]d4) {\(i_4\)};
        \node at ([shift={(0.2,0.5)}]d4) {\(j_4\)};
    \end{feynman}
\end{tikzpicture}

    \caption{A Feynman diagram which contributes to the calculation of $S_4$. The other 5 diagrams are given by permutations of the external legs.}
    \label{fig: appendix diagram s4} 
\end{figure}
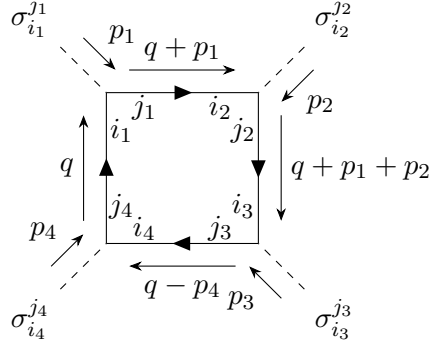

For $n=4$ there are $(4-1)!=6$ connected diagrams, corresponding to permutations of $\left(1,2,3,4\right)$ up to cyclic shifts. An example of one of the diagrams is given in figure \ref{fig: appendix diagram s4}. We have:
\begin{equation} \label{eq: critical boson s4 definition}
    \begin{split}
    & \left(S_{4}\right)_{j_{1},j_{2},j_{3},j_{4}}^{i_{1},i_{2},i_{3},i_{4}}\left(p_{1},p_{2},p_{3},p_{4}\right)=\left(\delta_{j_{2}}^{i_{1}}\delta_{j_{3}}^{i_{2}}\delta_{j_{4}}^{i_{3}}\delta_{j_{1}}^{i_{4}}\tilde{S}_{4}\left(p_{1},p_{2},p_{3},p_{4}\right)+\left(\text{permutations of \{2,3,4\}}\right)\right), \\
    & \tilde{S}_{4}\left(p_{1},p_{2},p_{3},p_{4}\right)=-N_c\intop\frac{d^3q}{\left(2\pi\right)^{3}}\frac{1}{q^{2}\left(q+p_{1}\right)^{2}\left(q+p_{1}+p_{2}\right)^{2}\left(q-p_{4}\right)^{2}} . 
    \end{split}
\end{equation}
The explicit expression of $\tilde{S}_{4}$ for arbitrary momenta is unknown.

A similar expression can be written for $n=5$, with $(5-1)!=24$ connected diagrams
\begin{equation}
\left(S_{5}\right)_{j_{1},j_{2},j_{3},j_{4},j_{5}}^{i_{1},i_{2},i_{3},i_{4},i_{5}}\left(p_{1},p_{2},p_{3},p_{4},p_{5}\right)=\left(\delta_{j_{2}}^{i_{1}}\delta_{j_{3}}^{i_{2}}\delta_{j_{4}}^{i_{3}}\delta_{j_{5}}^{i_{4}}\delta_{j_{1}}^{i_{5}}\tilde{S}_{5}\left(p_{1},p_{2},p_{3},p_{4},p_{5}\right)+\left(\text{permutations of \{2,3,4,5\}}\right)\right) ,
\end{equation}
with $\tilde{S}_{5}\left(p_{1},p_{2},p_{3},p_{4},p_{5}\right)$ containing the momentum part. We do not use the explicit form of $\tilde{S}_{5}$ in this work, and refer the reader to \cite{Aharony:2018pjn}.

\subsection{Amputated tree-level correlation functions for \texorpdfstring{$\sigma$}{sigma}} \label{subsec: App CB s4 s5}

In this section, we use \eqref{eq S eff J}, the effective action for $\sigma$,  to obtain the tree-level amputated correlation functions of $\sigma$, in the desired momenta limits.

First, we note that the coefficient of the quadratic term serves as the inverse of the $\sigma$ propagator. The $\sigma$ propagator (at tree-level) is thus\footnote{Note that here, and in the following, we ignore trivial factors of $\delta$ functions over the momenta.}
\begin{equation} \label{eq: green function two point tree level}
-\left(G_{2}\right)_{i_{1},i_{2}}^{j_{1},j_{2}}\left(p\right)\equiv\left\langle \sigma_{i_{1}}^{\ j_{1}}\left(p\right)\sigma_{i_{2}}^{\ j_{2}}\left(-p\right)\right\rangle =\left(S_{2}^{-1}\right)_{i_{_{1}},i_{2}}^{j_{1},j_{2}}\left(p\right)=-\frac{8\left|p\right|}{N_c}\delta_{i_{2}}^{j_{1}}\delta_{i_{1}}^{j_{2}} . 
\end{equation}

For the amputated three-point correlation function of $\sigma$, the only diagram at tree-level is that which contains only the 3-point vertex $S_3$, that is\footnote{Note that we flip the role of the $i$'s and $j$'s indices comparing to section \ref{subsec: App efective sigma action}. This is because when taking the amputated diagram, we have to multiply each external leg by the inverse of $S_2$, as in \eqref{eq: appendix, g3 from s3}.}
\begin{equation}
    \left\langle \sigma_{j_{1}}^{\ i_{1}}\left(p_{1}\right)\sigma_{j_{2}}^{\ i_{2}}\left(p_{2}\right)\sigma_{j_{3}}^{\ i_{3}}\left(p_{3}\right)\right\rangle _{\text{amp}}=-\left(S_{3}\right)_{j_{1},j_{2},j_{3}}^{i_{1},i_{2},i_{3}}\left(p_{1},p_{2},p_{3}\right).
\end{equation}

\begin{figure}[t]
    \centering
    \begin{equation*}
    \vcenter{\hbox{\begin{tikzpicture}[baseline=(c.base)]
    \begin{feynman}
        \node (c)  [dot];
        \vertex (i1) at (-1.8,1.8) {$\sigma^{j_1}_{i_1}$};
        \vertex (i2) at (1.8,1.8) {$\sigma^{j_2}_{i_2}$};
        \vertex (i3) at (1.8,-1.8) {$\sigma^{j_3}_{i_3}$};
        \vertex (i4) at (-1.8,-1.8) {$\sigma^{j_4}_{i_4}$};

        \vertex  (cn) at (0.4,0) {$S_4$};
        
        \diagram*{
            (i1) -- [scalar,momentum = {$ p_1 $}] (c),
            (i2) -- [scalar,momentum' = {$ p_2 $}] (c),
            (i3) -- [scalar,momentum = {$ p_3 $}] (c),
            (i4) -- [scalar,momentum = {$ p_4 $}] (c),  
        };
    \end{feynman}
\end{tikzpicture}}}\,+\,\vcenter{\hbox{\begin{tikzpicture}[baseline=(c.base)]
    \begin{feynman}
        \vertex (c);
        \node (d1) at (-1,0) [dot];
        \node (d2) at (1,0) [dot];
        
        \vertex (i1) at (-2,1.5) {$\sigma^{j_1}_{i_1}$};
        \vertex (i2) at (-2,-1.5) {$\sigma^{j_2}_{i_2}$};
        \vertex (i3) at (2,1.5) {$\sigma^{j_3}_{i_3}$};
        \vertex (i4) at (2,-1.5) {$\sigma^{j_4}_{i_4}$};

        \vertex  (d2n) at ([shift={(0.3,0.0)}]d2) {$S_3$};
        \vertex  (d1n) at ([shift={(-0.3,0.0)}]d1) {$S_3$};
        
        \diagram*{
            (i1) -- [scalar,momentum = {$ p_1 $}] (d1),
            (i2) -- [scalar,momentum' = {$ p_2 $}] (d1),
            (i3) -- [scalar,momentum' = {$ p_3 $}] (d2),
            (i4) -- [scalar,momentum = {$ p_4 $}] (d2),
            (d1) -- [scalar,momentum = {$ p_1+p_2 $}] (d2),  
        };
    \end{feynman}
\end{tikzpicture}}}
\end{equation*}
\begin{equation*}
    +\vcenter{\hbox{\begin{tikzpicture}[baseline=(c.base)]
    \begin{feynman}
        \vertex (c);
        \node (d1) at (-1,0) [dot];
        \node (d2) at (1,0) [dot];
        
        \vertex (i1) at (-2,1.5) {$\sigma^{j_1}_{i_1}$};
        \vertex (i2) at (-2,-1.5) {$\sigma^{j_3}_{i_3}$};
        \vertex (i3) at (2,1.5) {$\sigma^{j_2}_{i_2}$};
        \vertex (i4) at (2,-1.5) {$\sigma^{j_4}_{i_4}$};

        \vertex  (d2n) at ([shift={(0.3,0.0)}]d2) {$S_3$};
        \vertex  (d1n) at ([shift={(-0.3,0.0)}]d1) {$S_3$};
        
        \diagram*{
            (i1) -- [scalar,momentum = {$ p_1 $}] (d1),
            (i2) -- [scalar,momentum' = {$ p_3 $}] (d1),
            (i3) -- [scalar,momentum' = {$ p_2 $}] (d2),
            (i4) -- [scalar,momentum = {$ p_4 $}] (d2),
            (d1) -- [scalar,momentum = {$ p_1+p_3 $}] (d2),  
        };
    \end{feynman}
\end{tikzpicture}}}\,+\,\vcenter{\hbox{\begin{tikzpicture}[baseline=(c.base)]
    \begin{feynman}
        \vertex (c);
        \node (d1) at (-1,0) [dot];
        \node (d2) at (1,0) [dot];
        
        \vertex (i1) at (-2,1.5) {$\sigma^{j_1}_{i_1}$};
        \vertex (i2) at (-2,-1.5) {$\sigma^{j_4}_{i_4}$};
        \vertex (i3) at (2,1.5) {$\sigma^{j_3}_{i_3}$};
        \vertex (i4) at (2,-1.5) {$\sigma^{j_2}_{i_2}$};

        \vertex  (d2n) at ([shift={(0.3,0.0)}]d2) {$S_3$};
        \vertex  (d1n) at ([shift={(-0.3,0.0)}]d1) {$S_3$};
        
        \diagram*{
            (i1) -- [scalar,momentum = {$ p_1 $}] (d1),
            (i2) -- [scalar,momentum' = {$ p_4 $}] (d1),
            (i3) -- [scalar,momentum' = {$ p_3 $}] (d2),
            (i4) -- [scalar,momentum = {$ p_2 $}] (d2),
            (d1) -- [scalar,momentum = {$ p_1+p_4 $}] (d2),  
        };
    \end{feynman}
\end{tikzpicture}}}
\end{equation*}

    \caption{Feynman diagrams for calculating $S_4$.}
    \label{fig: feymann digarams for bosons g4} 
\end{figure}
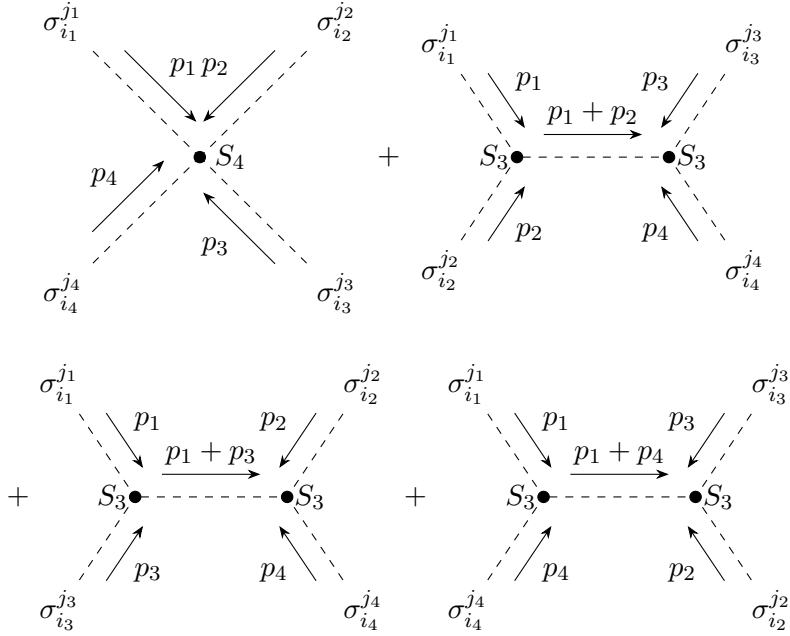

For the amputated four-point $\sigma$ correlation function,
there are four Feynman diagrams contributing to this correlation function, shown in figure \ref{fig: feymann digarams for bosons g4}. They are given by
\begin{gather} 
    \left\langle \sigma_{j_{1}}^{\ i_{1}}\left(p_{1}\right)\sigma_{j_{2}}^{\ i_{2}}\left(p_{2}\right)\sigma_{j_{3}}^{\ i_{3}}\left(p_{3}\right)\sigma_{j_{4}}^{\ i_{4}}\left(p_{4}\right)\right\rangle _{\text{amp}}=-\left(S_{4}\right)_{j_{1},j_{2},j_{3},j_{4}}^{i_{1},i_{2},i_{3},i_{4}}\left(p_{1},p_{2},p_{3},p_{4}\right) \nonumber \\
	+\left(S_{3}\right)_{j_{1},j_{2},j_{a}}^{i_{1},i_{2},i_{a}}\left(p_{1},p_{2},-p_{1}-p_{2}\right)\left(S_{3}\right)_{j_{3},j_{4},j_{b}}^{i_{3},i_{4},i_{b}}\left(p_{3},p_{4},p_{1}+p_{2}\right)\left(S_{2}^{-1}\right)_{i_{a},i_{b}}^{j_{a},j_{b}}\left(p_{1}+p_{2}\right) \label{eq: critical boson s4 calcualtion 1}\\
	+\left[\left(S_{3}\right)_{j_{1},j_{3},j_{a}}^{i_{1},i_{3},i_{a}}\left(p_{1},p_{3},-p_{1}-p_{3}\right)\left(S_{3}\right)_{j_{2},j_{4},j_{a}}^{i_{2},i_{4},i_{a}}\left(p_{2},p_{4},p_{1}+p_{3}\right)\left(S_{2}^{-1}\right)_{i_{a},i_{b}}^{j_{a},j_{b}}\left(p_{1}+p_{3}\right)+\left(3\leftrightarrow4\right)\right], \nonumber
\end{gather}
where we used \eqref{eq: green function two point tree level}, and we sum over $i_a,j_a,i_b,j_b$.

For the purpose of calculating the $\beta$ functions, we are interested in the limit where $p_1=-p_2=p,p_3=-p_4=k$ and $|p|\gg |k|$ (see section \ref{sec: Nf=1} in this paper, and section 3 in \cite{Aharony:2018pjn}). This limit cannot be plugged in directly, since both $\tilde{S}_4(p,-p,k,-k)$ and $S_3(p,-p.0)$ have an IR divergence, and so although the full tree-level 4-point function in this limit is well-defined, two of the Feynman diagrams contributing to \eqref{eq: critical boson s4 calcualtion 1} diverge separately.

To see that \eqref{eq: critical boson s4 calcualtion 1} for the single-flavor case $N_f=1$ is free from IR singularities, the authors in \cite{Aharony:2018pjn} used an IR regulator $\delta$. They calculated the four-point function with $p_1=p,p_2=p+\delta,p_3=k,p_4=-k-\delta$, in the $|p|\gg|k|\gg|\delta|$ limit, and showed that the resulting expression is well-defined, finite, and single-valued at $\delta\rightarrow0$. 

Adding flavors adds a subtlety to this procedure, since for specific indices, \eqref{eq: critical boson s4 calcualtion 1} with $p_1=-p_2=p,p_3=-p_4=k$ is not IR-finite. This can be fixed by symmetrizing over $p_3$ and $p_4$. We define
\begin{equation} \label{eq: appendix s4 effective}
    \begin{split}
        \left(S_{4}^{eff}\right)_{j_{1},j_{2},j_{3},j_{4}}^{i_{1},i_{2},i_{3},i_{4}}\left(p,-p,k,-k\right) \equiv - \lim_{\delta\rightarrow0}\frac{1}{2}\Big[&\left\langle \sigma_{j_{1}}^{\ i_{1}}\left(p\right)\sigma_{j_{2}}^{\ i_{2}}\left(-p+\delta\right)\sigma_{j_{3}}^{\ i_{3}}\left(k\right)\sigma_{j_{4}}^{\ i_{4}}\left(-k-\delta\right)\right\rangle_{\text{amp}} \\
        +&\left\langle \sigma_{j_{1}}^{\ i_{1}}\left(p\right)\sigma_{j_{2}}^{\ i_{2}}\left(-p+\delta\right)\sigma_{j_{3}}^{\ i_{3}}\left(-k-\delta\right)\sigma_{j_{4}}^{\ i_{4}}\left(k\right)\right\rangle_{\text{amp}} \Big] .
    \end{split}
\end{equation}
Then, \eqref{eq: appendix s4 effective} is free from IR divergences, even though each term on the right-hand side diverges. This should not concern us, since in the calculation involving the 4-point function in this limit, we need to include both the case that $p_3=k$ and the case $p_3=-k$.

The calculation of $S_{4}^{eff}$ follows the same lines as in \cite{Aharony:2018pjn}. The only difference is that one needs to compute the two different index structures (the near and far structures presented in \eqref{eq: g4 general case}) separately. We find that in the limit $|p|\gg|k|$, the effective amputated four-point function is
\begin{multline}
\left(S_{4}^{eff}\right)_{j_{1},j_{2},j_{3},j_{4}}^{i_{1},i_{2},i_{3},i_{4}}\left(p,-p,k,-k\right)\simeq
\\-\frac{N_{c}}{2}\left[\begin{array}{c}
\left(\left(\delta_{j_{2}}^{i_{1}}\delta_{j_{3}}^{i_{2}}\delta_{j_{4}}^{i_{3}}\delta_{j_{1}}^{i_{4}}+\left(3\leftrightarrow4\right)\right)+\left(1\leftrightarrow2\right)\right)\left(-\frac{1}{8}\frac{1}{p^{2}k^{2}\left|p-k\right|}-\frac{1}{8}\frac{1}{p^{2}k^{2}\left|p+k\right|}+\frac{1}{4p^{4}}\frac{1}{\left|k\right|}\frac{\left(k\cdot p\right)^{2}}{p^{2}k^{2}}\right)+\\
\left(\delta_{j_{3}}^{i_{1}}\delta_{j_{2}}^{i_{3}}\delta_{j_{4}}^{i_{2}}\delta_{j_{1}}^{i_{4}}+\delta_{j_{4}}^{i_{1}}\delta_{j_{2}}^{i_{4}}\delta_{j_{3}}^{i_{2}}\delta_{j_{1}}^{i_{3}}\right)\left(-\frac{1}{4}\frac{1}{p^{2}k^{2}\left|p-k\right|}-\frac{1}{4}\frac{1}{p^{2}k^{2}\left|p+k\right|}+\frac{1}{2\left|k\right|p^{4}}\right)
\end{array}\right] .
\end{multline}

The calculation of the amputated five-point function follows the same idea \cite{Aharony:2018pjn}. Besides the original 5 point interaction term $S_5$, we have two additional types of diagrams, one involving $S_3$ and $S_4$, and one involving three $S_3$ vertices (see figures 17, 18, and 19 in \cite{Aharony:2018pjn}). 

We will not attempt to calculate the final expression for the amputated five-point correlation function explicitly, nor will we show that for momenta in the limit of interest, it is free from IR singularities. Instead, we will use our knowledge of the full five-point correlation function $G_5$ in \eqref{eq: genral g5}, to construct the proper expression for $S_5$. 

For $|p|\gg|p_1|,|p_2|,|p_3|$, the form of $G_5$, which is the full correlation function, is given by \eqref{eq: genral g5}. To get the amputated correlation function, we remove the external legs by multiplying by $S_2 \left(k_i\right)$ for each external leg carrying a momentum $k_i$. The end result is 
\begin{equation} \label{eq: s5 effective RB lambda=0}
\left(S_{5}^{eff}\right)_{j_{1},j_{2},j_{3},j_{4},j_{5}}^{i_{1},i_{2},i_{3},i_{4},i_{5}}\left(p,-p,p_{1},p_{2},p_{3}\right)=\frac{N_{c}}{\left|p\right|^{4}\left|p_{1}\right|\left|p_{2}\right|\left|p_{3}\right|}\left(\begin{array}{c}
s_{5,N}\left(\left(\delta_{j_{2}}^{i_{1}}\delta_{j_{3}}^{i_{2}}\delta_{j_{4}}^{i_{3}}\delta_{j_{5}}^{i_{4}}\delta_{j_{1}}^{i_{5}}+\left(3,4,5\right)\right)+\left(1\leftrightarrow2\right)\right)+\\
s_{5,F}\left(\left(\delta_{j_{3}}^{i_{1}}\delta_{j_{2}}^{i_{3}}\delta_{j_{4}}^{i_{2}}\delta_{j_{5}}^{i_{4}}\delta_{j_{1}}^{i_{5}}+\left(3,4,5\right)\right)+\left(1\leftrightarrow2\right)\right)
\end{array}\right) ,
\end{equation}
where $s_{5,N}$ and $s_{5,F}$ are unknown coefficients, which do not scale with $N_c$, and we sum over permutations of $\{3,4,5\}$.

As a final comment, it should be clear that $S_n\propto N_c$ (at leading order at large $N_c$) for any $n$. As a result, the effective action \eqref{eq S eff J} is proportional to $N_c$, and we can think of $N_c$ as taking the role of $\hbar^{-1}$. Thus, the contribution of loop diagrams goes as $\frac{1}{N_c}$ to the power of the number of loops.

\subsection{Tree-level correlation functions} \label{subsec: App gn}

In the main text, we use the full (not amputated) $n$-point correlation functions of the $\sigma$ operators, which we denote by $G_n$ (up to minus sign)\footnote{Note that  we define $G_2$ and $G_3$ with minus the corresponding corelation functions, and $G_4$ and $G_5$ with a plus sign, to have compatibility with the notation in \eqref{eq: S zeta multiple flavor step 2}.}. To obtain them from the amputated correlation function, we multiply each external leg with momentum $p$ by the propagator $\left(S_{2}^{-1}\right)_{i_{a},i_{b}}^{j_{a},j_{b}}\left(p\right)$. This fixes the $N_c$ dependence of the tree-level contribution to $G_n$ to be $G_n\sim N_c^{-n+1}$.\footnote{Multiplying by $\left(S_{2}^{-1}\right)_{i_{a},i_{b}}^{j_{a},j_{b}}\left(p\right)$ also flips the role of the $j$ and $i$ indices in our notation.} The expression for the tree-level 2-point correlation function is given in \eqref{eq: green function two point tree level}, while for the 3-point function it is a contact term
\begin{equation} \label{eq: appendix, g3 from s3}
    \begin{split}\left(G_{3}\right)_{i_{1},i_{2},i_{3}}^{j_{1},j_{2},j_{3}}\left(p_{1},p_{2},p_{3}\right) & \equiv-\left\langle \sigma_{i_{1}}^{\ j_{1}}\left(p_{1}\right)\sigma_{i_{2}}^{\ j_{2}}\left(p_{2}\right)\sigma_{i_{3}}^{\ j_{3}}\left(p_{3}\right)\right\rangle \\
& = \left(S_{2}^{-1}\right)_{i_{1},i_{a}}^{j_{1},j_{a}}\left(p_{1}\right)\left(S_{2}^{-1}\right)_{i_{2},i_{b}}^{j_{2},j_{b}}\left(p_{2}\right)\left(S_{2}^{-1}\right)_{i_{3},i_{c}}^{j_{3},j_{c}}\left(p_{3}\right)\left(S_{3}\right)_{\left\{ j\right\} _{3}}^{\left\{ i\right\} _{3}}\left(p_{1},p_{2},p_{3}\right)\\
 & =-\frac{64}{N_{c}^{2}}\left(\delta_{i_{2}}^{j_{1}}\delta_{i_{3}}^{j_{2}}\delta_{i_{1}}^{j_{3}}+\delta_{i_{3}}^{j_{1}}\delta_{i_{2}}^{j_{3}}\delta_{i_{1}}^{j_{2}}\right),
\end{split}
\end{equation}
and for the four-point function in the desired momenta limit it is\footnote{See the discussion in section \ref{subsec: App CB s4 s5} for the meaning of the superscript $eff$.}
\begin{multline}
\left(G_{4}^{eff}\right)_{i_{1},i_{2},i_{3},i_{4}}^{j_{1},j_{2},j_{3},j_{4}}\left(p,-p,k,-k\right)\equiv\left\langle \sigma_{i_{1}}^{\ j_{1}}\left(p\right)\sigma_{i_{2}}^{\ j_{2}}\left(-p\right)\sigma_{i_{3}}^{\ j_{3}}\left(k\right)\sigma_{i_{4}}^{\ j_{4}}\left(-k\right)\right\rangle \\\simeq \frac{8^{4}}{N_{c}^{3}}\left[\begin{array}{c}
\left(\left(\delta_{i_{2}}^{j_{1}}\delta_{i_{3}}^{j_{2}}\delta_{i_{4}}^{j_{3}}\delta_{i_{1}}^{j_{4}}+\left(3\leftrightarrow4\right)\right)+\left(1\leftrightarrow2\right)\right)\left(-\frac{1}{16}\frac{1}{\left|p-k\right|}-\frac{1}{16}\frac{1}{\left|p+k\right|}+\frac{1}{8p^{4}}\frac{\left(k\cdot p\right)^{2}}{\left|k\right|}\right)+\\
\left(\delta_{i_{3}}^{j_{1}}\delta_{i_{2}}^{j_{3}}\delta_{i_{4}}^{j_{2}}\delta_{i_{1}}^{j_{4}}+\delta_{i_{4}}^{j_{1}}\delta_{i_{2}}^{j_{4}}\delta_{i_{3}}^{j_{2}}\delta_{i_{1}}^{j_{3}}\right)\left(-\frac{1}{8}\frac{1}{\left|p-k\right|}-\frac{1}{8}\frac{1}{\left|p+k\right|}+\frac{\left|k\right|}{4p^{2}}\right)
\end{array}\right]\\\approx\frac{8^{4}}{N_{c}^{3}}\left[\begin{array}{c}
\left(\left(\delta_{i_{2}}^{j_{1}}\delta_{i_{3}}^{j_{2}}\delta_{i_{4}}^{j_{3}}\delta_{i_{1}}^{j_{4}}+\left(3\leftrightarrow4\right)\right)+\left(1\leftrightarrow2\right)\right)\left(-\frac{1}{8}\frac{1}{p}+\frac{1}{8}\frac{\left(k\cdot p\right)^{2}}{p^{4}\left|k\right|}\right)+\\
\left(\delta_{i_{3}}^{j_{1}}\delta_{i_{2}}^{j_{3}}\delta_{i_{4}}^{j_{2}}\delta_{i_{1}}^{j_{4}}+\delta_{i_{4}}^{j_{1}}\delta_{i_{2}}^{j_{4}}\delta_{i_{3}}^{j_{2}}\delta_{i_{1}}^{j_{3}}\right)\left(-\frac{1}{4}\frac{1}{p}+\frac{\left|k\right|}{4p^{2}}\right)
\end{array}\right],
\end{multline}
where we assume $|p|\gg |k|$.

For the five-point function, we take the same result in \eqref{eq: genral g5}, which for completeness we copy here
\begin{equation}
\begin{split}\left(G_{5}\right)_{i_{1},i_{2},i_{3},i_{4},i_{5}}^{j_{1},j_{2},j_{3},j_{4},j_{5}}\left(p,-p,0,0,0\right) & \equiv-\left\langle \sigma_{i_{1}}^{\ j_{1}}\left(p\right)\sigma_{i_{2}}^{\ j_{2}}\left(-p\right)\sigma_{i_{3}}^{\ j_{3}}\left(0\right)\sigma_{i_{4}}^{\ j_{4}}\left(0\right)\sigma_{i_{5}}^{\ j_{5}}\left(0\right)\right\rangle \\
 & =\frac{1}{p^{2}}\left(\begin{array}{c}
G_{5,N}\left(\delta_{i_{2}}^{j_{1}}\delta_{i_{3}}^{j_{2}}\delta_{i_{4}}^{j_{3}}\delta_{i_{5}}^{j_{4}}\delta_{i_{1}}^{j_{5}}+\left(3,4,5\right)+\left(1\leftrightarrow2\right)\right)+\\
G_{5,F}\left(\delta_{i_{3}}^{j_{1}}\delta_{i_{2}}^{j_{3}}\delta_{i_{4}}^{j_{2}}\delta_{i_{5}}^{j_{4}}\delta_{i_{1}}^{j_{5}}+\left(3,4,5\right)+\left(1\leftrightarrow2\right)\right)
\end{array}\right).
\end{split}
\end{equation}
The connections between $G_{5,N},G_{5,F}$ and $s_{5,N},s_{5,F}$  in \eqref{eq: s5 effective RB lambda=0} are
\begin{equation} \label{eq: s5Nf}
    s_{5,F/N}=\frac{N_c^{4}}{8^{5}}G_{5,F/N} . 
\end{equation}

\subsection{First \texorpdfstring{$ \ln(\Lambda)$}{lnLambda} dependent sub-leading contribution to \texorpdfstring{$ G_2$}{G2}} \label{subsec: App subleading G2}

\begin{figure}[t]
    \centering
\begin{equation*}
    \vcenter{\hbox{\begin{tikzpicture}[baseline=(c.base)]
     \begin{feynman}
        \vertex (c);
        \vertex (i1) at (-3,0) {$\sigma^{j_1}_{i_1}$};
        \vertex (i2) at (3,0) {$\sigma^{j_2}_{i_2}$};
        
        \node (d1) at (-1.5,0) [dot];
        \node (d2) at (1.5,0)  [dot];

        \vertex  (d2n) at ([shift={(-0.3,0.0)}]d2) {$S_3$};
        \vertex  (d1n) at ([shift={(0.3,0.0)}]d1) {$S_3$};

        \diagram*{
            (i1) -- [scalar,momentum = {$ p $}] (d1),
            (i2) -- [scalar,reversed momentum' = {$ p $}] (d2),
            (d1) -- [scalar,half left,momentum = {$ q $}] (d2)
                 -- [scalar,half left,,momentum = {$ q - p $}] (d1),
        };
    \end{feynman}
\end{tikzpicture}}}\,-\,\vcenter{\hbox{\begin{tikzpicture}[baseline=(c.base)]
    \begin{feynman}
        \node (c) [dot];
        \vertex (h) at (0,2);
        \vertex (i1) at (-2,0) {$\sigma^{j_1}_{i_1}$};
        \vertex (i2) at (2,0) {$\sigma^{j_2}_{i_2}$};

        \node (d1) at (-1.5,0) [dot];
        \node (d2) at (1.5,0)  [dot];

        \vertex  (cn) at ([shift={(0.0,-0.3)}]c) {$S_4$};

        \diagram*{
            (i1) -- [scalar,momentum' = {$ p $}] (c),
            (i2) -- [scalar,reversed momentum = {$ p $}] (c),
            (c)--[scalar, out=0,in=0,looseness=0.7,,momentum' ={[arrow shorten=0.25] $ q $ } ](h), 
            (c)--[ scalar,out=180,in=180,,looseness=0.7](h)

        };
    \end{feynman}
\end{tikzpicture}}}
\end{equation*}

    \caption{Feynman diagrams for the first sub-leading correction to $G_2$. }
    \label{fig: diagrams for anomalus dimension with S} 
\end{figure}
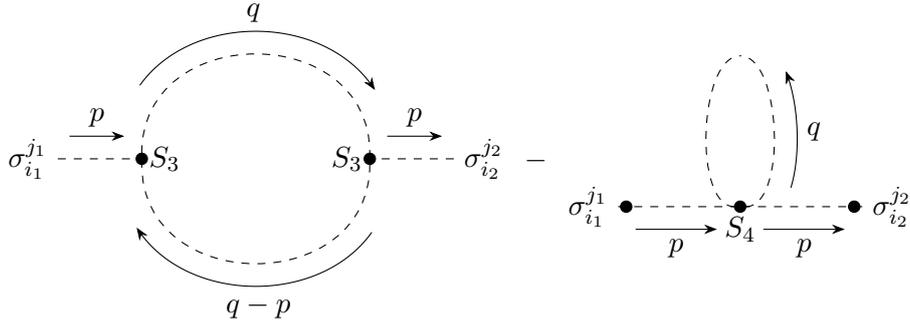

At first subleading order in $\frac{1}{N_c}$, there are two diagrams that contribute to the two-point function of the $\sigma$ operators, shown in figure \ref{fig: diagrams for anomalus dimension with S}. In the leading order calculation of the $\beta$ function, we use only the leading order dependence of the anomalous dimension (see sections \ref{sec: Nf=1} and \ref{sec: general case} in this paper and section 2 and Appendix B in \cite{Aharony:2018pjn}), which is extracted from the $\ln (\Lambda)$-dependent terms in the two-point function. Using the expressions for $S_n$, we evaluate the diagrams and find
\begin{equation}
    \begin{split}
        \left(\delta S_{2}\right)_{j_{1},j_{2}}^{i_{1},i_{2}}\left(p\right)&=\left\langle \sigma_{j_{1}}^{i_{1}}\left(p\right)\sigma_{j_{2}}^{i_{2}}\left(-p\right)\right\rangle _{amp,loop}\\&\supset\frac{1}{2}\intop\frac{d^{3}q}{\left(2\pi\right)^{3}}\lim_{\delta\rightarrow0}\left(S_{4}^{eff}\right)_{j_{3},j_{4},j_{1},j_{2}}^{i_{3},i_{4},i_{1},i_{2}}\left(q,-q+\delta,p,-p-\delta\right)\left(S_{2}^{-1}\right)_{i_{3},i_{4}}^{j_{3},j_{4}}\left(q\right)\\&\supset\left(\delta s_{20}\delta_{j_{2}}^{i_{1}}\delta_{j_{1}}^{i_{2}}+\delta s_{22}\delta_{j_{1}}^{i_{1}}\delta_{j_{2}}^{i_{2}}\right)\frac{1}{p}\ln\left(\frac{\Lambda}{\left|p\right|}\right) ,
    \end{split}
\end{equation}
where
\begin{equation}
    \delta s_{20}=\frac{N_{f}}{3\pi^{2}},\quad\delta s_{22}=\frac{1}{\pi^{2}} ,
\end{equation}
and\footnote{The inclusion symbol, instead of an equality symbol, is due to a subleading correction in $\frac{1}{N_c}$ to the $p$ term with no logarithm, which we ignore.}
\begin{equation}
\begin{split}\left(G_{2,tot}\right)_{i_{1},i_{2}}^{j_{1},j_{2}}\left(p\right)= & -\left(\left(S_{2}+\delta S_{2}\right)^{-1}\right)_{i_{1},i_{2}}^{j_{1},j_{2}}\left(p\right)\\
 & =-\left(S_{2}^{-1}\right)_{i_{1},i_{2}}^{j_{1},j_{2}}\left(p\right)+\left(S_{2}^{-1}\right)_{i_{1},i_{3}}^{j_{1},j_{3}}\left(p\right)\left(\delta S_{2}\right)_{j_{3},j_{4}}^{i_{3},i_{4}}\left(p\right)\left(S_{2}^{-1}\right)_{i_{4},i_{2}}^{j_{4},j_{2}}\left(p\right)\\
 & \supset+G_{2}p\delta_{i_{2}}^{j_{1}}\delta_{i_{1}}^{j_{2}}+G_{2}^{2}p\ln\left(\frac{\Lambda}{p}\right)\left(\delta s_{20}\delta_{i_{2}}^{j_{1}}\delta_{i_{1}}^{j_{2}}+\delta s_{22}\delta_{i_{1}}^{j_{1}}\delta_{i_{2}}^{j_{2}}\right).
\end{split}
\end{equation}

Note that, unlike the tree-level correlation function, the first subleading correction at order $ \frac{1}{N_c} $ introduces additional structures involving delta functions over the flavor indices, which do not correspond to a single-cycle permutation, specifically $ \delta_{i_{1}}^{j_{1}} \delta_{i_{2}}^{j_{2}} $.

To extract the anomalous dimensions of the singlet and adjoint structures, we multiply $\left(G_{2,tot}\right)_{i_{1},i_{2}}^{j_{1},j_{2}}$ by $\zeta_{j_{1}}^{i_{1}}\zeta_{j_{2}}^{i_{2}}$ and use \eqref{eq: projectio to SA to fields}. We find\footnote{Note that $G_{2,tot}$ and $\zeta$ have an implicit momentum dependence.}
\begin{equation}
\begin{split}
\left(G_{2,tot}\right)_{i_{1},i_{2}}^{j_{1},j_{2}}\zeta_{j_{1}}^{i_{1}}\zeta_{j_{2}}^{i_{2}} &=  pG_{2}\left(1-G_{2}\delta s_{20}\ln\left(\frac{p}{\Lambda}\right)\right)\left(\zeta_{A}\right)_{j}^{i}\left(\zeta_{A}\right)_{i}^{j}    \\
& +p\frac{1}{N_{f}}G_{2}\left(1-\left(G_{2}\delta s_{20}+N_{f}G_{2}\delta s_{22}\right)\ln\left(\frac{p}{\Lambda}\right)\right)\zeta_{S}\zeta_{S},
\end{split}
\end{equation}
and plugging in the numbers, we get
\begin{equation} \label{eq: anomalous dimension critical boson no CS}
    \gamma_{A}=-\frac{1}{2}G_{2}\delta s_{20}=-\frac{1}{N_{c}}\frac{4N_{f}}{3\pi^{2}},\quad \gamma_{S}=-\frac{1}{2}\left(G_{2}\delta s_{20}+N_{f}G_{2}\delta s_{22}\right)=-\frac{1}{N_{c}}\frac{16N_{f}}{3\pi^{2}}.
\end{equation}
For $N_f=1$, there is only a singlet structure, and $\gamma_S$ in \eqref{eq: anomalous dimension critical boson no CS} matches the result in the literature \cite{Aharony:2018pjn,Aharony:2012nh}.

\subsection{First sub-leading contribution to \texorpdfstring{$ G_3$}{G3}} \label{subsec: App subleading G3}

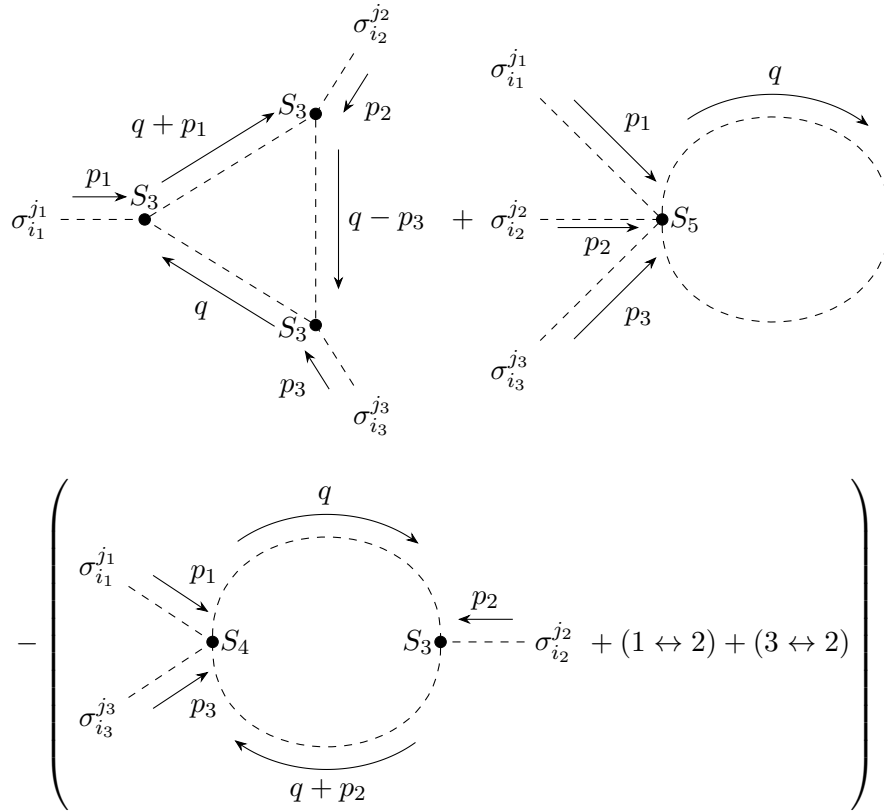
\begin{figure}[t]
    \centering
\begin{equation*}
    \vcenter{\hbox{\begin{tikzpicture}[baseline=(c.base)]
    \begin{feynman}
        \vertex (c);
        \vertex (i1) at (-3,0) {$\sigma^{j_1}_{i_1}$};
        \vertex (i2) at (1.5,2.6) {$\sigma^{j_2}_{i_2}$};
        \vertex (i3) at (1.5,-2.6) {$\sigma^{j_3}_{i_3}$};
        
        \node (d1) at (-1.5,0) [dot];
        \node (d2) at (0.75,1.4) [dot];
        \node (d3) at (0.75,-1.4) [dot];

        \vertex  (d1n) at ([shift={(0.0,0.3)}]d1) {$S_3$};
        \vertex  (d2n) at ([shift={(-0.3,0.1)}]d2) {$S_3$};
        \vertex  (d3n) at ([shift={(-0.3,-0.05)}]d3) {$S_3$};

        \diagram*{
            (i1) -- [scalar,momentum = {$ p_1 $}] (d1),
            (i2) -- [scalar, momentum = {$ p_2 $}] (d2),
            (i3) -- [scalar, momentum = {$ p_3 $}] (d3),
            (d1) -- [scalar,momentum = {$ q+p_1 $}] (d2)
                 -- [scalar,momentum = {$ q - p_3 $}] (d3)
                 -- [scalar,momentum = {$ q  $}] (d1),
            
        };
    \end{feynman}
\end{tikzpicture}
}}+\vcenter{\hbox{\begin{tikzpicture}[baseline=(c.base)]
    \begin{feynman}
        \node (c) [dot];
        \vertex (h) at (3,0);
        \vertex (i1) at (-2,2) {$\sigma^{j_1}_{i_1}$};
        \vertex (i2) at (-2,0) {$\sigma^{j_2}_{i_2}$};
        \vertex (i3) at (-2,-2) {$\sigma^{j_3}_{i_3}$};                
        
        \diagram*{
            (i1) -- [scalar,momentum = {$ p_1 $}] (c),
            (i3) -- [scalar, momentum' = {$ p_3 $}] (c),
            (i2) -- [scalar, momentum' = {[arrow distance=0.1] $ p_2 $}] (c),
            (c)--[scalar, out=90,in=90,looseness=1.5,momentum ={[arrow shorten=0.25] $ q $ } ](h), 
            (c)--[ scalar,out=270,in=270,looseness=1.5](h),
            
        };
        \vertex  (cn) at ([shift={(0.3,0.0)}]c) {$S_5$};
    \end{feynman}
\end{tikzpicture}}}
\end{equation*}
\begin{equation*}
-\left(\vcenter{\hbox{\begin{tikzpicture}[baseline=(c.base)]
    \begin{feynman}
        \vertex (c);
        \vertex (i1) at (-3,1) {$\sigma^{j_1}_{i_1}$};
        \vertex (i2) at (3,0) {$\sigma^{j_2}_{i_2}$};
        \vertex (i3) at (-3,-1) {$\sigma^{j_3}_{i_3}$};
        
        \node (d1) at (-1.5,0) [dot];
        \node (d2) at (1.5,0) [dot];

        \diagram*{
            (i1) -- [scalar,momentum = {$ p_1 $}] (d1),
            (i3) -- [scalar, momentum' = {$ p_3 $}] (d1),
            (i2) -- [scalar, momentum' = {$ p_2 $}] (d2),
            (d1) -- [scalar,half left,momentum ={[arrow shorten=0.25] $ q $ }] (d2)
                 -- [scalar,half left,,momentum ={[arrow shorten=0.25] $ q +p_2 $ }] (d1),
            
        };
        \vertex  (d1n) at ([shift={(0.3,0.0)}]d1) {$S_4$};
        \vertex  (d2n) at ([shift={(-0.3,0.0)}]d2) {$S_3$};

    \end{feynman}
\end{tikzpicture}}} +\left(1\leftrightarrow2\right)+\left(3\leftrightarrow2\right)\right)
\end{equation*}

    \caption{Feynman diagrams for calculating the first sub-leading contribution to $G_3$.}
    \label{fig: feynman diagrams for RB delta g3} 
\end{figure}

The first subleading contribution to $G_3$, which we denote in the main text as $\delta G_3$, comes from three diagrams, shown in figure \ref{fig: feynman diagrams for RB delta g3}. The calculations proceed in exactly the same manner as in section B.2.2 of \cite{Aharony:2018pjn}, and the only difference is adding flavors indices. We find that\footnote{Alternatively, and as in the $N_f=1$ case, these conditions can be found by requiring that \eqref{eq: beta multyflavour} vanish for $\lambda=\lambda_{n}=0$. This is true since the RB theory \eqref{eq: regular bosons} is free at this point, and thus the beta function should vanish.}
\begin{equation}
    \begin{split}
        \delta \bar{G}_{3,3} &= \frac{1}{N_{c}^{3}}\frac{512N_{f}\left(12s_{5,N}-1\right)}{\pi^{2}} ,  \\
        \delta \bar{G}_{3,2}&= \frac{1}{N_{c}^{3}}\frac{1024\left(4s_{5,F}-1\right)}{\pi^{2}}, \\ 
        \delta \bar{G}_{3,1} &= 0 .
    \end{split}
\end{equation}
Note that, at leading order in $ \frac{1}{N_c} $, there is no contribution to $ \delta \bar{G}_{3,1} $, as it involves three non-diagonal delta functions in the flavor indices. Such a structure requires diagrams with a different topology in order to be realized. We expect contributions to it only at order $\delta \bar{G}_{3,1} \sim \frac{1}{N_c^5}$ (that is three loops).

Transformation to the representation basis is done using \eqref{eq: transform to adjoint0}. The results of this appendix are summarized in table \ref{tab: beta coefficients}.

\section{Correlation functions for the free fermion theories} \label{sec: CF appendix}

In this appendix we present the calculations of correlation functions of the fermion bilinear mesons $M_{i}^{j}\equiv\frac{4\pi}{\kappa_F}\bar{\psi}_{c,i}\psi^{c,j}$ in the limit $\lambda_F=0$. We follow Appendix C of \cite{Aharony:2018pjn}, generalizing their results to multiple flavors.

The calculation of the two-point function is given by\footnote{Note that the minus sign from the fermion loop is canceled with the $i^2$ from the two fermion propagators in Euclidean signature.}
\begin{equation}
    \left\langle M_{i_{1}}^{\ j_{1}}\left(p\right)M_{i_{2}}^{\ j_{2}}\left(-p\right)\right\rangle =\delta_{i_{2}}^{j_{1}}\delta_{i_{1}}^{j_{2}}\left(\frac{4\pi}{\kappa_{F}}\right)^{2}N_{c}\intop\frac{d^{3}q}{\left(2\pi\right)^{3}}{\rm Tr}\left(\frac{\cancel{q}\left(\cancel{q}+\cancel{p}\right)}{q^{2}\left(q+p\right)^{2}}\right) .
\end{equation}
Solving the integral using ${\rm Tr}\left(\gamma^{\mu}\gamma^{\nu}\right)=2\eta^{\nu\mu}$, and ignoring terms which are linear in the cutoff (since we fine tune the relevant operator), we find
\begin{equation}
    \left(G_{2}\right)_{i_{1}i_{2}}^{j_{1}j_{2}}\left(p\right)=\frac{2\pi^{2}}{\kappa_{F}^{2}}N_{c}\left|q\right|\delta_{i_{2}}^{j_{1}}\delta_{i_{1}}^{j_{2}} .
\end{equation}

The four-point function is given by
\begin{equation}
\left(G_{4}\right)_{i_{1},i_{2},i_{3},i_{4}}^{j_{1},j_{2},j_{3},j_{4}}\left(p_{1},p_{2},p_{3},p_{4}\right)=\left(\frac{4\pi}{\kappa_{F}}\right)^{4}N_{c}\left(\delta_{i_{2}}^{j_{1}}\delta_{i_{3}}^{j_{2}}\delta_{i_{4}}^{j_{3}}\delta_{i_{1}}^{j_{4}}Y\left(p_{1},p_{2},p_{3},p_{4}\right)+\left(\text{permutations of \{2,3,4\}}\right)\right),
\end{equation}
where
\begin{equation}
    Y\left(p_{1},p_{2},p_{3},p_{4}\right)=-\intop\frac{d^{3}q}{\left(2\pi\right)^{3}}{\rm Tr}\left(\frac{\cancel{q}\left(\cancel{q}+\cancel{p_{1}}\right)\left(\cancel{q}+\cancel{p_{1}}+\cancel{p_{2}}\right)\left(\cancel{q}-\cancel{p_{4}}\right)}{q^{2}\left(q+p_{1}\right)^{2}\left(q+p_{1}+p_{2}\right)^{2}\left(q-p_{4}\right)^{2}}\right) .
\end{equation}
The trace over the $\gamma$ matrices can be simplified by
\begin{equation}
{\rm Tr}\left(\gamma^{\mu}\gamma^{\nu}\gamma^{\rho}\gamma^{\sigma}\right) = 2 \left(\eta^{\mu\nu}\eta^{\rho\sigma}-\eta^{\mu\rho}\eta^{\nu\sigma}+\eta^{\mu\sigma}\eta^{\nu\rho}\right) .
\end{equation}

We are interested in the kinematic limit where $ p_1 = -p_2 \equiv p $, $ p_3 = -p_4 \equiv k $, and $ |p| \gg |k| $. In this regime, it is sufficient to solve for $ Y(p, -p, k, -k) $ and $ Y(p, k, -p, -k) $. The remaining momentum configurations can be obtained either by replacing $ k $ with $ -k $, or by exploiting the symmetry
\begin{equation}
Y\left(p_{1},p_{2},p_{3},p_{4}\right)=Y\left(p_{1},p_{4},p_{3},p_{2}\right) .
\end{equation}
We find
\begin{equation}
\begin{split}- & Y\left(p,k,-p,-k\right)=\intop\frac{d^{3}q}{\left(2\pi\right)^{3}}\Bigg[-\frac{1}{2q^{2}\left(q+k\right)^{2}}-\frac{1}{2q^{2}\left(q+p\right)^{2}}-\frac{1}{2\left(q+p\right)^{2}\left(q+k\right)^{2}}\\
 & +\frac{k^{2}}{2q^{2}\left(q+k\right)^{2}\left(q+p\right)^{2}}+\frac{p^{2}}{2q^{2}\left(q+k\right)^{2}\left(q+p\right)^{2}}+\frac{3}{2q^{2}\left(q+p+k\right)^{2}}+\frac{1}{q^{2}\left(q+p\right)^{2}}\\
 & -\frac{k^{2}}{2q^{2}\left(q+k\right)^{2}\left(q+p+k\right)^{2}}-\frac{p^{2}}{2q^{2}\left(q+p\right)^{2}\left(q+p+k\right)^{2}}+\frac{3k^{2}}{2\left(q+k\right)^{2}\left(q+p\right)^{2}\left(q+p+k\right)^{2}}\\
 & +\frac{p^{2}}{2\left(q+k\right)^{2}\left(q+p\right)^{2}\left(q+p+k\right)^{2}}-\frac{k^{2}p^{2}}{2q^{2}\left(q+k\right)^{2}\left(q+p\right)^{2}\left(q+p+k\right)^{2}}+\frac{p\cdot k}{q^{2}\left(q+k\right)^{2}\left(q+p+k\right)^{2}}\\
 & +\frac{p\cdot k}{q^{2}\left(q+p\right)^{2}\left(q+p+k\right)^{2}}+\frac{2p\cdot k}{\left(q+k\right)^{2}\left(q+p\right)^{2}\left(q+p+k\right)^{2}}-\frac{k^{2}k\cdot p}{q^{2}\left(q+k\right)^{2}\left(q+p\right)^{2}\left(q+p+k\right)^{2}}\\
 & -\frac{p^{2}p\cdot k}{q^{2}\left(q+k\right)^{2}\left(q+p\right)^{2}\left(q+p+k\right)^{2}}+\frac{q\cdot k}{q^{2}\left(q+p\right)^{2}\left(q+p+k\right)^{2}}\\
 & +\frac{q\cdot p}{q^{2}\left(q+k\right)^{2}\left(q+p+k\right)^{2}}+\frac{2q\cdot k}{\left(q+k\right)^{2}\left(q+p\right)^{2}\left(q+p+k\right)^{2}}\Bigg],
\end{split}
\end{equation}
and
\begin{equation}
    \begin{split}-Y\left(p,-p,k,-k\right) & =\intop\frac{d^{3}q}{\left(2\pi\right)^{3}}\Bigg[\frac{1}{2q^{2}\left(q+p\right)^{2}}-\frac{k^{2}}{2q^{2}\left(q+k\right)^{2}\left(q+p\right)^{2}}+\frac{q\cdot p}{q^{2}\left(q+k\right)^{2}\left(q+p\right)^{2}}\\
 & +\frac{2\left(q\cdot p\right)\left(q\cdot k\right)}{q^{4}\left(q+k\right)^{2}\left(q+p\right)^{2}}+\frac{1}{2\left(q+k\right)^{2}\left(q+p\right)^{2}}-\frac{p\cdot k}{q^{2}\left(q+k\right)^{2}\left(q+p\right)^{2}}\Bigg] .
\end{split}
\end{equation}
Using the integral identities (where we denote $p_{0}=p_{1}-p_{2}$)
\begin{equation}
        \intop\frac{d^{3}q}{\left(2\pi\right)^{3}}\frac{1}{q^{2}\left(q+p\right)^{2}}=\frac{1}{8\left|p\right|},
\end{equation}
\begin{equation}
        \intop\frac{d^{3}q}{\left(2\pi\right)^{3}}\frac{1}{q^{2}\left(q+p_{1}\right)^{2}\left(q+p_{2}\right)^{2}}=\frac{1}{8}\frac{1}{\left|p_{1}\right|\left|p_{2}\right|\left|p_{0}\right|},
\end{equation}
\begin{equation}
    \intop\frac{d^{3}q}{\left(2\pi\right)^{3}}\frac{q_{\mu}}{q^{2}\left(q+p_{1}\right)^{2}\left(q+p_{2}\right)^{2}}=-\frac{1}{8\left|p_{0}\right|\left(\left|p_{1}\right|+\left|p_{2}\right|+\left|p_{0}\right|\right)}\left(\frac{1}{\left|p_{1}\right|}p_{1\mu}+\frac{1}{\left|p_{2}\right|}p_{2\mu}\right),
\end{equation}
\begin{equation}
    \begin{split}\intop\frac{d^{3}q}{\left(2\pi\right)^{3}}\frac{q_{\mu}q_{\nu}}{q^{4}\left(q+p_{1}\right)^{2}\left(q+p_{2}\right)^{2}}= & \frac{1}{16}\frac{1}{\left(p_{1}+p_{2}+p_{0}\right)}\Bigg[\frac{\eta_{\mu\nu}}{\left|p_{1}\right|\left|p_{2}\right|}\\
 & +\frac{\left(2\left|p_{1}\right|+\left|p_{2}\right|+\left|p_{0}\right|\right)p_{1,\mu}p_{1,\nu}}{\left|p_{1}\right|^{3}\left|p_{0}\right|\left(\left|p_{1}\right|+\left|p_{2}\right|+\left|p_{0}\right|\right)}+\frac{p_{1,\mu}p_{2,\nu}}{\left|p_{1}\right|\left|p_{2}\right|\left|p_{0}\right|\left(\left|p_{1}\right|+\left|p_{2}\right|+\left|p_{0}\right|\right)}\\
 & +\frac{p_{1,\nu}p_{2,\mu}}{\left|p_{1}\right|\left|p_{2}\right|\left|p_{0}\right|\left(\left|p_{1}\right|+\left|p_{2}\right|+\left|p_{0}\right|\right)}+\frac{\left(\left|p_{1}\right|+2\left|p_{2}\right|+\left|p_{0}\right|\right)p_{2,\nu}p_{2,\mu}}{\left|p_{2}\right|^{3}\left|p_{0}\right|\left(\left|p_{1}\right|+\left|p_{2}\right|+\left|p_{0}\right|\right)}\Bigg],
\end{split}
\end{equation}
and for $|p|\gg|k|$ the identity \cite{Aharony:2018pjn}
\begin{equation}
    \intop\frac{d^{3}p}{\left(2\pi\right)^{3}}\frac{1}{q^{2}\left(q+k\right)^{2}\left(q+p\right)^{2}\left(q+p+k\right)^{2}}\simeq \frac{1}{4\left|p\right|^{4}\left|k\right|} ,
\end{equation}
we find after Taylor expanding in $|k|/|p|$ that at the leading orders
\begin{equation}
    \begin{split}
        -Y\left(p,-p,k,-k\right)=&\frac{1}{8\left|p\right|}+\frac{k\cdot p}{8p^{3}}-\frac{k\cdot p}{8\left|k\right|p^{2}}-\frac{\left(k\cdot p\right)^{2}}{8\left|k\right|p^{4}}, \\
        -Y\left(p,k,-p,-k\right)=&\frac{1}{4\left|p\right|}-\frac{\left|k\right|}{4p^{2}} .
    \end{split}
\end{equation}
The four-point function is thus
\begin{equation}
    \left(G_{4}\right)_{i_{1},i_{2},i_{3},i_{4}}^{j_{1},j_{2},j_{3},j_{4}}\left(p,-p,k,-k\right)=-\left(\frac{4\pi}{\kappa}\right)^{4}N_{c}\left(\begin{array}{c}
\left(\delta_{i_{2}}^{j_{1}}\delta_{i_{3}}^{j_{2}}\delta_{i_{4}}^{j_{3}}\delta_{i_{1}}^{j_{4}}+\delta_{i_{4}}^{j_{1}}\delta_{i_{3}}^{j_{4}}\delta_{i_{2}}^{j_{3}}\delta_{i_{1}}^{j_{2}}\right)\left(\frac{1}{8\left|p\right|}+\frac{k\cdot p}{8p^{3}}-\frac{k\cdot p}{8\left|k\right|p^{2}}-\frac{\left(k\cdot p\right)^{2}}{8\left|k\right|p^{4}}\right)+\\
\left(\delta_{i_{2}}^{j_{1}}\delta_{i_{4}}^{j_{2}}\delta_{i_{3}}^{j_{4}}\delta_{i_{1}}^{j_{3}}+\delta_{i_{3}}^{j_{1}}\delta_{i_{4}}^{j_{3}}\delta_{i_{2}}^{j_{4}}\delta_{i_{1}}^{j_{2}}\right)\left(\frac{1}{8\left|p\right|}-\frac{k\cdot p}{8p^{3}}+\frac{k\cdot p}{8\left|k\right|p^{2}}-\frac{\left(k\cdot p\right)^{2}}{8\left|k\right|p^{4}}\right)+\\
\left(\delta_{i_{3}}^{j_{1}}\delta_{i_{2}}^{j_{3}}\delta_{i_{4}}^{j_{2}}\delta_{i_{1}}^{j_{4}}+\delta_{i_{4}}^{j_{1}}\delta_{i_{2}}^{j_{4}}\delta_{i_{3}}^{j_{2}}\delta_{i_{1}}^{j_{3}}\right)\left(\frac{1}{4\left|p\right|}-\frac{\left|k\right|}{4p^{2}}\right)
\end{array}\right).
\end{equation}

Note that when using the $G_4$ term in the main text, since we integrate over the large momenta when using $G_4$, we can ignore the $p\cdot k$ term, since it will give an antisymmetric integrand with respect to $p\leftrightarrow-p$, and so the full integral will vanish.

We summarize the results obtained in this appendix in table \ref{tab: beta coefficients}.

\bibliographystyle{JHEP}
\bibliography{biblio.bib}

\end{document}